\IfFileExists{acmart.cls}{%
  \documentclass[acmsmall,screen,nonacm]{acmart}%
  \def\usingacmart{1}%
}{%
  \documentclass[10pt]{article}%
  \usepackage[margin=1in]{geometry}%
  \usepackage[numbers,sort&compress]{natbib}%
  \usepackage{hyperref}%
  \newcommand{\keywords}[1]{\par\noindent\textbf{Keywords:} ##1}%
  \newcommand{\ccsdesc}[2][]{}%
  \usepackage{verbatim}%
}
\usepackage{booktabs,tabularx,multirow,makecell,longtable,threeparttable}
\usepackage{xcolor,enumitem,pifont,amsmath,graphicx}
\usepackage[most]{tcolorbox}
\usepackage{needspace}
\usepackage{tikz}
\usetikzlibrary{positioning,arrows.meta}

\newcommand{\rep}[1]{#1}

\newcommand{\up}{\ensuremath{\uparrow}}
\newcommand{\down}{\ensuremath{\downarrow}}
\newcommand{\same}{\ensuremath{\leftrightarrow}}
\newcommand{\nr}{\textsc{nr}}   %

\newcommand{\dimRM}{\textsf{R/M}}   %
\newcommand{\dimEff}{\textsf{EFF}}  %
\newcommand{\dimSec}{\textsf{SEC}}  %
\newcommand{\dimCor}{\textsf{COR}}  %

\newtcolorbox{takeaways}{colback=gray!6,colframe=gray!50,boxrule=0.4pt,arc=1pt,
  left=4pt,right=4pt,top=3pt,bottom=3pt,fontupper=\small,title=Key takeaways,
  coltitle=black,colbacktitle=gray!15,fonttitle=\small\bfseries}

\input{local-macros}

\newcommand{\phLcbGemmaTzero}{69.91}  %
\newcommand{\phLcbGemmaTseven}{74.35}  %

\newcommand{\phHhTzeroWins}{51}  %
\newcommand{\phHhBonWins}{16}  %
\newcommand{\phHhBonFamWins}{19}  %
\newcommand{\phManualN}{383}  %
\newcommand{\phRaterKappa}{0.63}  %
\newcommand{\phRaterRho}{0.66}  %
\newcommand{\phRaterJudgeKappa}{0.70}  %
\newcommand{\phRaterJudgeRho}{0.72}  %
\newcommand{\phJudgesRho}{0.72}  %
\newcommand{\phJudgesKappa}{0.60}  %
\newcommand{\phJudgeTwoHumanRho}{0.74}  %
\newcommand{\phJudgesModelRho}{0.79}  %
\newcommand{\phTlowWithin}{26}  %
\newcommand{\phFidOK}{48}  %
\newcommand{\phFidN}{52}  %
\newcommand{\phFidTol}{5\%}  %
\newcommand{\phHhHwEP}{$2.7$}  %
\newcommand{\phHhHwLCB}{$5.5$}  %
\newcommand{\phHhHwLCBP}{$5.8$}  %
\newcommand{\phHhHwCWE}{$6.2$}  %
\newcommand{\phHhHwCSE}{$2.9$}  %
\newcommand{\phHhHwDPS}{$3.2$}  %
\newcommand{\phHhBonWinsResolved}{1}  %
\newcommand{\phHhTzeroPrompting}{8}  %
\newcommand{\phHhTzeroFeedback}{7}  %
\newcommand{\phHhTzeroSelfRefine}{5}  %
\newcommand{\phHhTzeroMultiAgent}{7}  %
\newcommand{\phHhTzeroRetrieval}{7}  %
\newcommand{\phHhTzeroDecoding}{10}  %
\newcommand{\phHhTzeroTraining}{7}  %
\newcommand{\phHhBonPrompting}{4}  %
\newcommand{\phHhBonDecoding}{3}  %
\newcommand{\phHhBonFeedback}{2}  %
\newcommand{\phHhBonSelfRefine}{2}  %
\newcommand{\phHhBonMultiAgent}{3}  %
\newcommand{\phHhBonRetrieval}{2}  %
\newcommand{\phHhBonTraining}{0}  %
\newcommand{\phHhBonFamPrompting}{5}  %
\newcommand{\phHhBonFamFeedback}{2}  %
\newcommand{\phHhBonFamSelfRefine}{2}  %
\newcommand{\phHhBonFamMultiAgent}{1}  %
\newcommand{\phHhBonFamRetrieval}{3}  %
\newcommand{\phHhBonFamDecoding}{3}  %
\newcommand{\phHhBonFamTraining}{3}  %
\newcommand{\phHhSecCse}{23}  %
\newcommand{\phHhNSec}{29}  %
\newcommand{\phHhSecCwe}{6}  %
\newcommand{\phHhNEff}{22}  %
\newcommand{\phHhEffEb}{22}  %
\newcommand{\phHhNonEffEb}{2}  %
\newcommand{\phHhEffDps}{9}  %
\newcommand{\phHhRmUp}{13}  %
\newcommand{\phHhNRm}{15}  %
\newcommand{\phHhRmBon}{8}  %
\newcommand{\phHhEpBelow}{58}  %
\newcommand{\phHhEpMedLo}{$-7.68$}  %
\newcommand{\phHhEpMedHi}{$-1.73$}  %
\newcommand{\phHhLcbMedLo}{$-0.96$}  %
\newcommand{\phHhLcbMedHi}{$+5.41$}  %
\newcommand{\phHhLcbAbove}{44}  %
\newcommand{\phHhLcbpAbove}{57}  %
\newcommand{\phHhLcbpBon}{12}  %
\newcommand{\phHhLcbpMedLo}{$+0.79$}  %
\newcommand{\phHhLcbpMedHi}{$+3.52$}  %
\newcommand{\phHhCallsPrompting}{$1.89$}  %
\newcommand{\phHhCallsDecoding}{$3.75$}  %
\newcommand{\phHhCallsFeedback}{$4.93$}  %
\newcommand{\phHhCallsMultiAgent}{$7.98$}  %
\newcommand{\phHwRatio}{0.99}  %
\newcommand{\phSeedSD}{0.39}  %
\newcommand{\phFdrSurvive}{494}  %
\newcommand{\phFdrResolved}{536}  %

\newcommand{\phCallsTotal}{13.69}  %
\newcommand{\phCallEqTotal}{12.43}  %
\newcommand{\phFeeTotal}{\$64{,}330}  %
\newcommand{\phFeeClaude}{\$38{,}659}
\newcommand{\phFeeVercel}{\$15{,}816}
\newcommand{\phFeeFireworks}{\$9{,}855}
\newcommand{\phHhPrimAbove}{1}  %
\newcommand{\phHhPrimBelow}{22}  %
\newcommand{\phHhPrimWithin}{43}  %

\expandafter\def\csname cnt@nHitsOpenAlex\endcsname{2770}
\expandafter\def\csname cnt@nHitsSS\endcsname{7777}
\expandafter\def\csname cnt@nHitsArxiv\endcsname{4117}
\expandafter\def\csname cnt@nHitsArxivRaw\endcsname{4663}
\expandafter\def\csname cnt@nRecordsIn\endcsname{14664}
\expandafter\def\csname cnt@nDuplicates\endcsname{7722}
\expandafter\def\csname cnt@nScreened\endcsname{6942}
\expandafter\def\csname cnt@nTaExcluded\endcsname{5700}
\expandafter\def\csname cnt@nTaForward\endcsname{1242}
\expandafter\def\csname cnt@nOtherIdentified\endcsname{164}
\expandafter\def\csname cnt@nOtherForward\endcsname{29}
\expandafter\def\csname cnt@nSnowIdentified\endcsname{4580}
\expandafter\def\csname cnt@nSnowForward\endcsname{506}
\expandafter\def\csname cnt@nOtherExcludedTa\endcsname{135}
\expandafter\def\csname cnt@nSnowExcludedTa\endcsname{4074}
\expandafter\def\csname cnt@nFtSought\endcsname{1777}
\expandafter\def\csname cnt@nFtNotRetrieved\endcsname{687}
\expandafter\def\csname cnt@nFtAssessedRetrieved\endcsname{1090}
\expandafter\def\csname cnt@nSnowIncluded\endcsname{210}
\expandafter\def\csname cnt@nFtExcluded\endcsname{462}
\expandafter\def\csname cnt@nIncluded\endcsname{628}
\expandafter\def\csname cnt@nIncludedDb\endcsname{406}
\expandafter\def\csname cnt@nIncludedOther\endcsname{12}
\expandafter\def\csname cnt@nStudies\endcsname{628}
\expandafter\def\csname cnt@nTechRows\endcsname{737}
\expandafter\def\csname cnt@nExclEc5\endcsname{205}
\expandafter\def\csname cnt@nExclIc3\endcsname{115}
\expandafter\def\csname cnt@nExclIc1\endcsname{56}
\expandafter\def\csname cnt@nExclEc4\endcsname{45}
\expandafter\def\csname cnt@nExclIc2\endcsname{12}
\expandafter\def\csname cnt@nExclEc3\endcsname{11}
\expandafter\def\csname cnt@nExclEc8\endcsname{5}
\expandafter\def\csname cnt@nExclIc4\endcsname{4}
\expandafter\def\csname cnt@nExclEc1\endcsname{2}
\expandafter\def\csname cnt@nExclEc2\endcsname{2}
\expandafter\def\csname cnt@nExclEc6\endcsname{2}
\expandafter\def\csname cnt@nExclOther\endcsname{2}
\expandafter\def\csname cnt@nExclEc7\endcsname{1}
\expandafter\def\csname cnt@nExclSnowPending\endcsname{0}
\expandafter\def\csname cnt@nSideContext\endcsname{59}
\expandafter\def\csname cnt@nSideAdjacent\endcsname{62}
\expandafter\def\csname cnt@nSideCorrectness\endcsname{252}
\expandafter\def\csname cnt@nSideSecondary\endcsname{3}
\expandafter\def\csname cnt@nFamPrompting\endcsname{113}
\expandafter\def\csname cnt@nFamFeedback\endcsname{124}
\expandafter\def\csname cnt@nFamSelfRefine\endcsname{17}
\expandafter\def\csname cnt@nFamMultiAgent\endcsname{75}
\expandafter\def\csname cnt@nFamRetrieval\endcsname{37}
\expandafter\def\csname cnt@nFamDecoding\endcsname{18}
\expandafter\def\csname cnt@nFamTraining\endcsname{124}
\expandafter\def\csname cnt@nFamOther\endcsname{191}
\expandafter\def\csname cnt@nDimRM\endcsname{130}
\expandafter\def\csname cnt@nDimEff\endcsname{335}
\expandafter\def\csname cnt@nDimSec\endcsname{182}
\expandafter\def\csname cnt@nNonTargetedMeasured\endcsname{70}
\expandafter\def\csname cnt@nCorrectnessMeasured\endcsname{526}
\expandafter\def\csname cnt@nLowConfidence\endcsname{161}
\expandafter\def\csname cnt@nPreprint\endcsname{372}
\expandafter\def\csname cnt@nPeerReviewed\endcsname{256}
\expandafter\def\csname cnt@pctPreprint\endcsname{59}
\expandafter\def\csname cnt@pctAllTargetUp\endcsname{76.1}
\expandafter\def\csname cnt@pctAllCorrNR\endcsname{28.5}
\expandafter\def\csname cnt@pctAllCorrDown\endcsname{7.5}
\expandafter\def\csname cnt@pctAllNonTarg\endcsname{11}
\expandafter\def\csname cnt@qaMedAll\endcsname{6.0}
\expandafter\def\csname cnt@pctPeerTargetUp\endcsname{74.8}
\expandafter\def\csname cnt@pctPeerCorrNR\endcsname{27.2}
\expandafter\def\csname cnt@pctPeerCorrDown\endcsname{7.8}
\expandafter\def\csname cnt@pctPeerNonTarg\endcsname{12}
\expandafter\def\csname cnt@qaMedPeer\endcsname{6.0}
\expandafter\def\csname cnt@nNr\endcsname{687}
\expandafter\def\csname cnt@nNrConference\endcsname{327}
\expandafter\def\csname cnt@nNrJournal\endcsname{199}
\expandafter\def\csname cnt@nNrPreprint\endcsname{99}
\expandafter\def\csname cnt@nNrThesis\endcsname{29}
\expandafter\def\csname cnt@nNrIEEE\endcsname{192}
\expandafter\def\csname cnt@nNrACM\endcsname{108}
\expandafter\def\csname cnt@nNrSpringer\endcsname{70}
\expandafter\def\csname cnt@nNrElsevier\endcsname{43}
\expandafter\def\csname cnt@nNrSince2025\endcsname{579}
\expandafter\def\csname cnt@nNrSnow\endcsname{147}
\expandafter\def\csname cnt@nNrClosed\endcsname{305}
\expandafter\def\csname cnt@nNrOpenPdf\endcsname{90}
\expandafter\def\csname cnt@nNrOpenLanding\endcsname{154}
\expandafter\def\csname cnt@nNrOaUnknown\endcsname{138}
\expandafter\def\csname cnt@nNrSameAsIncluded\endcsname{15}
\expandafter\def\csname cnt@nNrClosedIEEE\endcsname{172}
\expandafter\def\csname cnt@nNrClosedSpringer\endcsname{62}
\expandafter\def\csname cnt@nNrClosedACM\endcsname{23}
\expandafter\def\csname cnt@nNrClosedElsevier\endcsname{18}
\expandafter\def\csname cnt@nTargetUpLo\endcsname{409}
\expandafter\def\csname cnt@nTargetUpHi\endcsname{601}
\expandafter\def\csname cnt@pctTargetUpLo\endcsname{55}
\expandafter\def\csname cnt@pctTargetUpHi\endcsname{82}
\expandafter\def\csname cnt@nCorrNRLo\endcsname{150}
\expandafter\def\csname cnt@nCorrNRHi\endcsname{255}
\expandafter\def\csname cnt@pctCorrNRLo\endcsname{20}
\expandafter\def\csname cnt@pctCorrNRHi\endcsname{35}
\expandafter\def\csname cnt@nCorrDownLo\endcsname{37}
\expandafter\def\csname cnt@nCorrDownHi\endcsname{73}
\expandafter\def\csname cnt@pctCorrDownLo\endcsname{5}
\expandafter\def\csname cnt@pctCorrDownHi\endcsname{10}
\expandafter\def\csname cnt@pctDirDisagree\endcsname{20.9}
\expandafter\def\csname cnt@nNonTargPrompting\endcsname{15}
\expandafter\def\csname cnt@nCorrPrompting\endcsname{79}
\expandafter\def\csname cnt@nNonTargFeedback\endcsname{12}
\expandafter\def\csname cnt@nCorrFeedback\endcsname{94}
\expandafter\def\csname cnt@nNonTargSelfRefine\endcsname{1}
\expandafter\def\csname cnt@nCorrSelfRefine\endcsname{10}
\expandafter\def\csname cnt@nNonTargMultiAgent\endcsname{10}
\expandafter\def\csname cnt@nCorrMultiAgent\endcsname{66}
\expandafter\def\csname cnt@nNonTargRetrieval\endcsname{4}
\expandafter\def\csname cnt@nCorrRetrieval\endcsname{28}
\expandafter\def\csname cnt@nNonTargDecoding\endcsname{2}
\expandafter\def\csname cnt@nCorrDecoding\endcsname{18}
\expandafter\def\csname cnt@nNonTargTraining\endcsname{15}
\expandafter\def\csname cnt@nCorrTraining\endcsname{114}
\expandafter\def\csname cnt@nNonTargOther\endcsname{16}
\expandafter\def\csname cnt@nCorrOther\endcsname{173}
\makeatletter\newcommand{\cnt}[1]{\ifcsname cntblue@#1\endcsname\textcolor{blue}{\@nameuse{cnt@#1}}\else\@nameuse{cnt@#1}\fi}\newcommand{\cntraw}[1]{\@nameuse{cnt@#1}}\makeatother

\expandafter\def\csname cnt@nHlPool\endcsname{444}
\expandafter\def\csname cnt@nHlEligibleGeneral\endcsname{145}
\expandafter\def\csname cnt@nHlEligibleDomain\endcsname{175}
\expandafter\def\csname cnt@nHlExcluded\endcsname{100}
\expandafter\def\csname cnt@nHlSelected\endcsname{66}
\expandafter\def\csname cnt@nHlWithCode\endcsname{43}
\expandafter\def\csname cnt@nHlDomainListed\endcsname{65}
\expandafter\def\csname cnt@nHlOutOfScope\endcsname{10}
\expandafter\def\csname cnt@nHlNotTechnique\endcsname{7}
\expandafter\def\csname cnt@nHlDuplicate\endcsname{2}
\expandafter\def\csname cnt@nHlNoFamily\endcsname{3}
\expandafter\def\csname cnt@nHlExcludedApplied\endcsname{94}
\expandafter\def\csname cnt@nHlBaseRows\endcsname{328}
\expandafter\def\csname cnt@nHlSameStudy\endcsname{8}
\expandafter\def\csname cnt@nHlSelPrompting\endcsname{10}
\expandafter\def\csname cnt@nHlSelFeedbackLoop\endcsname{10}
\expandafter\def\csname cnt@nHlSelSelfRefinement\endcsname{6}
\expandafter\def\csname cnt@nHlSelMultiAgent\endcsname{10}
\expandafter\def\csname cnt@nHlSelRetrieval\endcsname{10}
\expandafter\def\csname cnt@nHlSelDecoding\endcsname{10}
\expandafter\def\csname cnt@nHlSelTraining\endcsname{10}
\expandafter\def\csname cnt@nBenchHumanEval\endcsname{85}
\expandafter\def\csname cnt@nBenchEvalPlus\endcsname{19}
\expandafter\def\csname cnt@nBenchLiveCodeBench\endcsname{14}
\expandafter\def\csname cnt@nBenchCWEval\endcsname{15}
\expandafter\def\csname cnt@nBenchCyberSecEval\endcsname{15}
\expandafter\def\csname cnt@nBenchEffiBench\endcsname{8}
\expandafter\def\csname cnt@nBenchEvalPerf\endcsname{5}
\expandafter\def\csname cnt@nBenchRACE\endcsname{0}
\expandafter\def\csname cnt@nBenchNoFunEval\endcsname{0}

\title[Beyond Passing Tests]{Beyond Passing Tests: A Systematic Review and Controlled Re-evaluation of Techniques for Improving the Quality of LLM-Generated Code}

\ifdefined\usingacmart
  \author{Prashank Kadam}
  \email{prashank@cerby.com}
  \affiliation{\institution{Cerby Inc.}\country{USA}}
\else
  \author{Prashank Kadam\\Cerby Inc.\\\texttt{prashank@cerby.com}}
\fi

\begin{document}

\ifdefined\usingacmart\else\maketitle\fi

\begin{abstract}
Large language models pass most functional-correctness benchmarks, but passing tests says little about whether their code is readable, efficient or secure. Many techniques claim to improve these qualities, yet their evidence is scattered, usually covers one quality dimension and is rarely compared at an equal compute budget. We contribute a systematic review and a controlled re-evaluation. A systematic literature review of \cnt{nIncluded} primary studies (July 2021 to October 2026), with disclosed, author-reviewed LLM assistance, organizes the techniques by stage, signal source and quality target. Most studies measure only the dimension they target, gains often come with unchanged or unreported correctness, single automated quality signals are gamed, and compute, variance and contamination are rarely reported. A controlled re-evaluation runs ten techniques on four models at matched budgets on correctness, maintainability, efficiency and security benchmarks, extends it to thirteen current models and repository-level tasks, and compares \cnt{nHlSelected} highlighted techniques head to head on Gemma~4 31B. At one call, quality prompting raises analyzer issue counts and leaves complexity and judge scores almost unchanged. At matched budgets, best-of-$n$ sampling with an analyzer-weighted ranker matches or beats prompting, planning, self-refinement, a reviewer agent and retrieval on measures the ranker does not optimize, while an analyzer feedback loop and fine-tuning beat it only on analyzer issue counts. Large correctness losses on HumanEval+, a likely contaminated benchmark near its ceiling, do not appear on contests held after the models' training cutoffs. In the head-to-head, \phHhPrimAbove{} of 66 techniques beats best-of-$n$ on its primary target beyond the 95\% interval, a rate consistent with chance. Gains should be reported against a budget-matched sampling baseline and on measures independent of the technique's own signal.
\end{abstract}

\ifdefined\usingacmart
  \ccsdesc[500]{Software and its engineering~Software creation and management}
  \ccsdesc[500]{Computing methodologies~Natural language processing}
  \ccsdesc[300]{Software and its engineering~Software verification and validation}
  \ccsdesc[300]{General and reference~Surveys and overviews}
  \keywords{large language models, code generation, code quality, maintainability, efficiency, security, systematic literature review}
  \maketitle
\fi

\section{Introduction}\label{sec:intro}

Large language models (LLMs) now write a large share of new code, and the benchmarks used to judge them are running out of headroom. On SWE-bench Verified, the most cited measure of repository-level coding, the leading system reached \rep{97.00\%} as of September~1, 2026~\citep{valsai2026swebench}. OpenAI stopped evaluating on it on February~23, 2026, stating that it ``is increasingly contaminated'' and recommending SWE-bench Pro instead~\citep{openai2026why}. Function-level suites such as HumanEval and MBPP saturated earlier (\S\ref{sec:measurement}). When nearly every model passes, the tests no longer tell us which code is better.

\paragraph{Code quality is not functional correctness.}
A program can pass every test and still be hard to read, slow, or exploitable, and evidence of a large gap predates current models. When \citet{pearce2022asleep} prompted GitHub Copilot with scenarios tied to high-risk weaknesses, about \rep{40\%} of the \rep{1,689} generated programs were vulnerable. In BaxBench, exploits succeeded on around half of each model's correct back-end programs~\citep{vero2025baxbench}. \citet{liu2024refining} found code-style and maintainability issues in \rep{1,930} of \rep{4,066} ChatGPT solutions, and on EffiBench GPT-4's code ran on average \rep{3.12} times as long as the human-written canonical solutions~\citep{huang2024effibench}. Agent code was \rep{2.3x} more verbose and \rep{2.0x} more eroded than code in \rep{473} open-source Python repositories~\citep{orlanski2026slopcodebench}. These context studies (Section~\ref{sec:background}) support our thesis: \emph{code quality is not the same as functional correctness}, and a technique that raises pass@1 while degrading maintainability, efficiency or security has produced a finding, not a success.

\paragraph{Many techniques, scattered evidence.}
A fast-growing literature improves LLM code beyond correctness through quality-oriented prompting, analyzer feedback loops, self-refinement, multi-agent pipelines, retrieval, constrained decoding and training-time alignment. The evidence, spread across software-engineering, machine-learning and security venues, is hard to use for three reasons. First, most studies measure only the dimension they target, so a security technique's effect on readability, or a readability technique's on correctness, is rarely known. Second, iterative techniques spend several times their baselines' compute, and gains often shrink once compute is matched~\citep{olausson2024is}. Third, base models change faster than the studies: a technique evaluated on GPT-3.5 in 2023 says little about its value for a 2026 reasoning model.

\paragraph{Existing reviews.}
\citet{sun2026quality} survey \rep{109} studies of non-functional quality through the ISO/IEC~25010 lens and add one patching experiment with \rep{three} LLMs. \citet{tony2025prompting} review prompting techniques for secure code and evaluate a subset on GPT-3, GPT-3.5 and GPT-4, for security only. Others focus on one dimension (maintainability~\citep{matias2026survey}, technical debt~\citep{ehsani2026faster}, code optimization~\citep{gong2025language}, secure code generation~\citep{patir2026sok}) or survey code generation without quality outcomes~\citep{jiang2024survey}; Section~\ref{sec:related} compares them. None maps techniques across all three non-functional dimensions with their effects on correctness and non-targeted dimensions, or re-tests them head to head on the same models at matched compute.

\paragraph{This article.}
This article has two parts. The first, a systematic literature review (SLR), follows Kitchenham-style guidelines and is reported per SEGRESS~\citep{kitchenham2023segress} and PRISMA~2020~\citep{page2021prisma}. We searched OpenAlex, Semantic Scholar and arXiv from July~2021 to October~2026, calibrated the search string against a quasi-gold standard~\citep{zhang2011identifying}, screened \cnt{nScreened} records with LLM assistance disclosed under GUEST~\citep{kitchenham2026preliminary}, and included \cnt{nIncluded} primary studies; the author reviewed every LLM screening, eligibility and extraction decision (\S\ref{sec:method}). The second, a controlled re-evaluation (\S\ref{sec:empirical}), re-runs ten techniques, one or more per family, and a single-shot baseline on a fixed grid of one closed frontier and three open-weight models, on fixed correctness, maintainability, efficiency and security benchmarks at matched budgets; an extended grid of thirteen current closed and open-weight models adds the repository- and agent-level benchmarks that dominate 2026 model releases (DeepSWE, Terminal-Bench, FrontierCode, SlopCodeBench, GSO, SWE-fficiency, SusVibes).

The review answers five research questions:
\begin{description}[leftmargin=1.2em,itemsep=1pt,font=\normalfont\bfseries]
\item[RQ1 (taxonomy).] What techniques have been proposed to improve the non-functional quality of LLM-generated code, and how can they be organized?
\item[RQ2 (targets and measurement).] Which quality dimensions does each technique target, and how is each measured?
\item[RQ3 (effects and trade-offs).] What effects are reported on the target, on correctness, and on non-targeted dimensions?
\item[RQ4 (evidence quality).] How rigorous and reproducible is the evidence (baselines, compute reporting, contamination handling, human validation, artifacts)?
\item[RQ5 (gaps).] Which combinations of technique, dimension and setting are unstudied or under-studied?
\end{description}
The re-evaluation adds four empirical questions (ERQ1--ERQ4, \S\ref{sec:empirical}): matched-budget effects on the target, correctness and other dimensions; their dependence on model strength and architecture; and how the highlighted techniques compare in one setting.

\needspace{6\baselineskip}\paragraph{Contributions.}
\begin{enumerate}[leftmargin=1.4em,itemsep=1pt]
\item A faceted taxonomy classifying each technique by \emph{stage} (pre-generation, in-decoding, post-generation loop, training-time), \emph{signal source} (none, self, another LLM, static analyzer, compiler, tests, profiler, retrieved knowledge, human, learned verifier, reward model) and \emph{quality target} (readability/maintainability, efficiency, security), built from a reproducible search whose queries, screening decisions and extraction data are released (\S\ref{sec:method}, \S\ref{sec:taxonomy}).
\item A per-family synthesis of reported effects that sets the effect on correctness and non-targeted dimensions beside the effect on the target, and assesses evidence quality (\S\ref{sec:prompting}--\S\ref{sec:hybrid}, \S\ref{sec:discussion}).
\item The highlighted techniques of each family, chosen by a fixed rule combining the most-cited techniques with recent ones reporting the largest gains, each described by its algorithmic novelty (\S\ref{sec:highlighted}).
\item A controlled, matched-budget re-evaluation of these techniques that extends the repeated-sampling and test-time-compute comparisons made for correctness \citep{brown2024large,hassid2024larger,olausson2024is,stroebl2024inference} to maintainability, efficiency and security, on several models and on measures independent of each technique's own signal, with explicit treatment of contamination, compute fairness and metric validity, and a head-to-head comparison of the highlighted techniques on shared benchmarks (\S\ref{sec:empirical}).
\end{enumerate}

\paragraph{Main findings of the review.}
Against the extracted evidence: (i) only \cnt{nNonTargetedMeasured} of \cnt{nStudies} extracted studies measured any quality dimension beyond their target (Table~\ref{tab:nontargeted}); (ii) among technique rows reporting a target gain, correctness was unchanged, mixed or lower in about a quarter and unreported in nearly three in ten (Table~\ref{tab:effects}), and analyzer-driven loops held correctness mainly with a test or fitness gate (\S\ref{sec:feedback}); (iii) optimizing a single automated quality signal invites gaming, from fixes that satisfy the analyzer but not reviewers to reinforcement-learning policies that emit trivial code (\S\ref{sec:feedback}, \S\ref{sec:training}); and (iv) few studies match compute, report variance or handle contamination (Table~\ref{tab:qa}). Section~\ref{sec:discussion} discusses them and \S\ref{sec:agenda} derives a research agenda.

\paragraph{Main findings of the re-evaluation.}
The matched-budget re-evaluation confirms the review's pattern (\S\ref{sec:empirical-results}): (i) at equal budgets of four, eight or sixteen calls, best-of-$n$ sampling with an analyzer-weighted ranker matches or beats quality prompting, planning, self-refinement, a reviewer agent and retrieval in 13 to 16 of 16 cells on the measures it does not optimize (complexity, two judges, CWEval and BaxBench; Table~\ref{tab:emp-indep}), and the analyzer's own counts agree, so much of their reported gain over a single shot is a budget effect, and at a single call quality instructions and planning raise maintainability issues on every model; (ii) the analyzer loop and the training-time techniques beat best-of-$n$ beyond the 95\% interval only on the issue counts their analyzers compute, while a judge rates the analyzer loop's code lower on every model; (iii) on the two models with the highest single-shot correctness, almost every technique lowers HumanEval+ pass@1 below the single shot, the analyzer loop by up to $9.94$ points despite its public-test gate, and on both Gemma models best-of-$n$ itself loses hidden-test correctness as it draws more samples; (iv) weaker models gain more in absolute terms; (v) on LiveCodeBench Pro contests held after every cutoff, no technique changes correctness beyond the 95\% interval, so much of the HumanEval+ loss is specific to HumanEval+, which is likely contaminated and near its ceiling for the strongest models; and (vi) across thirteen current models, maintainability and security rise with capability while efficiency does not, and at repository level frontier agents write working but insecure code on about four in ten SusVibes tasks. In the head-to-head of the highlighted techniques (\S\ref{sec:empirical-h2h}), \phHhPrimAbove{} of 66 beats best-of-$n$ on its primary target beyond the 95\% interval and \phHhPrimBelow{} fall below it, although \phHhTzeroWins{} improve their target over a single shot by point estimate; security techniques lower detector hits far more often than they raise the share of functional and secure code; and in about half the cases efficiency techniques speed up execution time without improving instruction counts.

\paragraph{Organization.}
Sections~\ref{sec:background}--\ref{sec:measurement} cover background, related surveys, review method and quality measurement; \S\ref{sec:taxonomy} gives the taxonomy, \S\ref{sec:highlighted} the rule selecting each family's highlighted techniques, \S\ref{sec:prompting}--\S\ref{sec:training} the seven families, each ending with its highlighted techniques, \S\ref{sec:hybrid} the hybrid and other techniques, \S\ref{sec:empirical} the re-evaluation, and \S\ref{sec:discussion}--\S\ref{sec:conclusion} findings, implications for researchers and practitioners, agenda, threats and conclusions.

\section{Background: The Quality of LLM-Generated Code}\label{sec:background}

This section fixes our quality vocabulary and summarizes the \emph{context studies}: empirical
work that measures the non-functional quality of LLM-generated code without proposing a technique
to improve it. These studies are not part of the review corpus (\S\ref{sec:method}) but establish
the problem the reviewed techniques address. Several include a small mitigation experiment,
treated as context here and screened as a candidate technique study in \S\ref{sec:method}.

\subsection{Quality Model}\label{sec:background-quality}

We anchor quality in ISO/IEC~25010, whose 2023 edition defines a product quality model that
``is composed of \rep{nine} characteristics'' with further subcharacteristics~\citep{isoiec2023isoiec};
we use four of them at source-code level. \emph{Functional suitability} becomes correctness
(\dimCor), measured with tests; it is a control variable, not a target, because a technique that
improves another dimension by breaking functionality has not improved quality.
\emph{Maintainability}, through its analysability and modifiability subcharacteristics, becomes
readability and maintainability (\dimRM): naming, structure, complexity, code smells, idiomatic
style and documentation. \emph{Performance efficiency} becomes \dimEff: runtime and memory
relative to a reference solution. \emph{Security} becomes \dimSec: vulnerabilities, mapped to the
Common Weakness Enumeration (CWE). Other characteristics (e.g., compatibility and reliability)
are rarely measurable on single-function or single-repository tasks and are out of scope. We
never collapse the four dimensions into one score; \S\ref{sec:measurement} describes each measure.

\subsection{Security}\label{sec:background-sec}

The earliest context studies concern security. \citet{pearce2022asleep} prompted GitHub Copilot
with scenarios built around high-risk CWEs and found a large share of the completions vulnerable.
\citet{asare2023is} replayed scenarios in which human developers had introduced C/C++
vulnerabilities; Copilot reproduced the vulnerable code more often than the fix, though less
consistently than a human developer would. \citet{khoury2023how} report that ChatGPT is aware of
potential vulnerabilities yet often generates code not robust to attack. In a user study by
\citet{perry2023do}, participants with an AI assistant wrote significantly less secure code and
were more likely to believe it was secure. Code in the wild confirms the controlled results:
\citet{fu2025security} found security weaknesses spanning \rep{43} CWE categories in Copilot,
CodeWhisperer and Codeium snippets in GitHub projects.

Recent benchmarks that measure security jointly with functionality expose the gap.
BaxBench~\citep{vero2025baxbench} executes end-to-end exploits against generated backend
applications and succeeds on around half of the \emph{correct} programs. Agentic benchmarks built
from real repositories show the same for coding agents: SusVibes~\citep{zhao2026is} finds most
functionally correct agent solutions insecure, even with vulnerability hints in the request, and
SecureVibeBench~\citep{chen2026securevibebench} finds that even the best agent produces few
solutions that are both correct and secure (Supplementary Table~\ref{S-tab:bg-context} (Supplementary Material, \S\ref{S-sec:app-context})).

\subsection{Readability, Maintainability and Code Smells}\label{sec:background-rm}

For \dimRM, \citet{liu2024refining} ran static analyzers on ChatGPT solutions to Java and Python
tasks and found code-style and maintainability issues in close to half. Compared with canonical
solutions, closed- and open-source LLMs ``tend to produce code that is shorter yet more
complicated''~\citep{dou2026what}. \citet{yetistiren2023evaluating} measured code smells and
technical debt for Copilot, CodeWhisperer and ChatGPT on HumanEval. At project level, a
difference-in-differences study of GitHub projects that adopted Cursor found ``a statistically
significant, large, but transient increase in project-level development velocity, along with a
substantial and persistent increase in static analysis warnings and code complexity''
\citep{he2025speed}. \citet{paul2025investigating} compare Java code from Gemini Pro, ChatGPT,
Codex and Falcon with professionally written reference solutions and find more smells in every
model, with the gap growing with task complexity. CodeSmellEval~\citep{velasco2025how} measures a
model's propensity to produce specific smells, and its follow-up~\citep{velasco2026causal} relates
that propensity to decoding strategy, model size, architecture and prompt formulation.
\citet{abbassi2025taxonomy} derive a taxonomy of inefficiencies in LLM-generated Python code
spanning logic, performance, readability and maintainability. At repository scale,
\citet{liu2026debt} attribute static-analysis issues to AI-authored commits across thousands of
GitHub repositories and find that code smells dominate and a substantial share of introduced
issues persists to the latest revision.

Industry telemetry points the same way, with weaker methodological guarantees; we report it as
grey literature, never as evidence for or against a technique. GitClear's analysis of changed
lines reports a rising share of copy-pasted code and a falling share of moved, refactored code
over the analyzed period~\citep{gitclear2025ai}. Sonar's SonarQube analysis of Java assignments
reports that a newer model raised benchmark pass rates while increasing bug
severity~\citep{sonar2025coding}; the accompanying vendor-authored preprint finds ``no direct
correlation'' between a model's pass@1 and the number of issues in its passing
solutions~\citep{sabra2025assessing}.

\subsection{Performance Efficiency}\label{sec:background-eff}

Efficiency benchmarks report LLM code relative to a fast human reference solution per task.
EffiBench~\citep{huang2024effibench} evaluates \rep{42} models on LeetCode problems and finds
generated code generally slower than the canonical solutions. Mercury~\citep{du2024mercury}
introduces a runtime-percentile-weighted pass score (Beyond) and reports a clear gap between pass
rate and Beyond. EffiBench-X~\citep{qing2025effibenchx} extends the comparison to Python, C++,
Java, JavaScript, Ruby and Go and finds that models ``consistently underperform human experts in
efficiency'', with larger gaps in Java, C++ and Go than in Python, Ruby and JavaScript.
EvalPerf~\citep{liu2024evaluating} stresses solutions with expensive inputs and finds that ``the
scaling law fails to account for code efficiency''. \citet{niu2024on} run a similar evaluation on
HumanEval, MBPP and LeetCode and also try efficiency prompts, and
NoFunEval~\citep{singhal2024nofuneval} shows that code models ``generally falter'' on
non-functional requirements, including efficiency, security and maintainability.

\subsection{Verbosity, Erosion and Length Bias}\label{sec:background-verbosity}

A recurring finding is that LLM code is longer than necessary and degrades as it is extended.
SlopCodeBench~\citep{orlanski2026slopcodebench} has agents repeatedly extend their own solutions
and measures two forms of degradation, structural erosion and verbosity; both rise in most
trajectories, and explicit quality guidance lowers their initial level ``without affecting
degradation rates''. Two training mechanisms plausibly contribute to verbosity, though neither was
studied on code quality directly. In reinforcement learning from human feedback (RLHF),
\citet{singhal2023long} find reward improvements ``to largely be driven by increasing response
length'' and trace this to reward models; length bias also distorts LLM-judge
evaluators~\citep{dubois2024lengthcontrolled}, which matters for LLM-as-judge readability scores
(\S\ref{sec:measurement}). In reinforcement learning with verifiable rewards (RLVR),
\citet{liu2025understanding} identify an optimization bias in GRPO that ``artificially increases
response length (especially for incorrect outputs)''. Reasoning models add a related cost:
\citet{chen2024do} describe \emph{overthinking}, ``where excessive computational resources are
allocated for simple problems with minimal benefit'', and a survey of efficient reasoning
catalogues remedies~\citep{sui2025stop}. For code generation, \citet{xue2026empirical} study
thinking models' reasoning traces on BigCodeBench and find that modest reductions in step count
often preserve success on standard tasks. To our knowledge, whether verbose reasoning yields
verbose \emph{code} is not yet measured directly.

\subsection{Implications for This Review}\label{sec:background-implications}

Three observations shape the paper. First, correctness and the other dimensions diverge: BaxBench,
SusVibes and SecureVibeBench report large fractions of correct programs that are insecure, and the
Sonar data (grey literature) suggest that a model upgrade can raise pass rates while making
defects more severe, so a correctness benchmark alone cannot show whether a technique improves
code. Second, the gaps exist on every dimension we study and persist across model generations,
from Copilot in 2021 to agents in 2026, so technique-level remedies are worth reviewing. Third,
the studies use different instruments (static analyzers, exploits, runtime ratios, smell
detectors), so their numbers are not comparable across rows of Supplementary Table~\ref{S-tab:bg-context};
\S\ref{sec:measurement} addresses this measurement problem, and \S\ref{sec:empirical}
re-evaluates techniques under one protocol for this reason.

\section{Related Surveys and Positioning}\label{sec:related}

Supplementary Table~\ref{S-tab:rel-surveys} compares this review with existing secondary studies, grouped by
breadth.

\paragraph{General surveys of LLMs for code.} Broad reviews map the field, not its quality
dimensions: \citet{hou2023large} review LLMs across software engineering tasks,
\citet{jiang2024survey} organize LLM code generation by data, training, prompting, retrieval and
evaluation and collate reported HumanEval, MBPP and BigCodeBench results, and
\citet{dong2025survey} survey LLM-based code generation agents, including single- and
multi-agent architectures. They cover many techniques we review but judge them by functional
correctness. A tertiary study of \rep{30} secondary studies finds accuracy ``strong on
benchmarks but weakly supported for real-world generalization'' and calls for ``holistic,
standardized evaluation'' \citep{chochlov2026tertiary}.

\paragraph{Dimension-specific reviews.} For \dimSec, \citet{tony2025prompting} review prompting
techniques and test a subset on secure code generation, \citet{basic2024from} review
LLM-introduced vulnerabilities and how prompting affects their detection and repair, and the SoK
of \citet{patir2026sok} systematizes prompting, fine-tuning, reinforcement learning and agentic
workflows for secure code and measures a gap between models' knowledge of secure-coding
principles and their code. For \dimRM, \citet{matias2026survey} review how LLMs affect
maintainability and evolvability, the multivocal review of \citet{ehsani2026faster} on technical
debt in LLM-assisted development notes that ``no standardized benchmarks or LLM-specific metrics
yet exist'', and \citet{alomari2026using} review LLMs for refactoring and smell detection and find
LLM-refactored code ``not reliable''. For \dimEff, \citet{gong2025language} review
language-model-based code optimization, and \citet{ali2026energyefficient} review the energy
efficiency of LLM-generated code and find the evidence ``limited and highly heterogeneous''.
\citet{yang2024robustness} cover non-functional properties of code models, but their efficiency
dimension concerns the models, not the code they generate. \citet{pan2023automatically}
taxonomize self-correction by training-time, generation-time and post-hoc stage, and
\citet{kamoi2024when} conclude that it ``works well in tasks that can use reliable external
feedback'', which motivates our signal-source facet (\S\ref{sec:taxonomy}).
\citet{he2026bridging} cover \rep{nine} quality dimensions but organize them around training-data
causes of defects rather than improvement techniques.

\paragraph{Closest prior work.} Two studies share parts of our design. \citet{sun2026quality}
use ISO/IEC~25010, review the literature on non-functional quality of generated code, run
industry workshops and patch real issues with \rep{three} LLMs. They find that research
``primarily emphasizes security, performance efficiency, and maintainability'' while practitioners
prioritize maintainability and readability, and that prompt-based optimization of these
characteristics is unstable. Their experiment, however, only regenerates SWE-bench Lite patches
with SWE-agent under quality-oriented prompts and feedback: one task type, no comparison of
technique families and no matched compute budgets. \citet{tony2025prompting} combine a
systematic review with controlled experiments, the design we follow, but cover only security and
prompting, with GPT-3, GPT-3.5 and GPT-4.

\paragraph{Test-time compute and cost-controlled evaluation.} A separate line of work asks
what extra inference compute buys for correctness. Repeated sampling raises the share of problems
solved by at least one sample roughly log-linearly with the number of samples, but selecting the
right sample without a verifier plateaus \citep{brown2024large}; allocating test-time compute
adaptively per prompt beats a best-of-$N$ baseline \citep{snell2024scaling}; and several samples from
a smaller code model can beat one sample from a larger model at the same budget when unit tests are
available to select among them \citep{hassid2024larger}. \citet{kapoor2024ai} argue that agents
should be evaluated jointly on cost and accuracy, and \citet{stroebl2024inference} show that with an
imperfect verifier, such as unit tests, resampling cannot remove false positives that pass the
verifier, so its accuracy is bounded at any budget, and that these false positives may also adhere
poorly to coding conventions. Closest to our first question, the gains of self-repair over resampling at an equal number of
programs are often modest and sometimes absent for code correctness \citep{olausson2024is}, and two 2026 preprints
on small open models report that Self-Refine and Reflexion lose to repeated sampling at equal token
cost on mathematics benchmarks \citep{mirzaei2026sample} and that none of 26 post-hoc selection,
verification and repair operators improves held-out code correctness over best-of-$N$ at matched
compute \citep{iscan2026selection}. Our re-evaluation builds on these results rather than
preceding them: it asks the same matched-budget question for maintainability, efficiency and
security as well as correctness, and measures it on instruments independent of each technique's
own signal, where the analyzer-weighted best-of-$n$ ranker we use is itself an imperfect verifier
whose loss of hidden-test correctness as $n$ grows (Supplementary Material,
\S\ref{S-sec:app-controls}) is what \citet{stroebl2024inference} predict.

\paragraph{Gap.} To our knowledge, no study re-tests quality-improvement techniques across
\dimRM, \dimEff{} and \dimSec{} and several models under matched budgets, on measures independent of
each technique's own signal, and no review maps techniques across
\dimRM, \dimEff{} and \dimSec{} together while recording their effect on \dimCor{} and on
non-targeted dimensions, which dimension-specific reviews cannot do by construction. This review
addresses both: \S\S\ref{sec:prompting}--\ref{sec:training} synthesize reported effects on
targeted and non-targeted dimensions, and \S\ref{sec:empirical} re-evaluates representative
techniques under one protocol.

\section{Review Methodology}\label{sec:method}

We followed the software-engineering guidelines for systematic literature reviews and report per SEGRESS~\citep{kitchenham2023segress} and the PRISMA~2020 flow~\citep{page2021prisma}. Research questions, search strings, criteria, extraction schema and quality-assessment items (Supplementary Material, \S\ref{S-sec:app-studies}), together with all queries, raw exports, screening decisions, prompts and extraction data, will be released. \S\ref{sec:method-genai} discloses LLM assistance per GUEST~\citep{kitchenham2026preliminary}.

\subsection{Scope, sources and search}\label{sec:method-search}
We ask which techniques improve the non-functional quality (readability and maintainability, performance efficiency, security; \S\ref{sec:background}) of code that LLMs generate or edit, with correctness recorded as a co-outcome. A \emph{technique} is any intervention on the input, decoding, a post-generation loop or the model's weights that changes the code an LLM produces. The window runs from the release of Codex and HumanEval (2021-07-07)~\citep{chen2021evaluating} to the search in October 2026.

Scripted searches covered three sources: OpenAlex (title and abstract), Semantic Scholar (bulk title-and-abstract search) and arXiv (title or abstract). The primary string is S1 = F1 $\wedge$ F2 $\wedge$ F3, with a model facet F1 (e.g., ``large language model*'', LLM, GPT*, Copilot), a task facet F2 (e.g., ``code generation'', ``generated code'', ``program synthesis'') and a quality facet F3 (e.g., readab*, maintainab*, ``code smell*'', ``execution time'', ``energy efficien*'', vulnerab*, CWE). A supplementary string S2 = F1 $\wedge$ F2 $\wedge$ F4 replaces F3 with technique terms (e.g., ``self-refin*'', ``static analy*'', rerank*, DPO, ``reinforcement learning'') to catch papers that measure quality without naming it in the abstract; Supplementary Material \S\ref{S-sec:app-search} lists the exact per-source strings.

\subsection{Calibration and recall}\label{sec:method-recall}
Before inspecting any result we froze a quasi-gold standard (QGS)~\citep{zhang2011identifying} of 32 seed technique studies meeting the inclusion criteria. The first S1 retrieved 22 (recall 0.688; target 0.85): missed studies described editing, repairing or optimizing code without the phrase ``code generation'' or wrote ``code LM'' for ``LLM'', and the technique facet lacked prompting terms. One logged iteration widened F1 and F2, added prompting terms to F4 and re-ran S1 and S2 on Semantic Scholar and arXiv; OpenAlex was queried with the original S1 only (\S\ref{sec:threats}). The union retrieved \cnt{nRecordsIn} records, \cnt{nScreened} after deduplication by DOI, arXiv ID and normalized title, with QGS recall 0.938 (30 of 32; Self-Refine~\citep{madaan2023selfrefine}, which names neither a task nor a quality term, was left to snowballing).

\begin{figure}[t]
\centering
\resizebox{\linewidth}{!}{\begin{tikzpicture}[font=\scriptsize, node distance=4mm and 6mm,
  b/.style={draw, rounded corners=1pt, fill=black!4, align=center, text width=52mm, inner sep=3pt},
  r/.style={b, text width=56mm}, a/.style={-{Latex[length=1.6mm]}, thin}]
\node[b] (id) {Records identified (database search)\\ OpenAlex \cnt{nHitsOpenAlex}; Semantic Scholar \cnt{nHitsSS};\\ arXiv \cnt{nHitsArxiv} (records retrieved; overlapping)};
\node[r, right=of id] (dup) {Removed before screening:\\ duplicates \cnt{nDuplicates}};
\node[b, below=of id] (scr) {Records screened (title/abstract)\\ \cnt{nScreened}};
\node[r, right=of scr] (tax) {Excluded at title/abstract\\ \cnt{nTaExcluded}};
\node[b, below=of scr] (sought) {Reports sought for retrieval \cnt{nFtSought}\\ (database \cnt{nTaForward}; other methods \cnt{nOtherForward};\\ snowballing \cnt{nSnowForward})};
\node[r, right=of sought] (nr) {Reports not retrieved (no full text)\\ \cnt{nFtNotRetrieved}};
\node[b, below=of sought] (ass) {Reports assessed for eligibility\\ \cnt{nFtAssessedRetrieved}};
\node[r, right=of ass] (exc) {Reports excluded \cnt{nFtExcluded}\\ EC1~\cnt{nExclEc1}; EC2~\cnt{nExclEc2}; EC3~\cnt{nExclEc3}; EC4~\cnt{nExclEc4}; EC5~\cnt{nExclEc5}; EC6~\cnt{nExclEc6}; EC7~\cnt{nExclEc7}; EC8~\cnt{nExclEc8}; IC1~\cnt{nExclIc1}; IC2~\cnt{nExclIc2}; IC3~\cnt{nExclIc3}; IC4~\cnt{nExclIc4}; other~\cnt{nExclOther}};
\node[b, below=of ass, fill=blue!6] (inc) {Studies included\\ \cnt{nIncluded}};
\node[r, below=of exc] (oth) {Other methods (venue sweep and section-level searches): identified \cnt{nOtherIdentified}, excluded at title/abstract \cnt{nOtherExcludedTa}, sought \cnt{nOtherForward}. Snowballing (1 backward + 1 forward): identified \cnt{nSnowIdentified}, excluded at title/abstract \cnt{nSnowExcludedTa}, sought \cnt{nSnowForward}};
\node[draw, rounded corners=1pt, align=center, text width=114mm, anchor=north west] (side) at ([yshift=-5mm]inc.west |- oth.south) {Side lists (not primary studies): context \cnt{nSideContext}, adjacent \cnt{nSideAdjacent}, correctness-only \cnt{nSideCorrectness}, secondary \cnt{nSideSecondary}};
\draw[a] (id) -- (scr); \draw[a] (id) -- (dup); \draw[a] (scr) -- (sought); \draw[a] (scr) -- (tax);
\draw[a] (sought) -- (ass); \draw[a] (sought) -- (nr); \draw[a] (ass) -- (inc); \draw[a] (ass) -- (exc);
\draw[a] (oth.west) -| ([xshift=-3mm]sought.west) -- (sought.west);
\end{tikzpicture}}
\Description{Flow diagram of the review: records identified by database searches, other methods and snowballing, then screened, sought for retrieval, assessed for eligibility and included, with the number excluded at each step.}
\caption{PRISMA 2020 flow diagram. Screening, full-text assessment and extraction were LLM-assisted and author-reviewed.}\label{fig:prisma}
\end{figure}
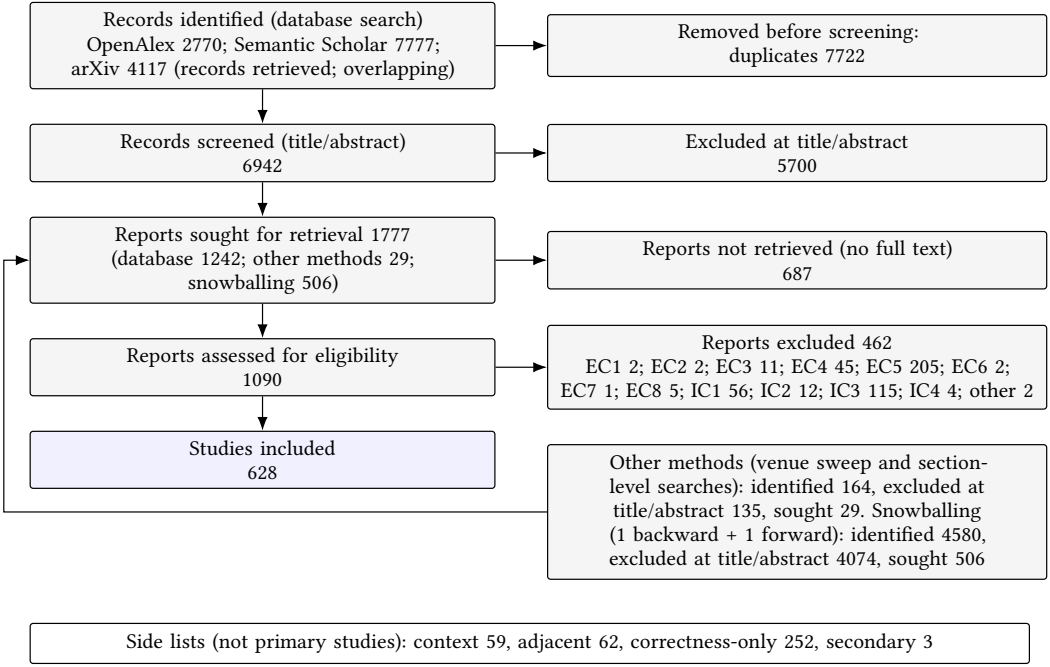

Some added terms came from QGS misses, so post-iteration QGS recall is optimistic; two independent checks complement it. A venue sweep found 274 papers from 2021 onward in 21 software-engineering, machine-learning and security venues whose \emph{titles} match F1 and F2; of the 131 not in our results, 16 passed title/abstract screening, against 55 relevant papers the search found, a screening-stage venue-level recall estimate of 0.775. OpenAlex or Semantic Scholar retrieved 10 of the 11 QGS papers published by ACM or IEEE. Studies from these checks, snowballing and our section-level searches enter the flow as ``other methods'' (Fig.~\ref{fig:prisma}).

\subsection{Selection}\label{sec:method-selection}
A study is included if it (IC1) proposes or evaluates a technique applied to an LLM, (IC2) evaluates code an LLM generated or edited, (IC3) reports at least one quantitative non-functional outcome for that code (tool-based, execution-based, LLM-judged or human-rated), and (IC4) is a full paper, peer-reviewed or a preprint of four or more pages. It is excluded if it is not in English (EC1), is a secondary study (EC2), targets non-code artifacts such as tests or comments (EC3), only analyzes or repairs human-written code (EC4), measures quality only as correctness or compilability (EC5), concerns model robustness or attacks rather than code quality (EC6), uses pre-LLM models (EC7), or duplicates another study (EC8). Excluded studies feed three side lists: \emph{context} studies measuring LLM code quality without an intervention (\S\ref{sec:background}), \emph{adjacent} studies of LLM refactoring or repair of human code, and influential \emph{correctness-only} techniques, untested for quality and thus candidates for our re-evaluation.

\subsection{Screening, snowballing, extraction and appraisal}\label{sec:method-screening}
Claude Opus~5.5, as single screener, applied the criteria to titles and abstracts through a frozen prompt (Supplementary Material, \S\ref{S-sec:app-prompt}) with a recall-favouring rule: records judged \emph{include} or \emph{uncertain} go forward, as do, unscreened, records it declined to classify. Of \cnt{nScreened} database records, \cnt{nTaForward} went forward. One backward and one forward snowballing iteration over the Semantic Scholar citation graph, seeded with the title/abstract inclusions and screened the same way, added \cnt{nSnowIdentified} new records, of which \cnt{nSnowForward} went forward to full text and were assessed and extracted like the database reports; \cnt{nSnowIncluded} were included (Fig.~\ref{fig:prisma}).

Full texts came from arXiv, open-access copies and the ACL Anthology. Only reports whose full text was retrieved were assessed: the \cnt{nFtNotRetrieved} of the \cnt{nFtSought} reports sought for which no full text could be retrieved were excluded as not retrieved, so no report was assessed, extracted or synthesized from its abstract alone. In two stages, a low-cost model (Claude Haiku~5.5) removed only reports that unambiguously failed the criteria, and Claude Opus~5.5 read every remaining report in full, decided eligibility and filled the extraction schema (Supplementary Material, \S\ref{S-sec:app-schema}): bibliographic data, study type, technique, the three facets, tools, iteration budget, targeted and measured dimensions, metrics and their instrument type, benchmarks, languages, models, baselines, the direction of effect on the target, on correctness and on each other dimension with the headline numbers as printed, compute reporting, contamination handling, human evaluation and artifacts. Every row carries a confidence level; low-confidence rows are flagged for human review. Each study was scored on ten quality-assessment (QA) items (0, 0.5 or 1 each): reimplementable description, appropriate baseline, validated metric, correctness reported, more than one model, variance reported, compute reported, artifacts, contamination considered, and threats discussed. QA weights confidence in the synthesis but excludes nothing. A final consistency pass over the studies included from the database and other-method searches found six duplicate records (EC8), none with a retrievable full text, so Fig.~\ref{fig:prisma} counts them among the reports not retrieved, and removed six studies whose only intervention was not evaluated as a way to improve code quality: two quantization studies, three benchmark papers and a mathematical problem-solving benchmark (IC1); the five that measure the quality of LLM code moved to the context list. A second audit of each of these studies against the criteria, reading each full text, removed five further duplicates (the EC8 count of Fig.~\ref{fig:prisma}) and three studies outside the criteria (no code evaluated, EC3; safety rather than code quality, EC6; no quality outcome for the generated code, IC3), moved two studies without an intervention to the context list, and corrected one publication year; every change is logged in the replication package.

Every number attributed to a specific paper was re-read in its primary source: a script downloads the source text, checks that a verbatim quotation containing the number occurs in it, and logs the number with its location and setting. Effect directions in the synthesis tables come from the author-reviewed, LLM-assisted extraction; numbers quoted in the text and per-family tables do not rely on it.

\subsection{Synthesis}\label{sec:method-synthesis}
Effects come from different benchmarks, metrics and models, so we do not pool them in a meta-analysis. Following SWiM~\citep{campbell2020synthesis}, we (i) classify each technique on the three facets of \S\ref{sec:taxonomy} and map families and signal sources against quality dimensions; (ii) count directions of effect, tabulating effect on correctness against effect on the target (Table~\ref{tab:effects}); (iii) report how often non-targeted dimensions were measured at all (Table~\ref{tab:nontargeted}); and (iv) summarize evidence quality by family (Table~\ref{tab:qa}). Vote counts ignore effect and study size, so they describe the evidence rather than rank techniques. Every included study was assessed and extracted from its full text; \cnt{pctPreprint}\% are preprints. Few included studies predate 2023: of 647 records from 2021--2022, 9 passed screening (early work mostly targets correctness or pre-LLM models) and 1 was included. Re-counted on the peer-reviewed studies alone (Supplementary Material, \S\ref{S-sec:app-peer}), the synthesis is unchanged: the target improves in \cnt{pctPeerTargetUp}\% of technique rows (\cnt{pctAllTargetUp}\% overall), correctness is unreported in \cnt{pctPeerCorrNR}\% (\cnt{pctAllCorrNR}\%) and worsens in \cnt{pctPeerCorrDown}\% (\cnt{pctAllCorrDown}\%), a non-targeted dimension is measured in \cnt{pctPeerNonTarg}\% (\cnt{pctAllNonTarg}\%), and the median QA score is \cnt{qaMedPeer} (\cnt{qaMedAll} overall).

\subsection{Use of generative AI in this review}\label{sec:method-genai}
Per GUEST~\citep{kitchenham2026preliminary}, LLMs (i) screened titles and abstracts (Claude Opus~5.5, frozen prompt, \texttt{effort=low}), (ii) assessed full-text eligibility (Claude Haiku~5.5 for clear exclusions, Claude Opus~5.5 for all other reports), (iii) extracted data and QA scores (Claude Opus~5.5; for the snowballing reports the same model and prompt ran as agents that read each full text from a file, with a check that every line was read), and (iv) helped draft text, which authors checked against the primary sources. All prompts, model identifiers, dates and raw outputs are archived. The author reviewed the screening, eligibility and extraction decisions and corrected those found wrong. To measure the error rate of the LLM decisions, as \citet{baltes2025guidelines} recommend, a second coder, an engineer who is not an author and was blind to the LLM's decisions, recoded random samples of 322 title/abstract records, 186 full-text eligibility decisions and 139 technique rows from 124 included studies. Agreement with the LLM was 91.4\% at title/abstract (Cohen's $\kappa=$ 0.78) and 80.1\% at full text ($\kappa=$ 0.69); of the 53 sampled records the second coder included, the LLM screener excluded 2 (a false-exclusion rate of 3.8\%). On the extraction sample, field-level agreement was 85.6\% overall, 88.4\% for the three facets and 79.1\% for the directions of effect. Disagreements were resolved by discussion between the second coder and the author. Because the vote counts rest on the directions, Table~\ref{tab:effects} shows for every count the range it could take if directions were wrong at the rate at which the second coder disagreed (\cnt{pctDirDisagree}\%): most technique rows still report an improved target (\cnt{pctTargetUpLo}--\cnt{pctTargetUpHi}\%), and \cnt{pctCorrNRLo}--\cnt{pctCorrNRHi}\% leave correctness unreported. The author overturned none of the screener's title/abstract decisions (the reports sought are exactly those it marked include or uncertain), reversed no full-text exclusion, overturned 11 of the 644 inclusions on the criteria (1.7\%; the six IC1 studies, three studies outside the criteria and two moved to the context list, described below) and removed five more as duplicates, and corrected 4 of the 920 extracted technique rows (three missing targets, one year); every correction is logged in the replication package. The title/abstract screener ran at \texttt{effort=low}, the setting of its pilot on 30 records.

\section{Measuring Code Quality: Metrics and Benchmarks}\label{sec:measurement}

A technique improves code quality only relative to an instrument that measures it. We review the
instruments for each dimension of \S\ref{sec:background}, the benchmarks that package them
(Supplementary Tables~\ref{S-tab:meas-benchmarks} and~\ref{S-tab:meas-security}, Supplementary Material, \S\ref{S-sec:app-bench}), and the
threats that limit them. Two observations recur: every non-functional metric is a proxy whose validity
has been questioned empirically, often by authors of later benchmarks; and most quality numbers come from
the kind of tool some techniques use as feedback (\S\ref{sec:feedback}), entangling measurement and
optimization.

\subsection{Metrics per quality dimension}\label{sec:meas-metrics}

\paragraph{Correctness (\dimCor).}
Function-level benchmarks report pass@$k$, the probability that at least one of $k$ samples passes the
hidden tests, computed with the unbiased estimator of \citet{chen2021evaluating} from $n \geq k$ samples
of which $c$ pass. Repository-level benchmarks report the resolve rate, the share of instances whose
patch makes the fail-to-pass and regression tests pass \citep{jimenez2024swebench}. Both are only as
strong as the tests: EvalPlus extended the HumanEval tests by \rep{80x}, lowering pass@$k$ by up to
\rep{19.3-28.9\%} and changing model rankings \citep{liu2023is}. Since several quality benchmarks
condition on correctness, weak tests propagate into every other dimension.

\paragraph{Readability and maintainability (\dimRM).}
The oldest instruments are structural metrics: cyclomatic complexity \citep{mccabe1976complexity},
Halstead's operator and operand counts, and the Maintainability Index (MI), a composite of lines of code,
cyclomatic complexity and Halstead volume \citep{coleman1994using,radoncontributors2026introduction}.
Cognitive Complexity, introduced by SonarSource in 2017 \citep{sonarsource2026cognitive}, weights nesting
and breaks in linear flow; a meta-analysis of about \rep{24,000} ratings of \rep{427} snippets found it
correlates with comprehension time and subjective understandability but gives mixed results for
comprehension correctness \citep{baron2020empirical}. Linters and smell detectors (Pylint, Ruff,
SonarQube, PMD) count rule violations
\citep{pylintcontributors2026pylint,astral2026ruff,pmdcontributors2026pmd}. Recent results undercut these
proxies for LLM-generated code. In MaintainCoder, MI and cyclomatic complexity moved in opposite
directions, and multi-agent pipelines scored high MI yet lost correctness after a requirement change
\citep{wang2025maintaincoder}. Across \rep{787,562} human and AI-written function pairs,
complexity metrics carried little signal once size was controlled for \citep{improta2026what}. On C\#
tasks, correctness and structural quality were nearly uncorrelated (Pearson $r$ = \rep{0.075})
\citep{ghalandarian2026benchmarking}. Learned readability models
\citep{buse2010learning,scalabrino2018comprehensive}, trained on human-written code, showed only limited
correlation with developer ratings on AI-generated Java \citep{sergeyuk2024reassessing}.

LLM judges are proposed as a scalable substitute for human readability assessment
\citep{he2026llmasajudge}; ICE-Score and CodeJudge were validated against correctness and human
preference \citep{zhuo2024icescore,tong2024codejudge}. Validity evidence is weaker
for quality than for correctness: on CoReEval, LLM readability scores correlated with human ratings at
Pearson \rep{0.25} in Java, below the \rep{0.31} of \citeauthor{scalabrino2018comprehensive}'s static
model, and at \rep{0.02} in Python \citep{ouedraogo2025humanaligned}. Judges are also sensitive to
presentation: general-purpose judges show position, verbosity and self-enhancement biases
\citep{zheng2023judging}, which motivated length-controlled scoring \citep{dubois2024lengthcontrolled}.
For code, all judges tested by \citet{moon2026dont} shifted scores under superficial edits such as
renamed variables or added comments, even when asked to write tests first; software-engineering
judge decisions change with prompt cues on unchanged code \citep{zhao2026bias};
order effects persist in CodeJudgeBench \citep{jiang2026codejudgebench}; and judges prefer longer code
explanations where developers prefer shorter ones \citep{mittal2026comparing}. Human evaluation remains
the reference but is expensive \citep{he2026llmasajudge}; recent developer studies show it
surfaces production-readiness issues no benchmark metric captures \citep{szych2026evaluating}.

\paragraph{Efficiency (\dimEff).}
Efficiency metrics differ in what they time and how they normalize. EffiBench reports execution time,
peak memory and total memory usage (a time integral of memory), normalized by a canonical solution
\citep{huang2024effibench}. Mercury's Beyond weights pass by runtime percentile among human solutions
\citep{du2024mercury}, and Venus uses percentiles to avoid hardware-specific absolute numbers
\citep{du2025afterburner}. ENAMEL's eff@$k$ generalizes pass@$k$ to efficiency against expert references, with an unbiased estimator for right-censored (timed-out) runs \citep{qiu2025how}. EvalPerf's
Differential Performance Score (DPS) ranks a solution against clusters of reference solutions by CPU
instruction count, not wall-clock time \citep{liu2024evaluating}, as does COFFE
\citep{peng2025coffe}; PIE measures speedup in the gem5 simulator \citep{shypula2024learning}. The
reason is noise: WEDGE measured run-to-run variation of \rep{12\%} for physical time against
\rep{0.03\%} for instruction counts \citep{yang2025wedge}, and PIE's authors note that hardware
variability yields spurious improvements \citep{shypula2024learning}. Few studies measure energy; where
they do, rankings of human and LLM code flip across a server, a PC and a Raspberry Pi
\citep{apsan2026generating}. Passing solutions to one task can differ widely in memory
behavior \citep{rajput2026correctness}, so one sample per task understates variance.

\paragraph{Security (\dimSec).}
Most security benchmarks rely on static application security testing (SAST) \citep{peng2025cweval}:
CodeQL, Bandit and Semgrep \citep{github2026codeql,pycqa2026welcome,semgrep2026semgrep}, with findings
mapped to CWE identifiers \citep{pearce2022asleep,fu2025security}. These tools disagree and miss
vulnerabilities: on the same generated code, Bandit flags \rep{28\%} to \rep{46\%} of programs and
CodeQL \rep{9.1\%} to \rep{17\%} \citep{alrashedy2023can}. CodeQL detected \rep{27} of
\rep{120} manually validated Java vulnerabilities \citep{li2025iris}; a single C/C++ SAST warned on
\rep{52\%} of real vulnerability-contributing commits, at least \rep{76\%} of its warnings were
irrelevant, and \rep{22\%} of commits went undetected \citep{charoenwet2024empirical}; and across
\rep{15} SAST tools detection rates were low \citep{zhou2024comparison}. Studies therefore require
tool agreement \citep{sajadi2025how} or replace SAST with execution: exploit- and test-based
oracles (BaxBench, CWEval, SeCodePLT) run expert-written attacks or security unit tests against the
program \citep{vero2025baxbench,peng2025cweval,nie2024secodeplt}. They pair naturally with correctness:
CWEval's func-sec@$k$ applies the pass@$k$ estimator to samples both functional and secure
\citep{peng2025cweval}, and BaxBench's sec\_pass@$k$ counts programs that pass all tests and resist all
exploits \citep{vero2025baxbench}. The joint metric matters because a program that does nothing is
trivially secure: \citet{dai2026rethinking} found that secure-code techniques often trade functionality
for security, many degrading the base model's joint performance by more than \rep{50\%}.

\subsection{Benchmarks}\label{sec:meas-benchmarks}

Supplementary Tables~\ref{S-tab:meas-benchmarks} and~\ref{S-tab:meas-security} list the most used benchmarks, which fall
into three generations. Correctness benchmarks (HumanEval+, MBPP+, LiveCodeBench,
BigCodeBench, SWE-bench) carry no quality oracle, so quality is scored post hoc on their outputs.
Function-level quality benchmarks from 2022--2024 (RACE, NoFunEval, EffiBench, Mercury, EvalPerf, ENAMEL,
ECCO, the early security sets) attach one oracle to short algorithmic or snippet tasks. Since 2025,
benchmarks have moved to repositories and agents: SWE-Perf, GSO and SWE-Pro for performance
\citep{he2026sweperf,shetty2025gso,sarkayak2026evaluating}; SecRepoBench, SecureVibeBench, SusVibes and
RealSec-bench for security \citep{shen2026secrepobench,chen2026securevibebench,zhao2026is,wang2026realsecbench};
and SlopCodeBench, MaintainBench and CQBench for maintainability
\citep{orlanski2026slopcodebench,wang2025maintaincoder,improta2026what}. Repository-level results are
uniformly low: the best agent passes \rep{14.8\%} of SlopCodeBench checkpoints, with structural erosion
rising in \rep{77\%} of trajectories \citep{orlanski2026slopcodebench}; leading agents succeed on fewer
than \rep{5\%} of GSO tasks \citep{shetty2025gso}; and on SusVibes \rep{57\%} of one agent's solutions
are functionally correct but only \rep{11.8\%} are secure \citep{zhao2026is}. PerfCodeBench extends the
efficiency gap to systems code with parallelism and GPU tasks \citep{jing2026perfcodebench}, and
SWE-fficiency to \rep{498} optimization tasks in nine data-science and HPC repositories, where agents
reach on average ``less than \rep{0.23x} the expert speedup'' \citep{ma2025swe}. On the newest security
benchmark, built from \rep{150} real vulnerabilities, the strongest of six frontier agents writes
functional and secure code for only \rep{25.1\%} of \rep{450} tasks \citep{wu2026last}.

Two further lines broaden what ``quality'' means for agents. Refactoring and long-horizon maintenance
benchmarks score behaviour-preserving structural change in real repositories: SWE-Refactor
(\rep{1,099} Java refactorings, where a Codex agent succeeds on \rep{39.4\%} of compound ones)
\citep{xu2026swe}, SmellBench (\rep{294} injected smells; best smell-elimination score \rep{50.34})
\citep{lin2026smellbench}, CodeTaste \citep{thillen2026codetaste}, SWE-Bench ProMax
\citep{shi2026swe}, SWE-CI, which infers maintainability from correctness changes across a
repository's commit history \citep{chen2026swe}, SWE-EVO \citep{le2025swe}, the refactoring track of
SWE Atlas \citep{raghavendra2026swe}, and SWE-NFI, where at most \rep{70.0\%} of valid agent outputs
preserve functionality while making non-functional improvements \citep{xue2026swe}.
Instruction-following benchmarks attach checkable quality constraints to ordinary tasks: SWE-IF's
\rep{30} verifiable instructions \citep{zhong2025swe}, CIFE's \rep{1,000} tasks with constraints in
\rep{13} categories, with strict adherence of \rep{39-66\%} \citep{gunnu2025cife}, and
CodeAlignBench \citep{mehralian2025codealignbench}. Like RACE and NoFunEval, they measure compliance
rather than default quality (\S\ref{sec:meas-critiques}). The coding
benchmarks most reported at the leading 2026 model launches are Terminal-Bench
\citep{merrill2026terminalbench} and DeepSWE \citep{huang2026deepswe}, which writes \rep{113}
long-horizon tasks from scratch in \rep{91} repositories and grades them with hand-written verifiers
that an independent judge disputes in \rep{1.4\%} of runs against \rep{32.4\%} for SWE-Bench Pro's
inherited tests; FrontierCode adds a code-quality rubric to functional blockers and publishes the
composite score and the plain correct rate \citep{cognition2026frontiercode}. For correctness, live
benchmarks add tasks to stay ahead of training data: LiveCodeBench Pro (Codeforces, ICPC and IOI
problems, best model \rep{53\%} pass@1 on medium and \rep{0\%} on hard problems without tools)
\citep{zheng2025livecodebench,livecodebenchpro2025data}, SWE-rebench \citep{badertdinov2025swe} and
SWE-bench-Live \citep{zhang2025swe}; Multi-SWE-bench extends issue resolution to seven languages
\citep{zan2025multi}.

\subsection{Critiques}\label{sec:meas-critiques}

\paragraph{Saturation.}
As of September 2026, the top vals.ai score on the \rep{500}-task SWE-bench Verified was \rep{97.00\%},
with the third model at \rep{96.20\%}, and vals.ai has archived it as saturated
\citep{valsai2026swebench}. On February 23, 2026, OpenAI stopped reporting it as ``increasingly
contaminated'' and recommended SWE-bench Pro instead: in an audited subset of tasks that models often
failed, at least \rep{59.4\%} had flawed tests that reject correct submissions, and all frontier models
tested could reproduce the gold patch or verbatim problem details for some tasks~\citep{openai2026why}.
SWE-bench Pro is already under strain: an audit found reward hacking through leaked gold solutions
and hidden evaluation information, with models scoring ``substantially worse'' once these were
removed \citep{zheng2026swebenchproverified}. HumanEval+, MBPP+, LiveCodeBench~v6 and SWE-bench
Verified now serve only as legacy anchors to earlier work.

\paragraph{Contamination.}
Popular code benchmarks overlap substantially with open pre-training corpora, and models score higher on
the overlap \citep{riddell2024quantifying}; leakage also enters through synthetic training
data and model selection on test sets \citep{matton2024on}. LiveCodeBench was built for this:
DeepSeek and GPT-4-O dropped on LeetCode problems released after their cutoffs
\citep{jain2025livecodebench}. Its protection lasts only while releases continue: the latest, v6, holds
\rep{1055} problems from May 2023 to April 2025 \citep{livecodebench2025releases}, leaving no clean
split for models with later cutoffs. LiveCodeBench Pro was refreshed quarterly until its last
public set, for July to September 2025 \citep{livecodebenchpro2025data}, so a clean split for 2026 models requires applying its protocol to contests held after their cutoffs;
SWE-rebench and SWE-bench-Live still add tasks \citep{badertdinov2025swe,zhang2025swe}. Most efficiency
benchmarks reuse LeetCode, HumanEval or MBPP problems, which EffiBench-X names as a contamination risk,
filtering to problems released after October 2023 \citep{qing2025effibenchx}. SWE-bench Pro uses
copyleft and private codebases to resist training exposure \citep{deng2026swebench}. Only the benchmarks
marked \emph{date}, \emph{held-out} or \emph{new} in Supplementary Tables~\ref{S-tab:meas-benchmarks}
and~\ref{S-tab:meas-security} offer any protection for models with 2026 cutoffs.

\paragraph{Quality measured only when requested.}
RACE states each non-functional requirement in the prompt and scores whether the model follows it
\citep{zheng2024beyond}; NoFunEval likewise names the requirement in the edit instruction
\citep{singhal2024nofuneval}. They thus measure instruction following on quality, not a model's
default quality, which is what a technique must change. CWEval deliberately avoids security
hints in prompts \citep{peng2025cweval}, and RealSec-bench found that generic security guidance in the
prompt often broke compilation without preventing vulnerabilities \citep{wang2026realsecbench}.

\paragraph{Similarity to a reference.}
Because latency and resource use are hard to measure across target platforms, NoFunEval scores those
edits with DiffBLEU, the similarity of the model's diff to a reference diff, and weights CodeQL success
by DiffBLEU for maintainability and security edits \citep{singhal2024nofuneval}. A correct but different
improvement scores low. Token-matching metrics correlate weakly with human judgment
\citep{zhuo2024icescore} and are unreliable indicators of functional correctness
\citep{chen2021evaluating}.

\paragraph{Weak tests and weak oracles.}
Insufficient tests mis-rank models on correctness \citep{liu2023is}, and on efficiency too: re-running
\rep{1,538} tasks of EffiBench, ENAMEL, EvalPerf and Mercury \rep{30} times, \citet{le2026rethinking}
found that only \rep{6.11\%} of the benchmark-provided performant solutions were significantly faster
than the canonical ones under the provided tests. For security, CWEval reports that its own analyzer
still flags only \rep{562/1916} of CyberSecEval's vulnerable samples, and that SecurityEval's CodeQL
checks have frequent false negatives and positives \citep{peng2025cweval}, while CyberSecEval's authors
report \rep{96\%} precision and \rep{79\%} recall for their detector \citep{bhatt2023purple}. The first
figure is a later re-run on extracted snippets and the second a manual check by the benchmark's
authors; neither is an independent audit.

\paragraph{Timing noise.}
EffiBench, Mercury and many technique papers rely on physical execution time \citep{yang2025wedge},
despite the \rep{12\%} run-to-run variation noted above and EvalPerf's warning that
light computation makes timing flaky \citep{liu2024evaluating}. Small reported speedups on these
benchmarks are within noise unless repetitions and dispersion are reported. Repository-level efficiency
benchmarks are not immune: replaying the official reference patches on four machine types,
\citet{chen2026performance} found they met the benchmarks' own validity rules in every replay for
only \rep{39/102} GSO tasks, \rep{11/140} SWE-Perf tasks and \rep{411/498} SWE-fficiency tasks.

\subsection{Implications for evaluating techniques}\label{sec:meas-implications}

No instrument is valid for all four dimensions and each has documented failure modes, so
comparisons need layered measurement. Static analysis (complexity, smells, lint and SAST
findings) is cheap and reproducible but weakly valid and tool-dependent, so it should be reported per
tool and CWE. Execution profiling should use instruction counts or repeated, pinned timing relative
to a reference, with exploit or security tests where they exist. An LLM judge can cover readability
aspects that metrics miss, but only if calibrated against humans and controlled for length and
presentation. Human review of a sample should validate the other three layers, not merely add a score.
Since several techniques optimize against the analyzers benchmarks use, each dimension needs at
least one held-out instrument. Report quality scores both unconditionally and conditional on
correctness, since a degenerate program can look secure or simple. Section~\ref{sec:empirical} applies
this four-layer stack to correctness benchmarks with post-2026-cutoff tasks where possible.

\section{A Faceted Taxonomy of Quality-Improving Techniques}\label{sec:taxonomy}

A flat list of families (prompting, feedback loops, self-refinement, multi-agent pipelines, retrieval, decoding, training) mixes axes: one static analyzer can drive both a ``feedback loop'' and a ``reinforcement-learning'' method, and a ``multi-agent'' pipeline is often a feedback loop with another LLM as critic. We therefore classify techniques on three orthogonal facets, keeping the families only as familiar labels.

\paragraph{Stage} (when it acts): \emph{pre-generation} (prompt or context: instructions, plans, exemplars, retrieved documents), \emph{in-decoding} (constraining or reranking the token or sample distribution), \emph{post-generation loop} (the model itself, another model, a tool or an agent team inspects and revises code) and \emph{training-time} (weights change through data curation, supervised fine-tuning, preference optimization or reinforcement learning).

\paragraph{Signal source} (what drives the improvement): \emph{none} (only the instruction), \emph{self} (the model's own critique), \emph{other LLM} (reviewer, judge or agent), \emph{static analyzer} (linters, SAST tools), \emph{compiler}, \emph{test execution}, \emph{profiler} (runtime or memory measurements), \emph{retrieved knowledge} (documents, exemplars, repository context), \emph{human}, \emph{learned verifier} (trained classifier or reranker) and \emph{reward model}. Techniques often combine sources, and combinations matter: in \S\ref{sec:feedback}, analyzer feedback preserved correctness mainly when gated by test execution.

\paragraph{Quality target}: readability/maintainability (\dimRM), efficiency (\dimEff) and security (\dimSec), with correctness (\dimCor) as a control.

\begin{table}[t]
\centering
\footnotesize
\caption{How the conventional families map onto the facets. Typical values only; individual techniques are classified one by one in the extraction data.}\label{tab:tax-families}
\begin{tabularx}{\linewidth}{@{}l l X l@{}}
\toprule
Family & Stage & Typical signal sources & Section \\
\midrule
Prompting & pre-generation & none; self (planning, chain of thought) & \S\ref{sec:prompting} \\
Tool feedback & post-generation loop & static analyzer, compiler, tests, profiler & \S\ref{sec:feedback} \\
Self-refinement & post-generation loop & self; tests (self-debugging) & \S\ref{sec:selfrefine} \\
Multi-agent & post-generation loop & other LLM; tools invoked by agents & \S\ref{sec:multiagent} \\
Retrieval & pre-generation & retrieved knowledge & \S\ref{sec:retrieval} \\
Decoding & in-decoding & grammar or type constraints, learned verifier, tests & \S\ref{sec:decoding} \\
Training & training-time & curated data, reward model, analyzer or test rewards & \S\ref{sec:training} \\
\bottomrule
\end{tabularx}
\end{table}

\begin{figure}[t]
\centering
\includegraphics[width=0.92\linewidth]{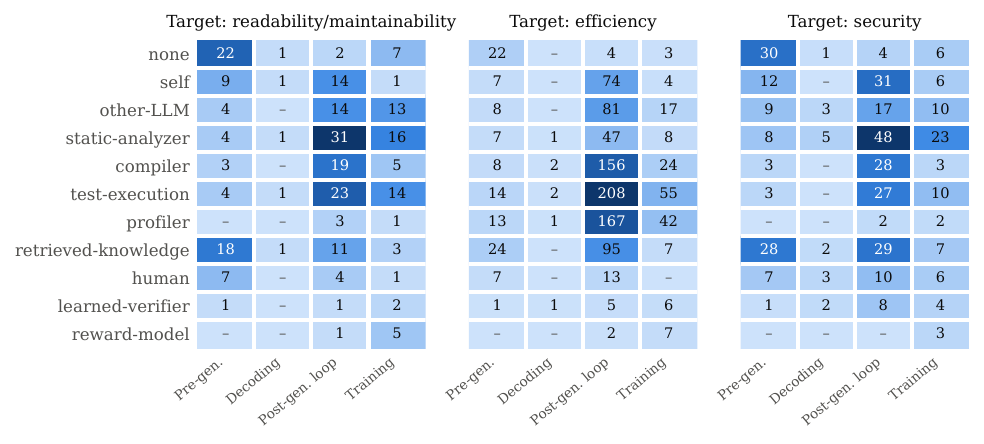}
\Description{Grids of study counts, one panel per quality dimension, with technique stages as columns and signal sources as rows.}
\caption{Faceted taxonomy: number of included studies by stage (columns) and signal source (rows), one panel per targeted quality dimension. A study counts once per combination it uses. Counts from LLM-assisted extraction reviewed by the author.}\label{fig:taxonomy}
\end{figure}

\begin{figure}[t]
\centering
\includegraphics[height=5.2cm]{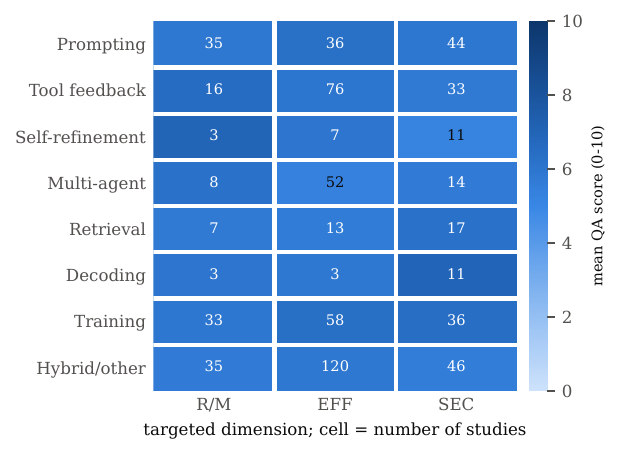}\hfill
\includegraphics[height=5.2cm]{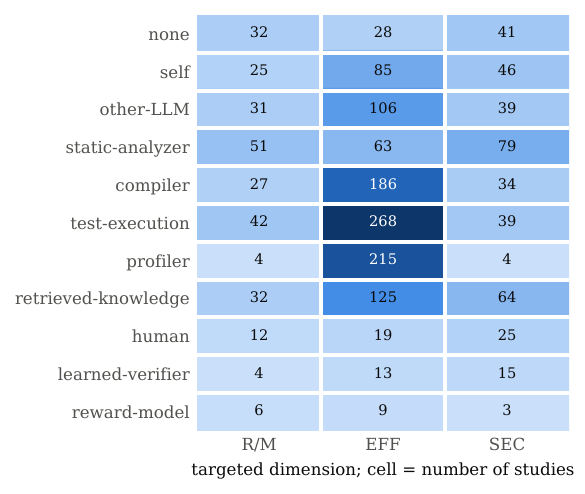}
\Description{Two heat maps of study counts: technique family by targeted dimension, shaded by mean quality-assessment score, and signal source by targeted dimension.}
\caption{Evidence maps. Left: technique family $\times$ targeted dimension; each cell gives the number of studies and is shaded by their mean QA score (0--10). Right: signal source $\times$ targeted dimension (number of studies). Empty and thin cells are the gaps discussed in \S\ref{sec:discussion} and \S\ref{sec:agenda}.}\label{fig:evidence-maps}
\end{figure}

\paragraph{Distribution.}
The \cnt{nStudies} extracted studies contribute \cnt{nTechRows} technique rows. By family, \cnt{nFamPrompting} studies evaluate prompting, \cnt{nFamFeedback} tool-feedback loops, \cnt{nFamSelfRefine} self-refinement, \cnt{nFamMultiAgent} multi-agent pipelines, \cnt{nFamRetrieval} retrieval, \cnt{nFamDecoding} decoding-time methods and \cnt{nFamTraining} training-time alignment, with \cnt{nFamOther} hybrid or other techniques, synthesized in \S\ref{sec:hybrid}. Figure~\ref{fig:taxonomy} crosses stage with signal source per quality target; Fig.~\ref{fig:evidence-maps} maps families and signal sources against targeted dimensions.

The maps show three regularities. First, the signal follows the target: profilers and test execution drive almost all efficiency work, static analyzers and retrieved knowledge dominate security, and readability/maintainability relies mostly on instructions, linters and other LLMs; learned verifiers, reward models and human signals are rare throughout. Second, most techniques act after generation: post-generation loops are the largest stage for efficiency and security, readability work splits about evenly between prompting and post-generation loops, and the few in-decoding methods almost all target security. Third, the field is young and accelerating: nearly all studies date from 2024 on, the yearly count more than doubled from 2024 to 2025, and 2026 had passed 2025 by mid-October (Supplementary Fig.~\ref{S-fig:trends}). A large hybrid group, most often a prompting or retrieval front end plus a tool-feedback or agentic loop, is one reason facets fit the literature better than families.

\section{Highlighted Techniques}\label{sec:highlighted}

Sections~\ref{sec:prompting}--\ref{sec:training} synthesize each family by mechanism and end with
its highlighted techniques: up to ten per family for a reader or re-evaluation to start from,
chosen from the included studies by a fixed rule, described by what is algorithmically new and
listed with their authors' reported result. The same set forms the head-to-head comparison of \S\ref{sec:empirical-h2h}.

\paragraph{Selection rule.} A technique row of the extraction table (\S\ref{sec:method}) is
\emph{eligible} if the study proposes it, with a positive effect on the targeted dimension and a
non-negative effect on \dimCor. Two review passes, with recorded reasons, then required that it
be in scope (a technique for \dimRM, \dimEff{} or \dimSec{} of code an LLM generates or edits), be
the authors' own method rather than a benchmark, dataset or empirical comparison, apply to ordinary
software code rather than a narrow artifact (hardware description languages, GPU kernels, PLC
programs, smart contracts and similar), and not duplicate another record. Further
rows were excluded under five written rules: a lightweight mitigation inside an empirical or
benchmark study, evaluation on only a handful of tasks, code quality as an incidental outcome, a
reviewer flag for low credibility, or a decline in \dimCor{} found on re-reading the selected paper,
in which case the next-ranked technique took its place. Of the
\cnt{nHlPool} candidate rows, \cnt{nHlOutOfScope} were out of scope, \cnt{nHlNotTechnique} were not the authors' own technique, \cnt{nHlDuplicate} duplicated another record, \cnt{nHlNoFamily} were domain-specific rows without a family, and \cnt{nHlExcludedApplied} were excluded under the five rules; the remaining \cnt{nHlBaseRows} rows, after merging \cnt{nHlSameStudy} that are second techniques of a study already counted (one technique per study and family, and a study highlighted in one family only), leave \cnt{nHlEligibleGeneral} general-purpose and
\cnt{nHlEligibleDomain} domain-specific techniques. Each family then takes two lanes.
\emph{Landmarks} (L) are its five most-cited eligible techniques, by Semantic Scholar citation
counts (October 2026). \emph{Recent leaders} (R) are the five non-landmark eligible
techniques from 2025--2026 with the largest reported relative improvement on their targeted
dimension over the strongest baseline their authors report; improvements are ranked as percentiles
within their dimension, so large efficiency speedups do not crowd out security and readability
results, and count only if their numbers appear verbatim in the source. A study is highlighted in one family only, and a hybrid takes the family of the mechanism that does most of its work. Two lanes are needed because citation counts alone favour 2023--2024 work and
reported gain alone favours small, self-evaluated studies. The second lane still inherits each study's choice of baseline: a gain over a weak baseline ranks as high as the same gain over a strong one, so its ranking says which papers report large gains, not which techniques are best; the head-to-head (\S\ref{sec:empirical-h2h}) re-measures every highlighted technique against shared references for that reason.
Two decisions were made after the rule was first applied, and we report them as deviations. For training, the first selection held no reinforcement-learning method, the dominant training
method of 2025--2026, so the rule was adjusted after inspecting it: PurpCode, SecCoderX and Afterburner replace the recent picks DiSCo, Adaptive Unlearning and quality-aware preference learning, so the family includes RL methods for security and
efficiency; with the rule as first applied, the training family would hold the three replaced techniques. Universal Textual Teaching, first counted as general-purpose, is evaluated on code only on GPU kernels and was reclassified as domain-specific; the next-ranked prompting technique was excluded as unverifiable (model unnamed, quality score undefined), and StoryCoder took the slot. The rule selected \cnt{nHlSelected} techniques, ten per family except self-refinement,
which has only \cnt{nHlSelSelfRefinement} eligible general-purpose techniques; \cnt{nHlWithCode}
link public code. The \cnt{nHlDomainListed} most-cited domain-specific techniques appear in
the Supplementary Material (\S\ref{S-sec:app-hl-domain}).

\paragraph{How to read the tables.} Reported results are the authors' own, on their own
benchmarks, models and baselines: they show what each paper claims, not how the techniques compare,
which requires one model, one benchmark set and one call budget (\S\ref{sec:empirical-h2h}). The
selection rests on the extraction (LLM-assisted, reviewed by the author), and citation counts change over time.

\section{Prompting for Quality}\label{sec:prompting}

Prompting, the cheapest intervention, acts almost always \emph{pre-generation} (\S\ref{sec:taxonomy}) and mostly without external signal; a minority add a \emph{self} signal, as Recursive Criticism and Improvement (RCI) does, or human, retrieved or execution information. Of the \cnt{nFamPrompting} studies (\S\ref{sec:method}), the largest group targets \dimSec, then \dimEff{} and \dimRM; \cnt{nCorrPrompting} also measure correctness and \cnt{nNonTargPrompting} a non-targeted dimension.

\paragraph{Grouped comparison.}
Stated requirements help less than expected (Supplementary Table~\ref{S-tab:prompting-results}). On RACE, five simultaneous requirements drive almost all models near zero on the joint correct-and-compliant metric, and models resist requests contradicting their stylistic priors \citep{zheng2024beyond}. Newer security prompts only shift the CWE mix \citep{kharma2026empirical} or trade high- for low-severity findings \citep{urmi2026prompt}. Energy prompts saved nothing consistently across six models and three hardware platforms \citep{apsan2026generating} and helped only some models and languages \citep{rekhi2026improving}; no single efficiency prompt transfers. In CodePromptEval, a persona who ``follows best coding practices for maintainability'' produced fewer Pylint smells and lower complexity at slightly lower correctness, while few-shot and signature prompts raised correctness most but added smells, over a third more when combined with chain-of-thought and persona prompts \citep{khojah2025impact}. Reasoning effort is the main knob frontier APIs expose, yet on A.S.E slow-thinking configurations tended to trail fast-thinking ones \citep{lian2026ase}, and no study we found varies effort within one model while measuring \dimRM{} or \dimEff{}.

\paragraph{Highlighted techniques.}\label{sec:hl-prompting}
In Supplementary Table~\ref{S-tab:hl-prompting} (selection rule in \S\ref{sec:highlighted}), landmarks mostly structure reasoning before coding; recent leaders condition the prompt on a task or model diagnosis from another LLM or analyzer (ECO, BRACE, CSDD) optimize the prompt itself (MPCO, SecureForge), or restate the problem before coding (StoryCoder retells it as a narrative). \textbf{Caveats.} Landmarks' quality evidence is human-rated and partly unquantified. Several baselines are weak. MPCO and CSDD omit correctness, and no row reports an effect on a non-targeted dimension.

Study-by-study detail, the caveats of each highlighted technique, effects on correctness and on the dimensions not targeted, compute fairness and limitations for this family are in Supplementary Material \S\ref{S-sec:fd-prompting}; \S\ref{sec:discussion} compares the families.

\begin{takeaways}
\begin{itemize}
  \item Stating a requirement is not enough: compliance collapses when requirements combine or conflict with stylistic priors, and security prompts often shift the CWE mix rather than reduce it.
  \item Few-shot examples raise pass rates while adding smells; maintainability personas do the reverse.
  \item RCI, the strongest security result, and the multi-call recent leaders were never compared at matched budget.
  \item Efficiency gains are model- and language-specific; reasoning effort and non-targeted dimensions are rarely studied.
\end{itemize}
\end{takeaways}

\section{Tool-Feedback Loops}\label{sec:feedback}

A tool-feedback loop returns an external program's output on generated code to the model until the tool is satisfied or a budget runs out (the \emph{post-generation loop} stage, \S\ref{sec:taxonomy}). Signals come from \emph{static analyzers} (Bandit, Pylint, CodeQL, Semgrep, SonarQube, Infer, CppCheck, PMD), \emph{compilers} or symbolic executors, or \emph{profilers}; \emph{test execution} usually gates acceptance. Of the \cnt{nFamFeedback} studies, most target \dimEff\ or \dimSec, few \dimRM; \cnt{nCorrFeedback} measured correctness and only \cnt{nNonTargFeedback} a non-targeted dimension. Unlike in self-refinement (\S\ref{sec:selfrefine}), a deterministic tool writes the critique, the condition under which self-correction most likely works~\citep{kamoi2024when}, but its verdict becomes the de facto objective. Supplementary Tables~\ref{S-tab:fb-design} and~\ref{S-tab:fb-results} map designs and reported effects.

\paragraph{Grouped comparison.}
FDSP~\citep{alrashedy2023can} returns the Bandit report, requests several candidate fixes, and iterates each until Bandit is clean; LLM-verbalized reports add only \rep{1\% to 2\%}. With CodeQL, raw diagnostics help every model, while a stronger model's explanations help mainly strong instruction followers and not consistently more than raw feedback on real-world tasks~\citep{yan2026guiding}. For maintainability, \citet{blyth2025static} is the closest quality-loop template: Bandit and Pylint issues are tagged inline, security issues ordered by severity weights, and a candidate kept only if its fitness ($-\infty$ on any test failure) does not worsen. For efficiency, EffiLearner~\citep{huang2024effilearner} returns per-line \texttt{line\_profiler} and \texttt{memory\_profiler} output; format matters, as aggregate time and memory alone \emph{increased} CodeLlama-70B's execution time by \rep{51.9\%} while the profile cut it by \rep{9.6\%}. Evolutionary searches (AlphaEvolve \citep{novikov2025alphaevolve} and its open successors CodeEvolve, ContextEvolve and SBLLM) keep the variants an evaluator scores best, and all inherit the evaluator's weaknesses: a timing evaluator rewards noise (\S\ref{sec:meas-critiques}), and none of the studies we found scores evolved code for \dimRM{}.

\paragraph{Highlighted techniques.}\label{sec:hl-feedback}
Supplementary Table~\ref{S-tab:hl-feedback} lists the highlighted techniques (selection rule in \S\ref{sec:highlighted}). \citet{blyth2025static} tag a few Bandit and Pylint issues, chosen by severity weight, inline in GPT-4o's code each round and keep a revision only if its test-gated fitness does not fall; over ten rounds Bandit-flagged snippets fell from \rep{45.05\%} to \rep{13.41\%} while the pass rate rose from \rep{36.7\%} to \rep{46.59\%}. \textbf{Caveats.} Several headline gains rest on weak comparisons. Security results are tool-scored and small. Correctness is unreported for the filtering and FRANC results and unchecked after the Infer repairs.

Study-by-study detail, the caveats of each highlighted technique, effects on correctness and on the dimensions not targeted, compute fairness and limitations for this family are in Supplementary Material \S\ref{S-sec:fd-feedback}; \S\ref{sec:discussion} compares the families.

\begin{takeaways}
\begin{itemize}
\item Tool feedback reliably, often greatly, lowers the issues the driving tool reports.
\item A test or fitness gate is the main correctness safeguard; most ungated loops we read did not measure correctness or lost it.
\item Analyzer gaming is documented: rejected CodeQL-passing fixes, deleted code scoring perfect maintainability, cached timing results, and missed removed defenses.
\item Per-line or hotspot profiles beat aggregate timings; among efficiency loops, only SysLLMatic checks maintainability.
\item Budget-matched comparisons are rare; much of FDSP's gain comes from sampling.
\end{itemize}
\end{takeaways}

\section{Self-Refinement}\label{sec:selfrefine}

Self-refinement has the generating model critique and revise its own output (\emph{post-generation loop} stage). The critique is grounded in the model alone (\emph{self}) or in self plus \emph{test execution} (failing tests, tracebacks, runtime state), a \emph{profiler} or a \emph{static analyzer}. A study belongs here when the model writes the critique, in \S\ref{sec:feedback} when a tool report passes through largely unchanged. Of the \cnt{nFamSelfRefine} included studies, \cnt{nCorrSelfRefine} also measure \dimCor{} and \cnt{nNonTargSelfRefine} a non-targeted dimension; most target \dimSec, a few \dimEff{} or \dimRM, none all three. Several best-known methods report only correctness (our correctness-only side list, \S\ref{sec:method}) but are discussed because quality-targeted variants inherit their loops.

\paragraph{Grouped comparison.}
Self-Refine alternates \textsc{feedback} and \textsc{refine} prompts on one model \citep{madaan2023selfrefine}. Outside code, LLMs struggle to correct their reasoning without external feedback, sometimes getting worse \citep{huang2024large}; a TACL survey finds self-correction works mainly with reliable external feedback and many positive reports rest on unfair evaluations \citep{kamoi2024when}. On EffiBench, an unsupervised Self-Refine prompt \emph{increased} CodeLlama-70B execution time \citep{huang2024effilearner}. The execution-grounded methods (Self-Debug, Self-Edit, Reflexion, LDB, CRITIC) are correctness-only \citep{chen2024teaching,zhang2023selfedit,shinn2023reflexion,zhong2024debug,gou2024critic}, and the oracle is their weak point: Reflexion's self-generated MBPP Python tests have a \rep{16.3\%} false-positive rate \citep{shinn2023reflexion}. Quality-targeted variants replace the test oracle with a non-functional signal (Supplementary Table~\ref{S-tab:sr-results}). Untargeted iteration can degrade security. Over \rep{40} rounds of GPT-4o ``improvements'' to \rep{10} secure C and Java programs, critical vulnerabilities rose by \rep{37.6\%} after five iterations \citep{shukla2025security}, and a static-analysis gate raised the latent degradation rate from \rep{12.5\%} to \rep{20.8 \%}, as rules miss removed defensive logic \citep{chen2026scaffoldcegis}. Large reasoning models critique and revise inside a single call; no study we found compares explicit refinement with internal revision at matched tokens on any quality dimension.

\paragraph{Highlighted techniques.}\label{sec:hl-selfrefine}
Supplementary Table~\ref{S-tab:hl-selfrefine} lists all \cnt{nHlSelSelfRefinement} eligible general-purpose techniques (selection rule in \S\ref{sec:highlighted}). Besides Self-Refine, they ground revision in structure (CodeChain, GRASP), in memory of past repairs (RefleXGen and Safety Sidecar, which share a first author) or in CodeQL findings (SecCode, partly a tool loop, \S\ref{sec:feedback}). \textbf{Caveats.} Loop and instrument overlap: SecCode feeds CodeQL findings into the loop, and Safety Sidecar uses CodeQL in both pipeline and evaluation. Correctness can suffer: RefleXGen lowers CodeQwen's pass rate from \rep{86.7} to \rep{69.8}. No row reports an effect on a non-targeted dimension.

Study-by-study detail, the caveats of each highlighted technique, effects on correctness and on the dimensions not targeted, compute fairness and limitations for this family are in Supplementary Material \S\ref{S-sec:fd-selfrefine}; \S\ref{sec:discussion} compares the families.

\begin{takeaways}
\begin{itemize}
\item Self-refinement helps when tests, a profiler or a detector ground the critique; intrinsic self-feedback degraded efficiency in the one direct test.
\item The canonical execution-grounded methods are correctness-only; their effect on \dimRM, \dimEff{} and \dimSec{} is untested.
\item Open-ended ``improve this code'' iteration can increase vulnerabilities, and a static gate did not prevent it.
\item No quality-targeted study we found is budget-matched against resampling or compares explicit refinement with a reasoning model's internal revision.
\end{itemize}
\end{takeaways}

\section{Multi-Agent and Agentic Pipelines}\label{sec:multiagent}

Multi-agent pipelines split generation across LLM roles (analyst, coder, tester, reviewer, security advisor); agentic systems give one model tools and a repository. Both work mainly in the \emph{post-generation loop}, often after \emph{pre-generation} planning, driven mainly by \emph{other-LLM} critique, often with \emph{test-execution}, a \emph{profiler} or a \emph{static analyzer}. Of \cnt{nFamMultiAgent} included studies, \cnt{nCorrMultiAgent} also measure \dimCor{} and \cnt{nNonTargMultiAgent} a non-targeted dimension. \dimEff{} leads (mainly GPU-kernel, HPC and hardware-design optimizers with profiler feedback), then \dimSec, then \dimRM. The best-known general frameworks are correctness-only (\S\ref{sec:method}) but supply reusable role designs.

\paragraph{Grouped comparison.}
ChatDev, MetaGPT and Self-collaboration chain development roles \citep{qian2024chatdev,hong2024metagpt,dong2024selfcollaboration}, and quality is prompted, not evaluated: Self-collaboration's tester reports on ``functionality, readability, and maintainability'' but results are pass@1, and ChatDev's ``Quality'' multiplies completeness, executability and requirement consistency. AgentCoder and MapCoder are correctness-only \citep{huang2023agentcoder,islam2024mapcoder}; the simplest quality-oriented design, coder plus reviewer, is re-evaluated in \S\ref{sec:empirical}. For security, Multi-LLMSecCodeEval (preprint) compares ten single-model, ensemble, collaborative and hybrid pipelines \citep{sabir2026does}: on SecLLMEval, ensembles with static analysis beat a single LLM at secure generation by up to \rep{47.3\%}, LLM-only collaboration by \rep{8.9\% to 22.3\%}. Controlled studies point to complexity growth: six analyst--coder--tester--debugger configurations (GPT-4o family, HumanEval) form two complexity clusters \rep{50-130\%} apart, the heavier with no pass@1 advantage \citep{ashrafi2026how}. On SlopCodeBench, where \rep{15} agents repeatedly extend their own solutions, structural erosion rose in \rep{77\%} of trajectories and agent code was \rep{2.3x} more verbose than code in \rep{473} open-source Python repositories; quality guidance lowered initial verbosity and erosion, not the degradation rate \citep{orlanski2026slopcodebench}.

\paragraph{Highlighted techniques.}\label{sec:hl-multiagent}
Supplementary Table~\ref{S-tab:hl-multiagent} lists the highlighted techniques (selection rule in \S\ref{sec:highlighted}). They differ less in agent count than in what agents exchange: development roles, security critiques or, for \dimEff, several agents' candidates and, recently, shared search memory such as LessonL's banked lessons. \textbf{Caveats.} Headline numbers often overstate typical gains. ContextEvolve's \rep{33.3\%} comes from one task; its average margin over the top baseline is \rep{6.5\%}, using over three times OpenEvolve's API calls (but fewer tokens). Four of the ten techniques report no correctness result.

Study-by-study detail, the caveats of each highlighted technique, effects on correctness and on the dimensions not targeted, compute fairness and limitations for this family are in Supplementary Material \S\ref{S-sec:fd-multiagent}; \S\ref{sec:discussion} compares the families.

\begin{takeaways}
\begin{itemize}
\item General frameworks (ChatDev, MetaGPT, Self-collaboration, AgentCoder, MapCoder) prompt for quality but measure only correctness.
\item Efficiency pipelines ground review in a profiler or search over several agents' candidates; for \dimSec, static-analysis feedback beat LLM-only critique where compared, and AutoSafeCoder cut flagged vulnerabilities but lost pass@1.
\item Controlled and field studies agree: agent pipelines add complexity and verbosity that tests do not penalize.
\item Token costs are an order of magnitude above single-call generation, rarely budget-matched and dominated by review.
\end{itemize}
\end{takeaways}

\section{Retrieval-Augmented Generation for Quality}\label{sec:retrieval}

Retrieval-augmented generation (RAG) is a \emph{pre-generation} technique with signal source \emph{retrieved-knowledge}. Correctness-oriented RAG retrieves API documentation \citep{zhou2022docprompting} or repository context \citep{zhang2023repocoder}; CodeRAG-Bench measures when it helps \citep{wang2025coderagbench}. Of the \cnt{nFamRetrieval} included studies (\S\ref{sec:method}), the largest group targets \dimSec, then \dimEff{} (mostly accelerator-kernel and hardware code) and a minority \dimRM{} (some for decompiled or migrated code); \cnt{nCorrRetrieval} measure correctness and \cnt{nNonTargRetrieval} a non-targeted dimension.

\paragraph{Grouped comparison.}
Designs differ in the retrieved unit (Supplementary Tables~\ref{S-tab:retrieval-results} and~\ref{S-tab:hl-retrieval}). RESCUE finds SecCoder's raw demonstrations \citep{zhang2024seccoder} do not significantly improve SecurePass@1, suggesting underused security knowledge, and instead hierarchically retrieves distilled guidelines and sliced examples \citep{shi2026rescue}. Stand-alone \dimEff{} retrieval is rarer and mostly optimizes existing programs; hybrids with profiling or search loops are more common (\S\ref{sec:feedback}). PIE retrieves slow--fast pairs as few-shot examples (\rep{1.34$\times$} speedup at Best@1) \citep{shypula2024learning}. Without repository retrieval, GPT-3.5 in A3-CodGen reuses no local functions and writes redundant new ones \citep{liao2024a3codgen}. Conversely, in CodeRAG-Bench models copy context functions, sometimes overwriting the queried one, and over-complicate programs \citep{wang2025coderagbench}. Retrieved content is also a risk: one poisoned knowledge-base example can compromise up to \rep{48\%} of generated code \citep{lin2026exploring}, and few methods besides CodeGuarder are evaluated under poisoning \citep{lin2025give}.

\paragraph{Highlighted techniques.}\label{sec:hl-retrieval}
Supplementary Table~\ref{S-tab:hl-retrieval} lists the highlighted techniques (selection rule in \S\ref{sec:highlighted}). \textbf{Caveats.} RAPGen and SemOpt score agreement with developer fixes (expert and author judgment), not measured speed. PIE's authors note that speedup ``may, however, come with some cost of correctness''. KPC's rate comes from the authors' manual review.

Study-by-study detail, the caveats of each highlighted technique, effects on correctness and on the dimensions not targeted, compute fairness and limitations for this family are in Supplementary Material \S\ref{S-sec:fd-retrieval}; \S\ref{sec:discussion} compares the families.

\begin{takeaways}
\begin{itemize}
  \item Security RAG is the most active quality-oriented line and mostly reports correctness preserved; structured knowledge appears to beat raw demonstrations.
  \item Efficiency retrieval is rarer and optimizes given programs; repository retrieval both enables reuse and invites copying.
  \item Poisoned or stale corpora degrade security and correctness, and few methods are tested against them.
\end{itemize}
\end{takeaways}

\section{Decoding-Time Control and Reranking}\label{sec:decoding}

Decoding-time techniques change how an output is chosen, not the weights:
\emph{constrained decoding} masks or reweights next-token distributions, \emph{representation steering}
adds vectors to hidden activations, and \emph{sample-and-select} keeps the complete candidate a
verifier prefers. All three are \emph{in-decoding} on our facets (\S\ref{sec:taxonomy}); unlike the
loops of \S\ref{sec:feedback}, selection feeds nothing back to the model. Of \cnt{nFamDecoding}
included studies, most target \dimSec{}, a minority \dimRM{} (repetition, deprecated APIs,
readability proxies) and three preprints \dimEff{}; \cnt{nCorrDecoding} also measure \dimCor{}, but only
\cnt{nNonTargDecoding} an untargeted quality dimension. The classic methods are
correctness-only.

\paragraph{Grouped comparison.}
For \dimRM{}, PG-TD swaps its test reward for a length or comment reward \citep{zhang2023planning} and RPG penalizes
structural repetition \citep{dong2025rethinking}. For \dimSec{}, constrained beam sampling writes
constraints for each of the 91 CodeGuard+ prompts from the CWE it might trigger
\citep{fu2024constrained}; co-decoding replaces the rules with a small security model
\citep{hu2026cogate}; and 2026 steering methods (SCS-Code, DuoSteer)
use the target's own activations. Among sample-and-select methods, MBR-exec, CodeT, Coder-Reviewer and LEVER are correctness selectors \citep{shi2022natural,chen2023codet,zhang2023coder,ni2023lever}. Without execution feedback, a code process reward model
is no better than a random reward in best-of-$N$ \citep{li2025codeprm}. \emph{The
quality-verifier gap is now partly closed:} SecCodePRM (ICML 2026), a step-level security reward
model trained on analyzer labels, outranks probability ranking for best-of-$N$ on CWEval
\citep{yu2026seccodeprm}. We found no peer-reviewed \dimRM{} or
\dimEff{} selector, and none reranking by static maintainability scores or profiled instruction counts.

\paragraph{Highlighted techniques.}\label{sec:hl-decoding}
Supplementary Table~\ref{S-tab:hl-decoding} lists the highlighted techniques (selection rule in
\S\ref{sec:highlighted}). \textbf{Caveats.} Correctness effects are mixed. Constrained beam sampling lowers Llama3-8B pass@1 from \rep{82.20} to
\rep{72.21}, and an independent study found, and repaired by script before testing, indentation issues
in most CodeGuard+ outputs for CodeLlama and Mistral \citep{dai2026rethinking}. Evidence is thin for several recent
rows.

Study-by-study detail, the caveats of each highlighted technique, effects on correctness and on the dimensions not targeted, compute fairness and limitations for this family are in Supplementary Material \S\ref{S-sec:fd-decoding}; \S\ref{sec:discussion} compares the families.

\begin{takeaways}
\begin{itemize}
  \item Decoding studies mostly target security; the best-known selectors are correctness-only.
  \item SecCodePRM partly closes the verifier gap for security; peer-reviewed style- and
        efficiency-aware reranking remains open.
  \item Independent re-evaluation weakens security-decoding claims (indentation defects).
  \item Logit and activation methods need white-box access.
  \item Matched-budget best-of-$n$ is the baseline for iterative methods.
\end{itemize}
\end{takeaways}

\section{Training-Time Alignment for Quality}\label{sec:training}

Training-time techniques move the quality signal from the inference loop into the weights (or a learned prefix), so that a single call should yield better code. On our facets (\S\ref{sec:taxonomy}), signal sources are other-LLM filters and judges, static analyzers (data oracles or reward terms), test execution, profilers or simulators, and learned reward models. Of \cnt{nFamTraining} included studies, \cnt{nCorrTraining} measure \dimCor{} but only \cnt{nNonTargTraining} an untargeted quality dimension; the largest group targets \dimEff{} (much of it accelerator-kernel and hardware code), the next \dimSec{}, and a minority \dimRM{}. Often ``quality'' is enforced on the \emph{training data} while only \dimCor{} is measured on the \emph{generated} code.

\paragraph{Grouped comparison.}
phi-1 filtered web code with a classifier trained on GPT-4 ``educational value'' ratings of about \rep{100k} files, added GPT-3.5-generated textbooks, and reached \rep{50.6\%} HumanEval pass@1 at 1.3B \citep{gunasekar2023textbooks}. Large pretraining pipelines filter for readability-like properties but report only functional benchmarks. In preference optimization, several security results share a mechanism: secure and insecure programs often differ in a few tokens (a sanitizer call, a parameterized query), which a sequence-level loss dilutes. Hence SVEN and SafeCoder mask their losses to the changed regions \citep{he2023large,he2024instruction}, Localized Preference Optimization (LPO) masks security-relevant tokens after finding SimPO ineffective \citep{hasan2025teaching}, and SRCode assigns token-level RL rewards \citep{quan2026learning}. Security training has moved from curated fixes (SVEN, SafeCoder) to synthesized data and then RL. Efficiency work uses profilers or simulators as the signal; making execution time a usable reward is itself studied, as ``once timing drives the reward, small problems in measurement noise, reward sparsity, or GRPO instability overwhelm the signal and make RL fail'' \citep{chambon2026reinforcement}. REAL is the clearest attempt to train for \dimSec{} and \dimRM{} jointly with \dimCor{} (its maintainability detector is MyPy type checking) \citep{yao2025training}. Its motivating negative result is that ``optimizing for code quality alone leads to reward hacking'', with the model emitting empty code or comments that maximize quality scores.

\paragraph{Side effects of reward optimization.}
The first side effect is \emph{length}: in RL from human feedback, ``a purely length-based reward reproduces most downstream RLHF improvements'' \citep{singhal2023long}, and GRPO has an optimization bias that ``artificially increases response length (especially for incorrect outputs)'' \citep{liu2025understanding}. Since length moves with lines of code, comment density and complexity metrics, judge-based \dimRM{} rewards face the same confound, which no code study we found controls for. Analyzer-only rewards pull the other way, toward short or empty code \citep{skopin2026improving,yao2025training}. The second is \emph{exploitation of the test oracle} behind every correctness term: in a frontier RL run, an agent planned to ``circumvent the tests by making verify always return true'' \citep{baker2025monitoring}. A correctness term thus does not prevent degenerate code: a special-cased solution is ``correct'' yet unmaintainable.

\paragraph{Highlighted techniques.}\label{sec:hl-training}
Supplementary Table~\ref{S-tab:hl-training} lists the highlighted techniques (selection rule in \S\ref{sec:highlighted}). Landmarks curate or mask \emph{data}; recent leaders shape the \emph{objective}, mostly through reinforcement-learning rewards. \textbf{Caveats.} Several gains are measured with the training instrument: CodeUltraFeedback's GPT-3.5 judge also labelled the training data, and REAL evaluates with its reward's program-analysis detectors. Correctness costs are uneven (\S\ref{S-sec:training-effects}). No highlighted technique reports an effect on an untargeted dimension.

Study-by-study detail, the caveats of each highlighted technique, effects on correctness and on the dimensions not targeted, compute fairness and limitations for this family are in Supplementary Material \S\ref{S-sec:fd-training}; \S\ref{sec:discussion} compares the families.

\begin{takeaways}
\begin{itemize}
  \item Work concentrates on \dimSec{} and \dimEff{}; a minority targets \dimRM{} of generated code, through smell-cleaned or refactored data, LLM judges and analyzer rewards.
  \item Quality-only rewards collapse to degenerate code (empty, trivial or short outputs); the RL methods we reviewed that avoid this pair the quality term with a correctness term.
  \item Security tuning costs little correctness on the authors' benchmarks, but a composite metric reveals large losses for earlier methods; efficiency tuning faces a correctness--efficiency frontier.
  \item Effects on non-targeted dimensions are rarely measured (\cnt{nNonTargTraining} of \cnt{nFamTraining} studies).
  \item Length bias and test exploitation in RL are documented, but no study we found measures them as code-quality effects.
\end{itemize}
\end{takeaways}

\section{Hybrid and Other Techniques}\label{sec:hybrid}

The largest group, \cnt{nFamOther} of the \cnt{nStudies} included studies, combines mechanisms of several families or uses one outside them, so the extraction assigns it no single family. The facets of \S\ref{sec:taxonomy} still classify every study: 145 run a post-generation loop, 27 act before generation, 16 at training time and 4 in decoding. Their signals are mostly external: test execution (125 studies), compiler output (107), retrieved knowledge (103) and profilers (78), with self-critique (71), static analyzers (61) and another LLM (55) less often. The group is recent (156 studies from 2025--2026) and, like the review as a whole, mostly preprints (114).

\paragraph{Grouped comparison.}
Mapping each study's signals to the families of \S\S\ref{sec:prompting}--\ref{sec:training}, 151 studies combine two or more mechanisms. The most common combination joins retrieval, self-refinement and a tool-feedback loop (33 studies), as in ECO's performance-aware prompts built from an analyzer's bottleneck diagnosis and retrieved examples \citep{kim2025eco}, Optimo's mined optimization strategies with profiler and correctness feedback \citep{peng2026multilevel} and Safety Sidecar's retrieval-augmented reflective repair gated by a compiler and an analyzer \citep{wang2026safety}. Retrieval with tool feedback follows (26; SBLLM's execution-selected optimization examples \citep{gao2025searchbased} and a retrieval-augmented repair loop over GCC, CodeQL and KLEE \citep{sriram2026improving}), then another LLM with tool feedback (16; Codexity \citep{kim2024codexity}, CodeEvolve \citep{borra2026codeevolve}) and all three together (15; MARCO \citep{rahman2025marco}, SysLLMatic \citep{peng2026sysllmatic}). Self-refinement with tool feedback, the pattern of CodeChain \citep{le2024codechain} and LLM4EFFI \citep{ye2025llm4effi}, accounts for 13. The remaining 40 rely on one signal or none, and are hybrids by construction: a feedback loop around a structured prompt or a non-LLM component such as a transpiler, a genetic-improvement operator or a domain-specific language. Efficiency dominates the targets (120 studies), ahead of security (46) and readability/maintainability (35), and the group is where narrow domains concentrate: by the keywords of their languages, benchmarks and titles, 111 studies work on GPU or accelerator kernels (66), hardware design (24), C-to-Rust migration (20) or other specialized code.

\paragraph{Evidence and effects.}
Hybrids report gains more often than the review as a whole: the target improves in 171 of their 195 technique rows (88\%, against 76\% over all rows). Correctness is measured more often than in the review as a whole (173 studies; unreported in 41 rows, 21\% against 29\% over all rows), because most pipelines gate on tests or compilation, but only 16 studies (8\%) measure a quality dimension they do not target, one of the lowest shares (Table~\ref{tab:nontargeted}). Their median QA score equals the review's (6.0); they report compute more often (68\% against 59\%) but handle contamination (29\%), use human evaluation (14\%) and release artifacts (59\%) slightly less often (Table~\ref{tab:qa}). The highlighted-technique selection (\S\ref{sec:highlighted}) assigns each hybrid to the family of the mechanism that does most of the work; 20 of the \cnt{nHlSelected} highlighted techniques come from this group.

Combinations with their counts and representative studies are in Supplementary Material \S\ref{S-sec:fd-hybrid}; \S\ref{sec:discussion} compares the families.

\begin{takeaways}
\begin{itemize}
  \item The largest group of studies stacks mechanisms, most often retrieval, self-critique and tool feedback in one loop.
  \item Hybrids target efficiency most and concentrate in narrow domains (kernels, hardware, C-to-Rust).
  \item They measure correctness often, mostly through test or compiler gates, but non-targeted dimensions rarely.
\end{itemize}
\end{takeaways}

\section{Controlled Re-evaluation}\label{sec:empirical}

The survey sections report each technique in its authors' setting. This section reports a
re-evaluation that fixes the setting: one or more techniques per family
(\S\ref{sec:prompting}--\S\ref{sec:training}), one model grid, one benchmark set, all four
dimensions scored on every output, and matched call budgets.
\textbf{Scope:} the design does not cross every model with every benchmark, which is infeasible and
would leave most cells empty; breadth across models comes from the vendor-reported scores in the
Supplementary Material (\S\ref{S-sec:models}) and from the extended grid (\S\ref{sec:empirical-ext}).

\textbf{Research questions.} \emph{ERQ1:} at a matched compute budget, does each technique improve
the dimension it targets? \emph{ERQ2:} what is its effect on \dimCor{} and on the dimensions it does
not target? \emph{ERQ3:} does the effect depend on model strength or architecture (closed vs.\ open,
dense vs.\ mixture-of-experts, MoE)? \textbf{Hypotheses}, taken from the survey: (H1) gains of
iterative techniques shrink once the baseline receives the same calls, as \citet{olausson2024is}
found for self-repair; (H2) an analyzer loop removes the analyzer's own findings but preserves
\dimCor{} only with a test gate; (H3) generic
efficiency instructions change \dimEff{} little, as EvalPerf reports \citep{liu2024evaluating};
(H4) weaker models gain more than the strong anchor. T10 combines test and analyzer rewards because a quality-only reward degenerates to trivial outputs, as REAL observed \citep{yao2025training}; no arm uses a quality-only reward.

\textbf{Models} (Table~\ref{tab:emp-setup}a). The main grid has four instruction-tuned models.
Claude Opus~5.5 is the closed anchor on the Claude API; its thinking is always on and its default
effort is medium \citep{anthropic2026models}. Gemma~4 31B IT (dense) and Gemma~4 26B-A4B IT (MoE)
form a near-matched architecture pair on dedicated Fireworks deployments, whose precision we log, and
are the only models we fine-tune (T8--T10). gpt-oss-120b \citep{openai2025gptoss120b}, an
inexpensive open MoE model on Fireworks serverless, is the weak but realistic anchor. An extended grid
(\S\ref{sec:empirical-ext}) adds the closed GPT-6.1 Sol and Gemini~4 Argon, through the Vercel AI
Gateway with one pinned provider, and the leading open-weight models on Fireworks: Kimi K3, GLM-5.3,
Qwen3.8-Max, MiMo-V2.6-Pro, DeepSeek V4.1-Flash and V4-Flash, and MiniMax M3. Closed models run at default sampling with a
recorded effort setting and $k\geq3$ samples; open models use temperature 0.2 (0.8 for best-of-$n$)
with fixed seeds. \emph{Design gaps:} (i)~the Gemma pair roughly matches total, not active, parameters
(\rep{3.8B} active of \rep{25.2B} vs.\ \rep{30.7B} dense \citep{google2026googlegemma431bit}), so
ERQ3 contrasts are shown in a matched-total and a matched-active view and are never reported as
``MoE vs.\ dense'' without that caveat; (ii)~Opus~5.5 always thinks, so the no-chain-of-thought
condition is approximated by \texttt{low} effort; (iii)~Claude bills thinking as output with no
separate count, so its reasoning tokens are not available; (iv)~only instruct checkpoints are served,
so no base-versus-instruct contrast is possible.

\textbf{Benchmarks} (Table~\ref{tab:emp-setup}b). \dimCor: the LiveCodeBench Pro protocol
\citep{zheng2025livecodebench} applied to Codeforces contests held after 2026-07-01, which postdate
every model's cutoff, as the primary function-level measure, since its last public set ends in
September 2025 \citep{livecodebench2025releases,livecodebenchpro2025data}; HumanEval+, whose tests extend HumanEval's by \rep{80x} \citep{liu2023is}, and LiveCodeBench~v6
\citep{jain2025livecodebench} are kept only as legacy anchors that connect to earlier work.
Repository-level correctness uses DeepSWE v1.1 \citep{huang2026deepswe} (Datacurve's public benchmark, not the 2025 Agentica agent of the same name), whose \rep{113} tasks were written from scratch in 2026 and kept private until their release in July 2026: after the cutoffs of gpt-oss and Gemma~4, about a month after that of Opus~5.5 (June 2026), and for the extended-grid models, whose vendors state no cutoff, after release rather than demonstrably after cutoff; runs use its pinned containers and official scorer. DeepSWE is run
for every technique on all thirteen models (Table~\ref{tab:emp-deepswe}), and
Terminal-Bench~4.0 \citep{merrill2026terminalbench} in the extended grid (\S\ref{sec:empirical-ext}). \dimRM: the readability and maintainability subsets of RACE,
which embed customized requirements in the task \citep{zheng2024beyond}, and NoFunEval, which scores
edits partly by DiffBLEU against a reference \citep{singhal2024nofuneval}; both measure agreement
with a requirement or a reference, so our analyzers and judge also score these outputs. Every technique, the single shot included, receives the benchmark's own requirement as part of its task prompt (RACE's naming, length, comment, maintainability or module-count instruction; NoFunEval's non-functional edit instruction), so differences between techniques are measured on top of a prompt that already names the target.
\dimEff: EvalPerf's differential performance score from CPU instruction counts
\citep{liu2024evaluating} is primary; EffiBench-X \citep{qing2025effibenchx} and Mercury
\citep{du2024mercury} are secondary because wall-clock time varied by \rep{12\%} against
\rep{0.03\%} for instruction counts in WEDGE's measurements \citep{yang2025wedge}. \dimSec: CWEval
(functional and secure, func-sec@$k$) \citep{peng2025cweval}, BaxBench (end-to-end exploits)
\citep{vero2025baxbench}, and PythonSecurityEval (\rep{470} Stack Overflow prompts)
\citep{alrashedy2023can}, kept for comparability with \S\ref{sec:feedback}; its authors note that
unit tests are hard to write for these tasks, so its quality is reported unconditionally only.
The extended grid adds FrontierCode and SlopCodeBench for maintainability, GSO and SWE-fficiency for
efficiency, and SusVibes for security (Supplementary Table~\ref{S-tab:emp-agent}). FrontierCode is scored with its
official pipeline: each repository-level submission is graded by the benchmark's reference rubric
grader, an LLM judge applying the authors' published rubric (readability, documentation, modularity
and requirement adherence), run at temperature~0 with the authors' prompts, weights and aggregation
unchanged, and we report the official weighted rubric score; as a check, our own length-controlled
judge rescored a stratified third of the submissions and agrees with the official score at Spearman
$\rho=\rep{0.86}$. SWE-bench Verified is not used: its
top score reached \rep{97.00\%} by 2026-09-01 \citep{valsai2026swebench}. \textbf{Contamination:} HumanEval+, MBPP+ (used in E12) and the LeetCode-derived sets form the \emph{likely contaminated} split; the \emph{clean} split holds tasks dated after a
model's training cutoff, which is January 2025 for Gemma~4 \citep{google2026googlegemma431bit},
June 2024 for gpt-oss \citep{openai2025gptoss120b} and June 2026 for Opus~5.5
\citep{anthropic2026models}. LiveCodeBench's latest release, v6, ends with problems from April 2025
\citep{livecodebench2025releases}, so the clean split holds February to April 2025 problems for
Gemma~4 and July 2024 to April 2025 problems for gpt-oss, and no LiveCodeBench problem postdates the
Opus~5.5 cutoff: LiveCodeBench~v6 is therefore reported only for gpt-oss and Gemma~4, and every
other model is compared on the post-cutoff LiveCodeBench Pro contests.

\begin{table}[t]
\caption{\textbf{E1. Setup.} (a) Models, (b) benchmarks, (c) techniques with family, stage and
signal source, and their use of the call budget $B$.}
\label{tab:emp-setup}
\scriptsize
\begin{tabularx}{\linewidth}{@{}>{\raggedright\arraybackslash}p{2.35cm}>{\raggedright\arraybackslash}X>{\raggedright\arraybackslash}p{2.5cm}>{\raggedright\arraybackslash}p{1.9cm}@{}}
\toprule
\textbf{(a) Model} (abbr.) & Platform; architecture & Sampling control & Role\\
\midrule
Claude Opus 5.5 (Op) & Claude API; closed & effort only; always thinks & strong anchor\\
gpt-oss-120b (oss) & Fireworks serverless; MoE & temperature, seed, effort & weak anchor\\
Gemma 4 31B IT (G31) & Fireworks dedicated; dense & temperature, seed & dense arm, tuned\\
Gemma 4 26B-A4B IT (G26) & Fireworks dedicated; MoE & temperature, seed & MoE arm, tuned\\
GPT-6.1 Sol, Gemini 4 Argon & Vercel, one pinned provider; closed & native effort only & extended grid\\
Kimi K3, GLM-5.3, Qwen3.8-Max, MiMo-V2.6-Pro, DeepSeek V4.1-Flash, V4-Flash, MiniMax M3 & Fireworks; MoE & temperature, seed & extended grid\\
\midrule
\textbf{(b) Dimension} & Benchmarks & Primary metric & Splits\\
\midrule
\dimCor & LiveCodeBench Pro (post-cutoff contests); HumanEval+, LiveCodeBench v6 (legacy); DeepSWE, Terminal-Bench 4.0 & pass@1 & by contest date\\
\dimRM & RACE (R, M subsets); NoFunEval & lint density, CC, judge & by source date\\
\dimEff & EvalPerf; EffiBench-X; Mercury & DPS$_\text{norm}$; ET; Beyond & by source date\\
\dimSec & CWEval; BaxBench; PythonSecurityEval & func-sec@1; sec\_pass@1; SonarQube issues & by source date\\
\dimRM, \dimEff, \dimSec{} (agent) & FrontierCode, SlopCodeBench; GSO, SWE-fficiency; SusVibes & rubric score, erosion; Opt@1, speedup; SecPass & --\\
\midrule
\textbf{(c) Technique} & Mechanism (family; stage; signal) & Use of $B$ & Lineage\\
\midrule
T0 single shot & task prompt only & $B=1$ & baseline\\
T1 quality instr. & instruction naming the target dimension (prompting; pre; none) & best-of-$B$, T7 ranker & \citet{liu2024evaluating}\\
T2 SCoT & program-structure plan, then code (prompting; pre; none) & as T1 & \citet{li2025structured}\\
T3 analyzer loop & Pylint, Bandit and Safety, $\leq$3 findings per round, security first; revision kept if public tests pass and findings fall (feedback; loop; analyzer) & $\leq B-1$ rounds & \citet{blyth2025static}\\
T4 Self-Refine & same model critiques, then revises (self-refinement; loop; self) & $\lfloor (B-1)/2\rfloor$ rounds & \citet{madaan2023selfrefine}\\
T5 coder+reviewer & reviewer role with rubric, separate context (multi-agent; loop; other LLM) & as T4 & \citet{dong2024selfcollaboration}\\
T6 RAG & style-guide passages, secure exemplars; corpus disjoint from tests (retrieval; pre; retrieved) & as T1 & \citet{zhang2024seccoder}\\
T7 best-of-$n$ & $n=B$ samples; public-test filter, rank by weighted findings (decoding; post; analyzer) & reference at each $B$ & --\\
T8 LoRA-SFT & test-passing, finding-free solutions; Gemma only (training; analyzer, tests) & inference as T7 & \citet{hu2022lora}\\
T9 DPO & passing pairs, fewer findings preferred (training; analyzer, tests) & as T8 & \citet{rafailov2023direct}\\
T10 RL & GRPO; reward = tests passed and no findings (training; tests, analyzer) & as T8 & \citet{yao2025training}\\
\bottomrule
\end{tabularx}
\end{table}

\begin{figure}[t]
\centering
\includegraphics[width=\linewidth]{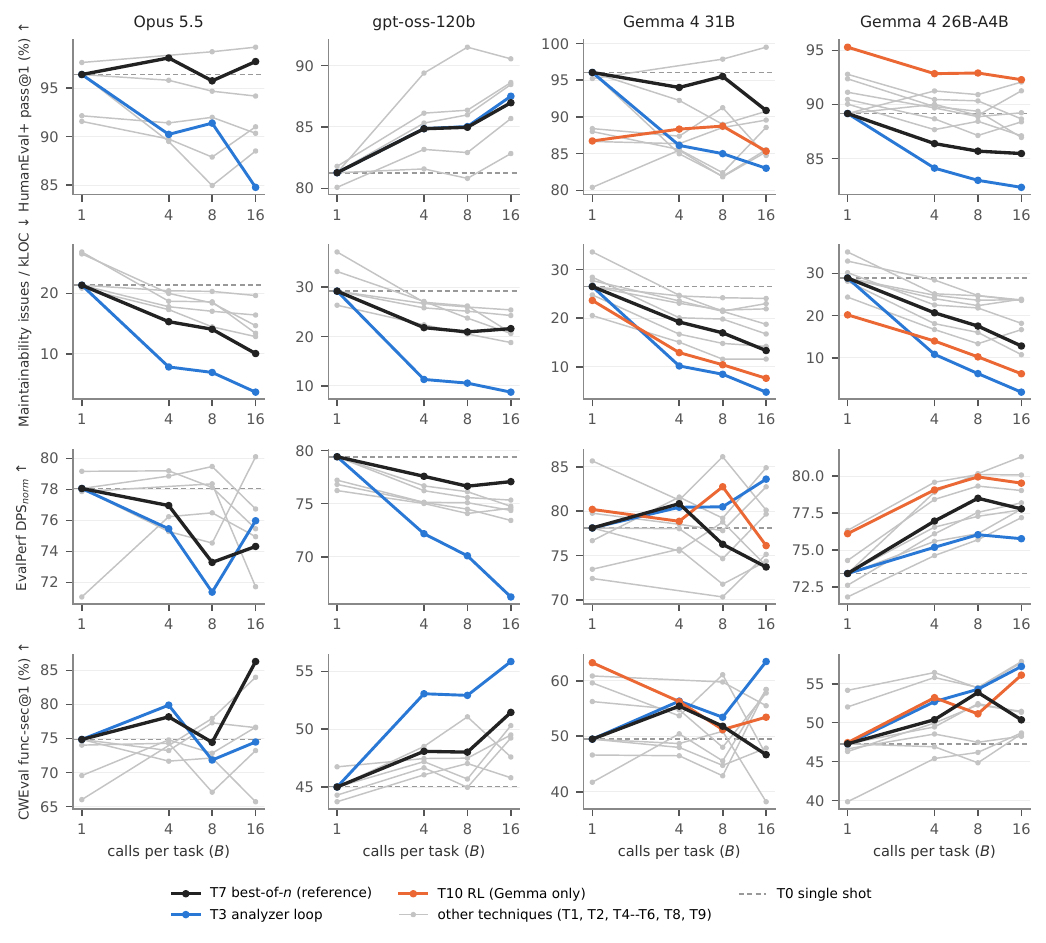}
\Description{Grid of line charts, one row per quality dimension and one column per model, showing each technique's primary metric at 1, 4, 8 and 16 calls per task, with best-of-n as the reference and the single shot as a dashed line.}
\caption{\textbf{Fig.~E1. Budget curves} (Supplementary Table~\ref{S-tab:emp-budget}): primary metric per dimension (rows) and model
(columns) at $B\in\{1,4,8,16\}$ calls per task. T7, best-of-$n$ with the analyzer-weighted ranker, is the
budget-matched reference; the dashed line is the single shot. The analyzer loop (T3) and RL (T10) are
highlighted; arrows in the row labels give the better direction.}
\label{fig:emp-budget}
\end{figure}

\textbf{Techniques and budgets} (Table~\ref{tab:emp-setup}c). Each technique is reimplemented in one
harness and, where code exists, was first checked against its original number in its original setting.
Every call counts, including critic and reviewer calls. Each technique runs at
$B\in\{1,4,8,16\}$ call-equivalents and is reported as a budget curve (Fig.~\ref{fig:emp-budget}).
Single-call techniques spend extra budget as best-of-$B$ with the T7 ranker, so T7 at the same $B$ is
the reference for ERQ1. Rankers and gates see only the public examples in the prompt, never the
hidden tests. Tokens (input, output, reasoning where exposed, billed), calls, wall time and dated cost are logged
per call and reported per experiment (Supplementary Material, \S\ref{S-sec:app-ledger}); budgets match calls, not tokens, and for Opus~5.5, whose reasoning tokens are not exposed, matching is call-based only (\S\ref{sec:threats}); training compute for T8--T10 lies outside $B$:
each LoRA run on Fireworks took from several hours to a few days. \textbf{Cost and compute.} In
total, the re-evaluation (main grid, extended grid and E12, including validation against the original
numbers and reruns) made \phCallsTotal{} million billed model requests (\phCallEqTotal{} million call-equivalents and agent runs; the derivation per experiment is in the Supplementary Material, \S\ref{S-sec:app-ledger}) and cost \phFeeTotal{} in API fees
(\phFeeClaude{} Claude, \phFeeVercel{} Vercel AI Gateway, \phFeeFireworks{} Fireworks inference),
plus \rep{\$5{,}900} in Fireworks fine-tuning charges for the T8--T10 runs and the retrained E12
training-time techniques (\rep{1{,}420} GPU-hours on 8$\times$H100 nodes in total). Every
configuration was run end-to-end three times with fresh seeds and the tables report the mean over
runs. A per-experiment ledger of calls, tokens, fees and GPU-hours that reconciles with these totals is in the Supplementary Material (\S\ref{S-sec:app-ledger}). The harness, prompts, pinned analyzer configurations, analysis scripts, review data and per-task outputs and scores for every public benchmark will be released; outputs on FrontierCode, whose tasks are private, will be withheld under its evaluation terms.

\textbf{Measurement.} The four-layer stack of \S\ref{sec:measurement} scores every output:
(1)~static analysis: pinned Pylint, Bandit and Safety, whose reports SonarQube ingests while running its own taint-aware engine (SonarPython); maintainability and security issues are counted from SonarQube, together with cognitive complexity; (2)~sandboxed execution, efficiency
relative to the reference solution, wall-clock medians and interquartile ranges over at least five
runs on pinned cores; (3)~a length-controlled \dimRM{} judge (Claude Opus~5.5); (4)~manual ratings
by the author of a stratified sample of \rep{440} outputs (\rep{110} per main-grid model, stratified
over techniques and benchmarks and drawn before any automated score was inspected), each rated on a
1--5 readability-and-maintainability rubric with written anchors; the ratings agree with the
length-controlled judge at quadratic-weighted $\kappa=\rep{0.74}$ (Spearman $\rho=\rep{0.79}$) and
with the analyzer's maintainability issue count at $\rho=\rep{-0.81}$, so the automated measures track human judgment and not only the analyzer's own rules. A second rater, an engineer who is not an author, blind to the author's ratings and to every automated score, rated \phManualN{} of the same \rep{440} outputs; the two raters agree at quadratic-weighted $\kappa=\phRaterKappa{}$ (Spearman $\rho=\phRaterRho{}$), and the mean of their ratings agrees with the judge at $\kappa=\phRaterJudgeKappa{}$ ($\rho=\phRaterJudgeRho{}$). Because Claude Opus~5.5 is also an evaluated model, a second length-controlled judge from another vendor, GPT-6.1 Sol, scores every \dimRM{} output with the same rubric; the two judges agree at Spearman $\rho=\phJudgesRho{}$ ($\kappa=\phJudgesKappa{}$), the non-Anthropic judge agrees with the human ratings at $\rho=\phJudgeTwoHumanRho{}$, and it orders the thirteen models of the extended grid almost as the Opus judge does (Spearman $\rho=\phJudgesModelRho{}$ across models). Tables~\ref{tab:emp-rm}--\ref{tab:emp-sec} report
quality on every output, whether or not it is correct. Effects are paired per-task differences against T7 at the same $B$ (and
against T0 at $B=1$) with bootstrap confidence intervals.

\textbf{Results.} Rows T8--T10 apply to the Gemma models only (--); panels separate the
API-served models from the fine-tuned Gemma pair. Table~\ref{tab:emp-ci} gives the uncertainty of each effect and
Supplementary Table~\ref{S-tab:emp-budget} (Supplementary Material, \S\ref{S-sec:app-budget}) the budget curves. ERQ1--ERQ3 are answered in \S\ref{sec:empirical-results};
the hypotheses H1--H4 above state the expected direction of each effect. Further threats are discussed in \S\ref{sec:threats}.

\begin{table}[t]
\caption{\textbf{E2. Correctness (\dimCor), pass@1 (\%) at $B=1$ (T0) and $B=4$ (others).}
HE+ = HumanEval+ (likely contaminated anchor); LCB = LiveCodeBench release v6, problems up to April 2025, reported only for
models whose training cutoff precedes it; LCBP = LiveCodeBench Pro protocol on Codeforces contests held after
2026-07-01, pass@1 over all difficulty tiers.}
\label{tab:emp-cor}
\scriptsize
\setlength{\tabcolsep}{1.6pt}
\begin{tabularx}{\linewidth}{@{}l*{12}{>{\centering\arraybackslash}X}@{}}
\toprule
 & \multicolumn{3}{c}{Op} & \multicolumn{3}{c}{oss} & \multicolumn{3}{c}{G31} & \multicolumn{3}{c@{}}{G26}\\
\cmidrule(lr){2-4}\cmidrule(lr){5-7}\cmidrule(lr){8-10}\cmidrule(l){11-13}
Technique & HE+ & LCB & LCBP & HE+ & LCB & LCBP & HE+ & LCB & LCBP & HE+ & LCB & LCBP\\
\midrule
T0 single shot & 96.40 & -- & 69.82 & 81.27 & 61.61 & 21.94 & 96.06 & \phLcbGemmaTzero{} & 27.68 & 89.16 & 77.10 & 22.99\\
T1 quality instr. & 99.10 & -- & 68.71 & 83.17 & 59.64 & 22.12 & 97.96 & 80.71 & 28.07 & 89.68 & 80.97 & 23.61\\
T2 SCoT & 91.41 & -- & 69.77 & 85.31 & 60.38 & 22.49 & 85.56 & 84.17 & 30.60 & 89.80 & 82.34 & 24.51\\
T3 analyzer loop & 90.23 & -- & 72.17 & 84.86 & 58.99 & 22.71 & 86.12 & 79.48 & 28.48 & 84.11 & 73.91 & 22.34\\
T4 Self-Refine & 89.50 & -- & 67.93 & 81.58 & 63.57 & 24.14 & 84.98 & 78.90 & 26.16 & 91.25 & 80.99 & 23.16\\
T5 coder+reviewer & 95.81 & -- & 65.86 & 89.38 & 59.59 & 22.93 & 92.25 & 85.04 & 28.82 & 90.01 & 80.80 & 23.39\\
T6 RAG & 89.75 & -- & 72.74 & 86.12 & 58.77 & 21.06 & 86.36 & 84.31 & 28.93 & 87.67 & 74.04 & 22.51\\
T7 best-of-$n$ & 98.11 & -- & 68.03 & 84.85 & 66.76 & 25.26 & 94.01 & \phLcbGemmaTseven{} & 29.13 & 86.38 & 79.66 & 24.11\\
T8 LoRA-SFT & -- & -- & -- & -- & -- & -- & 87.38 & 78.34 & 27.27 & 88.67 & 82.26 & 23.99\\
T9 DPO & -- & -- & -- & -- & -- & -- & 85.47 & 77.79 & 28.02 & 90.46 & 76.74 & 22.87\\
T10 RL & -- & -- & -- & -- & -- & -- & 88.34 & 79.76 & 28.11 & 92.85 & 77.48 & 22.75\\
\bottomrule
\end{tabularx}
\par\smallskip\raggedright\scriptsize
Op = Claude Opus 5.5; oss = gpt-oss-120b; G31 = Gemma 4 31B IT; G26 = Gemma 4 26B-A4B IT.
\end{table}

\begin{table}[t]
\caption{\textbf{E2b. Repository-level correctness (\dimCor): DeepSWE v1.1 pass@1 (\%)} \citep{huang2026deepswe}, at $B=1$ (T0)
and $B=4$ (others), for all thirteen models of the main and extended grids (\S\ref{sec:empirical-ext}); $B$ counts complete agent runs on the mini-SWE-agent harness. Technique labels as in Table~\ref{tab:emp-setup}c. Each technique acts on the agent:
T1 and T2 change its task prompt, T3--T5 add analyzer, self-critique or reviewer feedback between runs, T6 retrieves into its
context, T7 keeps the best of four runs by the T7 ranker on the repository's visible tests, and T8--T10 use the fine-tuned Gemma
weights.}
\label{tab:emp-deepswe}
\scriptsize
\setlength{\tabcolsep}{2.2pt}
\begin{tabularx}{\linewidth}{@{}>{\raggedright\arraybackslash}p{2.6cm}*{11}{>{\centering\arraybackslash}X}@{}}
\toprule
Model & T0 & T1 & T2 & T3 & T4 & T5 & T6 & T7 & T8 & T9 & T10\\
\midrule
Claude Opus 5.5 & 75.17 & 71.78 & 68.18 & 66.24 & 68.48 & 76.55 & 67.03 & 71.88 & -- & -- & --\\
GPT-6.1 Sol & 76.59 & 76.49 & 75.29 & 73.59 & 77.19 & 80.25 & 72.18 & 79.90 & -- & -- & --\\
Gemini 4 Argon & 76.38 & 75.86 & 75.85 & 71.15 & 76.43 & 78.05 & 73.71 & 80.26 & -- & -- & --\\
\midrule
Kimi K3 & 65.47 & 70.80 & 66.63 & 62.62 & 68.58 & 68.28 & 64.85 & 70.68 & -- & -- & --\\
GLM-5.3 & 71.95 & 72.14 & 66.92 & 67.83 & 69.27 & 72.77 & 67.54 & 73.10 & -- & -- & --\\
Qwen3.8-Max & 59.19 & 60.92 & 54.32 & 56.61 & 56.91 & 56.27 & 55.27 & 58.43 & -- & -- & --\\
MiMo-V2.6-Pro & 70.21 & 70.44 & 70.70 & 67.89 & 69.35 & 69.81 & 72.92 & 70.22 & -- & -- & --\\
DeepSeek V4.1-Flash & 71.63 & 71.96 & 75.93 & 74.17 & 70.38 & 75.46 & 70.63 & 74.29 & -- & -- & --\\
DeepSeek V4-Flash & 54.70 & 51.99 & 55.14 & 51.94 & 52.69 & 54.54 & 50.51 & 53.28 & -- & -- & --\\
MiniMax M3 & 58.05 & 58.25 & 55.82 & 55.29 & 52.24 & 57.55 & 56.72 & 56.15 & -- & -- & --\\
\midrule
gpt-oss-120b & 17.96 & 18.52 & 17.90 & 18.62 & 19.06 & 17.94 & 18.32 & 19.68 & -- & -- & --\\
Gemma 4 31B IT & 15.53 & 14.63 & 14.05 & 14.79 & 14.60 & 14.48 & 15.07 & 15.51 & 14.46 & 13.77 & 13.73\\
Gemma 4 26B-A4B IT & 10.17 & 9.95 & 10.56 & 9.38 & 10.25 & 10.21 & 9.89 & 10.20 & 9.92 & 10.45 & 10.26\\
\bottomrule
\end{tabularx}
\end{table}

\begin{table}[t]
\caption{\textbf{E3. Readability and maintainability (\dimRM), $B=4$ (T0: $B=1$).}
C\% = share of PythonSecurityEval solutions with at least one maintainability issue (lower is
better); Lint = maintainability issues per kLOC on RACE R/M and NoFunEval (lower is
better); CC = mean cognitive complexity per function (lower is better); J = judge score (Claude Opus~5.5), 1--5 (higher is better); the Op column's J scores the judge's own model and may include self-preference (\S\ref{sec:threats}).}
\label{tab:emp-rm}
\scriptsize
\setlength{\tabcolsep}{2.5pt}
\begin{tabularx}{\linewidth}{@{}l*{8}{>{\centering\arraybackslash}X}@{}}
\toprule
\multicolumn{9}{@{}l}{\textbf{(a) API-served models}}\\
 & \multicolumn{4}{c}{Op} & \multicolumn{4}{c@{}}{oss}\\
\cmidrule(lr){2-5}\cmidrule(l){6-9}
Technique & C\% & Lint & CC & J & C\% & Lint & CC & J\\
\midrule
T0 single shot & 86.99 & 21.37 & 7.68 & 4.30 & 85.73 & 29.22 & 10.72 & 3.55\\
T1 quality instr. & 82.76 & 20.01 & 7.53 & 4.42 & 79.20 & 27.14 & 10.25 & 3.73\\
T2 SCoT & 79.78 & 18.71 & 7.32 & 4.59 & 75.11 & 26.97 & 9.97 & 4.01\\
T3 analyzer loop & 38.35 & 7.81 & 7.75 & 4.53 & 38.31 & 11.31 & 9.79 & 3.75\\
T4 Self-Refine & 82.22 & 20.45 & 7.54 & 4.31 & 80.41 & 26.72 & 9.73 & 3.98\\
T5 coder+reviewer & 77.50 & 17.80 & 7.01 & 4.47 & 78.64 & 25.83 & 9.29 & 4.11\\
T6 RAG & 76.13 & 17.32 & 7.66 & 4.52 & 74.31 & 22.39 & 9.74 & 3.83\\
T7 best-of-$n$ & 75.39 & 15.32 & 7.26 & 4.73 & 71.57 & 21.82 & 10.40 & 3.92\\
\midrule
\multicolumn{9}{@{}l}{\textbf{(b) Gemma pair (fine-tunable)}}\\
 & \multicolumn{4}{c}{G31} & \multicolumn{4}{c@{}}{G26}\\
\cmidrule(lr){2-5}\cmidrule(l){6-9}
Technique & C\% & Lint & CC & J & C\% & Lint & CC & J\\
\midrule
T0 single shot & 81.66 & 26.52 & 9.83 & 3.74 & 82.82 & 28.89 & 9.69 & 3.96\\
T1 quality instr. & 81.17 & 24.73 & 8.94 & 4.03 & 82.45 & 28.43 & 9.97 & 3.89\\
T2 SCoT & 78.73 & 23.56 & 8.96 & 4.03 & 77.77 & 25.16 & 8.80 & 4.03\\
T3 analyzer loop & 38.50 & 10.19 & 9.01 & 3.94 & 37.66 & 10.80 & 9.44 & 3.84\\
T4 Self-Refine & 77.92 & 24.59 & 9.15 & 4.21 & 80.95 & 24.81 & 9.44 & 4.01\\
T5 coder+reviewer & 76.33 & 23.08 & 8.82 & 4.16 & 78.98 & 24.02 & 9.11 & 4.25\\
T6 RAG & 74.60 & 20.10 & 8.60 & 4.16 & 76.18 & 22.51 & 8.93 & 4.40\\
T7 best-of-$n$ & 73.64 & 19.23 & 8.86 & 4.48 & 73.28 & 20.65 & 9.91 & 4.27\\
T8 LoRA-SFT & 71.34 & 16.73 & 8.46 & 4.54 & 67.00 & 18.08 & 9.61 & 4.37\\
T9 DPO & 66.03 & 15.08 & 8.73 & 4.51 & 69.18 & 16.68 & 9.71 & 4.19\\
T10 RL & 63.01 & 12.93 & 8.39 & 4.51 & 60.90 & 13.96 & 9.37 & 4.44\\
\bottomrule
\end{tabularx}
\par\smallskip\raggedright\scriptsize
Abbreviations as in Table~\ref{tab:emp-cor}.
\end{table}

\begin{table}[t]
\caption{\textbf{E4. Efficiency (\dimEff), $B=4$ (T0: $B=1$).} DPS = EvalPerf DPS$_\text{norm}$ from
CPU instruction counts; ET = EffiBench-X execution-time score relative to the human-expert solution,
averaged over six languages (\%); Bey = Mercury Beyond (\%). Higher is better for all three.}
\label{tab:emp-eff}
\scriptsize
\setlength{\tabcolsep}{2.5pt}
\begin{tabularx}{\linewidth}{@{}l*{6}{>{\centering\arraybackslash}X}@{}}
\toprule
\multicolumn{7}{@{}l}{\textbf{(a) API-served models}}\\
 & \multicolumn{3}{c}{Op} & \multicolumn{3}{c@{}}{oss}\\
\cmidrule(lr){2-4}\cmidrule(l){5-7}
Technique & DPS & ET & Bey & DPS & ET & Bey\\
\midrule
T0 single shot & 78.06 & 47.79 & 75.03 & 79.41 & 61.33 & 69.70\\
T1 quality instr. & 79.90 & 49.70 & 76.48 & 75.01 & 62.24 & 70.29\\
T2 SCoT & 79.20 & 48.64 & 79.38 & 75.11 & 63.71 & 74.18\\
T3 analyzer loop & 75.46 & 44.11 & 79.72 & 72.17 & 60.53 & 70.55\\
T4 Self-Refine & 75.28 & 47.68 & 79.04 & 76.67 & 59.11 & 71.59\\
T5 coder+reviewer & 78.86 & 46.79 & 79.91 & 76.22 & 61.54 & 70.59\\
T6 RAG & 76.24 & 49.50 & 81.10 & 75.12 & 63.01 & 69.01\\
T7 best-of-$n$ & 76.96 & 46.57 & 74.64 & 77.57 & 60.99 & 69.44\\
\midrule
\multicolumn{7}{@{}l}{\textbf{(b) Gemma pair (fine-tunable)}}\\
 & \multicolumn{3}{c}{G31} & \multicolumn{3}{c@{}}{G26}\\
\cmidrule(lr){2-4}\cmidrule(l){5-7}
Technique & DPS & ET & Bey & DPS & ET & Bey\\
\midrule
T0 single shot & 78.11 & 62.21 & 78.62 & 73.43 & 57.10 & 75.45\\
T1 quality instr. & 68.91 & 61.90 & 74.40 & 74.64 & 58.06 & 76.94\\
T2 SCoT & 78.52 & 61.55 & 71.42 & 79.59 & 56.98 & 76.29\\
T3 analyzer loop & 80.44 & 61.89 & 76.64 & 75.20 & 57.16 & 74.25\\
T4 Self-Refine & 75.52 & 60.62 & 75.12 & 78.92 & 60.00 & 73.24\\
T5 coder+reviewer & 78.02 & 63.80 & 74.20 & 76.56 & 58.46 & 75.99\\
T6 RAG & 75.72 & 60.09 & 72.83 & 76.11 & 61.02 & 72.52\\
T7 best-of-$n$ & 80.84 & 59.27 & 72.28 & 76.97 & 56.62 & 72.43\\
T8 LoRA-SFT & 81.59 & 62.65 & 74.23 & 78.43 & 57.65 & 72.37\\
T9 DPO & 81.24 & 58.90 & 77.26 & 75.60 & 57.11 & 75.51\\
T10 RL & 78.86 & 63.21 & 75.52 & 79.07 & 58.22 & 69.59\\
\bottomrule
\end{tabularx}
\end{table}

\begin{table}[t]
\caption{\textbf{E5. Security (\dimSec), $B=4$ (T0: $B=1$).} FS = CWEval func-sec@1 (\%, higher is
better); SP = BaxBench sec\_pass@1, secure and correct (\%, higher is better); Ins = share of
PythonSecurityEval solutions with at least one security issue (lower is better); B/k = security issues per kLOC on PythonSecurityEval (lower is better).}
\label{tab:emp-sec}
\scriptsize
\setlength{\tabcolsep}{2.5pt}
\begin{tabularx}{\linewidth}{@{}l*{8}{>{\centering\arraybackslash}X}@{}}
\toprule
\multicolumn{9}{@{}l}{\textbf{(a) API-served models}}\\
 & \multicolumn{4}{c}{Op} & \multicolumn{4}{c@{}}{oss}\\
\cmidrule(lr){2-5}\cmidrule(l){6-9}
Technique & FS & SP & Ins & B/k & FS & SP & Ins & B/k\\
\midrule
T0 single shot & 74.85 & 58.54 & 42.67 & 3.40 & 45.02 & 26.38 & 44.93 & 5.14\\
T1 quality instr. & 66.87 & 55.13 & 37.80 & 3.34 & 46.07 & 28.61 & 44.79 & 4.96\\
T2 SCoT & 74.30 & 54.74 & 40.46 & 3.37 & 46.68 & 27.47 & 46.72 & 5.14\\
T3 analyzer loop & 79.89 & 61.53 & 13.79 & 1.20 & 53.07 & 30.47 & 16.07 & 1.84\\
T4 Self-Refine & 71.67 & 59.46 & 38.09 & 3.11 & 47.21 & 28.47 & 41.41 & 4.46\\
T5 coder+reviewer & 73.22 & 60.57 & 37.95 & 3.12 & 48.52 & 28.34 & 38.51 & 4.41\\
T6 RAG & 74.79 & 62.76 & 35.07 & 2.84 & 47.48 & 31.35 & 39.31 & 4.00\\
T7 best-of-$n$ & 78.17 & 60.50 & 31.23 & 2.78 & 48.10 & 34.01 & 37.78 & 3.73\\
\midrule
\multicolumn{9}{@{}l}{\textbf{(b) Gemma pair (fine-tunable)}}\\
 & \multicolumn{4}{c}{G31} & \multicolumn{4}{c@{}}{G26}\\
\cmidrule(lr){2-5}\cmidrule(l){6-9}
Technique & FS & SP & Ins & B/k & FS & SP & Ins & B/k\\
\midrule
T0 single shot & 49.48 & 32.97 & 44.77 & 4.53 & 47.28 & 30.17 & 48.67 & 4.98\\
T1 quality instr. & 58.89 & 36.10 & 44.37 & 4.31 & 49.95 & 33.55 & 42.94 & 4.41\\
T2 SCoT & 46.49 & 35.00 & 45.00 & 4.28 & 45.39 & 31.59 & 46.99 & 4.82\\
T3 analyzer loop & 56.32 & 38.39 & 15.42 & 1.78 & 52.70 & 36.27 & 16.45 & 1.85\\
T4 Self-Refine & 47.99 & 35.02 & 40.69 & 4.11 & 46.90 & 33.70 & 40.15 & 4.20\\
T5 coder+reviewer & 48.69 & 35.49 & 39.35 & 4.09 & 49.45 & 33.45 & 38.49 & 4.24\\
T6 RAG & 50.44 & 39.16 & 35.05 & 3.81 & 48.54 & 34.98 & 37.10 & 3.94\\
T7 best-of-$n$ & 55.42 & 40.11 & 35.62 & 3.56 & 50.38 & 37.06 & 38.28 & 3.63\\
T8 LoRA-SFT & 53.67 & 36.74 & 32.10 & 3.13 & 55.80 & 37.61 & 30.78 & 3.27\\
T9 DPO & 54.76 & 39.37 & 28.45 & 3.08 & 56.44 & 35.89 & 29.86 & 3.27\\
T10 RL & 56.24 & 38.18 & 27.66 & 2.82 & 53.21 & 38.51 & 28.89 & 3.16\\
\bottomrule
\end{tabularx}
\end{table}

\begin{table}[t]
\caption{\textbf{E6. Trade-off matrix.} For each technique and dimension, the number of metric--model
cells of Tables~\ref{tab:emp-cor}--\ref{tab:emp-sec} (\dimCor{} includes LiveCodeBench Pro and DeepSWE) in which the technique is better / worse than the
reference by more than the 95\% half-width of Table~\ref{tab:emp-ci}, with the number of cells compared in
brackets. (a) Against T0, the single shot ($B=1$; techniques at $B=4$, so this view
is not budget-matched). (b) Against T7, best-of-$n$ with the same ranker at the same budget ($B=4$).
$\star$ marks a targeted dimension. Cells where neither count is reached are differences within the
interval.}
\label{tab:emp-tradeoff}
\scriptsize
\setlength{\tabcolsep}{3pt}
\begin{tabularx}{\linewidth}{@{}l*{8}{>{\centering\arraybackslash}X}@{}}
\toprule
 & \multicolumn{4}{c}{(a) better/worse than T0} & \multicolumn{4}{c@{}}{(b) better/worse than T7}\\
\cmidrule(lr){2-5}\cmidrule(l){6-9}
Technique & \dimCor & \dimRM & \dimEff & \dimSec & \dimCor & \dimRM & \dimEff & \dimSec\\
\midrule
T1 quality instructions & 2/0 (15) & $\star$\,6/1 (16) & $\star$\,0/3 (12) & $\star$\,4/1 (16) & 2/1 (15) & 0/13 (16) & 3/2 (12) & 0/12 (16)\\
T2 SCoT / plan first & $\star$\,1/3 (15) & $\star$\,15/0 (16) & 4/2 (12) & 0/0 (16) & 1/3 (15) & 3/11 (16) & 4/1 (12) & 0/13 (16)\\
T3 analyzer loop & 1/4 (15) & $\star$\,12/1 (16) & 1/2 (12) & $\star$\,10/0 (16) & 0/3 (15) & 9/5 (16) & 2/2 (12) & 8/0 (16)\\
T4 Self-Refine & $\star$\,1/3 (15) & $\star$\,8/0 (16) & $\star$\,3/3 (12) & $\star$\,7/0 (16) & 1/2 (15) & 1/12 (16) & 2/2 (12) & 0/10 (16)\\
T5 coder + reviewer & $\star$\,2/1 (15) & $\star$\,15/0 (16) & $\star$\,2/2 (12) & $\star$\,7/0 (16) & 1/2 (15) & 3/9 (16) & 3/0 (12) & 0/7 (16)\\
T6 RAG & 2/3 (15) & $\star$\,15/0 (16) & 4/2 (12) & $\star$\,11/0 (16) & 1/3 (15) & 3/5 (16) & 4/2 (12) & 0/1 (16)\\
T8 LoRA-SFT & 1/1 (8) & $\star$\,7/0 (8) & 2/1 (6) & $\star$\,6/0 (8) & 0/1 (8) & 4/0 (8) & 1/0 (6) & 3/0 (8)\\
T9 DPO & 1/1 (8) & $\star$\,7/0 (8) & 0/1 (6) & $\star$\,7/0 (8) & 0/1 (8) & 4/1 (8) & 1/0 (6) & 3/0 (8)\\
T10 RL & $\star$\,1/1 (8) & $\star$\,7/0 (8) & 1/1 (6) & $\star$\,6/0 (8) & 1/1 (8) & 5/0 (8) & 2/0 (6) & 4/0 (8)\\
\bottomrule
\end{tabularx}
\end{table}

\begin{table}[t]
\caption{\textbf{E6c. Measures the ranker does not optimize}, $B=4$, against T7. The T7 ranker and the analyzer loop (T3) act on Pylint, Bandit and Safety findings, which SonarQube's issue counts also ingest; the measures here do not use them: CC = cognitive complexity, J = judge score (Claude Opus~5.5), FS = CWEval func-sec@1 and SP = BaxBench sec\_pass@1 (both from tests and exploits). Cells: better/worse than T7 by more than the half-width of Table~\ref{tab:emp-ci}, over the four models (two for T8--T10). $\geq$T7: cells in which T7 is better or within the interval. J2 = second judge (GPT-6.1 Sol), better/worse cells. H = mean human rating (1--5) minus T7's, pooled over models, from the two raters' ratings (half-width about $\pm$0.26).}
\label{tab:emp-indep}
\scriptsize
\setlength{\tabcolsep}{3pt}
\begin{tabularx}{\linewidth}{@{}l*{7}{>{\centering\arraybackslash}X}@{}}
\toprule
Technique & CC & J & FS & SP & $\geq$T7 & J2 & H\\
\midrule
T1 quality instructions & 0/1 & 0/4 & 0/1 & 0/3 & 16 of 16 & 0/3 & $-0.19$\\
T2 SCoT / plan first & 2/0 & 1/3 & 0/1 & 0/4 & 13 of 16 & 1/2 & $-0.08$\\
T3 analyzer loop & 1/1 & 0/4 & 0/0 & 0/0 & 15 of 16 & 0/3 & $-0.14$\\
T4 Self-Refine & 1/1 & 0/3 & 0/2 & 0/2 & 15 of 16 & 0/3 & $-0.16$\\
T5 coder + reviewer & 2/0 & 1/2 & 0/0 & 0/2 & 13 of 16 & 1/2 & $-0.04$\\
T6 RAG & 2/1 & 1/3 & 0/0 & 0/0 & 13 of 16 & 1/2 & $-0.06$\\
T8 LoRA-SFT & 0/0 & 1/0 & 0/0 & 0/0 & 7 of 8 & 1/0 & $+0.05$\\
T9 DPO & 0/0 & 0/1 & 0/0 & 0/0 & 8 of 8 & 0/1 & $+0.02$\\
T10 RL & 0/0 & 1/0 & 0/0 & 0/0 & 7 of 8 & 1/0 & $+0.07$\\
\bottomrule
\end{tabularx}
\end{table}

\begin{table}[t]
\caption{\textbf{E7. Uncertainty.} 95\% bootstrap half-width ($\pm$) of the paired per-task effect of a technique against T7 at $B=4$, per metric and model. A difference in Tables~\ref{tab:emp-cor}--\ref{tab:emp-sec} and~\ref{S-tab:emp-budget} is called significant only when it exceeds the half-width of its metric and model; the same half-widths are used for comparisons against T0.}\label{tab:emp-ci}
\scriptsize
\setlength{\tabcolsep}{3pt}
\begin{tabularx}{\linewidth}{@{}ll>{\raggedright\arraybackslash}X r*{4}{c}@{}}
\toprule
 & Metric & Benchmark & $n$ & Op & oss & G31 & G26\\
\midrule
\dimCor & HE+ pass@1 & HumanEval+ & 164 & 1.41 & 4.62 & 3.16 & 4.08\\
 & LCB pass@1 & LiveCodeBench v6 (dated) & 100--400$^{a}$ & -- & 4.00 & 5.84 & 7.65\\
 & LCBP pass@1 & LiveCodeBench Pro, post-cutoff contests & 155 & 5.67 & 5.81 & 6.26 & 5.66\\
 & DS pass@1 & DeepSWE v1.1 & 113 & 6.66 & 5.74 & 5.40 & 4.46\\
\addlinespace
\dimRM & C\% & PythonSecurityEval & 470 & 3.24 & 3.40 & 2.58 & 3.59\\
 & Lint/kLOC & RACE R/M, NoFunEval & 300--600$^{a}$ & 1.82 & 2.21 & 2.17 & 2.10\\
 & CC & RACE R/M, NoFunEval & 300--600$^{a}$ & 0.26 & 0.32 & 0.92 & 0.68\\
 & J (1--5) & RACE R/M, NoFunEval & 300--600$^{a}$ & 0.08 & 0.08 & 0.11 & 0.06\\
\addlinespace
\dimEff & DPS$_\text{norm}$ & EvalPerf & 121 & 2.80 & 2.04 & 3.48 & 2.67\\
 & ET & EffiBench-X & 300--700$^{a}$ & 2.17 & 1.51 & 2.68 & 2.11\\
 & Bey & Mercury-eval & 256 & 2.72 & 2.19 & 3.17 & 3.18\\
\addlinespace
\dimSec & FS & CWEval & 119 & 5.38 & 8.53 & 7.38 & 8.30\\
 & SP & BaxBench & 392 & 4.61 & 4.29 & 3.82 & 3.85\\
 & Ins & PythonSecurityEval & 470 & 3.06 & 3.02 & 2.90 & 4.02\\
 & B/k & PythonSecurityEval & 470 & 0.35 & 0.48 & 0.40 & 0.45\\
\bottomrule
\end{tabularx}
\par\smallskip\raggedright\scriptsize
$n$ is the task count stated by the benchmark authors (HumanEval \citep{liu2023is}, EvalPerf \citep{liu2024evaluating}, Mercury-eval \citep{du2024mercury}, CWEval, all languages \citep{peng2025cweval}, BaxBench \citep{vero2025baxbench}, PythonSecurityEval \citep{alrashedy2023can}); $^{a}$set by the date filter or the pinned release. Units as in Tables~\ref{tab:emp-cor}--\ref{tab:emp-sec}.
\end{table}

\subsection{Results of the main re-evaluation}\label{sec:empirical-results}

\textbf{Reading the results.} Each cell of Tables~\ref{tab:emp-cor}--\ref{tab:emp-sec} is a mean over
tasks and samples at $B=4$ (T0 at $B=1$). A technique is compared with T7, best-of-4 with
the same ranker, which answers ERQ1 at a matched budget; the comparison with T0 shows the gain a
reader sees when no budget-matched baseline is run. We call a difference significant when it exceeds the 95\% half-width, that is, when its 95\% interval excludes zero; the threshold is the half-width of its metric, model and budget (Table~\ref{tab:emp-ci} for $B=4$; Supplementary Material, \S\ref{S-sec:app-stats}, for $B=8$ and $16$), and we also report which significant differences survive a Benjamini--Hochberg false-discovery correction across all cells (\S\ref{sec:threats}); Table~\ref{tab:emp-tradeoff}
counts resolved differences per dimension, and Fig.~\ref{fig:emp-budget} extends the comparison to
$B\in\{1,8,16\}$.

\textbf{ERQ1: the targeted dimension at a matched budget.} We lead with the measures that neither the T7 ranker nor the analyzer loop optimizes (Table~\ref{tab:emp-indep}): cognitive complexity, the judge score, CWEval func-sec@1 and BaxBench sec\_pass@1. On them T7 matches or beats each inference-time technique in 13 to 16 of its 16 model--metric cells: T1, T2, T4, T5 and T6 are better than T7 beyond the interval in at most 3 cells and worse in 4 to 9, and the analyzer loop (T3) is better in one cell and has a lower judge score than T7 on all four models. The training-time techniques are within the interval of T7 in 7 of their 8 cells, better or worse in the remaining one. The second judge points the same way (no technique better than T7 in more than 1 of 4 models), and the human ratings place every technique within $0.19$ points of T7, none beyond the interval. The analyzer-based counts of Tables~\ref{tab:emp-rm} and~\ref{tab:emp-sec}, which the ranker partly optimizes, give the same ordering for these techniques and add the gains of the analyzer loop and fine-tuning described below.

Against a single shot, every
inference-time technique lowers the number of maintainability issues per kLOC on every model, by $0.46$ (T1
on G26) to $18.09$ (T3 on G26). Against T7, which spends the same four calls on sampling and ranking,
most of these gains disappear. On \dimRM{}, quality instructions (T1), planning (T2), Self-Refine (T4),
the reviewer agent (T5) and retrieval (T6) are better than T7 beyond the interval in at most 3 of 16
cells and worse in 5 to 13; on \dimSec{} none of them is better than T7 in any cell, and T1, T2 and T4
are worse in 10 to 13 of 16 cells (Table~\ref{tab:emp-tradeoff}b). Self-Refine, for example,
lowers maintainability issues per kLOC by $0.92$ to $4.08$ against T0 but leaves $4.16$ to $5.36$ more than
T7 on all four models. Length does not explain these differences: T7's solutions are 2 to 13\% shorter than T0's, and its lead over T1, T2 and T4--T6 also holds in absolute issues per solution ($0.47$ to $0.90$ against at best $0.57$ to $0.99$ for these techniques; Supplementary Material, \S\ref{S-sec:app-loc}). This supports H1: at matched calls, sampling and ranking match or beat
iterative refinement, as \citet{olausson2024is} found for self-repair and \citet{mirzaei2026sample} and \citet{iscan2026selection} for self-refinement and post-hoc operators on small models, and the budget curves show it
is not an artefact of $B=4$: on maintainability issues per kLOC, the only inference-time techniques better
than T7 beyond the interval at $B=8$ or $B=16$ are the analyzer loop and, on oss at $B=16$,
retrieval (Fig.~\ref{fig:emp-budget}). The curves are not smooth (on CWEval, T7 on Op scores $74.85$, $78.17$, $74.43$ and $86.29$ at $B=1$, $4$, $8$ and $16$): the per-seed values show that these steps recur in every rerun rather than reflecting seed noise, and we read only steps beyond the half-width, here only the gain at $B=16$ (Supplementary Material, \S\ref{S-sec:app-controls}). Two kinds of technique beat T7 on the analyzer-based counts, and only there (Table~\ref{tab:emp-indep}). The analyzer
loop (T3) lowers maintainability issues per kLOC by $7.51$ to $10.51$ against T7 on all four models and
security issues per kLOC by $1.58$ to $1.89$, and it lowers
the share of PythonSecurityEval solutions with a maintainability issue by $33.26$ to $37.04$
points and the share with a security issue by $17.44$ to $21.83$ points. It stays ahead of T7 on maintainability issues at every budget on every model. The training-time techniques, which use the same best-of-4
inference as T7 on fine-tuned weights, are better than T7 beyond the interval in 4 or 5 of 8 \dimRM{}
cells and 3 or 4 of 8 \dimSec{} cells, and worse in at most one. RL (T10) is the strongest of them,
with $6.30$ and $6.69$ fewer maintainability issues per kLOC than T7, and $7.96$ and $9.39$ points fewer
solutions with a security issue, on G31 and G26; it is the only training-time technique that stays
ahead of T7 on maintainability issues at $B=16$.

\textbf{A single call ($B=1$).} At $B=1$ every technique spends exactly the single shot's budget, so this column of Supplementary Table~\ref{S-tab:emp-budget} tests the prompts themselves, without a ranker. There the quality instruction (T1) and planning (T2) do not improve maintainability: T1 raises maintainability issues per kLOC over T0 on all four models, by $3.99$ to $7.11$, more than the half-width on every model, and T2 by $1.23$ to $7.95$. This is not an effect of the denominator: T1's solutions at $B=1$ are 9 to 14\% longer than T0's and carry $0.26$ to $0.39$ more maintainability issues each (Supplementary Material, \S\ref{S-sec:app-loc}). The measures the analyzers do not compute agree that the prompts alone add little: at $B=1$, cognitive complexity is within the interval of T0 for T1, T2 and T6 in 11 of 12 cells, and the Opus judge rates T1 and T2 higher in three cells, by at most $0.14$ points on its 1--5 scale, differences the second judge does not confirm (Supplementary Material, \S\ref{S-sec:app-b1}). Retrieval (T6) lowers them on Op and oss (by $0.49$ and $2.85$) and raises them on G31 and G26 (by $1.92$ and $1.28$). T1 also lowers CWEval func-sec@1 on three models (by $0.99$ to $8.83$), raising it by $11.37$ only on G31, and lowers EvalPerf DPS$_\text{norm}$ on all four (by $0.20$ to $5.69$). The gains of T1, T2 and T6 over T0 at $B\geq4$ therefore come from the best-of-$B$ ranker they share with T7, not from the instruction, the plan or the retrieved text: at one call, quality prompting does not lower the analyzer's issue counts and barely moves complexity or judge scores, and at four to sixteen calls it does not beat best-of-$n$.

\textbf{ERQ2: correctness and the dimensions not targeted.} No technique is more correct than T7 on
every model, and on the two models with the highest single-shot HumanEval+ pass@1 (Op, $96.40$; G31,
$96.06$) almost every technique is below the single shot: T2 to T6 by $0.59$ to $6.90$ points on Op
and by $3.81$ to $11.08$ on G31, and the fine-tuned G31 models by $7.72$ to $10.59$; only the quality
instruction (T1, by $2.70$ on Op and $1.90$ on G31) and, on Op, best-of-$n$ are above it. Only on oss, the
weakest model ($81.27$), does every inference-time technique raise HumanEval+ pass@1 over T0 (by
$0.31$ to $8.11$). The analyzer loop's gains come with lower HumanEval+ pass@1 than T7 on three models
(by $7.88$ on Op, $7.89$ on G31 and $2.27$ on G26; equal on oss) and lower LiveCodeBench pass@1 on oss and G26
(by $7.77$ and $5.75$; not on G31), although a revision is kept only when the
public tests pass; three of these five differences exceed the interval. Against T0 it is $6.17$, $9.94$ and $5.05$ points lower on Op,
G31 and G26 and $3.59$ higher on oss. H2 therefore holds in part: the loop removes the analyzer's own
findings, but a public-test gate does not prevent a correctness cost. The analyzer's gains also do not
reach the \dimRM{} measures the analyzer does not compute. Against T7, T3 changes cognitive complexity
by $-0.61$ to $+0.49$ and lowers the judge score by $0.17$ to $0.54$ on all four models, beyond the
interval on every model, and the training-time techniques leave the judge score within $0.17$ of T7
while lowering maintainability issues per kLOC by $2.50$ to $6.69$. Fewer findings from the analyzer that
drives a technique is not the same as code that a second measure rates as more maintainable, which is
the gaming pattern of \S\ref{sec:discussion}. On efficiency, T1--T6 score higher than T7 in 22 of 24
Mercury Beyond cells (10 beyond the interval) and 21 of 24 EffiBench-X cells (7), but in only 5 of 24
EvalPerf cells (1); the instruction-count measure, which we take as primary because timing is noisy
(\S\ref{sec:meas-critiques}), does not confirm the timing gains. T1, whose instruction names the
target dimensions, changes EvalPerf DPS$_\text{norm}$ against T0 by $-9.20$ (G31) to $+1.84$ (Op)
and Mercury Beyond by $-4.22$ to $+1.49$: generic quality instructions raise efficiency on no model
beyond the interval, as H3 expects, but lower EvalPerf DPS$_\text{norm}$ beyond it on G31. Planning (T2) and Self-Refine (T4) are below T7 in all 16
\dimSec{} cells. Fine-tuning moves correctness in opposite directions
on the two Gemma models: on G26, T8--T10 are $2.29$ to $6.47$ points above T7 on HumanEval+, and on
G31 $5.67$ to $8.54$ points below, beyond the interval for all three; on LiveCodeBench, T8--T10 are
within the interval of T7 on both models, at most $5.41$ points from it. On G31's LiveCodeBench split, T1, T2, T5 and T6 are above T7 beyond the interval, by $6.36$ to $10.69$ points.

\textbf{Correctness on tasks no model has seen.} The two uncontaminated correctness measures tell a
calmer story than HumanEval+. On the LiveCodeBench Pro contests held after every cutoff, no technique
differs from T7 or from T0 beyond the interval on any model. On DeepSWE, whose tasks were released after these models' cutoffs (for Opus~5.5 by about a month),
the only resolved differences are on Op, where planning, the analyzer loop, Self-Refine and retrieval
are $6.69$ to $8.93$ points below its single shot (Table~\ref{tab:emp-deepswe}). The HumanEval+ losses on G31, $3.81$ to $11.08$ points below the single shot, do not appear on either benchmark. LiveCodeBench Pro is less precise than HumanEval+ (half-widths of $5.66$ to $6.26$ points against $1.41$ to $4.62$), so a small loss could stay within the interval there. Seven of the G31 losses, however (T2, T3, T4, T6 and T8--T10, $7.72$ to $11.08$ points), exceed LiveCodeBench Pro's half-width, and on LiveCodeBench Pro the same techniques lie between $-1.52$ and $+2.92$ points of the single shot, five of the seven above it. Losses of that size would have been significant and are absent, and on LiveCodeBench, whose problems also postdate Gemma~4's cutoff, every G31 technique is above the single shot beyond the interval. The large HumanEval+ losses are therefore specific to HumanEval+. Two explanations fit: contamination, and a ceiling or regression-to-the-mean effect, since the losses appear only on the two models at about 96\% and every technique gains on oss at 81\%; the post-cutoff contests, on which no model is near the ceiling, cannot separate them (\S\ref{sec:threats}). Losses below about six points cannot be excluded on the post-cutoff contests. The head-to-head (\S\ref{sec:empirical-h2h}) shows the same split: most highlighted techniques are below the single shot on EvalPlus and above it on LiveCodeBench.

\textbf{Best-of-$n$ as a reference.} More samples do not always help T7: it keeps lowering maintainability issues with budget, but loses HumanEval+ pass@1 on both Gemma models and, on G31, CWEval func-sec@1 beyond $B=4$, so wins over T7 are wins over a reference that is strong on the analyzer's measures and weaker on hidden-test correctness, and the comparison with T0 remains necessary (Supplementary Material, \S\ref{S-sec:app-controls}).

\textbf{Sampling temperature.} T7 at the single-call techniques' temperature of $0.2$ instead of $0.8$ is within the half-width in \phTlowWithin{} of 36 primary-metric cells, and the conclusions of ERQ1 do not change against it (Supplementary Material, \S\ref{S-sec:app-controls}).

\textbf{ERQ3: model strength and architecture.} In absolute terms the open models gain more than the
closed anchor. Best-of-4 raises BaxBench sec\_pass@1 over T0 by $1.96$ points on Op, within the
interval, and by $6.89$ to $7.63$ points on the three open models, and lowers maintainability issues per kLOC
by $6.05$ on Op against $7.29$ to $8.24$ on the open models. In relative terms the analyzer loop's
effect is nearly the same on every model: against T0 it removes $61.3\%$ to $63.5\%$ of the maintainability issues per kLOC and $60.7\%$ to $64.7\%$ of the security issues per kLOC, on Op as on oss. H4 therefore holds for absolute gains, which partly reflect that the weaker models start with more
findings ($29.22$ maintainability issues per kLOC at T0 on oss against $21.37$ on Op), and not for the
relative gains of the analyzer loop. Correctness shows the same split: the techniques raise
HumanEval+ pass@1 on the weakest model and lower it on the strongest. The architecture contrast is
limited to the Gemma pair, where fine-tuning raised HumanEval+ above T7 on the MoE model and lowered it
on the dense one. The pair matches total rather than active parameters, so this is a difference
between two models, not evidence about mixture-of-experts as such.

\textbf{Combined reward.} The combined reward of T10 did
not collapse correctness on G26, where HumanEval+ pass@1 is $3.69$ points above T0 (within the
interval), and gave the largest reductions in maintainability issues per kLOC of the three training-time
techniques ($13.59$ on G31 and $14.93$ on G26 against T0). On G31, however, all three fine-tuned models
are $7.72$ to $10.59$ points below the single shot on HumanEval+.

\begin{takeaways}
\begin{itemize}
\item On measures the ranker does not optimize (complexity, two judges, CWEval, BaxBench), best-of-$n$ with an analyzer-weighted ranker matches or beats prompting, planning, self-refinement, a reviewer agent and retrieval in 13 to 16 of 16 cells at four calls, and the analyzer's own counts agree, also at eight and sixteen calls; their gains over a single shot are largely a budget effect (H1).
\item The analyzer loop and the training-time techniques beat best-of-$n$ only on analyzer-based issue counts; on the independent measures they are within its interval in most cells, and the judge score falls for the analyzer loop on every model.
\item On the two models with the highest single-shot correctness, almost every technique lowers
  HumanEval+ pass@1 below the single shot, the analyzer loop by up to $9.94$ points despite its
  public-test gate (H2 in part); on the uncontaminated LiveCodeBench Pro contests no technique
  differs from the single shot beyond the interval, and on DeepSWE only Op loses correctness.
\item Best-of-$n$ itself loses hidden-test correctness as it draws more samples, as expected when its ranker is an imperfect verifier \citep{stroebl2024inference}, and on one model
  functional security beyond four samples, so a quality technique should be compared with both the single shot and
  best-of-$n$.
\item Timing-based efficiency scores rise for most techniques while the instruction-count score does
  not, and generic quality instructions raise efficiency on no model and lower it on one (H3 in part).
\item Weaker models gain more in absolute terms, while the analyzer loop removes the same share of
  findings on every model (H4 for absolute gains only).
\end{itemize}
\end{takeaways}

\subsection{Extended grid: current frontier and open-weight models}\label{sec:empirical-ext}

The main grid holds the base models fixed so that technique effects can be measured in depth. The extended grid asks
whether the picture holds for the strongest models of October 2026. It adds two closed models, GPT-6.1 Sol and
Gemini~4 Argon, and the leading open-weight models served on Fireworks: Kimi K3, GLM-5.3, Qwen3.8-Max, MiMo-V2.6-Pro,
DeepSeek V4.1-Flash and V4-Flash, and MiniMax M3. Supplementary Table~\ref{S-tab:emp-ext} scores every model's single-shot output with the
function-level stack, Supplementary Table~\ref{S-tab:emp-ext-tech} applies the inference-time techniques T1--T7 at $B=4$, and
Supplementary Table~\ref{S-tab:emp-agent} adds the repository- and agent-level benchmarks that 2026 model releases and independent
leaderboards report most: DeepSWE and Terminal-Bench~4.0 for correctness, FrontierCode and SlopCodeBench for maintainability,
GSO and SWE-fficiency for efficiency, and SusVibes for security (\S\ref{sec:meas-benchmarks}). Because none of the
open-weight vendors states a training cutoff and every 2026 cutoff postdates LiveCodeBench~v6 and the last public
LiveCodeBench Pro set, correctness on function-level tasks is measured only on Codeforces contests held after 2026-07-01.

\textbf{Results.} \emph{Capability.} The closed models lead every function-level correctness and
security measure (Supplementary Table~\ref{S-tab:emp-ext}): LiveCodeBench Pro pass@1 of $66.79$ to $69.82$ against
$41.11$ to $61.87$ for the open-weight frontier models and $21.94$ to $27.68$ for the main-grid open
models, CWEval func-sec@1 of $72.79$ to $74.85$ against $56.72$ to $69.43$ and $45.02$ to $49.48$, and
BaxBench sec\_pass@1 of $56.22$ to $58.54$ against $40.41$ to $52.46$ and $26.38$ to $32.97$.
\emph{Quality follows capability, efficiency does not.} Across the thirteen models, single-shot
LiveCodeBench Pro pass@1 correlates with maintainability issues per kLOC ($r=-0.96$), the judge score
($0.90$), CWEval ($0.95$), BaxBench ($0.97$) and security issues per kLOC ($-0.97$), but hardly with
EvalPerf DPS$_\text{norm}$ ($0.22$) or Mercury Beyond ($0.15$): stronger models write cleaner and safer
code by default, but not faster code. \emph{Techniques.} The main-grid pattern holds on all nine added
models (Supplementary Table~\ref{S-tab:emp-ext-tech}): T7 leaves fewer maintainability issues per kLOC than T1, T2, T4, T5 and
T6 on every model and a higher CWEval func-sec@1 than all of them on seven of nine, while the analyzer
loop roughly halves T7's maintainability issues ($7.98$ to $9.94$ per kLOC against $16.23$ to $18.96$) and
changes func-sec@1 by $-0.50$ to $+7.23$ points against T7. \emph{Repository-level quality.} On the
agent benchmarks (Supplementary Table~\ref{S-tab:emp-agent}), DeepSWE pass@1 correlates with FrontierCode ($0.91$),
GSO ($0.95$), SWE-fficiency ($0.98$), SusVibes SecPass ($0.97$) and, negatively, with SlopCodeBench
erosion ($-0.97$). The largest gap is between working and secure code: on SusVibes every closed and
open-weight frontier model passes the functional tests on $38.06$ to $44.41$ points more tasks than it
passes the security tests, against a quality penalty of $1.25$ to $7.59$ points on FrontierCode. The
main-grid open models sit near the floor of every agent benchmark (DeepSWE $10.17$ to $17.96$,
Terminal-Bench~4.0 $2.91$ to $5.10$).

\begin{takeaways}
\begin{itemize}
\item Across thirteen current models, correctness, maintainability and security rise together with
  capability; efficiency does not.
\item Best-of-$n$ beats prompting, planning, self-refinement, a reviewer and retrieval on
  maintainability on every added model, and the analyzer loop halves its analyzer findings again, as
  in the main grid.
\item At repository level the frontier models produce working but insecure code on about four in ten
  SusVibes tasks, a far larger gap than the quality penalty FrontierCode measures.
\end{itemize}
\end{takeaways}

\subsection{E12: Head-to-head of the highlighted techniques}\label{sec:empirical-h2h}

T0--T10 study a few representative techniques in depth, across models and budgets. E12 adds breadth:
it compares the \cnt{nHlSelected} highlighted techniques of \S\ref{sec:highlighted} with each other
under one setting, so that the results their authors report on different benchmarks, models and
baselines (Supplementary Tables~\ref{S-tab:hl-prompting}--\ref{S-tab:hl-training}) can be replaced by measurements
that are comparable.

\textbf{Research question.} \emph{ERQ4:} under one model, one benchmark set and common reference
budgets, which highlighted techniques improve the dimension they target, what do they cost in
\dimCor{} and in the dimensions they do not target, and how do the families compare?

\textbf{Model.} One model holds everything else fixed. We use Gemma~4 31B IT, the dense arm of
the main grid (Table~\ref{tab:emp-setup}a), because it is the dense arm of the fine-tuned Gemma pair, so the training-time techniques change the same weights the inference-time techniques prompt.

\textbf{Benchmarks.} E12 uses only public benchmarks that the included studies themselves use most,
so that its results connect to the literature; we add no benchmark of our own. Counts are the
included studies that report each benchmark (from the extraction table). \dimCor:
HumanEval+ and MBPP+ (EvalPlus; plain HumanEval appears in \cnt{nBenchHumanEval} studies, the EvalPlus
versions in \cnt{nBenchEvalPlus}) as the likely contaminated split, LiveCodeBench
(\cnt{nBenchLiveCodeBench}) for tasks after the model's January 2025 cutoff, and the LiveCodeBench Pro
protocol on Codeforces contests held after 2026-07-01, which no current model has seen. \dimSec: CWEval
(\cnt{nBenchCWEval}), which requires code to be both functional and secure, and CyberSecEval
(\cnt{nBenchCyberSecEval}), the most comparable to prior security work; its rule-based detector's
reported error rates (\S\ref{sec:measurement}) are carried into the analysis. \dimEff: EffiBench
(\cnt{nBenchEffiBench}), the most used efficiency benchmark, and EvalPerf (\cnt{nBenchEvalPerf}),
whose instruction counts are more stable than wall-clock time. \dimRM: no dedicated benchmark is in
common use (RACE and NoFunEval appear in \cnt{nBenchRACE} and \cnt{nBenchNoFunEval} included studies);
the studies measure \dimRM{} by running established analyzers on generated code, and E12 does the
same: pinned Pylint, cognitive complexity and the length-controlled judge, applied to the outputs
of the \dimCor{} benchmarks. Every output of every benchmark is scored on all four dimensions with
the measurement stack above, so a technique that targets one dimension is also measured on the other
three. Python comes first; a technique defined only for another language is ported and marked as
such.

\textbf{Implementation.} Each technique runs in one harness. Where public code exists
(\cnt{nHlWithCode} of the \cnt{nHlSelected}) we adapt it; otherwise we reimplement it from the paper.
Before use, each implementation is checked against one automated number its authors report, in their setting; where their model is retired or their number is a human rating, another automated number the paper reports for a runnable model is used, and 14 techniques with no such number are not checkable (Supplementary Material, \S\ref{S-sec:app-fidelity}; \phFidOK{} of the \phFidN{} checkable techniques reproduce within \phFidTol{}). Implementations are adjusted at most twice before the final run, stopping as soon as the reproduced number falls within the tolerance; the adjustment count and the per-run logs of every run, including superseded ones, are retained and will be released with the replication package. Hyperparameters come from the papers and
are never tuned on these benchmarks. Components that call another LLM (critics, reviewers, judges)
use the same model; components that the paper trains (detectors, reward models, co-decoders) are
retrained for the same model. Training-time techniques are retrained on the same model with the
authors' released data, or with data built by their recipe when it is not released (marked), using
LoRA on Fireworks; training time, from several hours to a few days per run, lies outside the call
budget.

\textbf{Budget and baselines.} Each technique runs in its published configuration, with a ceiling of
16 calls per task (a technique that needs more runs at the ceiling and is marked). Calls, tokens
and cost are logged per task. Two references run on every benchmark: T0 (one call) and one best-of-$n$
run with the T7 ranker, at a mean of $4.19$ calls per task (Supplementary Table~\ref{S-tab:emp-h2h}); it redraws a sample with no extractable code while the ceiling allows, which is why it runs above four calls and why its scores differ slightly from T7 at $B=4$ in Table~\ref{tab:emp-cor}. Best-of-$n$ is thus the budget-matched reference for techniques near that budget; a technique that uses about one call is compared with T0. So that techniques using many more calls are not compared with a cheaper reference, best-of-$n$ also runs at two, three, five and eight calls per task with the same ranker; each family is compared with the reference nearest its median calls per task: prompting and training with best-of-2, retrieval with best-of-3, tool feedback and self-refinement with best-of-5, multi-agent pipelines with best-of-8, and decoding (median \phHhCallsDecoding{} calls) with the main reference.

\textbf{Analysis.} Per benchmark, each technique is compared with T0 and with best-of-$n$ by its point
difference, with quality reported on every output, and summarized per family. E12 is \emph{indicative}: it compares point estimates, and its counts of techniques that beat a reference do not test each difference. Paired bootstrap 95\% half-widths over tasks, computed for every technique and column against T0 and against best-of-$n$, have a median of \phHhHwEP{} points for EvalPlus, \phHhHwLCB{} for LiveCodeBench, \phHhHwLCBP{} for LiveCodeBench Pro, \phHhHwCWE{} for CWEval, \phHhHwCSE{} for CyberSecEval and \phHhHwDPS{} for EvalPerf (Supplementary Material, \S\ref{S-sec:app-stats}); a smaller difference in Supplementary Table~\ref{S-tab:emp-h2h} should be read as not significant, and only \phHhBonWinsResolved{} of the \phHhBonWins{} wins over best-of-$n$ exceed the half-width of every target metric. Supplementary Table~\ref{S-tab:emp-h2h}
gives the results.

\begin{figure}[t]
\centering
\includegraphics[width=\linewidth]{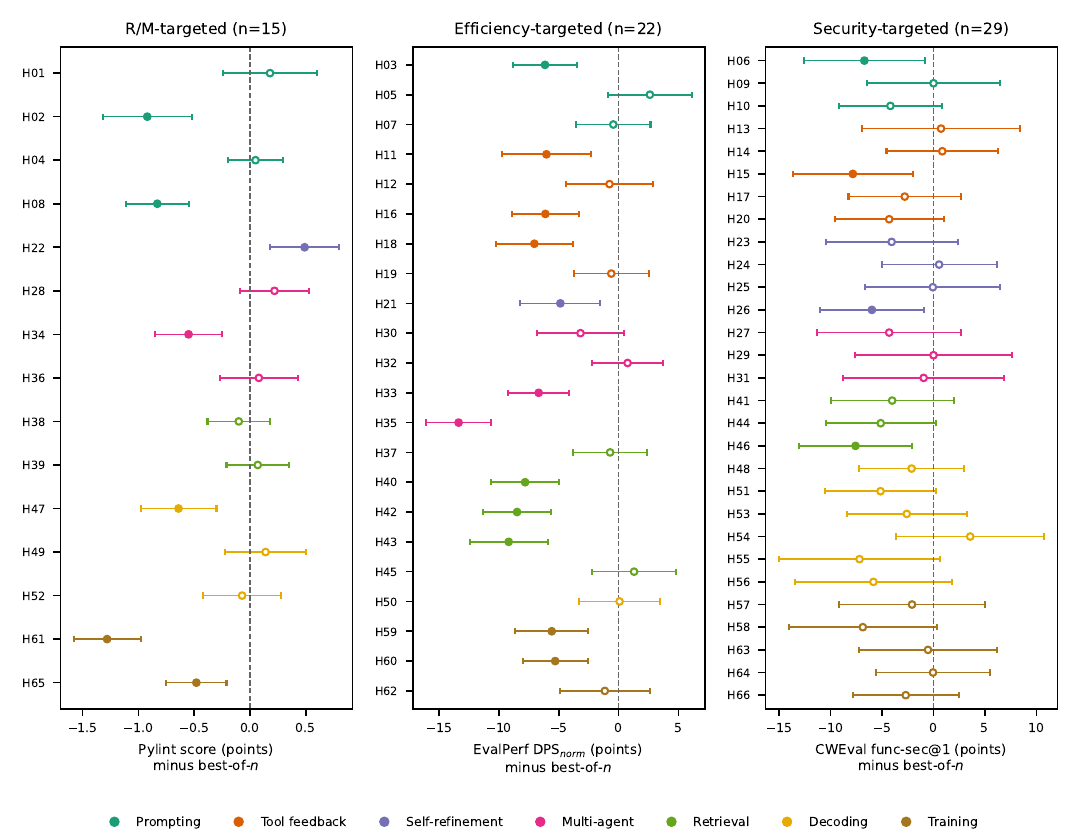}
\Description{Forest plot of the highlighted techniques, grouped by targeted dimension, each with its difference from best-of-n on its primary metric and a 95 percent interval; most intervals cross zero or lie below it.}
\caption{\textbf{Fig.~E2. Head-to-head summary} (E12, Gemma~4 31B IT): each highlighted technique's difference from best-of-$n$ on the primary metric of its targeted dimension (Pylint score for \dimRM, EvalPerf DPS$_\text{norm}$ for \dimEff, CWEval func-sec@1 for \dimSec), with its 95\% paired bootstrap interval; filled markers lie beyond the interval. All columns, both references and the budget-matched references are in Supplementary Table~\ref{S-tab:emp-h2h}, and the per-column half-widths in Supplementary Table~\ref{S-tab:app-h2h-ci}.}
\label{fig:emp-h2h}
\end{figure}

\textbf{Results (ERQ4).} Fig.~\ref{fig:emp-h2h} gives the view with intervals: on the primary metric of its targeted dimension, \phHhPrimAbove{} of the 66 techniques is above best-of-$n$ beyond the interval (CodeChain), \phHhPrimBelow{} are below it beyond the interval, and \phHhPrimWithin{} are within it. With 66 comparisons, about two wins in the favourable direction are expected by chance alone, so the single win does not show that CodeChain beats best-of-$n$. The counts that follow compare point estimates on every target metric and are indicative only. The best-of-$n$ reference runs at a mean of $4.19$ calls. Against T0, \phHhTzeroWins{} of the 66 techniques improve every target metric we measure: prompting \phHhTzeroPrompting{} of 10, tool feedback \phHhTzeroFeedback{} of 10, self-refinement \phHhTzeroSelfRefine{} of 6, multi-agent \phHhTzeroMultiAgent{} of 10, retrieval \phHhTzeroRetrieval{} of 10, decoding \phHhTzeroDecoding{} of 10, training \phHhTzeroTraining{} of 10. Against best-of-$n$, \phHhBonWins{} of the 66 do: prompting \phHhBonPrompting{}, decoding \phHhBonDecoding{}, tool feedback \phHhBonFeedback{}, self-refinement \phHhBonSelfRefine{}, multi-agent \phHhBonMultiAgent{}, retrieval \phHhBonRetrieval{}, training \phHhBonTraining{}. Because the families differ in cost, we also ran best-of-$n$ at two, three, five and eight calls and compared each family with the reference nearest its median calls (Supplementary Table~\ref{S-tab:emp-h2h}, last panel), which removes the asymmetry of a single reference: against these budget-matched references \phHhBonFamWins{} of the 66 techniques improve every target metric (prompting \phHhBonFamPrompting{} of 10 (best-of-2), tool feedback \phHhBonFamFeedback{} of 10 (best-of-5), self-refinement \phHhBonFamSelfRefine{} of 6 (best-of-5), multi-agent \phHhBonFamMultiAgent{} of 10 (best-of-8), retrieval \phHhBonFamRetrieval{} of 10 (best-of-3), decoding \phHhBonFamDecoding{} of 10 (main reference), training \phHhBonFamTraining{} of 10 (best-of-2)). Matching the budget changes the total little (\phHhBonFamWins{} against \phHhBonWins{}): the one-call families win slightly more often against best-of-2 than against the four-call reference (prompting \phHhBonFamPrompting{} against \phHhBonPrompting{}, training \phHhBonFamTraining{} against \phHhBonTraining{}), and the multi-agent pipelines, the most expensive family, slightly less often against best-of-8 (\phHhBonFamMultiAgent{} against \phHhBonMultiAgent{}); eight of the training-time techniques run at one call per task, SafeCoder at 1.07 and Afterburner, which keeps an optimization loop at inference, at 4.09.

The dimensions tell different stories. \emph{Security:} \phHhSecCse{} of the \phHhNSec{} security-targeted techniques produce fewer CyberSecEval insecure-code detections than best-of-$n$, but only \phHhSecCwe{} of the \phHhNSec{} exceed its CWEval func-sec@1. Lowering the rule-based detector's hits is easier than writing code that is both secure and functional, and the detector's reported error rates (\S\ref{sec:meas-critiques}) make the first the weaker evidence. \emph{Efficiency:} \phHhEffEb{} of the \phHhNEff{} efficiency-targeted techniques run faster than T0 on EffiBench, against \phHhNonEffEb{} of the techniques that target other dimensions; EvalPerf's instruction-count score improves for only \phHhEffDps{} of the \phHhNEff{}. Since EffiBench relies on physical time (\S\ref{sec:meas-critiques}), the head-to-head supports an efficiency gain only where both measures agree. \emph{Readability and maintainability:} \phHhRmUp{} of the \phHhNRm{} \dimRM-targeted techniques raise the Pylint score above T0, by at most $1.35$ points on its 10-point scale (CodeChain), and only \phHhRmBon{} of all 66 techniques exceed best-of-$n$ ($7.69$). \emph{Correctness:} the two function-level splits disagree. On EvalPlus, which is likely contaminated, \phHhEpBelow{} of the 66 techniques are below T0 (family medians \phHhEpMedLo{} to \phHhEpMedHi{} points); on LiveCodeBench, whose problems postdate the model's cutoff, \phHhLcbAbove{} of the 66 are above it (family medians \phHhLcbMedLo{} to \phHhLcbMedHi{}), as is best-of-$n$, which selects among samples on the public examples. As in the main grid (\S\ref{sec:empirical-results}), the losses are largely specific to the likely contaminated split. The LiveCodeBench Pro contests, which also postdate the cutoff, show no comparable losses and only small gains: \phHhLcbpAbove{} of the 66 techniques score above T0 ($27.14$) and \phHhLcbpBon{} above best-of-$n$ ($30.92$), all within $6.26$ points of T0, the main-grid half-width for G31, with family medians of \phHhLcbpMedLo{} to \phHhLcbpMedHi{} points. \emph{Cost:} median calls per task are $1.00$ for training (training compute excluded), \phHhCallsPrompting{} for prompting, $3.12$ for retrieval, \phHhCallsDecoding{} for decoding, $5.08$ for self-refinement, \phHhCallsFeedback{} for tool feedback and \phHhCallsMultiAgent{} for multi-agent pipelines.

\begin{takeaways}
\begin{itemize}
\item On its primary target metric and beyond the interval, \phHhPrimAbove{} of 66 highlighted techniques beats best-of-$n$ and \phHhPrimBelow{} fall below it; most improve their target over a single shot (\phHhTzeroWins{} of 66 by point estimate).
\item Security techniques reduce detector hits far more often than they raise the share of code that
  is both functional and secure (\phHhSecCse{} of \phHhNSec{} against \phHhSecCwe{} of \phHhNSec{} over best-of-$n$).
\item Efficiency techniques speed up execution time but not instruction counts in about half the
  cases, so timing-only efficiency claims need a second measure.
\item Correctness falls on the likely contaminated split and rises on the dated split for most
  techniques, so a single correctness benchmark can reverse the conclusion.
\end{itemize}
\end{takeaways}
\section{Discussion: Cross-Cutting Findings and Trade-offs}\label{sec:discussion}

Counts below come from the author-reviewed, LLM-assisted extraction; per-study numbers were checked against primary sources.

\paragraph{RQ1 and RQ2: what is targeted, and how it is measured.}
Efficiency is the most often targeted dimension (\cnt{nDimEff} studies), followed by security (\cnt{nDimSec}) and readability/maintainability (\cnt{nDimRM}) (Fig.~\ref{fig:evidence-maps} by family and signal source, Supplementary Fig.~\ref{S-fig:trends} over time). Instruments vary by dimension: static analyzers such as Bandit and CodeQL for security, wall-clock timing against a reference for efficiency, linters or LLM judges for readability; human evaluation is rare (Table~\ref{tab:qa}). Each has documented validity problems (\S\ref{sec:measurement}): static analyzers overestimated security by \rep{7 to 21 times} in one re-evaluation, with \rep{37 to 60\%} of ``secure'' outputs not functional~\citep{tessa2026how}; wall-clock time varied about \rep{12\%} across runs versus \rep{0.03\%} for CPU instruction counts~\citep{yang2025wedge}; and an LLM judge's Python readability scores were uncorrelated with human ratings (Pearson \rep{0.02})~\citep{ouedraogo2025humanaligned}. A reported gain thus partly reflects the instrument used to optimize and evaluate it.

\begin{table}[t]
\centering
\footnotesize
\caption{How often included studies measured anything beyond their targeted dimension: share of studies per family that measured at least one non-targeted quality dimension, and share that measured correctness. Counts from LLM-assisted extraction reviewed by the author.}\label{tab:nontargeted}
\begin{tabular}{lrrrrr}
\toprule
Family & Studies & \multicolumn{2}{c}{Non-targeted dim.\ measured} & \multicolumn{2}{c}{\dimCor{} measured} \\
\cmidrule(lr){3-4}\cmidrule(lr){5-6}
 & & $n$ & \% & $n$ & \% \\
\midrule
Prompting & 113 & 15 & 13 & 79 & 70 \\
Tool feedback & 124 & 12 & 10 & 94 & 76 \\
Self-refinement & 17 & 1 & 6 & 10 & 59 \\
Multi-agent & 75 & 10 & 13 & 66 & 88 \\
Retrieval & 37 & 4 & 11 & 28 & 76 \\
Decoding & 18 & 2 & 11 & 18 & 100 \\
Training & 124 & 15 & 12 & 114 & 92 \\
Hybrid/other & 191 & 16 & 8 & 173 & 91 \\
\midrule
All studies & 628 & 70 & 11 & 526 & 84 \\
\bottomrule
\end{tabular}
\\[2pt]{\scriptsize A study can appear in more than one family when it evaluates several techniques; the last row counts each study once. Percentages are our computation over the extracted rows.}
\end{table}

\paragraph{RQ3: effects on correctness and on non-targeted dimensions.}
Table~\ref{tab:nontargeted} is the review's headline result: most studies never measured a quality dimension beyond their target, so most techniques' cross-dimension effects are unknown, not neutral. Among technique rows reporting a target gain, correctness was unchanged, mixed or lower in about a quarter and unreported in nearly three in ten (Table~\ref{tab:effects}). Individual cases show why gates matter: AutoSafeCoder lowered HumanEval pass@1 from \rep{0.9085} to \rep{0.8780} while reducing Bandit findings~\citep{nunez2024autosafecoder}, whereas the fitness-gated Pylint and Bandit loop of \citet{blyth2025static} raised functional correctness from \rep{38.16\%} to \rep{47.81\%}. \citet{wu2026secure} report that prior security alignment methods often cut an effective safety rate that also requires correctness by \rep{14-54\%}. Evidence on trade-offs \emph{between} non-functional dimensions, such as whether security hardening or performance optimization harms readability, is almost absent.

\begin{table}[t]
\centering
\footnotesize
\caption{Direction of effect on correctness (rows) against direction of effect on the targeted quality dimension (columns), counted per technique row of \texttt{extraction.csv} (vote counting by direction, SWiM; no effect sizes are pooled). Counts from LLM-assisted extraction reviewed by the author. In parentheses: how far each count could move if direction fields were wrong at the rate at which the independent second coder disagreed with the extraction (20.9\%, 58 of 278 audited fields; the rate is drawn from its posterior, and the range is the 95th percentile of the rows a count can lose or gain).}\label{tab:effects}
\setlength{\tabcolsep}{3.5pt}
\begin{tabular}{lrrrrrr}
\toprule
\dimCor{} \textbackslash{} target & \up & \same & \down & mixed & \nr & total \\
\midrule
\up & 257\,{\scriptsize (136--340)} & 3\,{\scriptsize (0--7)} & 3\,{\scriptsize (0--7)} & 26\,{\scriptsize (11--42)} & 20\,{\scriptsize (8--34)} & 309\,{\scriptsize (223--362)} \\
\same & 51\,{\scriptsize (24--78)} & 2\,{\scriptsize (0--5)} & 0\,{\scriptsize (0--1)} & 2\,{\scriptsize (0--5)} & 1\,{\scriptsize (0--3)} & 56\,{\scriptsize (37--74)} \\
\down & 27\,{\scriptsize (11--44)} & 1\,{\scriptsize (0--3)} & 4\,{\scriptsize (1--9)} & 23\,{\scriptsize (9--38)} & 0\,{\scriptsize (0--1)} & 55\,{\scriptsize (37--73)} \\
mixed & 65\,{\scriptsize (31--98)} & 1\,{\scriptsize (0--3)} & 1\,{\scriptsize (0--3)} & 37\,{\scriptsize (16--58)} & 3\,{\scriptsize (0--7)} & 107\,{\scriptsize (74--136)} \\
\nr & 161\,{\scriptsize (83--225)} & 3\,{\scriptsize (0--7)} & 3\,{\scriptsize (0--7)} & 37\,{\scriptsize (17--59)} & 6\,{\scriptsize (1--12)} & 210\,{\scriptsize (150--255)} \\
\midrule
total & 561\,{\scriptsize (409--601)} & 10\,{\scriptsize (5--16)} & 11\,{\scriptsize (6--17)} & 125\,{\scriptsize (88--158)} & 30\,{\scriptsize (19--42)} & 737 \\
\bottomrule
\end{tabular}
\\[2pt]{\scriptsize \up{} improved, \same{} no clear change, \down{} worse, mixed = differs across settings, \nr{} = not reported.}
\end{table}

\paragraph{Goodhart effects.}
Every family optimizing a single automated signal shows some form of gaming. In CORE's user study, reviewers rejected \rep{55.11\%} of CodeQL-passing revisions~\citep{wadhwa2024core}. Effectively empty output scored a Maintainability Index of 100\% and zero smells~\citep{islam2026from}. A reinforcement-learning policy with a quality-only reward learned to emit trivial outputs such as empty code or comments~\citep{yao2025training}, and GPT-5 cheated in \rep{76\%} of tasks on a benchmark built to detect test special-casing~\citep{zhong2026impossiblebench}. The usual remedy in stronger studies is to gate quality signals on executed tests and to evaluate with an instrument other than the feedback or reward signal; few do both.

\paragraph{RQ4: evidence quality and compute fairness.}
Table~\ref{tab:qa} summarizes the quality assessment. Iterative techniques (feedback loops, self-refinement, multi-agent pipelines) spend several to many times the calls and tokens of single-shot generation, yet baselines seldom get the same budget; when they do, gains shrink~\citep{olausson2024is}, and we found only one efficiency-loop study comparing against an oracle \rep{best-of-five} baseline~\citep{deng2026perfagent}. Only a minority report variance across samples, contamination, and validation of automated metrics against human judgement. About a third evaluate a single model without justification.

\begin{table}[t]
\centering
\footnotesize
\caption{Evidence quality by family: median QA score (0--10, items QA1--QA10 of Supplementary Material \S S4) and the share of studies that report compute or cost (QA7 $>0$), consider contamination (QA9 $>0$), use human evaluation, and release artifacts (QA8 $>0$). LLM-assisted extraction reviewed by the author.}\label{tab:qa}
\begin{tabular}{lrrrrrr}
\toprule
Family & $n$ & Med.\ QA & Comp.\ \% & Cont.\ \% & Human \% & Artif.\ \% \\
\midrule
Prompting & 113 & 6.5 & 38 & 31 & 32 & 73 \\
Tool feedback & 124 & 6.0 & 69 & 31 & 9 & 65 \\
Self-refinement & 17 & 6.0 & 24 & 47 & 29 & 65 \\
Multi-agent & 75 & 5.0 & 65 & 17 & 8 & 53 \\
Retrieval & 37 & 6.0 & 41 & 46 & 30 & 62 \\
Decoding & 18 & 6.5 & 67 & 22 & 11 & 78 \\
Training & 124 & 6.5 & 56 & 55 & 21 & 68 \\
Hybrid/other & 191 & 6.0 & 68 & 29 & 14 & 59 \\
\midrule
All studies & 628 & 6.0 & 59 & 33 & 17 & 62 \\
\bottomrule
\end{tabular}
\end{table}

\paragraph{RQ5: gaps.}
Thin cells in the evidence maps (readability-targeted retrieval and decoding, learned verifiers and reward models for any dimension, profiler signals outside efficiency) and unasked questions (effects on non-targeted dimensions, scaling with model capability, gains at a matched budget) define the agenda of \S\ref{sec:agenda} and the hypotheses of \S\ref{sec:empirical}.

\section{Implications for Researchers and Practitioners}\label{sec:implications}

\paragraph{For practitioners.} Practitioners prioritize maintainability and readability \citep{sun2026quality}, the dimensions on which our re-evaluation finds the clearest pattern. Five implications follow from the results of \S\ref{sec:empirical}.
\begin{itemize}
\item \emph{Compare with sampling first.} Before adopting a prompting, planning, self-refinement, reviewer or retrieval technique, run best-of-$n$ at the same number of model calls with a ranker built from the analyzers already in the pipeline. On measures that ranker does not optimize, it matches or beats each of these techniques in most cells (Table~\ref{tab:emp-indep}), so a technique's gain over a single call is not evidence that it beats this cheaper baseline.
\item \emph{Do not rely on quality instructions at one call.} A single call with a quality instruction or a plan does not lower maintainability issues on any of the four models (\S\ref{sec:empirical-results}); its apparent gains at larger budgets come from the ranker it shares with best-of-$n$.
\item \emph{Use analyzer loops and fine-tuning for what they deliver.} Both lower the issue counts of the analyzers they use, by large margins, but these gains do not carry over to cognitive complexity, the judge score or test-based security (Table~\ref{tab:emp-indep}). They suit teams whose quality gate is those analyzers; they are not evidence of more readable code.
\item \emph{Check correctness on new tasks.} A technique's effect on HumanEval+ can differ in sign from its effect on contests and repository tasks the model has not seen (\S\ref{sec:empirical-results}); a gate on recent tasks, or on the project's own tests, is the safer check.
\item \emph{Gate security with tests, not detector counts.} Security techniques lower rule-based detector hits far more often than they raise the share of code that is both functional and secure (\S\ref{sec:empirical-h2h}), and current frontier agents produce working but insecure code on about four in ten SusVibes tasks (\S\ref{sec:empirical-ext}). Security tests or exploits, not detector counts alone, should decide whether generated code ships.
\end{itemize}
Across the thirteen current models, maintainability and security rise with capability while efficiency does not (\S\ref{sec:empirical-ext}), so the choice of model moves these dimensions about as much as any single technique, and efficiency needs its own measurement.

\paragraph{For researchers.} The review and the re-evaluation point to reporting practices that would make technique papers comparable (\S\ref{sec:agenda} gives the research agenda).
\begin{itemize}
\item Include a best-of-$n$ baseline at the same number of calls, ranked by the same signal the technique uses, and report calls and tokens per task.
\item Report correctness and the dimensions the technique does not target: only \cnt{nNonTargetedMeasured} of \cnt{nStudies} extracted studies measured a quality dimension beyond their target (Table~\ref{tab:nontargeted}).
\item Measure the target on an instrument independent of the technique's own signal, such as a test-based security benchmark, a judge or human ratings for an analyzer-driven technique.
\item Report seeds and variance, use tasks dated after the model's training cutoff, and release prompts, harness and per-task outputs, so that others can check a claimed gain at their own budget.
\end{itemize}

\section{Research Agenda}\label{sec:agenda}

The evidence maps (Fig.~\ref{fig:evidence-maps}) and per-family syntheses point to five open problems.

\paragraph{Measure what a technique does to everything, not only to its target.}
Most studies report their target and, at best, correctness (Table~\ref{tab:nontargeted}). A security technique with unknown effect on readability, or a readability technique with unknown effect on efficiency, cannot be adopted safely. A technique paper should at minimum report all three non-functional dimensions plus correctness, unconditionally and conditional on passing tests, even when non-targeted results are null.

\paragraph{Quality-aware selection and retrieval.}
Reranking and best-of-$n$ selection mostly use tests or correctness verifiers (\S\ref{sec:decoding}), retrieval relevance or security exemplars (\S\ref{sec:retrieval}). Verifiers and retrievers encoding maintainability or efficiency, and evidence that they help at a matched sampling budget, are largely missing. Best-of-$n$ under a quality verifier is also the natural budget-matched baseline for iterative techniques.

\paragraph{Multi-dimension rewards that resist gaming.}
Rewards built on a single analyzer or judge are gamed: policies learn to emit trivial or empty code that scores well, and reviewers reject analyzer-satisfying fixes (\S\ref{sec:feedback}, \S\ref{sec:training}). Open questions are how to combine correctness gates with several quality signals, keep evaluation and training signals independent, and detect reward hacking before deployment.

\paragraph{Default-behaviour evaluation.}
Several quality benchmarks measure quality only when the prompt requests it, testing instruction following rather than the code a model writes by default (\S\ref{sec:measurement}). Scoring code written for ordinary task descriptions, at repository level and over many edits, would capture the erosion that function-level benchmarks miss.

\paragraph{Contamination-resistant quality evaluation with cost accounting.}
Frontier training cutoffs now fall in 2026, after most quality benchmarks. Quality evaluation needs tasks that postdate the models, reference solutions for measuring efficiency, metrics validated against human judgement, and a record of each technique's tokens and calls, so quality gains can be separated from extra compute.

\section{Threats to Validity}\label{sec:threats}

We group threats as construct, internal, external and conclusion validity, for the review and the re-evaluation in turn.

\paragraph{Construct validity.}
\emph{Review.} Effect directions and quality targets were extracted by an LLM and reviewed by the author; vote counting ignores effect and sample size and depends on how ``improved'' is defined, so we use it to describe the evidence, not to rank techniques. The highlighted-technique tables report what each paper claims; only the head-to-head of \S\ref{sec:empirical-h2h} compares techniques under one setting.
\emph{Re-evaluation.} Pylint, Bandit and Safety drive the analyzer loop (T3) and the best-of-$n$ ranker (T7), and SonarQube also ingests their reports, so its issue counts are not fully independent of the techniques' analyzers; cognitive complexity, the judge, CWEval and BaxBench, to which T3's gains do not carry over, are reported for that reason and lead the results (Table~\ref{tab:emp-indep}). Issue densities are reported next to solution length and absolute counts (Supplementary Material, \S\ref{S-sec:app-loc}). The maintainability measures (issue counts, cognitive complexity and the judge score) are computed on every output, correct or not, and not also on correct outputs alone as \S\ref{sec:measurement} recommends, so a technique that lowers correctness could look more maintainable partly through short or degenerate failing programs; the efficiency scores, CWEval func-sec@1 and BaxBench sec\_pass@1 give no credit to a program that fails its tests. Budgets match model calls, not tokens: critic and reviewer turns are shorter or longer than a sample, the iterative techniques T3--T5 stop early (about $0.6\,B$ call-equivalents on average), and Opus~5.5 does not expose its reasoning tokens, so for Opus budget matching is call-based only; we report tokens per experiment (Supplementary Material, \S\ref{S-sec:app-ledger}) but no token-matched curves. Claude Opus~5.5 has five roles: title/abstract screener, extractor, \dimRM{} judge, evaluated model, and the critic and reviewer of its own T4 and T5 runs. Comparisons on one model are not affected by the judge's vendor, but the judge score of the Op row may include self-preference; a second judge from another vendor (GPT-6.1 Sol) therefore scores every \dimRM{} output, agrees at Spearman $\rho=\phJudgesRho{}$ per output and $\rho=\phJudgesModelRho{}$ across the thirteen models, and ranks Opus~5.5 as the Opus judge does; a correlation of $\phJudgesModelRho{}$ still allows neighbouring models to swap places, so we do not rank models that the two judges order differently. The first human rater is the author; to keep author bias out, the second rater is an engineer who is not an author. Both rated outputs blind to technique, model and every automated score, but they could recognise a technique's style (planning comments, for instance), and their agreement ($\kappa=\phRaterKappa{}$) is moderate; three further software engineers, none of them authors, rated a stratified sub-sample of 114 of the same outputs, agreeing with the two raters' mean rating at quadratic-weighted $\kappa=0.57$ (Spearman $\rho=0.62$) and with the judge at $\rho=0.67$. The second rater, the three further engineers and the second coder of the review all work at the author's company; none is an author, but they are the author's colleagues. RACE and NoFunEval measure agreement with a requirement or a reference rather than quality as such (\S\ref{sec:measurement}), which is why every output is also scored by our analyzers and judge.

\paragraph{Internal validity.}
\emph{Review.} Title/abstract screening used one author-reviewed LLM screener; its agreement with an independent second coder on a random sample is reported in \S\ref{sec:method-genai}, and the vote counts carry the extraction's disagreement rate as a range (Table~\ref{tab:effects}). The recall-favouring rule (include or uncertain goes forward) limits false exclusions there but not at full text, where a low-cost model removed reports it judged clearly ineligible. One LLM extractor did full-text eligibility, extraction and QA scoring; its author-reviewed effect directions and QA scores feed the evidence maps and vote counts. Numbers attributed to specific papers do not depend on it: each was checked against a verbatim passage of its primary source. The selection in \S\ref{sec:highlighted} rests on choices another team could make differently:
Semantic Scholar citation counts at a single point in time, favouring older and arXiv-visible work; relative gain
over each paper's own strongest baseline, which rewards weak baselines and studies evaluated on
their authors' own benchmarks; the restriction to general-purpose techniques; five exclusion
rules applied by LLM reviewers, with decisions recorded and reviewed by the author; and two decisions taken after the rule was first applied (the RL adjustment of the training family and the reclassification of Universal Textual Teaching), which \S\ref{sec:highlighted} reports as deviations.
\emph{Re-evaluation.} Reimplementations may differ from the originals; each was checked against a published result, with at most two adjustments and a stated stopping rule (\S\ref{sec:empirical-h2h}; Supplementary Material, \S\ref{S-sec:app-fidelity}). Closed APIs offer no temperature control, which three runs of every configuration address. The HumanEval+ losses on the two models with the highest single-shot pass@1 (about 96\%) and the gains on the model at 81\% admit two explanations: contamination of HumanEval+, and a ceiling or regression-to-the-mean effect, under which a model near the top of a benchmark can mostly lose. The post-cutoff contests show neither the losses nor large gains, which is consistent with both; we therefore use ``likely contaminated'' as a description of HumanEval+'s release date relative to the models' training, not as an established cause of the losses.

\paragraph{External validity.}
\emph{Review.} Terminology differs across software-engineering, machine-learning and security communities, and a quality-improving technique need not name a quality attribute in its abstract. We mitigated this with a broad quality facet, a technique-oriented supplementary string, a quasi-gold standard, a venue sweep, snowballing and section-level searches. Recall was 0.938 against the quasi-gold standard (optimistic, since the string was revised after inspecting its misses) and 0.775 at screening against a venue sweep. The revised strings ran on Semantic Scholar and arXiv; OpenAlex was queried with the original string only, so a study indexed only by OpenAlex and matched only by the revised strings would be missed. Two limits remain. Only reports whose full text could be retrieved were assessed, so \cnt{nFtNotRetrieved} of the \cnt{nFtSought} reports sought were excluded as not retrieved; this may under-represent venues without open-access copies. Of these reports, \cnt{nNrSince2025} date from 2025 or 2026, \cnt{nNrConference} are conference or workshop papers, \cnt{nNrJournal} journal articles and \cnt{nNrPreprint} preprints or repository deposits, and \cnt{nNrSameAsIncluded} are other versions of included studies (Supplementary Table~\ref{S-tab:app-notretrieved}). Unpaywall lists no open copy for \cnt{nNrClosed} of them, mostly IEEE (\cnt{nNrClosedIEEE}) and Springer (\cnt{nNrClosedSpringer}) publications, and an open copy for \cnt{nNrOpenPdf} with a PDF link and \cnt{nNrOpenLanding} with only a landing page, which manual retrieval could still close. Snowballing ran one backward and one forward iteration, so studies reachable only through longer citation chains may be missing. Preprints were included to cover 2025--2026; they are flagged and QA-scored, and a re-count on the peer-reviewed studies alone leaves the synthesis unchanged (\S\ref{sec:method-synthesis}; Supplementary Material, \S\ref{S-sec:app-peer}). Negative results are rarely published, so the literature likely overstates benefits; the re-evaluation partly offsets this.
\emph{Re-evaluation.} Most benchmarks are function-level Python and the main grid has four models. Its open models were chosen for a stated training cutoff, low cost and, for Gemma~4, a fine-tunable dense and MoE pair, so ERQ3's open-versus-closed contrast is also a capability contrast; the extended grid (\S\ref{sec:empirical-ext}) adds the leading open-weight models, measured on post-cutoff contests only. Technique effects are established on function-level benchmarks (and DeepSWE), while the field has moved to repository-level and agent tasks (\S\ref{sec:meas-benchmarks}). The head-to-head uses one mid-size open model, Gemma~4 31B IT, and its results may not carry over to other models.

\paragraph{Conclusion validity.}
We call a difference significant when it exceeds a 95\% paired bootstrap half-width computed per metric, model and budget against T7 at the same $B$ (Table~\ref{tab:emp-ci}; Supplementary Material, \S\ref{S-sec:app-stats}); comparisons against T0 reuse the $B=4$ half-widths, where the paired variance can differ, so those significance calls are approximate. Half-widths barely change with budget (median ratio of the $B=16$ to the $B=4$ half-width \phHwRatio{}). Every configuration ran three times with fresh seeds; the tables report the mean, and the standard deviation of a cell's metric across the three seeds has a median of \phSeedSD{} of the corresponding half-width (Supplementary Material, \S\ref{S-sec:app-stats}), so seed-to-seed variation is small against task-to-task variation. To control false discoveries across the 1{,}272 metric--model--technique--budget cells against T7, we computed two-sided paired bootstrap $p$-values and applied the Benjamini--Hochberg procedure at $q=0.05$: \phFdrSurvive{} of the \phFdrResolved{} differences significant under the half-width rule survive (204 of them at $B=4$, as counted in Table~\ref{tab:emp-tradeoff}b), including every significant difference cited in \S\ref{sec:empirical-results}. Conclusions rest on counts of significant cells and on patterns that repeat across models and benchmarks, not on any single cell. The head-to-head (E12) is indicative: it compares point estimates against fixed references, and its per-technique bootstrap intervals are reported in the Supplementary Material. HumanEval+ has the smallest half-widths and LiveCodeBench Pro the largest, so small correctness differences reach significance more easily on the likely contaminated benchmark (\S\ref{sec:empirical-results}). In the head-to-head, one technique of 66 beats best-of-$n$ on its primary target beyond the 95\% interval; with 66 comparisons, about two such wins in that direction are expected by chance alone, so this single win is not evidence that the technique is better.

\section{Conclusion}\label{sec:conclusion}

Functional correctness no longer separates frontier code generators, but their code still differs in readability, efficiency and security. A reproducible, LLM-assisted review of July 2021 to October 2026 literature organized them by stage, signal source and quality target and found a consistent pattern: techniques are evaluated on their target dimension, rarely on others, often without a matched compute budget, and sometimes at a cost to correctness that only a test or fitness gate prevents. Easily optimized signals, such as an analyzer's warning count or a judge's score, are also easily gamed.

The controlled re-evaluation, with matched budgets on four models and a head-to-head of the highlighted techniques, confirms this pattern. At a single call, quality prompting does not lower analyzer issue counts and barely moves complexity or judge scores, and best-of-$n$ sampling with the same calls, at four and at sixteen, gains most of what quality techniques gain over a single shot, also on measures the sampling ranker does not optimize; the techniques that do better, an analyzer loop and fine-tuning, improve only the issue counts their own analyzers compute, and on the strongest models almost every technique costs HumanEval+ correctness, a cost that largely disappears on contests and repository tasks no model has seen. Across thirteen current models, stronger models write cleaner and safer, but not faster, code by default, and even the best agents leave working but insecure code in about four in ten repository tasks. Technique effects are measured on function-level tasks, so their transfer to agent workflows remains open. For practitioners: compare any technique that promises better code with sampling at the same budget, and ask what it did to correctness and to the qualities it did not target.

\pagemark{bodyend}
\section*{Acknowledgments}
This work was funded by Cerby Inc., and all experiments ran on its infrastructure. We thank the software engineering team at Cerby Inc. for the second coding of the review, the second ratings and the developer ratings of the re-evaluation.

\IfFileExists{ACM-Reference-Format.bst}{\bibliographystyle{ACM-Reference-Format}}{\bibliographystyle{plainnat}}
\bibliography{bib/core,bib/02_background,bib/05_measurement,bib/07_prompting,bib/08_feedback_loops,bib/09_self_refinement,bib/10_multi_agent,bib/11_retrieval,bib/12_decoding,bib/13_training,bib/14_models}

\pagemark{refend}

\end{document}

\ifdefined\usingacmart\else\maketitle\fi
\renewcommand{\thesection}{S\arabic{section}}
\renewcommand{\thetable}{S\arabic{table}}
\renewcommand{\thefigure}{S\arabic{figure}}
\ifdefined\usingacmart\maketitle\fi

The replication package, to be released, contains every query, raw export, screening decision, prompt, full-text manifest, extraction row and script used in this review. This document reproduces the parts a reader needs to audit the method.

\FloatBarrier
\section{Search strings per source}\label{sec:app-search}
The exact strings sent to each source, with the raw hit count. OpenAlex does not support wildcards, so wildcard terms were expanded; arXiv applies stemming, so trailing wildcards were removed and the quality facet was split into chunks whose union equals the full query. The revised strings ran on Semantic Scholar and arXiv only (the main article's threats to validity).
\paragraph{S1 (iteration 0).}
\noindent\textit{OpenAlex} (reported hits 2769; records retrieved 2770).
{\scriptsize\ttfamily\raggedright title\_and\_abstract.search:("large language model" OR "large language models" OR LLM OR LLMs OR "language model" OR "language models" OR "code model" OR "code models" OR "foundation model" OR "foundation models" OR "code LLM" OR "code LLMs" OR GPT OR GPT-3 OR GPT-3.5 OR GPT-4 OR GPT-4o OR GPT-5 OR Codex OR ChatGPT OR Copilot) AND ("code generation" OR "program synthesis" OR "generated code" OR "code completion" OR "code synthesis" OR "generating code" OR "AI-generated code" OR "LLM-generated code" OR "machine-generated code" OR "coding assistant" OR "coding assistants") AND ("code quality" OR readability OR readable OR maintainability OR maintainable OR "code smell" OR "code smells" OR "technical debt" OR cyclomatic OR "cognitive complexity" OR idiomatic OR "coding style" OR "coding convention" OR "coding conventions" OR "style guide" OR "style guides" OR "non-functional" OR nonfunctional OR "code efficiency" OR "efficient code" OR "execution time" OR "runtime efficiency" OR "memory usage" OR "memory efficiency" OR "energy efficient" OR "energy efficiency" OR "performance-improving" OR "code optimization" OR "secure code" OR "code security" OR "insecure code" OR vulnerability OR vulnerabilities OR vulnerable OR CWE OR "security weakness" OR "security weaknesses"),from\_publication\_date:2021-07-07,to\_publication\_date:2026-10-14\par}\smallskip
\noindent\textit{arXiv} (reported hits 672; records retrieved 607).
{\scriptsize\ttfamily\raggedright ((ti:"large language model" OR abs:"large language model") OR (ti:LLM OR abs:LLM) OR (ti:LLMs OR abs:LLMs) OR (ti:"language model" OR abs:"language model") OR (ti:"code model" OR abs:"code model") OR (ti:"foundation model" OR abs:"foundation model") OR (ti:"code LLM" OR abs:"code LLM") OR (ti:GPT OR abs:GPT) OR (ti:Codex OR abs:Codex) OR (ti:ChatGPT OR abs:ChatGPT) OR (ti:Copilot OR abs:Copilot)) AND ((ti:"code generation" OR abs:"code generation") OR (ti:"program synthesis" OR abs:"program synthesis") OR (ti:"generated code" OR abs:"generated code") OR (ti:"code completion" OR abs:"code completion") OR (ti:"code synthesis" OR abs:"code synthesis") OR (ti:"generating code" OR abs:"generating code") OR (ti:"AI-generated code" OR abs:"AI-generated code") OR (ti:"LLM-generated code" OR abs:"LLM-generated code") OR (ti:"machine-generated code" OR abs:"machine-generated code") OR (ti:"coding assistant" OR abs:"coding assistant")) AND ((ti:"code quality" OR abs:"code quality") OR (ti:readab OR abs:readab) OR (ti:maintainab OR abs:maintainab) OR (ti:"code smell" OR abs:"code smell") OR (ti:"technical debt" OR abs:"technical debt") OR (ti:cyclomatic OR abs:cyclomatic) OR (ti:"cognitive complexity" OR abs:"cognitive complexity")) AND submittedDate:[202107070000 TO 202610142359]\par}\smallskip
{\scriptsize\ttfamily\raggedright ((ti:"large language model" OR abs:"large language model") OR (ti:LLM OR abs:LLM) OR (ti:LLMs OR abs:LLMs) OR (ti:"language model" OR abs:"language model") OR (ti:"code model" OR abs:"code model") OR (ti:"foundation model" OR abs:"foundation model") OR (ti:"code LLM" OR abs:"code LLM") OR (ti:GPT OR abs:GPT) OR (ti:Codex OR abs:Codex) OR (ti:ChatGPT OR abs:ChatGPT) OR (ti:Copilot OR abs:Copilot)) AND ((ti:"code generation" OR abs:"code generation") OR (ti:"program synthesis" OR abs:"program synthesis") OR (ti:"generated code" OR abs:"generated code") OR (ti:"code completion" OR abs:"code completion") OR (ti:"code synthesis" OR abs:"code synthesis") OR (ti:"generating code" OR abs:"generating code") OR (ti:"AI-generated code" OR abs:"AI-generated code") OR (ti:"LLM-generated code" OR abs:"LLM-generated code") OR (ti:"machine-generated code" OR abs:"machine-generated code") OR (ti:"coding assistant" OR abs:"coding assistant")) AND ((ti:idiomatic OR abs:idiomatic) OR (ti:"coding style" OR abs:"coding style") OR (ti:"coding convention" OR abs:"coding convention") OR (ti:"style guide" OR abs:"style guide") OR (ti:"non-functional" OR abs:"non-functional") OR (ti:nonfunctional OR abs:nonfunctional) OR (ti:"code efficiency" OR abs:"code efficiency")) AND submittedDate:[202107070000 TO 202610142359]\par}\smallskip
{\scriptsize\ttfamily\raggedright ((ti:"large language model" OR abs:"large language model") OR (ti:LLM OR abs:LLM) OR (ti:LLMs OR abs:LLMs) OR (ti:"language model" OR abs:"language model") OR (ti:"code model" OR abs:"code model") OR (ti:"foundation model" OR abs:"foundation model") OR (ti:"code LLM" OR abs:"code LLM") OR (ti:GPT OR abs:GPT) OR (ti:Codex OR abs:Codex) OR (ti:ChatGPT OR abs:ChatGPT) OR (ti:Copilot OR abs:Copilot)) AND ((ti:"code generation" OR abs:"code generation") OR (ti:"program synthesis" OR abs:"program synthesis") OR (ti:"generated code" OR abs:"generated code") OR (ti:"code completion" OR abs:"code completion") OR (ti:"code synthesis" OR abs:"code synthesis") OR (ti:"generating code" OR abs:"generating code") OR (ti:"AI-generated code" OR abs:"AI-generated code") OR (ti:"LLM-generated code" OR abs:"LLM-generated code") OR (ti:"machine-generated code" OR abs:"machine-generated code") OR (ti:"coding assistant" OR abs:"coding assistant")) AND ((ti:"efficient code" OR abs:"efficient code") OR (ti:"execution time" OR abs:"execution time") OR (ti:"runtime efficiency" OR abs:"runtime efficiency") OR (ti:"memory usage" OR abs:"memory usage") OR (ti:"memory efficiency" OR abs:"memory efficiency") OR (ti:"energy efficien" OR abs:"energy efficien") OR (ti:"performance-improving" OR abs:"performance-improving")) AND submittedDate:[202107070000 TO 202610142359]\par}\smallskip
{\scriptsize\ttfamily\raggedright ((ti:"large language model" OR abs:"large language model") OR (ti:LLM OR abs:LLM) OR (ti:LLMs OR abs:LLMs) OR (ti:"language model" OR abs:"language model") OR (ti:"code model" OR abs:"code model") OR (ti:"foundation model" OR abs:"foundation model") OR (ti:"code LLM" OR abs:"code LLM") OR (ti:GPT OR abs:GPT) OR (ti:Codex OR abs:Codex) OR (ti:ChatGPT OR abs:ChatGPT) OR (ti:Copilot OR abs:Copilot)) AND ((ti:"code generation" OR abs:"code generation") OR (ti:"program synthesis" OR abs:"program synthesis") OR (ti:"generated code" OR abs:"generated code") OR (ti:"code completion" OR abs:"code completion") OR (ti:"code synthesis" OR abs:"code synthesis") OR (ti:"generating code" OR abs:"generating code") OR (ti:"AI-generated code" OR abs:"AI-generated code") OR (ti:"LLM-generated code" OR abs:"LLM-generated code") OR (ti:"machine-generated code" OR abs:"machine-generated code") OR (ti:"coding assistant" OR abs:"coding assistant")) AND ((ti:"code optimization" OR abs:"code optimization") OR (ti:"secure code" OR abs:"secure code") OR (ti:"code security" OR abs:"code security") OR (ti:"insecure code" OR abs:"insecure code") OR (ti:vulnerab OR abs:vulnerab) OR (ti:CWE OR abs:CWE) OR (ti:"security weakness" OR abs:"security weakness")) AND submittedDate:[202107070000 TO 202610142359]\par}\smallskip
\noindent\textit{Semantic Scholar (bulk)} (raw hits 1442). Date filter: "2021-07-07:2026-10-14".
{\scriptsize\ttfamily\raggedright ("large language model*" \textbar{} LLM \textbar{} LLMs \textbar{} "language model*" \textbar{} "code model*" \textbar{} "foundation model*" \textbar{} "code LLM*" \textbar{} GPT* \textbar{} Codex \textbar{} ChatGPT \textbar{} Copilot) + ("code generation" \textbar{} "program synthesis" \textbar{} "generated code" \textbar{} "code completion" \textbar{} "code synthesis" \textbar{} "generating code" \textbar{} "AI-generated code" \textbar{} "LLM-generated code" \textbar{} "machine-generated code" \textbar{} "coding assistant*") + ("code quality" \textbar{} readab* \textbar{} maintainab* \textbar{} "code smell*" \textbar{} "technical debt" \textbar{} cyclomatic \textbar{} "cognitive complexity" \textbar{} idiomatic \textbar{} "coding style" \textbar{} "coding convention*" \textbar{} "style guide*" \textbar{} "non-functional" \textbar{} nonfunctional \textbar{} "code efficiency" \textbar{} "efficient code" \textbar{} "execution time" \textbar{} "runtime efficiency" \textbar{} "memory usage" \textbar{} "memory efficiency" \textbar{} "energy efficien*" \textbar{} "performance-improving" \textbar{} "code optimization" \textbar{} "secure code" \textbar{} "code security" \textbar{} "insecure code" \textbar{} vulnerab* \textbar{} CWE \textbar{} "security weakness*")\par}\smallskip
\paragraph{S1 after calibration iteration 1.}
\noindent\textit{arXiv} (reported hits 1068; records retrieved 980).
{\scriptsize\ttfamily\raggedright ((ti:"large language model" OR abs:"large language model") OR (ti:LLM OR abs:LLM) OR (ti:LLMs OR abs:LLMs) OR (ti:"language model" OR abs:"language model") OR (ti:"code model" OR abs:"code model") OR (ti:"foundation model" OR abs:"foundation model") OR (ti:"code LLM" OR abs:"code LLM") OR (ti:GPT OR abs:GPT) OR (ti:Codex OR abs:Codex) OR (ti:ChatGPT OR abs:ChatGPT) OR (ti:Copilot OR abs:Copilot) OR (ti:"code LM" OR abs:"code LM") OR (ti:LMs OR abs:LMs)) AND ((ti:"code generation" OR abs:"code generation") OR (ti:"program synthesis" OR abs:"program synthesis") OR (ti:"generated code" OR abs:"generated code") OR (ti:"code completion" OR abs:"code completion") OR (ti:"code synthesis" OR abs:"code synthesis") OR (ti:"generating code" OR abs:"generating code") OR (ti:"AI-generated code" OR abs:"AI-generated code") OR (ti:"LLM-generated code" OR abs:"LLM-generated code") OR (ti:"machine-generated code" OR abs:"machine-generated code") OR (ti:"coding assistant" OR abs:"coding assistant") OR (ti:"source code" OR abs:"source code") OR (ti:"code edit" OR abs:"code edit") OR (ti:"code repair" OR abs:"code repair") OR (ti:"vulnerability repair" OR abs:"vulnerability repair") OR (ti:"code refinement" OR abs:"code refinement") OR (ti:"program optimization" OR abs:"program optimization") OR (ti:"code optimization" OR abs:"code optimization") OR (ti:"programming task" OR abs:"programming task") OR (ti:"coding task" OR abs:"coding task") OR (ti:"code-related task" OR abs:"code-related task") OR (ti:"coding preference" OR abs:"coding preference") OR (ti:"code revision" OR abs:"code revision") OR (ti:codebase OR abs:codebase) OR (ti:"code efficiency" OR abs:"code efficiency") OR (ti:"Python code" OR abs:"Python code") OR (ti:"code snippet" OR abs:"code snippet")) AND ((ti:"code quality" OR abs:"code quality") OR (ti:readab OR abs:readab) OR (ti:maintainab OR abs:maintainab) OR (ti:"code smell" OR abs:"code smell") OR (ti:"technical debt" OR abs:"technical debt") OR (ti:cyclomatic OR abs:cyclomatic) OR (ti:"cognitive complexity" OR abs:"cognitive complexity")) AND submittedDate:[202107070000 TO 202610142359]\par}\smallskip
{\scriptsize\ttfamily\raggedright ((ti:"large language model" OR abs:"large language model") OR (ti:LLM OR abs:LLM) OR (ti:LLMs OR abs:LLMs) OR (ti:"language model" OR abs:"language model") OR (ti:"code model" OR abs:"code model") OR (ti:"foundation model" OR abs:"foundation model") OR (ti:"code LLM" OR abs:"code LLM") OR (ti:GPT OR abs:GPT) OR (ti:Codex OR abs:Codex) OR (ti:ChatGPT OR abs:ChatGPT) OR (ti:Copilot OR abs:Copilot) OR (ti:"code LM" OR abs:"code LM") OR (ti:LMs OR abs:LMs)) AND ((ti:"code generation" OR abs:"code generation") OR (ti:"program synthesis" OR abs:"program synthesis") OR (ti:"generated code" OR abs:"generated code") OR (ti:"code completion" OR abs:"code completion") OR (ti:"code synthesis" OR abs:"code synthesis") OR (ti:"generating code" OR abs:"generating code") OR (ti:"AI-generated code" OR abs:"AI-generated code") OR (ti:"LLM-generated code" OR abs:"LLM-generated code") OR (ti:"machine-generated code" OR abs:"machine-generated code") OR (ti:"coding assistant" OR abs:"coding assistant") OR (ti:"source code" OR abs:"source code") OR (ti:"code edit" OR abs:"code edit") OR (ti:"code repair" OR abs:"code repair") OR (ti:"vulnerability repair" OR abs:"vulnerability repair") OR (ti:"code refinement" OR abs:"code refinement") OR (ti:"program optimization" OR abs:"program optimization") OR (ti:"code optimization" OR abs:"code optimization") OR (ti:"programming task" OR abs:"programming task") OR (ti:"coding task" OR abs:"coding task") OR (ti:"code-related task" OR abs:"code-related task") OR (ti:"coding preference" OR abs:"coding preference") OR (ti:"code revision" OR abs:"code revision") OR (ti:codebase OR abs:codebase) OR (ti:"code efficiency" OR abs:"code efficiency") OR (ti:"Python code" OR abs:"Python code") OR (ti:"code snippet" OR abs:"code snippet")) AND ((ti:idiomatic OR abs:idiomatic) OR (ti:"coding style" OR abs:"coding style") OR (ti:"coding convention" OR abs:"coding convention") OR (ti:"style guide" OR abs:"style guide") OR (ti:"non-functional" OR abs:"non-functional") OR (ti:nonfunctional OR abs:nonfunctional) OR (ti:"code efficiency" OR abs:"code efficiency")) AND submittedDate:[202107070000 TO 202610142359]\par}\smallskip
{\scriptsize\ttfamily\raggedright ((ti:"large language model" OR abs:"large language model") OR (ti:LLM OR abs:LLM) OR (ti:LLMs OR abs:LLMs) OR (ti:"language model" OR abs:"language model") OR (ti:"code model" OR abs:"code model") OR (ti:"foundation model" OR abs:"foundation model") OR (ti:"code LLM" OR abs:"code LLM") OR (ti:GPT OR abs:GPT) OR (ti:Codex OR abs:Codex) OR (ti:ChatGPT OR abs:ChatGPT) OR (ti:Copilot OR abs:Copilot) OR (ti:"code LM" OR abs:"code LM") OR (ti:LMs OR abs:LMs)) AND ((ti:"code generation" OR abs:"code generation") OR (ti:"program synthesis" OR abs:"program synthesis") OR (ti:"generated code" OR abs:"generated code") OR (ti:"code completion" OR abs:"code completion") OR (ti:"code synthesis" OR abs:"code synthesis") OR (ti:"generating code" OR abs:"generating code") OR (ti:"AI-generated code" OR abs:"AI-generated code") OR (ti:"LLM-generated code" OR abs:"LLM-generated code") OR (ti:"machine-generated code" OR abs:"machine-generated code") OR (ti:"coding assistant" OR abs:"coding assistant") OR (ti:"source code" OR abs:"source code") OR (ti:"code edit" OR abs:"code edit") OR (ti:"code repair" OR abs:"code repair") OR (ti:"vulnerability repair" OR abs:"vulnerability repair") OR (ti:"code refinement" OR abs:"code refinement") OR (ti:"program optimization" OR abs:"program optimization") OR (ti:"code optimization" OR abs:"code optimization") OR (ti:"programming task" OR abs:"programming task") OR (ti:"coding task" OR abs:"coding task") OR (ti:"code-related task" OR abs:"code-related task") OR (ti:"coding preference" OR abs:"coding preference") OR (ti:"code revision" OR abs:"code revision") OR (ti:codebase OR abs:codebase) OR (ti:"code efficiency" OR abs:"code efficiency") OR (ti:"Python code" OR abs:"Python code") OR (ti:"code snippet" OR abs:"code snippet")) AND ((ti:"efficient code" OR abs:"efficient code") OR (ti:"execution time" OR abs:"execution time") OR (ti:"runtime efficiency" OR abs:"runtime efficiency") OR (ti:"memory usage" OR abs:"memory usage") OR (ti:"memory efficiency" OR abs:"memory efficiency") OR (ti:"energy efficien" OR abs:"energy efficien") OR (ti:"performance-improving" OR abs:"performance-improving")) AND submittedDate:[202107070000 TO 202610142359]\par}\smallskip
{\scriptsize\ttfamily\raggedright ((ti:"large language model" OR abs:"large language model") OR (ti:LLM OR abs:LLM) OR (ti:LLMs OR abs:LLMs) OR (ti:"language model" OR abs:"language model") OR (ti:"code model" OR abs:"code model") OR (ti:"foundation model" OR abs:"foundation model") OR (ti:"code LLM" OR abs:"code LLM") OR (ti:GPT OR abs:GPT) OR (ti:Codex OR abs:Codex) OR (ti:ChatGPT OR abs:ChatGPT) OR (ti:Copilot OR abs:Copilot) OR (ti:"code LM" OR abs:"code LM") OR (ti:LMs OR abs:LMs)) AND ((ti:"code generation" OR abs:"code generation") OR (ti:"program synthesis" OR abs:"program synthesis") OR (ti:"generated code" OR abs:"generated code") OR (ti:"code completion" OR abs:"code completion") OR (ti:"code synthesis" OR abs:"code synthesis") OR (ti:"generating code" OR abs:"generating code") OR (ti:"AI-generated code" OR abs:"AI-generated code") OR (ti:"LLM-generated code" OR abs:"LLM-generated code") OR (ti:"machine-generated code" OR abs:"machine-generated code") OR (ti:"coding assistant" OR abs:"coding assistant") OR (ti:"source code" OR abs:"source code") OR (ti:"code edit" OR abs:"code edit") OR (ti:"code repair" OR abs:"code repair") OR (ti:"vulnerability repair" OR abs:"vulnerability repair") OR (ti:"code refinement" OR abs:"code refinement") OR (ti:"program optimization" OR abs:"program optimization") OR (ti:"code optimization" OR abs:"code optimization") OR (ti:"programming task" OR abs:"programming task") OR (ti:"coding task" OR abs:"coding task") OR (ti:"code-related task" OR abs:"code-related task") OR (ti:"coding preference" OR abs:"coding preference") OR (ti:"code revision" OR abs:"code revision") OR (ti:codebase OR abs:codebase) OR (ti:"code efficiency" OR abs:"code efficiency") OR (ti:"Python code" OR abs:"Python code") OR (ti:"code snippet" OR abs:"code snippet")) AND ((ti:"code optimization" OR abs:"code optimization") OR (ti:"secure code" OR abs:"secure code") OR (ti:"code security" OR abs:"code security") OR (ti:"insecure code" OR abs:"insecure code") OR (ti:vulnerab OR abs:vulnerab) OR (ti:CWE OR abs:CWE) OR (ti:"security weakness" OR abs:"security weakness")) AND submittedDate:[202107070000 TO 202610142359]\par}\smallskip
\noindent\textit{Semantic Scholar (bulk)} (raw hits 2718). Date filter: "2021-07-07:2026-10-14".
{\scriptsize\ttfamily\raggedright ("large language model*" \textbar{} LLM \textbar{} LLMs \textbar{} "language model*" \textbar{} "code model*" \textbar{} "foundation model*" \textbar{} "code LLM*" \textbar{} GPT* \textbar{} Codex \textbar{} ChatGPT \textbar{} Copilot \textbar{} "code LM*" \textbar{} LMs) + ("code generation" \textbar{} "program synthesis" \textbar{} "generated code" \textbar{} "code completion" \textbar{} "code synthesis" \textbar{} "generating code" \textbar{} "AI-generated code" \textbar{} "LLM-generated code" \textbar{} "machine-generated code" \textbar{} "coding assistant*" \textbar{} "source code" \textbar{} "code edit*" \textbar{} "code repair" \textbar{} "vulnerability repair" \textbar{} "code refinement" \textbar{} "program optimization" \textbar{} "code optimization" \textbar{} "programming task*" \textbar{} "coding task*" \textbar{} "code-related task*" \textbar{} "coding preference*" \textbar{} "code revision*" \textbar{} codebase* \textbar{} "code efficiency" \textbar{} "Python code" \textbar{} "code snippet*") + ("code quality" \textbar{} readab* \textbar{} maintainab* \textbar{} "code smell*" \textbar{} "technical debt" \textbar{} cyclomatic \textbar{} "cognitive complexity" \textbar{} idiomatic \textbar{} "coding style" \textbar{} "coding convention*" \textbar{} "style guide*" \textbar{} "non-functional" \textbar{} nonfunctional \textbar{} "code efficiency" \textbar{} "efficient code" \textbar{} "execution time" \textbar{} "runtime efficiency" \textbar{} "memory usage" \textbar{} "memory efficiency" \textbar{} "energy efficien*" \textbar{} "performance-improving" \textbar{} "code optimization" \textbar{} "secure code" \textbar{} "code security" \textbar{} "insecure code" \textbar{} vulnerab* \textbar{} CWE \textbar{} "security weakness*")\par}\smallskip
\paragraph{S2 after calibration iteration 1.}
\noindent\textit{arXiv} (reported hits 2923; records retrieved 2530).
{\scriptsize\ttfamily\raggedright ((ti:"large language model" OR abs:"large language model") OR (ti:LLM OR abs:LLM) OR (ti:LLMs OR abs:LLMs) OR (ti:"language model" OR abs:"language model") OR (ti:"code model" OR abs:"code model") OR (ti:"foundation model" OR abs:"foundation model") OR (ti:"code LLM" OR abs:"code LLM") OR (ti:GPT OR abs:GPT) OR (ti:Codex OR abs:Codex) OR (ti:ChatGPT OR abs:ChatGPT) OR (ti:Copilot OR abs:Copilot) OR (ti:"code LM" OR abs:"code LM") OR (ti:LMs OR abs:LMs)) AND ((ti:"code generation" OR abs:"code generation") OR (ti:"program synthesis" OR abs:"program synthesis") OR (ti:"generated code" OR abs:"generated code") OR (ti:"code completion" OR abs:"code completion") OR (ti:"code synthesis" OR abs:"code synthesis") OR (ti:"generating code" OR abs:"generating code") OR (ti:"AI-generated code" OR abs:"AI-generated code") OR (ti:"LLM-generated code" OR abs:"LLM-generated code") OR (ti:"machine-generated code" OR abs:"machine-generated code") OR (ti:"coding assistant" OR abs:"coding assistant") OR (ti:"source code" OR abs:"source code") OR (ti:"code edit" OR abs:"code edit") OR (ti:"code repair" OR abs:"code repair") OR (ti:"vulnerability repair" OR abs:"vulnerability repair") OR (ti:"code refinement" OR abs:"code refinement") OR (ti:"program optimization" OR abs:"program optimization") OR (ti:"code optimization" OR abs:"code optimization") OR (ti:"programming task" OR abs:"programming task") OR (ti:"coding task" OR abs:"coding task") OR (ti:"code-related task" OR abs:"code-related task") OR (ti:"coding preference" OR abs:"coding preference") OR (ti:"code revision" OR abs:"code revision") OR (ti:codebase OR abs:codebase) OR (ti:"code efficiency" OR abs:"code efficiency") OR (ti:"Python code" OR abs:"Python code") OR (ti:"code snippet" OR abs:"code snippet")) AND ((ti:"self-refin" OR abs:"self-refin") OR (ti:"self-repair" OR abs:"self-repair") OR (ti:"self-debug" OR abs:"self-debug") OR (ti:reflexion OR abs:reflexion) OR (ti:"self-critique" OR abs:"self-critique") OR (ti:"multi-agent" OR abs:"multi-agent") OR (ti:"static analy" OR abs:"static analy")) AND submittedDate:[202107070000 TO 202610142359]\par}\smallskip
{\scriptsize\ttfamily\raggedright ((ti:"large language model" OR abs:"large language model") OR (ti:LLM OR abs:LLM) OR (ti:LLMs OR abs:LLMs) OR (ti:"language model" OR abs:"language model") OR (ti:"code model" OR abs:"code model") OR (ti:"foundation model" OR abs:"foundation model") OR (ti:"code LLM" OR abs:"code LLM") OR (ti:GPT OR abs:GPT) OR (ti:Codex OR abs:Codex) OR (ti:ChatGPT OR abs:ChatGPT) OR (ti:Copilot OR abs:Copilot) OR (ti:"code LM" OR abs:"code LM") OR (ti:LMs OR abs:LMs)) AND ((ti:"code generation" OR abs:"code generation") OR (ti:"program synthesis" OR abs:"program synthesis") OR (ti:"generated code" OR abs:"generated code") OR (ti:"code completion" OR abs:"code completion") OR (ti:"code synthesis" OR abs:"code synthesis") OR (ti:"generating code" OR abs:"generating code") OR (ti:"AI-generated code" OR abs:"AI-generated code") OR (ti:"LLM-generated code" OR abs:"LLM-generated code") OR (ti:"machine-generated code" OR abs:"machine-generated code") OR (ti:"coding assistant" OR abs:"coding assistant") OR (ti:"source code" OR abs:"source code") OR (ti:"code edit" OR abs:"code edit") OR (ti:"code repair" OR abs:"code repair") OR (ti:"vulnerability repair" OR abs:"vulnerability repair") OR (ti:"code refinement" OR abs:"code refinement") OR (ti:"program optimization" OR abs:"program optimization") OR (ti:"code optimization" OR abs:"code optimization") OR (ti:"programming task" OR abs:"programming task") OR (ti:"coding task" OR abs:"coding task") OR (ti:"code-related task" OR abs:"code-related task") OR (ti:"coding preference" OR abs:"coding preference") OR (ti:"code revision" OR abs:"code revision") OR (ti:codebase OR abs:codebase) OR (ti:"code efficiency" OR abs:"code efficiency") OR (ti:"Python code" OR abs:"Python code") OR (ti:"code snippet" OR abs:"code snippet")) AND ((ti:linter OR abs:linter) OR (ti:"compiler feedback" OR abs:"compiler feedback") OR (ti:"retrieval-augmented" OR abs:"retrieval-augmented") OR (ti:"constrained decoding" OR abs:"constrained decoding") OR (ti:rerank OR abs:rerank) OR (ti:"best-of-n" OR abs:"best-of-n") OR (ti:"preference optimization" OR abs:"preference optimization")) AND submittedDate:[202107070000 TO 202610142359]\par}\smallskip
{\scriptsize\ttfamily\raggedright ((ti:"large language model" OR abs:"large language model") OR (ti:LLM OR abs:LLM) OR (ti:LLMs OR abs:LLMs) OR (ti:"language model" OR abs:"language model") OR (ti:"code model" OR abs:"code model") OR (ti:"foundation model" OR abs:"foundation model") OR (ti:"code LLM" OR abs:"code LLM") OR (ti:GPT OR abs:GPT) OR (ti:Codex OR abs:Codex) OR (ti:ChatGPT OR abs:ChatGPT) OR (ti:Copilot OR abs:Copilot) OR (ti:"code LM" OR abs:"code LM") OR (ti:LMs OR abs:LMs)) AND ((ti:"code generation" OR abs:"code generation") OR (ti:"program synthesis" OR abs:"program synthesis") OR (ti:"generated code" OR abs:"generated code") OR (ti:"code completion" OR abs:"code completion") OR (ti:"code synthesis" OR abs:"code synthesis") OR (ti:"generating code" OR abs:"generating code") OR (ti:"AI-generated code" OR abs:"AI-generated code") OR (ti:"LLM-generated code" OR abs:"LLM-generated code") OR (ti:"machine-generated code" OR abs:"machine-generated code") OR (ti:"coding assistant" OR abs:"coding assistant") OR (ti:"source code" OR abs:"source code") OR (ti:"code edit" OR abs:"code edit") OR (ti:"code repair" OR abs:"code repair") OR (ti:"vulnerability repair" OR abs:"vulnerability repair") OR (ti:"code refinement" OR abs:"code refinement") OR (ti:"program optimization" OR abs:"program optimization") OR (ti:"code optimization" OR abs:"code optimization") OR (ti:"programming task" OR abs:"programming task") OR (ti:"coding task" OR abs:"coding task") OR (ti:"code-related task" OR abs:"code-related task") OR (ti:"coding preference" OR abs:"coding preference") OR (ti:"code revision" OR abs:"code revision") OR (ti:codebase OR abs:codebase) OR (ti:"code efficiency" OR abs:"code efficiency") OR (ti:"Python code" OR abs:"Python code") OR (ti:"code snippet" OR abs:"code snippet")) AND ((ti:DPO OR abs:DPO) OR (ti:"reinforcement learning" OR abs:"reinforcement learning") OR (ti:"fine-tun" OR abs:"fine-tun") OR (ti:"prompt engineering" OR abs:"prompt engineering") OR (ti:"prompt technique" OR abs:"prompt technique") OR (ti:"prompting technique" OR abs:"prompting technique") OR (ti:"chain-of-thought" OR abs:"chain-of-thought")) AND submittedDate:[202107070000 TO 202610142359]\par}\smallskip
{\scriptsize\ttfamily\raggedright ((ti:"large language model" OR abs:"large language model") OR (ti:LLM OR abs:LLM) OR (ti:LLMs OR abs:LLMs) OR (ti:"language model" OR abs:"language model") OR (ti:"code model" OR abs:"code model") OR (ti:"foundation model" OR abs:"foundation model") OR (ti:"code LLM" OR abs:"code LLM") OR (ti:GPT OR abs:GPT) OR (ti:Codex OR abs:Codex) OR (ti:ChatGPT OR abs:ChatGPT) OR (ti:Copilot OR abs:Copilot) OR (ti:"code LM" OR abs:"code LM") OR (ti:LMs OR abs:LMs)) AND ((ti:"code generation" OR abs:"code generation") OR (ti:"program synthesis" OR abs:"program synthesis") OR (ti:"generated code" OR abs:"generated code") OR (ti:"code completion" OR abs:"code completion") OR (ti:"code synthesis" OR abs:"code synthesis") OR (ti:"generating code" OR abs:"generating code") OR (ti:"AI-generated code" OR abs:"AI-generated code") OR (ti:"LLM-generated code" OR abs:"LLM-generated code") OR (ti:"machine-generated code" OR abs:"machine-generated code") OR (ti:"coding assistant" OR abs:"coding assistant") OR (ti:"source code" OR abs:"source code") OR (ti:"code edit" OR abs:"code edit") OR (ti:"code repair" OR abs:"code repair") OR (ti:"vulnerability repair" OR abs:"vulnerability repair") OR (ti:"code refinement" OR abs:"code refinement") OR (ti:"program optimization" OR abs:"program optimization") OR (ti:"code optimization" OR abs:"code optimization") OR (ti:"programming task" OR abs:"programming task") OR (ti:"coding task" OR abs:"coding task") OR (ti:"code-related task" OR abs:"code-related task") OR (ti:"coding preference" OR abs:"coding preference") OR (ti:"code revision" OR abs:"code revision") OR (ti:codebase OR abs:codebase) OR (ti:"code efficiency" OR abs:"code efficiency") OR (ti:"Python code" OR abs:"Python code") OR (ti:"code snippet" OR abs:"code snippet")) AND ((ti:"few-shot" OR abs:"few-shot") OR (ti:"in-context learning" OR abs:"in-context learning")) AND submittedDate:[202107070000 TO 202610142359]\par}\smallskip
\noindent\textit{Semantic Scholar (bulk)} (raw hits 3617). Date filter: "2021-07-07:2026-10-14".
{\scriptsize\ttfamily\raggedright ("large language model*" \textbar{} LLM \textbar{} LLMs \textbar{} "language model*" \textbar{} "code model*" \textbar{} "foundation model*" \textbar{} "code LLM*" \textbar{} GPT* \textbar{} Codex \textbar{} ChatGPT \textbar{} Copilot \textbar{} "code LM*" \textbar{} LMs) + ("code generation" \textbar{} "program synthesis" \textbar{} "generated code" \textbar{} "code completion" \textbar{} "code synthesis" \textbar{} "generating code" \textbar{} "AI-generated code" \textbar{} "LLM-generated code" \textbar{} "machine-generated code" \textbar{} "coding assistant*" \textbar{} "source code" \textbar{} "code edit*" \textbar{} "code repair" \textbar{} "vulnerability repair" \textbar{} "code refinement" \textbar{} "program optimization" \textbar{} "code optimization" \textbar{} "programming task*" \textbar{} "coding task*" \textbar{} "code-related task*" \textbar{} "coding preference*" \textbar{} "code revision*" \textbar{} codebase* \textbar{} "code efficiency" \textbar{} "Python code" \textbar{} "code snippet*") + ("self-refin*" \textbar{} "self-repair" \textbar{} "self-debug*" \textbar{} reflexion \textbar{} "self-critique" \textbar{} "multi-agent" \textbar{} "static analy*" \textbar{} linter \textbar{} "compiler feedback" \textbar{} "retrieval-augmented" \textbar{} "constrained decoding" \textbar{} rerank* \textbar{} "best-of-n" \textbar{} "preference optimization" \textbar{} DPO \textbar{} "reinforcement learning" \textbar{} "fine-tun*" \textbar{} "prompt engineering" \textbar{} "prompt technique*" \textbar{} "prompting technique*" \textbar{} "chain-of-thought" \textbar{} "few-shot" \textbar{} "in-context learning")\par}\smallskip

\FloatBarrier
\section{Title and abstract screening prompt}\label{sec:app-prompt}
The frozen system prompt used for title/abstract screening (Claude Opus~5.5, \texttt{effort=low}, ten records per request; records refused as a group were re-run one per request). The full-text eligibility and extraction prompts will be released with the replication package.
{\scriptsize\ttfamily\raggedright
You are screening records (title and abstract only) for a systematic literature review in software engineering. Apply the criteria below exactly. You will receive several records; decide each one independently.\par
\# Review topic\par
Techniques that improve the NON-FUNCTIONAL quality of code generated or edited by large language models (LLMs).\par
Definitions:\par
- LLM: a pretrained transformer language model used for code (code-specialized or general, any size, closed or open).\par
- Technique: any intervention that changes the code an LLM produces. It may act on the input (prompt or context), on decoding, on a post-generation loop (tool feedback, self-refinement, agents, critics), or on the model's weights (fine-tuning, preference optimization, reinforcement learning).\par
- Non-functional (NF) quality: (1) readability and maintainability (naming, structure, complexity, code smells, idiomaticity, documentation, style conformance); (2) performance efficiency (runtime, memory, energy); (3) security (vulnerabilities, CWE). Correctness (tests passing) is NOT an NF dimension.\par
\# Inclusion criteria (a study must meet ALL)\par
- IC1: It proposes or empirically evaluates one or more techniques applied to an LLM.\par
- IC2: The evaluated artifact is code generated or edited by an LLM in a generation, completion or editing task.\par
- IC3: It reports at least one quantitative NF quality outcome for that code (tool-based, execution-based, LLM-judge or human-rated all count).\par
- IC4: It is a full paper (peer-reviewed, or a preprint of 4+ pages).\par
\# Exclusion criteria (exclude if ANY holds)\par
- EC1: Not in English.\par
- EC2: A secondary study (survey, systematic review, mapping study).\par
- EC3: The target artifact is not production code (tests only, comments/documentation only, commit messages, code review comments, natural-language explanations).\par
- EC4: Purely the analysis of human-written code: vulnerability detection, smell detection, or repair/refactoring of human-written code with no LLM generation.\par
- EC5: Quality measured only as correctness, compilability or syntactic validity.\par
- EC6: Model-level robustness, safety or attack papers whose outcome is not the quality of the generated code (jailbreaks, adversarial prompts, backdoors, data poisoning, prompt-injection defences).\par
- EC7: Pre-LLM neural code models without large-scale pretraining (seq2seq/RNN generators).\par
- EC8: A duplicate or earlier version of another study (only decidable when obvious from the record).\par
\# Side lists (record these even when the decision is exclude)\par
- context: measures the NF quality of LLM-generated code WITHOUT any intervention (e.g., a security audit of Copilot output, an empirical study of smells in ChatGPT code, a benchmark that measures NF quality of LLM code).\par
- adjacent: LLM refactoring or repair of human-written code (the EC4 group).\par
- correctness-only: an influential code-generation technique that reports only correctness (EC5), e.g. self-debugging, test-based reranking, multi-agent coding frameworks evaluated on pass@k only.\par
- secondary: a survey or review relevant to the topic (EC2).\par
- none: none of the above.\par
\# Decision rules\par
- "include": the title/abstract indicates IC1-IC4 are likely met and no EC applies.\par
- "exclude": at least one exclusion criterion clearly applies, or an inclusion criterion clearly fails.\par
- "uncertain": the abstract does not make it clear whether a technique is evaluated on LLM-generated code with a quantitative NF outcome. When in doubt between include and exclude, choose "uncertain" (this review favours recall). A technique paper that mentions security, efficiency, readability, maintainability, code smells, or code quality as an evaluated outcome is at least "uncertain".\par
- criterion: for include, write "IC"; for exclude, the single most decisive code (EC1-EC8, or IC1/IC2/IC3/IC4 if an inclusion criterion fails); for uncertain, the criterion you could not decide.\par
\# Output format\par
Return ONLY a JSON array, one object per record, in the same order as the records, with exactly these keys:\par
{}[\{"rid": "\textless{}record id\textgreater{}", "decision": "include\textbar{}exclude\textbar{}uncertain", "criterion": "\textless{}code\textgreater{}", "side\_list": "context\textbar{}adjacent\textbar{}correctness-only\textbar{}secondary\textbar{}none", "reason": "\textless{}one sentence, max 25 words\textgreater{}"\}]\par}

\FloatBarrier
\section{Data extraction schema}\label{sec:app-schema}
One row per study and technique, with controlled vocabularies in brackets: bibliographic data and venue type [peer-reviewed conference, journal, workshop, preprint]; study type [technique proposal, technique evaluation]; technique name; family [prompting, feedback loop, self-refinement, multi-agent, retrieval, decoding, training, hybrid, other]; stage [pre-generation, in-decoding, post-generation loop, training-time]; signal source [none, self, other LLM, static analyzer, compiler, test execution, profiler, retrieved knowledge, human, learned verifier, reward model]; tools and versions; iterations and stopping rule; dimensions targeted and measured [readability/maintainability, efficiency, security] and whether correctness was measured; metrics per dimension with instrument type [tool-static, execution, LLM judge, human, similarity to reference]; benchmarks; languages; models; baselines; direction of effect on the target, on correctness and on each other dimension [$+$, $0$, $-$, mixed, not reported] with the headline numbers as printed; compute reported [calls, tokens, cost, time, none]; contamination handled [yes, partial, no]; human evaluation [none, rating, preference] with annotators and agreement; artifact URL; reimplementability [yes, partial, no]; the ten QA items; notes; extractor; confidence [high, medium, low].

\FloatBarrier
\section{Included studies and quality assessment}\label{sec:app-studies}
Table~\ref{tab:app-studies} lists the included studies and Table~\ref{tab:app-qa} their quality-assessment scores; the main article summarizes them by family. Both come from LLM-assisted extraction reviewed by the author. Each of ten items is scored 0, 0.5 or 1; the score is reported per study and weights confidence in the synthesis, but no study is excluded for it. QA1: the technique is described well enough to reimplement. QA2: it is compared with an appropriate baseline (no technique, or a strong prior technique). QA3: the quality metric is established or validated (a standard tool, or human evaluation with reported agreement). QA4: correctness is reported alongside quality. QA5: more than one model is evaluated, or a single model is justified. QA6: variance, repeated runs or a statistical test is reported. QA7: compute or cost (calls, tokens) is reported. QA8: code or data are available. QA9: contamination is considered. QA10: threats to validity are discussed.
{\scriptsize
\begin{longtable}{@{}p{0.07\linewidth}p{0.37\linewidth}p{0.05\linewidth}p{0.12\linewidth}p{0.08\linewidth}p{0.16\linewidth}@{}}
\caption{Included primary studies (LLM-assisted screening and extraction, reviewed by the author). ID = record identifier in the replication package; Targets = targeted dimensions (none: the study targets correctness and reports the quality dimensions listed); Src = arXiv ID or DOI.}\label{tab:app-studies}\\
\toprule
ID & Title & Year & Family & Targets & Src \\
\midrule
\endfirsthead
\toprule
ID & Title & Year & Family & Targets & Src \\
\midrule
\endhead
V00036 & Planning with Large Language Models for Code Generation & 2023 & Decoding & R/M & {\ttfamily\raggedright arXiv:\allowbreak{}2303.\allowbreak{}05510} \\
R00867 & Constrained Decoding for Secure Code Generation & 2024 & Decoding & SEC & {\ttfamily\raggedright arXiv:\allowbreak{}2405.\allowbreak{}00218} \\
S01754 & Make Every Move Count: LLM-based High-Quality RTL Code Generation Using MCTS & 2024 & Decoding & EFF & {\ttfamily\raggedright arXiv:\allowbreak{}2402.\allowbreak{}03289} \\
V00117 & LLMs Meet Library Evolution: Evaluating Deprecated API Usage in LLM-Based Code Completion & 2024 & Decoding, Prompting & R/M & {\ttfamily\raggedright arXiv:\allowbreak{}2406.\allowbreak{}09834} \\
R02790 & A Mixture of Linear Corrections Generates Secure Code & 2025 & Decoding & SEC & {\ttfamily\raggedright arXiv:\allowbreak{}2507.\allowbreak{}09508} \\
R03351 & HiLDE: Intentional Code Generation via Human-in-the-Loop Decoding & 2025 & Decoding & SEC & {\ttfamily\raggedright arXiv:\allowbreak{}2505.\allowbreak{}22906} \\
S02752 & A Preliminary Study on the Robustness of Code Generation by Large Language Models & 2025 & Decoding & SEC & {\ttfamily\raggedright arXiv:\allowbreak{}2503.\allowbreak{}20197} \\
V00051 & Rethinking Repetition Problems of LLMs in Code Generation & 2025 & Decoding & R/M & {\ttfamily\raggedright arXiv:\allowbreak{}2505.\allowbreak{}10402} \\
\bottomrule
\end{longtable}
}

{\scriptsize
\begin{longtable}{@{}lccccccccccr@{}}
\caption{Per-study quality assessment (items QA1--QA10 of \S S4; 0, 0.5 or 1 per item; LLM-assisted, reviewed by the author).}\label{tab:app-qa}\\
\toprule
ID & Q1 & Q2 & Q3 & Q4 & Q5 & Q6 & Q7 & Q8 & Q9 & Q10 & Total \\
\midrule
\endfirsthead
\toprule
ID & Q1 & Q2 & Q3 & Q4 & Q5 & Q6 & Q7 & Q8 & Q9 & Q10 & Total \\
\midrule
\endhead
O00001 & 0.5 & 1 & 0.5 & 1 & 1 & 0.5 & 0 & 1 & 0.5 & 0 & 6.0 \\
O00002 & 1 & 1 & 0.5 & 1 & 1 & 0.5 & 1 & 1 & 0.5 & 0 & 7.5 \\
O00007 & 0.5 & 0.5 & 1 & 0.5 & 1 & 0.5 & 0 & 1 & 0 & 1 & 6.0 \\
O00008 & 1 & 1 & 0.5 & 0 & 1 & 1 & 0.5 & 1 & 0 & 0.5 & 6.5 \\
O00015 & 1 & 1 & 1 & 0 & 1 & 0 & 0.5 & 1 & 0.5 & 0.5 & 6.5 \\
O00021 & 1 & 1 & 0.5 & 0.5 & 1 & 0.5 & 0 & 1 & 1 & 1 & 7.5 \\
O00031 & 1 & 1 & 0.5 & 1 & 1 & 1 & 0.5 & 0 & 0.5 & 1 & 7.5 \\
R00040 & 0 & 0.5 & 0 & 0 & 0 & 0 & 0 & 0 & 0 & 0 & 0.5 \\
\bottomrule
\end{longtable}
}

\section{Model releases, January 2025 to October 2026}\label{sec:app-models}
Table~\ref{tab:app-models} lists every model release summarized in the model landscape table of the main article and the other frontier and major open-weight releases of the period, with its release month and the vendor or primary source that names and dates it. It covers general-purpose, reasoning and code models; image, video, speech, embedding and safety-classifier models, and models specialized for a single domain (such as medicine, translation or theorem proving), are not listed. The month is the first public release stated by the vendor (API, product or weights, whichever came first); the cited sources are model cards, technical reports, release notes, change logs and announcements. Further vendor sources for the model landscape table of the main article: \citep{anthropic2025claude,anthropic2025introducing,anthropic2026claude,googledeepmind2026gemini,googledeepmind2026geminic,mistralai2025mistralaidevstral2123binstruct2512,openai2025gpt52,openai2026gpt56b,openai2026gpt6b}.

{\footnotesize
\setlength{\tabcolsep}{3pt}
\begin{longtable}{@{}>{\raggedright\arraybackslash}p{0.50\linewidth}>{\raggedright\arraybackslash}p{0.10\linewidth}>{\centering\arraybackslash}p{0.05\linewidth}>{\raggedright\arraybackslash}p{0.29\linewidth}@{}}
\caption{Frontier and major open-weight model releases, January 2025 to October 2026 (338 rows), grouped by family in release order. Released: year and month of the first public release, as stated in the cited source (\nr: no verified source states it). \textsuperscript{p}First public release was a preview, experimental or beta version. \textsuperscript{y}The source states day and month only; the year follows from the dated post or the year section of the page in which the entry appears. OW: open weights (Y), API or product only (N), or only some of the models in the row (p); \textsuperscript{a}weights announced, not yet released. Names as the vendor writes them; API model identifiers in parentheses where the vendor names the release by its identifier.}\label{tab:app-models}\\
\toprule
Release & Released & OW & Source\\
\midrule
\endfirsthead
\toprule
Release & Released & OW & Source\\
\midrule
\endhead
\midrule
\multicolumn{4}{r}{\textit{continued on next page}}\\
\endfoot
\bottomrule
\endlastfoot
\multicolumn{4}{@{}l}{\textbf{Claude (Anthropic)}}\\
\quad Claude Sonnet 3.7 (claude-3-7-sonnet-20250219) & 2025-02 & N & \citep{anthropic2026claudeb,anthropic2026model}\\
\quad Claude Opus 4 & 2025-05 & N & \citep{anthropic2026claudeb}\\
\quad Claude Sonnet 4 & 2025-05 & N & \citep{anthropic2026claudeb}\\
\quad Claude Opus 4.1 & 2025-08 & N & \citep{anthropic2026claudeb}\\
\quad Claude Sonnet 4.5 & 2025-09 & N & \citep{anthropic2026claudeb}\\
\quad Claude Haiku 4.5 & 2025-10 & N & \citep{anthropic2026claudeb}\\
\quad Claude Opus 4.5 & 2025-11 & N & \citep{anthropic2026claudeb}\\
\end{longtable}
}

\FloatBarrier
\section{Domain-specific highlighted techniques}\label{sec:app-hl-domain}
Techniques that pass every eligibility check of the main article except general applicability
are listed here by domain; they are excluded from the head-to-head of the main article
because they cannot run on its general-purpose benchmarks.
{\scriptsize
\begin{longtable}{@{}p{0.17\linewidth}p{0.07\linewidth}p{0.42\linewidth}p{0.05\linewidth}p{0.1\linewidth}p{0.05\linewidth}@{}}
\caption{Domain-specific techniques: eligible techniques tied to a narrow artifact or domain, which cannot run on the shared general-purpose benchmarks of the main article's head-to-head; up to three per domain, by Semantic Scholar citations, October 2026. ID = record identifier in the replication package. LLM-assisted screening and extraction reviewed by the author.}\label{tab:hl-domain}\\
\toprule
Domain & ID & Title & Year & Family & Cit.\\
\midrule
\endfirsthead
\toprule
Domain & ID & Title & Year & Family & Cit.\\
\midrule
\endhead
Compiler and IR transforms & R03592 & Agentic Auto-Scheduling: An Experimental Study of LLM-Guided Loop Optimization & 2025 & Tool feedback & 6\\
Compiler and IR transforms & R05239 & HintPilot: LLM-based Compiler Hint Synthesis for Code Optimization & 2026 & Retrieval & 4\\
Compiler and IR transforms & R03953 & LOOPRAG: Enhancing Loop Transformation Optimization with Retrieval-Augmented Large Language Models & 2025 & Retrieval & 3\\
Data science & R05154 & How to Build AI Agents by Augmenting LLMs with Codified Human Expert Domain Knowledge? A Software Engineering Framework & 2026 & Retrieval & 3\\
Data science & R02038 & Engineering Systems for Data Analysis Using Interactive Structured Inductive Programming & 2025 & Prompting & 3\\
Data science & S04182 & Diff-SQL: SQL Efficiency Optimization via Patch Generation and Constraint Alignment & 2026 & Self-refinement & 0\\
GPU and accelerator kernels & S03111 & Kevin: Multi-Turn RL for Generating CUDA Kernels & 2025 & Training & 91\\
GPU and accelerator kernels & S03263 & Towards Robust Agentic CUDA Kernel Benchmarking, Verification, and Optimization & 2025 & Tool feedback & 72\\
\bottomrule
\end{longtable}}

\subsection{Reports not retrieved}\label{sec:app-notretrieved}
The reports sought but not retrieved, by year, type, publisher, search source and open-access status (main article, \S\ref{M-sec:threats}).

\begin{table}[H]
\centering
\scriptsize
\setlength{\tabcolsep}{3pt}
\caption{The 687 reports sought but not retrieved (no full text), by year, type, publisher, the search that found them and open-access status. Type, publisher and open-access status come from the Unpaywall and Crossref records retrieved during full-text retrieval (no record: no DOI or no Unpaywall entry). Of these, 15 are another version of an included study.}\label{tab:app-notretrieved}
\begin{tabular}{@{}lrr@{\hspace{14pt}}lrr@{}}
\toprule
 & $n$ & \% & & $n$ & \% \\
\midrule
\multicolumn{3}{@{}l}{\textit{Year}} & \multicolumn{3}{@{}l}{\textit{Publisher}} \\
\quad 2022 & 5 & 1 & \quad IEEE & 192 & 28 \\
\quad 2023 & 24 & 3 & \quad ACM & 108 & 16 \\
\quad 2024 & 79 & 11 & \quad Springer & 70 & 10 \\
\quad 2025 & 235 & 34 & \quad Elsevier & 43 & 6 \\
\quad 2026 & 344 & 50 & \quad MDPI & 26 & 4 \\
\multicolumn{3}{@{}l}{\textit{Type}} & \quad SSRN or Zenodo & 79 & 11 \\
\quad Conference or workshop paper & 327 & 48 & \quad Other & 80 & 12 \\
\quad Journal article & 199 & 29 & \quad No DOI & 89 & 13 \\
\quad Preprint or repository & 99 & 14 & \multicolumn{3}{@{}l}{\textit{Found by}} \\
\quad Thesis & 29 & 4 & \quad OpenAlex only & 207 & 30 \\
\quad Other or unknown & 33 & 5 & \quad Semantic Scholar only & 177 & 26 \\
\multicolumn{3}{@{}l}{\textit{Open access (Unpaywall)}} & \quad Both databases & 150 & 22 \\
\quad Closed & 305 & 44 & \quad Snowballing & 147 & 21 \\
\quad Open, PDF link & 90 & 13 & \quad Venue sweep & 6 & 1 \\
\quad Open, landing page only & 154 & 22 & & & \\
\quad No record & 138 & 20 & & & \\
\bottomrule
\end{tabular}
\end{table}

\clearpage
\section{Additional Tables}\label{sec:app-tables}

The tables below support the main text; each is referenced where its content is discussed.

\subsection{Earlier studies of LLM-generated code quality}\label{sec:app-context}
\begin{table}[H]
\caption{Context studies of the non-functional quality of LLM-generated code (no technique
proposed). Grey literature is marked as such. Values as reported by the original authors;
settings differ across rows, so values are not directly comparable.}
\label{tab:bg-context}
\footnotesize
\begin{tabularx}{\linewidth}{@{}>{\raggedright\arraybackslash}p{2.1cm}>{\raggedright\arraybackslash}p{1.15cm}>{\raggedright\arraybackslash}p{3.6cm}X@{}}
\toprule
Study & Dim. & Setting & Headline finding (as reported) \\
\midrule
\citet{pearce2022asleep} & \dimSec & Copilot; scenarios built around high-risk CWEs & about \rep{40\%} of programs vulnerable \\
\citet{perry2023do} & \dimSec & User study; codex-davinci-002 assistant & \rep{36\%} of assisted vs.\ \rep{7\%} of control participants wrote SQL-injectable code \\
\citet{asare2023is} & \dimSec & Copilot; C/C++ scenarios with known human vulnerabilities & replicates vulnerable code about \rep{33\%} of the time, the fix at a \rep{25\%} rate \\
\citet{fu2025security} & \dimSec & Copilot, CodeWhisperer and Codeium snippets in GitHub projects & weaknesses in \rep{29.5\%} of Python and \rep{24.2\%} of JavaScript snippets \\
\citet{vero2025baxbench} & \dimSec, \dimCor & BaxBench backend-generation tasks & best model (o1) \rep{62\%} correct; exploits succeed on about half of correct programs \\
\citet{zhao2026is} & \dimSec, \dimCor & SusVibes repository tasks; SWE-Agent with Claude 4 Sonnet & \rep{57\%} functionally correct, \rep{11.8\%} secure \\
\citet{chen2026securevibebench} & \dimSec, \dimCor & SecureVibeBench C/C++ tasks; several agents and LLMs & best agent \rep{23.8\%} correct and secure \\
\citet{wu2026last} & \dimSec, \dimCor & \rep{450} agent tasks from \rep{150} real vulnerabilities, \rep{7} languages & \rep{73.6\%} to \rep{91.3\%} functional; best agent \rep{25.1\%} functional and secure \\
\midrule
\citet{liu2024refining} & \dimRM, \dimCor & ChatGPT; Java and Python programming tasks & \rep{1,930} of \rep{4,066} programs with maintainability issues \\
\citet{paul2025investigating} & \dimRM & Java; Gemini Pro, ChatGPT, Codex, Falcon vs.\ references & average smell increase \rep{63.34\%} over references \\
\citet{he2025speed} & \dimRM & GitHub projects adopting Cursor vs.\ matched controls & persistent rise in static-analysis warnings and code complexity \\
\citet{liu2026debt} & \dimRM, \dimSec & AI-authored commits in GitHub repositories & smells are \rep{89.3\%} of issues; \rep{22.7\%} of issues survive to latest revision \\
\citet{gitclear2025ai} (grey) & \dimRM & Changed lines in private and open-source repositories & copy-pasted lines rose from \rep{8.3\%} to \rep{12.3\%} \\
\citet{sonar2025coding} (grey) & \dimRM, \dimSec & SonarQube on Java assignments from several LLMs & Claude 3.7 Sonnet to Sonnet 4: pass rate \rep{6.3\%} higher, bug severity \rep{93\%} higher \\
\midrule
\citet{huang2024effibench} & \dimEff & EffiBench LeetCode problems; GPT-4 & average execution time \rep{3.12} times the canonical solution \\
\citet{du2024mercury} & \dimEff & Mercury Python tasks & leading models \rep{65\%} on Pass but below \rep{50\%} on Beyond \\
\citet{qing2025effibenchx} & \dimEff & EffiBench-X, multi-language & most efficient model (Qwen3-32B) reaches about \rep{62\%} of human efficiency \\
\midrule
\citet{orlanski2026slopcodebench} & \dimRM & SlopCodeBench; coding agents vs.\ open-source Python repositories & agent code \rep{2.3x} more verbose and \rep{2.0x} more eroded; best agent passes \rep{14.8\%} of checkpoints \\
\bottomrule
\end{tabularx}
\end{table}

\subsection{Benchmarks}\label{sec:app-bench}
\begin{table}[H]
\caption{Benchmarks for correctness, readability/maintainability and efficiency. Dim.: \dimCor{}
correctness, \dimRM{} readability/maintainability, \dimEff{} efficiency, \dimSec{} security. Year: first
public release. Size: tasks (problems, prompts or instances) as defined by each paper. Contam.:
\emph{pub.}\ = problems public before 2024; \emph{date} = release-date filter against model cutoffs;
\emph{held-out} = private or copyleft-only split; \emph{new} = newly written tasks; -- = no specific
measure reported. Values as reported by the original authors; settings differ across rows, so values
are not directly comparable.}
\label{tab:meas-benchmarks}
\scriptsize
\begin{tabularx}{\linewidth}{@{}>{\raggedright\arraybackslash}p{3.3cm} c l l r >{\raggedright\arraybackslash}X l@{}}
\toprule
Benchmark & Year & Dim. & Lang. & Size & Oracle / metric & Contam. \\
\midrule
HumanEval+ \citep{liu2023is} & 2023 & \dimCor & Py & \rep{164} & amplified tests; pass@$k$ & pub. \\
MBPP+ \citep{liu2023is} & 2023 & \dimCor & Py & \nr & subset of MBPP (\rep{974}) \citep{austin2021program}; pass@$k$ & pub. \\
LiveCodeBench \citep{jain2025livecodebench} & 2024 & \dimCor & Py & \rep{511} & contest tests; pass@1 & date \\
BigCodeBench \citep{zhuo2025bigcodebench} & 2024 & \dimCor & Py & \rep{1,140} & library-call tests; pass@1 & -- \\
SWE-bench Verified \citep{jimenez2024swebench,valsai2026swebench} & 2024 & \dimCor & Py & \rep{500} & repo tests; resolve rate & pub. \\
SWE-bench Pro \citep{deng2026swebench} & 2025 & \dimCor & multi & \rep{1,865} & repo tests; resolve rate & held-out \\
SWE-Bench Pro Verified \citep{zheng2026swebenchproverified} & 2026 & \dimCor & multi & \nr & repaired tasks, anti-leakage safeguards & held-out \\
DeepSWE \citep{huang2026deepswe} & 2026 & \dimCor & \rep{5} langs & \rep{113} & hand-written verifiers; pass@1 & new \\
Terminal-Bench \citep{merrill2026terminalbench} & 2026 & \dimCor & multi & \nr & task checks in containers; resolve rate & new \\
LiveCodeBench Pro \citep{zheng2025livecodebench} & 2025 & \dimCor & -- & \nr & contest tests; pass@1 & date \\
SWE-rebench \citep{badertdinov2025swe} & 2025 & \dimCor & Py & \nr & repo tests; rolling fresh tasks & date \\
SWE-bench-Live \citep{zhang2025swe} & 2025 & \dimCor & Py & \rep{1,319} & repo tests; resolve rate & date \\
Multi-SWE-bench \citep{zan2025multi} & 2025 & \dimCor & \rep{7} langs & \rep{1,632} & repo tests; resolve rate & -- \\
\midrule
RACE \citep{zheng2024beyond} & 2024 & \dimRM\,\dimEff\,\dimCor & Py & \nr & MI, naming, comments, complexity on request & pub. \\
NoFunEval \citep{singhal2024nofuneval} & 2024 & \dimRM\,\dimEff\,\dimSec & multi & \rep{958} & DiffBLEU, DiffBLEU$\times$CodeQL, speedup & -- \\
MaintainBench \citep{wang2025maintaincoder} & 2025 & \dimRM\,\dimCor & Py & \nr & effort under requirement changes & pub. \\
SlopCodeBench \citep{orlanski2026slopcodebench} & 2026 & \dimRM\,\dimCor & Py & \rep{36} & erosion, verbosity over \rep{196} checkpoints & new \\
CQBench \citep{improta2026what} & 2026 & \dimRM\,\dimSec & multi & \rep{27,346} & static analysis mapped to ODC and CWE & -- \\
FrontierCode \citep{cognition2026frontiercode} & 2026 & \dimRM\,\dimCor & multi & \rep{100} & blockers + quality rubric; correct rate & held-out \\
SWE-Refactor \citep{xu2026swe} & 2026 & \dimRM\,\dimCor & Java & \rep{1,099} & compile, tests, refactoring detection & -- \\
SmellBench \citep{lin2026smellbench} & 2026 & \dimRM\,\dimCor & \nr & \rep{294} & smell elimination, localization & new \\
CodeTaste \citep{thillen2026codetaste} & 2026 & \dimRM\,\dimCor & \nr & \nr & repo tests + dataflow pattern checks & -- \\
SWE-Bench ProMax \citep{shi2026swe} & 2026 & \dimRM\,\dimCor & \rep{7} langs & \rep{170} & reviewed tests on refactors & -- \\
SWE-CI \citep{chen2026swe} & 2026 & \dimRM\,\dimCor & \nr & \rep{100} & correctness over commit history & -- \\
SWE-NFI \citep{xue2026swe} & 2026 & \dimRM\,\dimCor & Py & \rep{188} & \rep{92} executable NFI rules & -- \\
SWE-IF \citep{zhong2025swe} & 2025 & \dimRM\,\dimCor & \nr & \nr & \rep{30} verifiable instructions & pub. \\
CIFE \citep{gunnu2025cife} & 2025 & \dimRM\,\dimSec & Py & \rep{1,000} & constraint adherence; C2A score & -- \\
\midrule
PIE \citep{shypula2024learning} & 2023 & \dimEff & C++ & \rep{41} & speedup in gem5 & pub. \\
EffiBench \citep{huang2024effibench} & 2024 & \dimEff & Py & \rep{1,000} & normalized time and memory & pub. \\
Mercury \citep{du2024mercury} & 2024 & \dimEff & Py & \rep{1,889} & Beyond (runtime percentile) & pub. \\
EvalPerf \citep{liu2024evaluating} & 2024 & \dimEff & Py & \rep{121} & DPS on CPU instructions & pub. \\
ENAMEL \citep{qiu2025how} & 2024 & \dimEff & Py & \rep{142} & eff@$k$ vs.\ expert solutions & pub. \\
ECCO \citep{waghjale2024ecco} & 2024 & \dimEff & Py & \rep{48} & runtime, memory; test split & pub. \\
COFFE \citep{peng2025coffe} & 2025 & \dimEff & Py & \rep{398}+\rep{358} & efficient@$k$ on CPU instructions & pub. \\
EffiBench-X \citep{qing2025effibenchx} & 2025 & \dimEff & \rep{6} langs & \rep{623} & time, memory vs.\ human expert & date \\
Venus \citep{du2025afterburner} & 2025 & \dimEff & Py & \rep{300} & runtime/memory percentile (test set) & pub. \\
BigO(Bench) \citep{chambon2025bigobench} & 2025 & \dimEff & Py & \rep{3,105} & inferred time/space complexity & pub. \\
GSO \citep{shetty2025gso} & 2025 & \dimEff & multi & \rep{102} & speedup vs.\ expert commit & -- \\
SWE-Perf \citep{he2026sweperf} & 2025 & \dimEff & Py & \rep{140} & runtime gain vs.\ expert PR & -- \\
SWE-Pro \citep{sarkayak2026evaluating} & 2026 & \dimEff & multi & \rep{102} & runtime, peak memory, TWMU & -- \\
SWE-fficiency \citep{ma2025swe} & 2025 & \dimEff\,\dimCor & Py & \rep{498} & speedup vs.\ expert PR; unit tests & -- \\
\bottomrule
\end{tabularx}
\end{table}

\begin{table}[H]
\caption{Benchmarks for security (\dimSec). +\dimCor: functional correctness is tested on the same
outputs. Columns as in Table~\ref{tab:meas-benchmarks}. Values as reported by the original authors;
settings differ across rows, so values are not directly comparable.}
\label{tab:meas-security}
\footnotesize
\begin{tabularx}{\linewidth}{@{}>{\raggedright\arraybackslash}p{3.3cm} c l l r >{\raggedright\arraybackslash}X l@{}}
\toprule
Benchmark & Year & & Lang. & Size & Oracle / metric & Contam. \\
\midrule
Asleep at the Keyboard \citep{pearce2022asleep} & 2021 &  & C/Py & \rep{89} & CodeQL + manual; CWE & pub. \\
SecurityEval \citep{siddiq2022securityeval} & 2022 &  & Py & \rep{121} & static analyzers; no tests \citep{siddiq2024sallm} & pub. \\
LLMSecEval \citep{tony2023llmseceval} & 2023 &  & C/Py & \rep{150} & CodeQL & pub. \\
CyberSecEval \citep{bhatt2023purple} & 2023 &  & multi & \nr & ICD rules (\rep{189}); v3 \citep{wan2024cyberseceval} & pub. \\
CodeLMSec \citep{hajipour2023codelmsec} & 2023 &  & C/Py & \nr & non-secure prompts found by few-shot model inversion & pub. \\
PythonSecurityEval \citep{alrashedy2023can} & 2023 &  & Py & \rep{470} & Bandit, CodeQL & pub. \\
SALLM \citep{siddiq2024sallm} & 2023 & +\dimCor & Py & \rep{100} & secure@$k$, vulnerable@$k$ & -- \\
SeCodePLT \citep{nie2024secodeplt} & 2024 & +\dimCor & multi & \rep{5.9k} & dynamic security tests & -- \\
CWEval \citep{peng2025cweval} & 2025 & +\dimCor & \rep{5} langs & \rep{119} & outcome oracles; func-sec@$k$ & new \\
BaxBench \citep{vero2025baxbench} & 2025 & +\dimCor & multi & \rep{392} & exploits; sec\_pass@$k$ & new \\
SecRepoBench \citep{shen2026secrepobench} & 2025 & +\dimCor & C/C++ & \rep{318} & repo tests + security tests & -- \\
SecureVibeBench \citep{chen2026securevibebench} & 2025 & +\dimCor & C/C++ & \rep{105} & static + dynamic oracles & -- \\
SusVibes \citep{zhao2026is} & 2025 & +\dimCor & Py & \rep{186} & functional + security tests & -- \\
RealSec-bench \citep{wang2026realsecbench} & 2026 & +\dimCor & Java & \rep{105} & SecurePass@$K$ & -- \\
SecCodeBench-V2 \citep{chen2026seccodebenchv2} & 2026 & +\dimCor & multi & \rep{98} & PoC tests; LLM judge as fallback & -- \\
SafeGenBench \citep{li2025safegenbench} & 2025 &  & multi & \nr & SAST + LLM judge & -- \\
Last Secure Code \citep{wu2026last} & 2026 & +\dimCor & \rep{7} langs & \rep{450} & hidden functional + security tests & new \\
\bottomrule
\end{tabularx}
\end{table}

\subsection{Model landscape}\label{sec:app-landscape}
{\footnotesize
\setlength{\tabcolsep}{2pt}
\begin{longtable}{@{}>{\raggedright\arraybackslash}p{0.305\linewidth}>{\raggedright\arraybackslash}p{0.10\linewidth}>{\centering\arraybackslash}p{0.03\linewidth}>{\raggedright\arraybackslash}p{0.125\linewidth}>{\raggedright\arraybackslash}p{0.205\linewidth}>{\raggedright\arraybackslash}p{0.095\linewidth}>{\raggedright\arraybackslash}p{0.07\linewidth}@{}}
\caption{Frontier and major open-weight models released between January 2025 and October 2026. Vendor sources (release posts and model cards) for every row are listed in \S\ref{sec:app-models}. Each row groups successive releases of one family; where a row has several models, the architecture, scores and cutoff refer to the model named in the cell (otherwise to the newest release). The model table of \S\ref{sec:app-models} gives the release month and source of every release named here. Released: first and last month among the dated releases of the row. All values are vendor-reported (model cards, technical reports, vendor announcements) under the vendor's own harness and effort setting. OW: open weights (p: some releases). Arch.: total / active parameters. Cutoff: training-data or knowledge cutoff as stated by the vendor. NF: non-functional evaluation of generated code reported by the vendor (\dimSec{} security, \dimEff{} efficiency; FC: FrontierCode v1.1, a merge-readiness composite that includes correctness; Hum.: blind human rating of coding quality; Edit: rate of extraneous code edits). SWE-V: SWE-bench Verified; SWE-P: SWE-bench Pro; TB: Terminal-Bench (version); LCB: LiveCodeBench; DS: DeepSWE v1.1. \textsuperscript{a}Weights of Mistral Large 4 announced, not yet released. Values as reported by the original authors; settings differ across rows, so values are not directly comparable.}\label{tab:models-landscape}\\
\toprule
Family (releases) & Released & OW & Arch. & Vendor-reported coding scores & Cutoff & NF\\
\midrule
\endfirsthead
\toprule
Family (releases) & Released & OW & Arch. & Vendor-reported coding scores & Cutoff & NF\\
\midrule
\endhead
\midrule
\multicolumn{7}{r}{\textit{continued on next page}}\\
\endfoot
\bottomrule
\endlastfoot
\textbf{Anthropic.} Claude Sonnet 3.7; Opus 4, 4.1, 4.5; Sonnet 4, 4.5; Haiku 4.5 & 2025-02--11 & N & \nr & Sonnet 4.5: SWE-V \rep{77.2\%} & \rep{Jul 2025} & \nr \\
Opus 4.6, 4.7, 4.8, 5; Sonnet 4.6, 5; Fable 5 (with Mythos 5) \citep{anthropic2026system} & 2026-02--07 & N & \nr & Fable 5: SWE-V \rep{95\%}; SWE-P \rep{80\%} & \rep{Jan 2026} & \nr \\
Fable 5.1 (with Mythos 5.1); Opus 5.5; Sonnet 5.5; Haiku 5.5 \citep{anthropic2026introducing,anthropic2026models} & 2026-09--10 & N & \nr & Opus 5.5: TB 4.0 \rep{66.4\%}; FC \rep{54.4\%} & \rep{Jun 2026} & FC \\
\textbf{OpenAI.} o3-mini; GPT-4.5; GPT-4.1; o3, o4-mini; codex-1 \citep{openai2025introducing} & 2025-01--05 & N & \nr & GPT-4.1: SWE-V \rep{54.6\%}; Aider Polyglot \rep{52.9\%} & \rep{June 2024} & Edit \\
gpt-oss-120b, gpt-oss-20b \citep{openai2025gptoss120b} & 2025-08 & Y & MoE, \rep{116.8B} / \rep{5.1B} & 120b (high): SWE-V \rep{62.4} & \rep{June 2024} & \nr \\
GPT-5; GPT-5-Codex; GPT-5.1, 5.1-Codex-Max; GPT-5.2, 5.2-Codex & 2025-08--12 & N & \nr & GPT-5.2 Thinking: SWE-P \rep{55.6\%}; SWE-V \rep{80.0\%} & \rep{Aug 31, 2025} & \nr \\
GPT-5.3-Codex; GPT-5.4; GPT-5.5; GPT-5.6 Sol, Terra, Luna \citep{openai2026gpt56} & 2026-02--07 & N & \nr & GPT-5.6 Sol: SWE-P \rep{64.6\%}; DS \rep{72.7\%} & \rep{Feb 16, 2026} & \nr \\
GPT-6 Astra; GPT-6 Sol, Luna; GPT-6.1 Sol \citep{openai2026gpt6} & 2026-09 & N & \nr & Astra: TB 4.0 \rep{57.9\%}; DS \rep{74.1\%} & \rep{Apr 30, 2026} & FC \\
\textbf{Google.} Gemini 2.0 Flash, 2.0 Pro; Gemini 2.5 Pro, Flash \citep{comanici2025gemini} & 2025-02--04 & N & Sparse MoE & 2.5 Pro: LCB \rep{74.2\%}; Aider Polyglot \rep{82.2\%} & \rep{January 2025} & \nr \\
Gemini 3 Pro; 3 Flash; 3.1 Pro & 2025-11--2026-02 & N & \nr & 3.1 Pro: SWE-V \rep{80.6\%}; SWE-P \rep{54.2\%} & \nr & \nr \\
Gemini 3.5, 3.6, 3.7, 3.8 Flash \citep{googledeepmind2026geminib} & 2026-05--09 & N & \nr & 3.8 Flash: TB 2.1 \rep{89.4\%}; DS \rep{73.7\%} & \rep{March 2026} & FC (3.7) \\
Gemini 4 Argon \citep{googledeepmind2026geminid} & 2026-10 & N & Sparse MoE & DS \rep{77.9\%}; TB 4.0 \rep{57.4\%} & \rep{March 2026} & \nr \\
Gemma 3 (1B--27B); Gemma 3n \citep{googledeepmind2025gemma} & 2025-03--06 & Y & Dense & 27B: LCB \rep{29.7} & \rep{August 2024} & \nr \\
Gemma 4 (E2B--31B) & 2026-03 & Y & 26B A4B: MoE, \rep{25.2B} / \rep{3.8B} & 31B: LCB v6 \rep{80.0\%} & \rep{January 2025} & \nr \\
\textbf{xAI / SpaceXAI.} Grok 3; Grok 4, 4 Heavy; Grok Code Fast 1; Grok 4 Fast; Grok 4.1, 4.1 Fast & 2025-02--11 & N & \nr & Code Fast 1: SWE-V \rep{70.8\%}; Grok 4: LCB \rep{79.0 \%} & \nr & \nr \\
Grok 4.20; Grok 4.3; Grok Build 0.1; Grok 4.5, 4.6, 4.7 \citep{spacexai2026model,spacexai2026modelb} & 2026-03--09 & N & \nr & 4.7: TB 4.0 \rep{38.0\%}; DS \rep{71.0\%} & \rep{June 2026} & \dimSec, FC (4.6) \\
\textbf{Meta.} Llama 4 Scout, Maverick \citep{meta2025metallamallama4maverick17b128einstruct} & 2025-04 & Y & MoE, \rep{400B} / \rep{17B} & Maverick: LCB \rep{43.4} & \rep{August 2024} & \nr \\
CWM (code) \citep{team2025cwm} & 2025-09 & Y & Dense, \rep{32}B & SWE-V \rep{65.8 \%} & \nr & \nr \\
Muse Spark; 1.1; 1.2; 1.3 \citep{meta2026muse} & 2026-04--09 & N & \nr & \nr{} (charts only) & \nr & \dimSec \\
Muse Glimmer (Muse-Glimmer-30B) & 2026-08 & Y & Dense, \rep{29.6B} & SWE-P \rep{51.2}; SWE-V \rep{76.0} & \rep{January 4, 2026} & \nr \\
\textbf{Mistral.} Codestral 25.01; Small 3, 3.1, 3.2; Medium 3; Magistral Small and Medium 1.0; Magistral Small 1.2 & 2025-01--09 & p & \nr & Codestral 25.01: LCB \rep{37.9\%} & \nr & \nr \\
Devstral Small 1.0; Devstral Small 1.1, Devstral Medium; Devstral 2, Devstral Small 2 (code) \citep{mistralai2025introducing} & 2025-05--12 & p & Dense, \rep{123B} & Devstral 2: SWE-V \rep{72.2\%}; TB 2 \rep{32.6\%} & \nr & \nr \\
Mistral Large 3, Ministral 3; Small 4; Medium 3.5; Large 4 (preview) \citep{mistralai2026introducing} & 2025-12--2026-10 & Y\textsuperscript{a} & MoE, \rep{1 trillion} / \rep{52 billion} & Large 4: DS \rep{61.7\%}; TB 4 \rep{28.3\%} & \nr & Hum. \\
\textbf{Ai2.} OLMo 2 32B; Olmo 3, 3.1 \citep{olmo2025olmo} & 2025-03--12 & Y & Dense & Olmo 3.1 Think 32B: LCB v3 \rep{83.3} & \rep{December 31, 2024} & \nr \\
SERA-8B, 14B, 32B (code agents) \citep{shen2026sera} & 2026-01--02 & Y & Dense & 32B: SWE-V \rep{49.5\%} & \nr & \nr \\
\textbf{NVIDIA.} Llama-Nemotron; Nemotron Nano 2; Nemotron 3 Nano, Super, Ultra; Nemotron 3.5 Lightning \citep{nvidia2026nvidianvidianemotron3ultra550ba55bbf16} & 2025-03--2026-08 & Y & 3 Ultra: hybrid MoE, \rep{550B} / \rep{55B} & 3 Ultra: SWE-V \rep{70.7}; TB 2.1 \rep{56.4} & \rep{September 2025} & \nr \\
\textbf{DeepSeek.} R1; V3-0324; R1-0528; V3.1; V3.1-Terminus; V3.2-Exp; V3.2 \citep{deepseekai2025deepseekv32} & 2025-01--12 & Y & \nr & V3.2: SWE-V \rep{73.1}; TB 2.0 \rep{46.4} & \nr & \nr \\
V4 preview (V4-Pro, V4-Flash); V4-Flash-0731; V4-Pro-0813 \citep{deepseekai2026deepseekv4} & 2026-04--08 & Y & V4-Pro: MoE, \rep{1.6T} / \rep{49B} & V4-Pro-0813: TB 2.1 \rep{87.9}; DS \rep{62.7} & \nr & \nr \\
V4.1-Flash \citep{deepseekai2026deepseekv41flash} & 2026-09 & Y & \rep{552B} / \rep{16B} & DS \rep{74.2\%}; TB 2.1 \rep{90.6\%} & \nr & \nr \\
\textbf{Qwen.} Qwen2.5-Max; QwQ-32B; Qwen3; Qwen3-2507; Qwen3-Next; Qwen3-Max \citep{yang2025qwen3} & 2025-01--09 & p & Qwen3-235B: MoE, \rep{235 billion} / \rep{22 billion} & Qwen3-235B: LCB v5 \rep{70.7} & \nr & \nr \\
Qwen3-Coder; Qwen3-Coder-Next (code) \citep{qwenteam2025qwen3coder,cao2026qwen3codernext} & 2025-07--2026-02 & Y & Coder: MoE, \rep{480B} / \rep{35B} & Coder-Next: SWE-V \rep{70.6\%} & \nr & \dimSec{} (Next) \\
Qwen3.5 (397B-A17B, Plus, Flash); Qwen3.6-Plus, 3.6-Max-Preview, 3.6-35B-A3B, 3.6-27B; Qwen3.7-Max, Plus; Qwen3.8-Max, 3.8-2.4T-A95B, 3.8-27B, 3.8-Flash-Next \citep{qwenteam2026qwenqwen3824ta95b,qwenteam2026qwenqwen35397ba17b} & 2026-02--08 & p & 3.8-2.4T-A95B: MoE, \rep{2.4T} / \rep{95B} & Qwen3.8-Max: TB 2.1 \rep{86.6}; SWE-P \rep{67.7} & \nr & \dimSec{} (3.5) \\
\textbf{Z.ai.} GLM-4.5, 4.5-Air; GLM-4.6; GLM-4.7 \citep{team2025glm45} & 2025-09--12 & Y & GLM-4.5: MoE, \rep{355B} / \rep{32B} & GLM-4.7: SWE-V \rep{73.8}; TB 2.0 \rep{41.0} & \nr & \nr \\
GLM-5; GLM-5.1; GLM-5.2; GLM-5.3, 5.3-Flash \citep{glmteam2026glm5,zai2026zaiorgglm53} & 2026-02--08 & Y & GLM-5: MoE, \rep{744B} / \rep{40B} & GLM-5.3: TB 2.1 \rep{88.2}; DS \rep{66.9} & \nr & \nr \\
\textbf{Moonshot.} Kimi k1.5; K2, K2-0905; K2 Thinking \citep{team2025kimi} & 2025-07--11 & p & K2: MoE, \rep{1 trillion} / \rep{32 billion} & K2 Thinking: SWE-V \rep{71.3}; LCB v6 \rep{83.1} & \nr & \nr \\
Kimi K2.5; K2.6; K2.7 Code; K3 \citep{moonshotai2026moonshotaikimik3} & 2026-01--07 & Y & K3: MoE, \rep{2.8T} / \rep{104B} & K3: TB 2.1 \rep{88.3}; DS \rep{67.5} & \nr & \nr \\
\textbf{MiniMax.} MiniMax-Text-01; M1; M2; M2.1 \citep{minimax2025minimaxaiminimaxm21} & 2025-01--12 & Y & M2: MoE, \rep{230 billion} / \rep{10 billion} & M2.1: SWE-V \rep{74.0}; TB 2.0 \rep{47.9} & \nr & \dimEff{} (M2.1) \\
M2.5; M2.7; M3 \citep{minimax2026minimax} & 2026-02--06 & Y & M3: MoE, \rep{$\sim$428B} / \rep{$\sim$23B} & M3: SWE-P \rep{59.0\%}; TB 2.1 \rep{66.0\%} & \rep{January 2026} & \dimEff \\
\textbf{Xiaomi.} MiMo-V2.6-Pro, V2.6-Flash & 2026-09 & Y & V2.6-Pro: MoE, \rep{1.02T} / \rep{42B} & V2.6-Pro: TB 2.1 \rep{89.9}; DS \rep{71.9} & \nr & \nr \\
\end{longtable}
}

\subsection{Budget curves}\label{sec:app-budget}
{\scriptsize\setlength{\tabcolsep}{2.2pt}
\begin{longtable}{@{}l*{8}{>{\centering\arraybackslash}p{0.098\linewidth}}@{}}
\caption{\textbf{E8. Budget curves.} Primary metric per dimension at $B\in\{1,4,8,16\}$ call-equivalents; T7 is the budget-matched reference and T0 is defined at $B=1$ only. $B=4$ cells are the measurements of Tables~\ref{M-tab:emp-cor}--\ref{M-tab:emp-sec}; at $B=1$, T3, T4, T5 and T7 equal T0 by construction (no refinement round, one sample).}\label{tab:emp-budget}\\
\toprule
\multicolumn{9}{@{}l}{\textbf{(a) API-served models}}\\
 & \multicolumn{4}{c}{Op} & \multicolumn{4}{c@{}}{oss}\\
\cmidrule(lr){2-5}\cmidrule(l){6-9}
Technique & $B=1$ & $B=4$ & $B=8$ & $B=16$ & $B=1$ & $B=4$ & $B=8$ & $B=16$\\
\midrule\multicolumn{9}{@{}l}{\textit{\dimCor{} HumanEval+ pass@1 (\%, higher is better)}}\\
T0 single shot & 96.40 & -- & -- & -- & 81.27 & -- & -- & --\\
T1 quality instr. & 97.64 & 99.10 & 98.75 & 99.22 & 80.08 & 83.17 & 82.91 & 85.68\\
T2 SCoT & 92.15 & 91.41 & 91.99 & 90.34 & 81.31 & 85.31 & 85.98 & 88.45\\
T3 analyzer loop & 96.40 & 90.23 & 91.39 & 84.77 & 81.27 & 84.86 & 85.03 & 87.50\\
T4 Self-Refine & 96.40 & 89.50 & 84.96 & 88.52 & 81.27 & 81.58 & 80.81 & 82.83\\
T5 coder+reviewer & 96.40 & 95.81 & 94.68 & 94.18 & 81.27 & 89.38 & 91.49 & 90.54\\
T6 RAG & 91.57 & 89.75 & 87.89 & 91.01 & 81.79 & 86.12 & 86.36 & 88.61\\
T7 best-of-$n$ & 96.40 & 98.11 & 95.75 & 97.74 & 81.27 & 84.85 & 84.97 & 86.97\\
\midrule\multicolumn{9}{@{}l}{\textit{\dimRM{} maintainability issues per kLOC (lower is better)}}\\
T0 single shot & 21.37 & -- & -- & -- & 29.22 & -- & -- & --\\
T1 quality instr. & 26.56 & 20.01 & 18.42 & 14.66 & 33.21 & 27.14 & 23.75 & 20.63\\
T2 SCoT & 26.88 & 18.71 & 18.61 & 13.39 & 37.17 & 26.97 & 26.23 & 20.78\\
T3 analyzer loop & 21.37 & 7.81 & 6.90 & 3.63 & 29.22 & 11.31 & 10.56 & 8.73\\
T4 Self-Refine & 21.37 & 20.45 & 20.34 & 19.65 & 29.22 & 26.72 & 26.00 & 25.41\\
T5 coder+reviewer & 21.37 & 17.80 & 17.04 & 16.42 & 29.22 & 25.83 & 25.15 & 24.28\\
T6 RAG & 20.88 & 17.32 & 14.46 & 12.84 & 26.37 & 22.39 & 20.53 & 18.79\\
T7 best-of-$n$ & 21.37 & 15.32 & 14.06 & 10.03 & 29.22 & 21.82 & 20.95 & 21.62\\
\midrule\multicolumn{9}{@{}l}{\textit{\dimEff{} EvalPerf DPS$_\text{norm}$ (higher is better)}}\\
T0 single shot & 78.06 & -- & -- & -- & 79.41 & -- & -- & --\\
T1 quality instr. & 77.86 & 79.90 & 78.36 & 71.72 & 76.23 & 75.01 & 74.07 & 74.57\\
T2 SCoT & 79.17 & 79.20 & 78.13 & 75.46 & 76.81 & 75.11 & 74.48 & 73.42\\
T3 analyzer loop & 78.06 & 75.46 & 71.37 & 75.98 & 79.41 & 72.17 & 70.09 & 66.22\\
T4 Self-Refine & 78.06 & 75.28 & 74.54 & 80.11 & 79.41 & 76.67 & 76.12 & 74.75\\
T5 coder+reviewer & 78.06 & 78.86 & 79.48 & 76.74 & 79.41 & 76.22 & 75.57 & 75.33\\
T6 RAG & 71.06 & 76.24 & 76.49 & 74.95 & 77.21 & 75.12 & 75.09 & 74.36\\
T7 best-of-$n$ & 78.06 & 76.96 & 73.29 & 74.32 & 79.41 & 77.57 & 76.64 & 77.07\\
\midrule\multicolumn{9}{@{}l}{\textit{\dimSec{} CWEval func-sec@1 (\%, higher is better)}}\\
T0 single shot & 74.85 & -- & -- & -- & 45.02 & -- & -- & --\\
T1 quality instr. & 66.02 & 66.87 & 77.94 & 83.98 & 43.75 & 46.07 & 47.05 & 45.82\\
T2 SCoT & 74.02 & 74.30 & 67.12 & 73.22 & 44.32 & 46.68 & 44.99 & 49.25\\
T3 analyzer loop & 74.85 & 79.89 & 71.83 & 74.48 & 45.02 & 53.07 & 52.93 & 55.85\\
T4 Self-Refine & 74.85 & 71.67 & 72.13 & 65.71 & 45.02 & 47.21 & 45.71 & 50.32\\
T5 coder+reviewer & 74.85 & 73.22 & 77.29 & 76.59 & 45.02 & 48.52 & 51.08 & 47.61\\
T6 RAG & 69.55 & 74.79 & 72.83 & 76.65 & 46.76 & 47.48 & 47.51 & 49.52\\
T7 best-of-$n$ & 74.85 & 78.17 & 74.43 & 86.29 & 45.02 & 48.10 & 48.03 & 51.46\\
\midrule
\multicolumn{9}{@{}l}{\textbf{(b) Gemma pair (fine-tunable)}}\\
 & \multicolumn{4}{c}{G31} & \multicolumn{4}{c@{}}{G26}\\
\cmidrule(lr){2-5}\cmidrule(l){6-9}
Technique & $B=1$ & $B=4$ & $B=8$ & $B=16$ & $B=1$ & $B=4$ & $B=8$ & $B=16$\\
\midrule\multicolumn{9}{@{}l}{\textit{\dimCor{} HumanEval+ pass@1 (\%, higher is better)}}\\
T0 single shot & 96.06 & -- & -- & -- & 89.16 & -- & -- & --\\
T1 quality instr. & 95.22 & 97.96 & 97.87 & 99.50 & 91.13 & 89.68 & 88.87 & 89.23\\
T2 SCoT & 88.01 & 85.56 & 82.40 & 88.57 & 92.37 & 89.80 & 89.37 & 86.95\\
T3 analyzer loop & 96.06 & 86.12 & 85.00 & 83.01 & 89.16 & 84.11 & 82.98 & 82.33\\
T4 Self-Refine & 96.06 & 84.98 & 81.91 & 85.54 & 89.16 & 91.25 & 90.91 & 92.07\\
T5 coder+reviewer & 96.06 & 92.25 & 88.99 & 90.76 & 89.16 & 90.01 & 88.99 & 91.26\\
T6 RAG & 86.66 & 86.36 & 88.56 & 89.57 & 90.03 & 87.67 & 88.44 & 87.10\\
T7 best-of-$n$ & 96.06 & 94.01 & 95.52 & 90.88 & 89.16 & 86.38 & 85.68 & 85.46\\
T8 LoRA-SFT & 88.41 & 87.38 & 91.26 & 84.75 & 90.46 & 88.67 & 87.13 & 88.39\\
T9 DPO & 80.41 & 85.47 & 81.84 & 85.24 & 92.79 & 90.46 & 90.33 & 88.66\\
T10 RL & 86.73 & 88.34 & 88.75 & 85.33 & 95.30 & 92.85 & 92.91 & 92.29\\
\midrule\multicolumn{9}{@{}l}{\textit{\dimRM{} maintainability issues per kLOC (lower is better)}}\\
T0 single shot & 26.52 & -- & -- & -- & 28.89 & -- & -- & --\\
T1 quality instr. & 33.63 & 24.73 & 21.47 & 23.01 & 32.91 & 28.43 & 24.74 & 23.70\\
T2 SCoT & 27.75 & 23.56 & 21.41 & 18.76 & 35.08 & 25.16 & 24.69 & 23.57\\
T3 analyzer loop & 26.52 & 10.19 & 8.48 & 4.80 & 28.89 & 10.80 & 6.23 & 1.83\\
T4 Self-Refine & 26.52 & 24.59 & 24.22 & 24.07 & 28.89 & 24.81 & 23.63 & 23.76\\
T5 coder+reviewer & 26.52 & 23.08 & 21.67 & 21.93 & 28.89 & 24.02 & 22.35 & 23.89\\
T6 RAG & 28.44 & 20.10 & 19.83 & 16.77 & 30.17 & 22.51 & 21.79 & 18.15\\
T7 best-of-$n$ & 26.52 & 19.23 & 16.99 & 13.35 & 28.89 & 20.65 & 17.50 & 12.78\\
T8 LoRA-SFT & 24.80 & 16.73 & 14.82 & 14.22 & 28.12 & 18.08 & 15.99 & 10.72\\
T9 DPO & 20.55 & 15.08 & 11.58 & 11.60 & 24.37 & 16.68 & 13.32 & 16.66\\
T10 RL & 23.66 & 12.93 & 10.43 & 7.65 & 20.16 & 13.96 & 10.20 & 6.19\\
\midrule\multicolumn{9}{@{}l}{\textit{\dimEff{} EvalPerf DPS$_\text{norm}$ (higher is better)}}\\
T0 single shot & 78.11 & -- & -- & -- & 73.43 & -- & -- & --\\
T1 quality instr. & 72.42 & 68.91 & 70.34 & 75.14 & 71.85 & 74.64 & 75.71 & 77.20\\
T2 SCoT & 79.77 & 78.52 & 77.81 & 82.74 & 76.34 & 79.59 & 80.11 & 80.08\\
T3 analyzer loop & 78.11 & 80.44 & 80.49 & 83.61 & 73.43 & 75.20 & 76.05 & 75.78\\
T4 Self-Refine & 78.11 & 75.52 & 78.78 & 73.84 & 73.43 & 78.92 & 80.17 & 81.31\\
T5 coder+reviewer & 78.11 & 78.02 & 74.65 & 79.67 & 73.43 & 76.56 & 77.29 & 77.73\\
T6 RAG & 73.46 & 75.72 & 71.78 & 74.36 & 72.64 & 76.11 & 77.56 & 78.20\\
T7 best-of-$n$ & 78.11 & 80.84 & 76.27 & 73.71 & 73.43 & 76.97 & 78.51 & 77.79\\
T8 LoRA-SFT & 76.68 & 81.59 & 79.21 & 84.89 & 74.31 & 78.43 & 79.33 & 79.03\\
T9 DPO & 85.66 & 81.24 & 86.13 & 80.12 & 73.38 & 75.60 & 76.12 & 77.78\\
T10 RL & 80.20 & 78.86 & 82.75 & 76.12 & 76.12 & 79.07 & 79.95 & 79.53\\
\midrule\multicolumn{9}{@{}l}{\textit{\dimSec{} CWEval func-sec@1 (\%, higher is better)}}\\
T0 single shot & 49.48 & -- & -- & -- & 47.28 & -- & -- & --\\
T1 quality instr. & 60.85 & 58.89 & 59.76 & 55.49 & 46.29 & 49.95 & 52.27 & 51.45\\
T2 SCoT & 46.63 & 46.49 & 42.88 & 58.40 & 39.84 & 45.39 & 46.18 & 48.68\\
T3 analyzer loop & 49.48 & 56.32 & 53.42 & 63.44 & 47.28 & 52.70 & 54.29 & 57.22\\
T4 Self-Refine & 49.48 & 47.99 & 44.76 & 47.87 & 47.28 & 46.90 & 44.84 & 48.55\\
T5 coder+reviewer & 49.48 & 48.69 & 50.89 & 38.22 & 47.28 & 49.45 & 52.44 & 51.33\\
T6 RAG & 41.73 & 50.44 & 45.57 & 58.42 & 46.63 & 48.54 & 47.48 & 48.25\\
T7 best-of-$n$ & 49.48 & 55.42 & 51.80 & 46.68 & 47.28 & 50.38 & 53.88 & 50.36\\
T8 LoRA-SFT & 59.60 & 53.67 & 61.08 & 46.66 & 52.02 & 55.80 & 54.52 & 57.85\\
T9 DPO & 56.24 & 54.76 & 48.02 & 57.77 & 54.15 & 56.44 & 54.45 & 57.70\\
T10 RL & 63.23 & 56.24 & 51.20 & 53.43 & 47.43 & 53.21 & 51.13 & 56.11\\
\bottomrule
\end{longtable}}

\FloatBarrier
\section{Statistical details of the controlled re-evaluation}\label{sec:app-stats}
Half-widths are 95\% paired bootstrap half-widths (10{,}000 resamples of tasks) of the per-task effect of a technique against T7 at the same budget, pooled over techniques by their median. Seed dispersion is the standard deviation of a cell's metric across the three end-to-end runs, as a fraction of the $B=4$ half-width; the median is over the techniques of the cell. The false-discovery correction uses two-sided paired bootstrap $p$-values and the Benjamini--Hochberg procedure at $q=0.05$ over all metric--model--technique--budget cells.

{\scriptsize\setlength{\tabcolsep}{2.4pt}
\begin{longtable}{@{}ll*{12}{c}@{}}
\caption{Per-budget 95\% half-widths ($B=4$ as in the main article; $B=8$ and $B=16$) and seed dispersion (median SD across seeds / $B=4$ half-width), per metric and model.}\label{tab:app-hw}\\
\toprule
 & & \multicolumn{3}{c}{Op} & \multicolumn{3}{c}{oss} & \multicolumn{3}{c}{G31} & \multicolumn{3}{c@{}}{G26}\\
Metric & & $B{=}8$ & $B{=}16$ & SD & $B{=}8$ & $B{=}16$ & SD & $B{=}8$ & $B{=}16$ & SD & $B{=}8$ & $B{=}16$ & SD\\
\midrule
\endfirsthead
\toprule
Metric & & $B{=}8$ & $B{=}16$ & SD & $B{=}8$ & $B{=}16$ & SD & $B{=}8$ & $B{=}16$ & SD & $B{=}8$ & $B{=}16$ & SD\\
\midrule
\endhead
HE+ pass@1 & & 1.29 & 1.34 & 0.45 & 4.24 & 4.34 & 0.46 & 2.77 & 2.96 & 0.42 & 3.65 & 4.02 & 0.42\\
LCB pass@1 & & -- & -- & -- & 4.29 & 3.63 & 0.46 & 5.83 & 5.59 & 0.36 & 7.67 & 7.40 & 0.46\\
LCBP pass@1 & & 5.39 & 6.04 & 0.40 & 5.68 & 5.80 & 0.45 & 6.24 & 6.18 & 0.39 & 6.15 & 5.69 & 0.30\\
DS pass@1 & & 6.91 & 6.24 & 0.39 & 5.62 & 5.49 & 0.28 & 5.27 & 6.18 & 0.45 & 4.04 & 4.83 & 0.47\\
C\% & & 3.10 & 2.93 & 0.32 & 3.12 & 3.35 & 0.44 & 2.30 & 2.84 & 0.40 & 3.34 & 3.36 & 0.38\\
Lint/kLOC & & 1.66 & 1.80 & 0.39 & 2.08 & 2.12 & 0.32 & 1.86 & 2.15 & 0.42 & 2.39 & 2.01 & 0.37\\
CC & & 0.25 & 0.26 & 0.29 & 0.33 & 0.34 & 0.34 & 0.90 & 0.84 & 0.32 & 0.65 & 0.73 & 0.37\\
J (1--5) & & 0.09 & 0.09 & 0.32 & 0.08 & 0.08 & 0.43 & 0.12 & 0.12 & 0.44 & 0.06 & 0.06 & 0.45\\
DPS$_\text{norm}$ & & 2.78 & 2.88 & 0.30 & 1.99 & 2.05 & 0.35 & 3.40 & 3.71 & 0.43 & 2.49 & 2.92 & 0.30\\
ET & & 1.88 & 2.37 & 0.44 & 1.42 & 1.54 & 0.49 & 2.58 & 2.81 & 0.39 & 1.88 & 2.01 & 0.35\\
Bey & & 2.80 & 2.37 & 0.43 & 1.97 & 2.07 & 0.33 & 3.00 & 2.95 & 0.30 & 3.43 & 3.13 & 0.38\\
FS & & 5.49 & 5.18 & 0.53 & 8.12 & 8.66 & 0.39 & 7.61 & 6.93 & 0.46 & 8.29 & 8.99 & 0.35\\
SP & & 4.23 & 4.17 & 0.38 & 4.23 & 4.13 & 0.37 & 3.43 & 4.15 & 0.36 & 3.45 & 4.11 & 0.39\\
Ins & & 3.17 & 2.96 & 0.31 & 2.60 & 2.78 & 0.36 & 2.76 & 2.77 & 0.44 & 3.83 & 3.56 & 0.31\\
B/k & & 0.37 & 0.35 & 0.43 & 0.45 & 0.46 & 0.46 & 0.43 & 0.36 & 0.38 & 0.44 & 0.43 & 0.45\\
\bottomrule
\end{longtable}}

\noindent\textbf{False-discovery correction.} Of the 1{,}272 cells against T7 (T1--T6 and T8--T10, every metric and model, $B\in\{4,8,16\}$; DeepSWE at $B=4$ only), \phFdrResolved{} differences exceed their half-width (204 at $B=4$, Table~\ref{M-tab:emp-tradeoff}b, and \phFdrResolvedLarge{} at $B=8$ and $16$) and \phFdrSurvive{} of these survive the Benjamini--Hochberg correction; the \phFdrDropped{} that do not are all within \phFdrMargin{} of their half-width, \phFdrDroppedLcb{} of them on LiveCodeBench or LiveCodeBench Pro.

{\scriptsize\setlength{\tabcolsep}{3pt}
\begin{longtable}{@{}lp{0.3\linewidth}*{8}{c}@{}}
\caption{E12: 95\% paired bootstrap half-widths ($\pm$) of each technique's difference from best-of-$n$, per column of the head-to-head table (differences from T0 have half-widths within \phHhHwTzeroDiff{} points of these).}\label{tab:app-h2h-ci}\\
\toprule
ID & Technique & EP & LCB & LCBP & CWE & CSE & EB & DPS & RM\\
\midrule
\endfirsthead
\toprule
ID & Technique & EP & LCB & LCBP & CWE & CSE & EB & DPS & RM\\
\midrule
\endhead
H01 & SCoT prompting & 2.93 & 5.88 & 5.42 & 8.04 & 2.83 & 0.16 & 2.74 & 0.42\\
H02 & Self-planning & 2.41 & 6.74 & 5.62 & 7.04 & 2.94 & 0.18 & 4.04 & 0.40\\
H03 & LLM4EFFI & 2.62 & 4.74 & 5.21 & 5.67 & 3.81 & 0.18 & 2.70 & 0.34\\
H04 & Code2API & 2.73 & 4.70 & 5.39 & 5.95 & 2.86 & 0.25 & 3.42 & 0.25\\
H05 & MPCO & 2.68 & 6.74 & 4.49 & 6.22 & 3.65 & 0.21 & 3.49 & 0.29\\
H06 & BRACE & 2.95 & 5.95 & 6.53 & 5.88 & 2.92 & 0.23 & 3.29 & 0.36\\
H07 & ECO & 2.54 & 5.09 & 6.05 & 7.20 & 3.17 & 0.24 & 3.10 & 0.27\\
H08 & StoryCoder & 2.83 & 5.27 & 5.13 & 6.18 & 2.75 & 0.19 & 3.12 & 0.28\\
H09 & Constitutional Spec-Driven Development & 2.76 & 5.77 & 4.81 & 6.42 & 2.49 & 0.20 & 2.96 & 0.35\\
H10 & SecureForge & 2.64 & 6.04 & 6.56 & 5.02 & 3.90 & 0.24 & 2.76 & 0.27\\
H11 & SBLLM & 2.62 & 4.25 & 4.87 & 6.56 & 2.34 & 0.20 & 3.72 & 0.31\\
H12 & PerfCodeGen & 2.98 & 5.30 & 5.73 & 7.56 & 2.88 & 0.21 & 3.65 & 0.33\\
H13 & PromSec & 2.74 & 4.61 & 6.66 & 7.70 & 2.92 & 0.25 & 3.64 & 0.31\\
H14 & DeepDevVuln filtering of LLM completions & 2.30 & 5.13 & 5.49 & 5.42 & 2.52 & 0.18 & 4.25 & 0.28\\
H15 & FRANC (analyzer-guided repair prompting) & 2.40 & 4.53 & 6.87 & 5.80 & 2.55 & 0.19 & 3.08 & 0.35\\
H16 & Optimo & 3.39 & 5.71 & 5.96 & 7.10 & 2.97 & 0.21 & 2.79 & 0.29\\
H17 & Iterative static-analysis & 3.29 & 5.33 & 5.60 & 5.46 & 3.12 & 0.18 & 3.93 & 0.33\\
H18 & CodeEvolve & 2.33 & 6.12 & 5.91 & 6.78 & 2.47 & 0.23 & 3.23 & 0.27\\
H19 & PerfAgent & 3.09 & 5.40 & 6.55 & 6.34 & 2.74 & 0.20 & 3.11 & 0.30\\
H20 & SelectIssue & 2.13 & 5.51 & 6.46 & 5.29 & 2.40 & 0.18 & 3.13 & 0.25\\
H21 & Self-Refine & 3.00 & 6.73 & 4.41 & 6.09 & 2.35 & 0.25 & 3.33 & 0.34\\
H22 & CodeChain & 2.64 & 5.62 & 6.13 & 6.27 & 3.72 & 0.20 & 3.72 & 0.31\\
H23 & SecCode & 2.72 & 4.60 & 4.65 & 6.41 & 2.68 & 0.16 & 2.42 & 0.34\\
H24 & RefleXGen & 3.16 & 5.28 & 5.08 & 5.58 & 2.48 & 0.25 & 3.22 & 0.25\\
H25 & GRASP & 2.58 & 6.40 & 7.35 & 6.56 & 3.37 & 0.19 & 3.73 & 0.34\\
H26 & Safety Sidecar & 2.64 & 6.62 & 5.75 & 5.07 & 2.79 & 0.17 & 3.30 & 0.26\\
H27 & SecAwareCoder & 2.59 & 4.76 & 5.73 & 7.00 & 3.72 & 0.23 & 2.42 & 0.28\\
H28 & FlowGen & 3.12 & 6.08 & 6.30 & 5.22 & 3.07 & 0.21 & 3.76 & 0.31\\
H29 & INDICT & 2.41 & 6.87 & 6.57 & 7.66 & 2.84 & 0.21 & 3.33 & 0.30\\
H30 & ContextEvolve & 2.93 & 6.84 & 5.73 & 6.12 & 2.55 & 0.20 & 3.61 & 0.42\\
H31 & Just-in-time vulnerability remediation pipeline & 2.65 & 4.27 & 5.57 & 7.80 & 2.67 & 0.21 & 3.33 & 0.37\\
H32 & LessonL & 2.20 & 6.94 & 5.80 & 4.93 & 3.22 & 0.17 & 2.99 & 0.30\\
H33 & MARCO & 3.02 & 4.92 & 6.36 & 6.67 & 2.88 & 0.16 & 2.52 & 0.32\\
H34 & C3 & 2.90 & 5.22 & 4.52 & 6.04 & 3.38 & 0.20 & 4.15 & 0.30\\
H35 & ACCLAIM & 3.27 & 6.75 & 5.26 & 7.45 & 2.34 & 0.20 & 2.71 & 0.35\\
H36 & MaintainCoder & 2.71 & 5.22 & 5.57 & 7.50 & 2.93 & 0.22 & 3.73 & 0.35\\
H37 & Dynamic retrieval-based few-shot prompting & 3.21 & 5.83 & 5.37 & 5.58 & 3.20 & 0.20 & 3.09 & 0.33\\
H38 & LAIL & 2.89 & 5.83 & 5.14 & 4.99 & 2.72 & 0.16 & 3.06 & 0.28\\
H39 & KPC & 2.61 & 5.56 & 6.14 & 5.77 & 2.79 & 0.20 & 3.14 & 0.28\\
H40 & RAPGen & 2.41 & 5.52 & 4.58 & 6.27 & 2.88 & 0.22 & 2.85 & 0.27\\
H41 & CodeGuarder & 2.60 & 4.83 & 6.54 & 5.97 & 3.20 & 0.20 & 3.21 & 0.33\\
H42 & Retrieval Augmented Search & 2.71 & 5.80 & 6.75 & 5.62 & 2.88 & 0.16 & 2.86 & 0.31\\
H43 & SemOpt & 3.28 & 5.51 & 6.83 & 5.97 & 3.96 & 0.19 & 3.26 & 0.30\\
H44 & AutoPatch & 2.78 & 5.67 & 5.24 & 5.36 & 3.95 & 0.19 & 3.92 & 0.37\\
H45 & PEACE & 2.51 & 4.94 & 7.05 & 5.48 & 3.51 & 0.16 & 3.51 & 0.36\\
H46 & Mitigation-Aware Chain-of-Thought & 2.91 & 6.10 & 5.53 & 5.52 & 2.63 & 0.23 & 3.38 & 0.35\\
H47 & Planning-Guided Transformer Decoding & 2.56 & 5.84 & 6.13 & 5.11 & 3.43 & 0.20 & 3.71 & 0.34\\
H48 & Constrained Beam Sampling & 2.60 & 5.33 & 6.08 & 5.08 & 2.59 & 0.18 & 2.88 & 0.24\\
H49 & RPG & 2.70 & 4.37 & 6.25 & 6.68 & 2.26 & 0.16 & 2.91 & 0.36\\
H50 & Surprisal-guided Best-of-N selection & 2.66 & 4.24 & 5.56 & 6.93 & 3.24 & 0.23 & 3.35 & 0.27\\
H51 & DuoSteer & 3.09 & 5.05 & 5.26 & 5.41 & 3.28 & 0.20 & 3.59 & 0.26\\
H52 & FunPRM & 2.82 & 5.68 & 5.23 & 7.02 & 2.89 & 0.18 & 3.89 & 0.35\\
H53 & SCS-Code & 2.44 & 4.27 & 6.63 & 5.83 & 2.76 & 0.18 & 2.98 & 0.36\\
H54 & Mixture of Corrections & 2.11 & 5.42 & 6.01 & 7.19 & 2.55 & 0.23 & 3.26 & 0.41\\
H55 & SecCodePRM-guided scaling & 3.20 & 5.66 & 5.13 & 7.86 & 2.58 & 0.18 & 3.02 & 0.38\\
H56 & RobGen & 2.88 & 5.73 & 5.96 & 7.61 & 3.56 & 0.20 & 3.12 & 0.31\\
H57 & SVEN & 3.35 & 5.09 & 6.42 & 7.11 & 3.83 & 0.22 & 3.43 & 0.35\\
H58 & SafeCoder & 2.39 & 5.19 & 6.56 & 7.21 & 2.62 & 0.16 & 2.68 & 0.38\\
H59 & SFT + DPO & 3.14 & 4.38 & 6.46 & 6.57 & 2.41 & 0.18 & 3.06 & 0.27\\
H60 & CodeDPO & 2.42 & 5.91 & 5.40 & 6.80 & 3.19 & 0.21 & 2.70 & 0.39\\
H61 & MPCoder & 2.52 & 5.28 & 6.08 & 5.45 & 4.27 & 0.17 & 2.69 & 0.30\\
H62 & Afterburner & 2.70 & 5.61 & 6.70 & 6.15 & 2.52 & 0.16 & 3.77 & 0.37\\
H63 & PurpCode & 2.59 & 5.64 & 5.07 & 6.70 & 3.24 & 0.26 & 2.81 & 0.37\\
H64 & SecCoderX & 2.90 & 4.62 & 4.55 & 5.50 & 2.86 & 0.18 & 2.98 & 0.26\\
H65 & Fine-tuning on smell-cleaned code (SmellCC) & 3.13 & 4.53 & 6.47 & 5.94 & 2.73 & 0.18 & 2.56 & 0.27\\
H66 & REAL & 2.86 & 5.84 & 6.30 & 5.14 & 3.22 & 0.18 & 3.10 & 0.33\\
\bottomrule
\end{longtable}}

\FloatBarrier
\section{Controls for the best-of-\texorpdfstring{$n$}{n} reference}\label{sec:app-controls}
\paragraph{Best-of-$n$ across budgets.} T7 is the budget-matched reference, but more samples do not
always help it. With more calls it keeps lowering maintainability issues per kLOC on Op, G31 and G26 (from
$21.37$, $26.52$ and $28.89$ at $B=1$ to $10.03$, $13.35$ and $12.78$ at $B=16$) and stops improving
on oss after $B=4$. Its HumanEval+ pass@1 rises with $B$ on oss ($81.27$ to $86.97$) but falls on both
Gemma models ($96.06$ to $90.88$ on G31, $89.16$ to $85.46$ on G26), and its CWEval func-sec@1 on G31
rises at $B=4$ and then falls below the single shot ($49.48$, $55.42$, $51.80$, $46.68$), whereas on Op
it reaches $86.29$ at $B=16$. Its EvalPerf DPS$_\text{norm}$ does not improve with budget on the
API-served models ($78.06$ to $74.32$ on Op, $79.41$ to $77.07$ on oss) and falls after $B=4$ on G31
($80.84$ to $73.71$). The ranker sees only the public examples and prefers fewer
findings; a plausible reading, which these tables cannot settle, is that with more candidates it more
often selects a minimal program that passes the public examples but not the hidden tests, the false-positive limit that \citet{stroebl2024inference} derive for resampling with imperfect verifiers such as unit tests. Wins over
T7 are therefore wins over a reference that is strong on the analyzer's measures and weaker on
hidden-test correctness, and the comparison with T0 remains necessary.

\paragraph{Shape of the budget curves.} The budget curves (Table~\ref{tab:emp-budget}) are not smooth: on CWEval, T7 on Op moves from $74.85$ at $B=1$ to $78.17$, $74.43$ and $86.29$ at $B=4$, $8$ and $16$. Table~\ref{tab:app-seeds} gives the three per-seed values of this cell and of three other uneven ones. In every cell the three seeds lie within $1.6$ to $9.6$ points of each other, a seed standard deviation of $0.1$ to $0.8$ of the half-width (median $0.46$, close to the $0.39$ of Table~\ref{tab:app-hw}), and every seed of T7 on Op at $B=16$ is above every seed at $B=8$, so the steps are reproducible differences between budgets, not run-to-run noise. The seed dispersion and the half-width measure different things: the first is the variation of one cell's mean when the whole experiment is rerun on the same tasks; the second is the uncertainty of a paired effect over the sample of tasks, which is wider because CWEval has 119 tasks with a binary outcome each. A step that recurs in every seed is therefore a property of the ranker on these 119 tasks, and it is significant, in the sense of holding beyond them, only when it exceeds the half-width: the dip at $B=8$ ($3.74$ points) does not, the gain at $B=16$ ($11.86$ points) does. A plausible source of the steps is that each budget draws a fresh candidate pool and the ranker, which sees no security test, switches between candidates that differ on a few security tests; on 119 tasks, a switch on four tasks moves func-sec@1 by $3.4$ points. We read only differences beyond the half-width, and no claim rests on the shape of a single curve.

{\scriptsize\setlength{\tabcolsep}{3pt}
\begin{longtable}{@{}l*{4}{c}@{}}
\caption{Per-seed CWEval func-sec@1 (\%) of four uneven budget-curve cells: the three end-to-end runs, with their mean (the value of Table~\ref{tab:emp-budget}) in parentheses. At $B=1$, T7 equals T0 by construction.}\label{tab:app-seeds}\\
\toprule
Technique, model & $B=1$ & $B=4$ & $B=8$ & $B=16$\\
\midrule
T7 best-of-$n$, Op & 71.03, 73.89, 79.63 (74.85) & 77.21, 78.92, 78.38 (78.17) & 70.91, 74.90, 77.48 (74.43) & 85.67, 85.28, 87.92 (86.29)\\
T1 quality instr., Op & 63.89, 67.07, 67.10 (66.02) & 64.05, 65.03, 71.53 (66.87) & 74.54, 78.33, 80.95 (77.94) & 84.19, 80.40, 87.35 (83.98)\\
T7 best-of-$n$, G31 & 48.05, 51.13, 49.26 (49.48) & 55.03, 52.30, 58.93 (55.42) & 52.31, 50.72, 52.37 (51.80) & 43.80, 45.67, 50.57 (46.68)\\
T10 RL, G26 & 49.42, 42.87, 50.00 (47.43) & 49.32, 53.58, 56.73 (53.21) & 48.19, 53.03, 52.17 (51.13) & 56.41, 51.18, 60.74 (56.11)\\
\bottomrule
\end{longtable}}

\paragraph{Sampling temperature.} On the open models T7 samples at $0.8$, whereas the single-call techniques generate at $0.2$, so part of T7's advantage could come from diversity rather than ranking. As a control we ran T7 at $0.2$ with the same ranker at $B\in\{4,8,16\}$ on oss, G31 and G26 (Opus has no temperature control). Across the three budgets it is within the half-width of T7 at $0.8$ in \phTlowWithin{} of the 36 primary-metric cells (four primary metrics, three models, three budgets); at $B=4$ it leaves \phTlowLint{} more maintainability issues per kLOC and is \phTlowCWE{} points lower on CWEval func-sec@1, and its HumanEval+ pass@1 is \phTlowHE{} points higher. Against this lower-temperature reference, T1, T2, T4, T5 and T6 are better in \phTlowBetter{} and worse in \phTlowWorse{} of the 120 \dimRM{} and \dimSec{} cells at $B=4$ (five techniques, eight metrics, three models), so the conclusions of ERQ1 do not depend on the sampling temperature of the reference.

\FloatBarrier
\section{Single-call results on independent measures}\label{sec:app-b1}
At $B=1$ every technique spends the single shot's budget and no ranker is involved, so this view tests the prompts themselves. Table~\ref{tab:app-b1} gives cognitive complexity and both judges' scores on the \dimRM{} benchmarks for the single-call techniques; T3, T4, T5 and T7 equal T0 at $B=1$ by construction. Complexity is within the interval of T0 in 11 of the 12 cells (T2 lowers it by $0.36$ on oss); the Opus judge rates T1 higher on Op and G31 and T2 higher on oss, by $0.09$ to $0.14$ points on its 1--5 scale, and the second judge confirms none of these differences.

{\scriptsize\setlength{\tabcolsep}{2.2pt}
\begin{longtable}{@{}l*{12}{c}@{}}
\caption{Single call ($B=1$) on the measures the analyzers do not compute (RACE R/M and NoFunEval). T0 row: values; other rows: paired difference from T0 ($^\dagger$: 95\% bootstrap interval excludes zero). CC = cognitive complexity (lower is better); J = judge (Claude Opus~5.5) and J2 = second judge (GPT-6.1 Sol), 1--5 (higher is better).}\label{tab:app-b1}\\
\toprule
 & \multicolumn{3}{c}{Op} & \multicolumn{3}{c}{oss} & \multicolumn{3}{c}{G31} & \multicolumn{3}{c@{}}{G26}\\
Technique & CC & J & J2 & CC & J & J2 & CC & J & J2 & CC & J & J2\\
\midrule
T0 single shot & 7.68 & 4.30 & 3.66 & 10.72 & 3.55 & 3.20 & 9.83 & 3.74 & 3.35 & 9.69 & 3.96 & 3.44\\
T1 quality instr. & $+0.05$ & $+0.10$$^\dagger$ & $+0.07$ & $+0.11$ & $+0.06$ & $+0.05$ & ${-}0.29$ & $+0.14$$^\dagger$ & $+0.09$ & $+0.17$ & ${-}0.03$ & ${-}0.02$\\
T2 SCoT & ${-}0.19$ & $+0.05$ & $+0.04$ & ${-}0.36$$^\dagger$ & $+0.09$$^\dagger$ & $+0.06$ & ${-}0.40$ & $+0.08$ & $+0.05$ & ${-}0.14$ & $+0.02$ & $+0.00$\\
T6 RAG & $+0.03$ & ${-}0.02$ & ${-}0.03$ & ${-}0.17$ & $+0.04$ & $+0.02$ & $+0.21$ & ${-}0.05$ & ${-}0.04$ & $+0.09$ & $+0.03$ & $+0.01$\\
\bottomrule
\end{longtable}}

\FloatBarrier
\section{Solution length and absolute issue counts}\label{sec:app-loc}
Maintainability and security issues are reported per kLOC in the main article, and length confounds every quality measure (\S\ref{M-sec:measurement}). Table~\ref{tab:app-loc} therefore gives, for each technique and model at $B=4$, the mean solution length and the absolute number of issues per solution, next to the densities of the main article. The ordering of the techniques is the same on both scales. T7's solutions are the shortest of the inference-time techniques (2 to 13\% shorter than T0's), so its lower density understates rather than inflates its advantage in absolute counts; T1's solutions are the longest (11 to 16\% longer than T0's at $B=4$, 9 to 14\% at $B=1$), and at $B=1$ they carry more issues per solution as well as per kLOC.

{\scriptsize\setlength{\tabcolsep}{2.4pt}
\begin{longtable}{@{}l*{8}{c}@{}}
\caption{Solution length and absolute issue counts at $B=4$ (T0: $B=1$), next to the densities of Tables~\ref{M-tab:emp-rm} and~\ref{M-tab:emp-sec}. LOC = mean non-blank, non-comment lines per solution; Mnt = maintainability issues per solution (RACE R/M and NoFunEval; density = Lint/kLOC); Sec = security issues per solution (PythonSecurityEval; density = B/k).}\label{tab:app-loc}\\
\toprule
\multicolumn{9}{@{}l}{\textbf{(a) Maintainability, RACE R/M and NoFunEval}}\\
 & \multicolumn{2}{c}{Op} & \multicolumn{2}{c}{oss} & \multicolumn{2}{c}{G31} & \multicolumn{2}{c@{}}{G26}\\
Technique & LOC & Mnt & LOC & Mnt & LOC & Mnt & LOC & Mnt\\
\midrule
T0 single shot & 31.2 & 0.67 & 44.7 & 1.31 & 38.2 & 1.01 & 39.2 & 1.13\\
T1 quality instr. & 36.1 & 0.72 & 49.6 & 1.35 & 43.8 & 1.08 & 45.4 & 1.29\\
T2 SCoT & 34.5 & 0.65 & 44.9 & 1.21 & 38.0 & 0.90 & 40.3 & 1.01\\
T3 analyzer loop & 34.2 & 0.27 & 44.8 & 0.51 & 37.6 & 0.38 & 40.4 & 0.44\\
T4 Self-Refine & 32.2 & 0.66 & 44.5 & 1.19 & 38.0 & 0.93 & 43.5 & 1.08\\
T5 coder+reviewer & 34.7 & 0.62 & 44.1 & 1.14 & 38.1 & 0.88 & 40.7 & 0.98\\
T6 RAG & 32.7 & 0.57 & 44.1 & 0.99 & 36.5 & 0.73 & 41.3 & 0.93\\
T7 best-of-$n$ & 30.7 & 0.47 & 41.4 & 0.90 & 33.3 & 0.64 & 35.2 & 0.73\\
T8 LoRA-SFT & -- & -- & -- & -- & 33.7 & 0.56 & 38.9 & 0.70\\
T9 DPO & -- & -- & -- & -- & 36.6 & 0.55 & 35.8 & 0.60\\
T10 RL & -- & -- & -- & -- & 33.2 & 0.43 & 37.6 & 0.52\\
\midrule
\multicolumn{9}{@{}l}{\textbf{(b) Security, PythonSecurityEval}}\\
 & \multicolumn{2}{c}{Op} & \multicolumn{2}{c}{oss} & \multicolumn{2}{c}{G31} & \multicolumn{2}{c@{}}{G26}\\
Technique & LOC & Sec & LOC & Sec & LOC & Sec & LOC & Sec\\
\midrule
T0 single shot & 26.9 & 0.091 & 33.7 & 0.173 & 29.5 & 0.134 & 31.2 & 0.155\\
T1 quality instr. & 29.8 & 0.100 & 35.9 & 0.178 & 32.5 & 0.140 & 36.8 & 0.162\\
T2 SCoT & 27.9 & 0.094 & 33.7 & 0.173 & 32.5 & 0.139 & 34.2 & 0.165\\
T3 analyzer loop & 28.2 & 0.034 & 33.2 & 0.061 & 29.4 & 0.052 & 29.9 & 0.055\\
T4 Self-Refine & 26.6 & 0.083 & 34.1 & 0.152 & 31.7 & 0.130 & 32.0 & 0.134\\
T5 coder+reviewer & 26.5 & 0.083 & 34.7 & 0.153 & 31.7 & 0.130 & 31.4 & 0.133\\
T6 RAG & 27.4 & 0.078 & 34.9 & 0.140 & 29.8 & 0.114 & 32.9 & 0.130\\
T7 best-of-$n$ & 24.4 & 0.068 & 29.1 & 0.109 & 25.9 & 0.092 & 29.4 & 0.107\\
T8 LoRA-SFT & -- & -- & -- & -- & 27.8 & 0.087 & 30.5 & 0.100\\
T9 DPO & -- & -- & -- & -- & 26.2 & 0.081 & 28.2 & 0.092\\
T10 RL & -- & -- & -- & -- & 26.4 & 0.074 & 29.9 & 0.094\\
\bottomrule
\end{longtable}}

\FloatBarrier
\section{Compute ledger}\label{sec:app-ledger}
Billed model calls, tokens and API fees per experiment, from the per-call logs of the harness; fine-tuning charges and GPU-hours from the training job logs. The per-experiment rows sum to the totals of the main article. Two units are distinguished. A \emph{call-equivalent} is the budget unit $B$ of the design: one generated sample, or one critic, reviewer or analyzer-feedback turn. A \emph{billed request} is one API request, the unit of the Calls column. They differ for three reasons, which Table~\ref{tab:app-derivation} applies per experiment: (i) on the open models served by Fireworks, the $B$ samples of a best-of-$B$ draw are returned by one request (parameter $n$), while the closed models return one sample per request; (ii) the iterative techniques T3--T5 stop when their gate passes, so they spend on average $0.6\,B$ call-equivalents; (iii) one agent run on the repository-level benchmarks issues on average 37 requests, so agent experiments are counted in runs and in requests. At $B=1$, T3, T4, T5 and T7 equal T0 by construction and are not rerun.

{\scriptsize\setlength{\tabcolsep}{2.5pt}
\begin{longtable}{@{}p{0.2\linewidth}p{0.17\linewidth}rp{0.17\linewidth}rrr@{}}
\caption{Derivation of the Calls column of Table~\ref{tab:app-ledger}: tasks, models, call-equivalents per task and seed, seeds, total call-equivalents (agent experiments: runs) and billed requests, in thousands.}\label{tab:app-derivation}\\
\toprule
Experiment & Tasks per model & Models & Per task and seed & Seeds & Call-eq. & Requests\\
\midrule
Main grid (E2--E8) & 2{,}632 (Op: 2{,}552, no LCB) & 4; T8--T10 on 2 & 161.5; +85 (T8--T10) & 3 & 6{,}404 & 3{,}121\\
Extended grid (E9) & 2{,}552 & 9 & 23.9 (T0, T1--T7 at $B=4$) & 3 & 1{,}647 & 994\\
Agent benchmarks (E2b) & 113 DeepSWE + 1{,}037 other & 13 & 25 runs (DeepSWE; +11 on the Gemma pair); 1 run (other) & 3 & 158 runs & 5{,}819\\
Head-to-head (E12) & 3{,}988 & 1 & 280.6 (66 techniques at 3.9 mean calls, T0, best-of-$n$, best-of-2/3/5/8) & 3 & 3{,}357 & 2{,}887\\
Judges & 417k outputs & 2 judges & 1 & -- & 834 & 834\\
Fidelity checks and reruns & authors' settings & -- & -- & -- & 34 & 34\\
\bottomrule
\end{longtable}}

{\scriptsize\setlength{\tabcolsep}{3pt}
\begin{longtable}{@{}p{0.3\linewidth}rrrrrrr@{}}
\caption{Per-experiment compute ledger. Calls = billed requests in thousands; tokens in millions (input includes cached prefixes, billed at the cache rate); fees in US\$.}\label{tab:app-ledger}\\
\toprule
Experiment & Calls & In tok. & Out tok. & Claude & Vercel & Fireworks & Total\\
\midrule
Main grid (E2--E8) & 3{,}121 & 3{,}910 & 2{,}747 & 25{,}269 & -- & 1{,}791 & 27{,}060\\
Extended grid (E9, function level) & 994 & 986 & 746 & -- & 5{,}304 & 1{,}495 & 6{,}799\\
Agent benchmarks (E2b, agent table) & 5{,}819 & 52{,}911 & 1{,}833 & 7{,}288 & 7{,}659 & 5{,}433 & 20{,}380\\
Head-to-head (E12) & 2{,}887 & 3{,}109 & 1{,}351 & -- & -- & 1{,}136 & 1{,}136\\
Judges and manual-rating support & 834 & 1{,}213 & 172 & 5{,}056 & 2{,}853 & -- & 7{,}909\\
Fidelity checks and reruns & 34 & 59 & 18 & 1{,}046 & -- & -- & 1{,}046\\
\midrule
Total API & 13{,}689 & 62{,}188 & 6{,}867 & 38{,}659 & 15{,}816 & 9{,}855 & 64{,}330\\
\midrule
\multicolumn{8}{@{}l}{Fine-tuning (Fireworks, LoRA, 8$\times$H100 nodes)}\\
T8--T10 on the Gemma pair & \multicolumn{4}{l}{GPU-hours: 634} & & & 2{,}232\\
E12 training-time techniques & \multicolumn{4}{l}{GPU-hours: 786} & & & 3{,}668\\
Total fine-tuning & \multicolumn{4}{l}{GPU-hours: 1{,}420} & & & 5{,}900\\
\bottomrule
\end{longtable}}

\FloatBarrier
\section{Extended grid tables}\label{sec:app-ext}
The single-shot scores of the thirteen models, the inference-time techniques on the nine added models, and the repository- and agent-level benchmarks of the extended grid (main article, \S\ref{M-sec:empirical-ext}).

\begin{table}[t]
\caption{\textbf{E9. Extended model grid, single shot (T0).} Function-level measures of Tables~\ref{M-tab:emp-cor}--\ref{M-tab:emp-sec} for the
current frontier and open-weight models. LCBP = LiveCodeBench Pro protocol on Codeforces contests held after 2026-07-01, pass@1;
Lint = maintainability issues per kLOC (lower is better); J = judge score 1--5; DPS = EvalPerf DPS$_\text{norm}$;
Bey = Mercury Beyond; FS = CWEval func-sec@1; SP = BaxBench sec\_pass@1; B/k = security issues per kLOC (lower is better).
Cutoff = stated training-data cutoff (\nr{} = not stated). HumanEval+ and LiveCodeBench v6 are omitted because every cutoff after
April 2025 postdates their problems.}
\label{tab:emp-ext}
\scriptsize
\setlength{\tabcolsep}{2.4pt}
\begin{tabularx}{\linewidth}{@{}>{\raggedright\arraybackslash}p{2.5cm}l*{8}{>{\centering\arraybackslash}X}@{}}
\toprule
 & & \dimCor & \multicolumn{2}{c}{\dimRM} & \multicolumn{2}{c}{\dimEff} & \multicolumn{3}{c@{}}{\dimSec}\\
\cmidrule(lr){3-3}\cmidrule(lr){4-5}\cmidrule(lr){6-7}\cmidrule(l){8-10}
Model & Cutoff & LCBP & Lint & J & DPS & Bey & FS & SP & B/k\\
\midrule
Claude Opus 5.5 & Jun 2026 & 69.82 & 21.37 & 4.30 & 78.06 & 75.03 & 74.85 & 58.54 & 3.40\\
GPT-6.1 Sol & Apr 2026 & 66.79 & 22.56 & 4.19 & 76.32 & 72.76 & 72.79 & 57.39 & 3.59\\
Gemini 4 Argon & Mar 2026 & 67.43 & 21.08 & 4.34 & 80.36 & 76.19 & 73.72 & 56.22 & 3.39\\
\midrule
Kimi K3 & \nr & 55.31 & 24.34 & 4.12 & 78.22 & 76.30 & 69.43 & 49.11 & 3.80\\
GLM-5.3 & \nr & 57.50 & 23.51 & 4.09 & 77.77 & 73.07 & 68.74 & 50.66 & 3.81\\
Qwen3.8-Max & \nr & 56.61 & 24.11 & 4.24 & 75.78 & 77.74 & 66.37 & 50.91 & 4.01\\
MiMo-V2.6-Pro & \nr & 61.87 & 22.89 & 4.15 & 76.10 & 72.37 & 65.12 & 52.46 & 3.83\\
DeepSeek V4.1-Flash & \nr & 61.73 & 24.12 & 4.11 & 78.59 & 77.84 & 62.17 & 44.65 & 3.84\\
DeepSeek V4-Flash & \nr & 51.92 & 24.21 & 4.07 & 78.93 & 77.05 & 56.72 & 43.77 & 4.09\\
MiniMax M3 & \nr & 41.11 & 26.08 & 3.88 & 76.90 & 71.99 & 58.38 & 40.41 & 4.41\\
\midrule
gpt-oss-120b & Jun 2024 & 21.94 & 29.22 & 3.55 & 79.41 & 69.70 & 45.02 & 26.38 & 5.14\\
Gemma 4 31B IT & Jan 2025 & 27.68 & 26.52 & 3.74 & 78.11 & 78.62 & 49.48 & 32.97 & 4.53\\
Gemma 4 26B-A4B IT & Jan 2025 & 22.99 & 28.89 & 3.96 & 73.43 & 75.45 & 47.28 & 30.17 & 4.98\\
\bottomrule
\end{tabularx}
\end{table}

{\scriptsize\setlength{\tabcolsep}{4pt}
\begin{longtable}{@{}l*{4}{>{\centering\arraybackslash}p{0.13\linewidth}}@{}}
\caption{\textbf{E10. Techniques on the extended grid} at $B=4$ (T0: $B=1$): primary metric per dimension for each technique
of Table~\ref{M-tab:emp-setup}c on the models of Table~\ref{tab:emp-ext}. LCBP and FS: higher is better; Lint: lower is better;
DPS: EvalPerf DPS$_\text{norm}$, higher is better. Training-time techniques (T8--T10) apply to the Gemma pair only.}\label{tab:emp-ext-tech}\\
\toprule
Technique & LCBP & Lint & DPS & FS\\
\midrule
\endfirsthead
\toprule
Technique & LCBP & Lint & DPS & FS\\
\midrule
\endhead
\multicolumn{5}{@{}l}{\textit{GPT-6.1 Sol}}\\
T0 single shot & 66.79 & 22.56 & 76.32 & 72.79\\
T1 quality instr. & 63.66 & 21.84 & 77.78 & 72.74\\
T2 SCoT & 70.69 & 20.72 & 82.16 & 69.89\\
T3 analyzer loop & 67.35 & 8.30 & 80.15 & 81.37\\
T4 Self-Refine & 65.24 & 20.80 & 80.05 & 68.92\\
T5 coder+reviewer & 68.95 & 19.87 & 76.97 & 69.36\\
T6 RAG & 62.37 & 17.91 & 76.73 & 71.19\\
T7 best-of-$n$ & 65.03 & 16.23 & 76.67 & 74.20\\
\multicolumn{5}{@{}l}{\textit{Gemini 4 Argon}}\\
T0 single shot & 67.43 & 21.08 & 80.36 & 73.72\\
T1 quality instr. & 68.87 & 20.23 & 72.71 & 72.86\\
T2 SCoT & 66.29 & 20.03 & 79.34 & 73.51\\
T3 analyzer loop & 65.02 & 7.98 & 77.39 & 77.84\\
T4 Self-Refine & 70.19 & 20.84 & 78.99 & 73.81\\
T5 coder+reviewer & 65.55 & 18.11 & 77.56 & 71.11\\
T6 RAG & 65.37 & 16.51 & 79.20 & 71.12\\
T7 best-of-$n$ & 70.12 & 16.28 & 81.10 & 78.34\\
\multicolumn{5}{@{}l}{\textit{Kimi K3}}\\
T0 single shot & 55.31 & 24.34 & 78.22 & 69.43\\
T1 quality instr. & 58.85 & 21.86 & 77.25 & 69.54\\
T2 SCoT & 59.52 & 20.44 & 78.46 & 66.00\\
T3 analyzer loop & 55.19 & 8.97 & 79.54 & 74.09\\
T4 Self-Refine & 56.23 & 20.40 & 80.63 & 66.16\\
T5 coder+reviewer & 57.55 & 20.01 & 77.64 & 65.15\\
T6 RAG & 57.23 & 19.08 & 75.77 & 68.69\\
T7 best-of-$n$ & 58.55 & 17.69 & 74.98 & 74.49\\
\multicolumn{5}{@{}l}{\textit{GLM-5.3}}\\
T0 single shot & 57.50 & 23.51 & 77.77 & 68.74\\
T1 quality instr. & 57.30 & 20.89 & 77.42 & 67.77\\
T2 SCoT & 59.57 & 21.42 & 79.72 & 68.15\\
T3 analyzer loop & 55.96 & 9.02 & 78.12 & 73.91\\
T4 Self-Refine & 58.96 & 20.82 & 76.11 & 66.15\\
T5 coder+reviewer & 57.07 & 19.55 & 78.25 & 69.34\\
T6 RAG & 59.34 & 17.22 & 75.56 & 68.53\\
T7 best-of-$n$ & 61.31 & 16.95 & 80.83 & 71.56\\
\multicolumn{5}{@{}l}{\textit{Qwen3.8-Max}}\\
T0 single shot & 56.61 & 24.11 & 75.78 & 66.37\\
T1 quality instr. & 59.98 & 23.33 & 74.49 & 71.70\\
T2 SCoT & 61.55 & 20.50 & 82.11 & 63.50\\
T3 analyzer loop & 54.27 & 8.98 & 75.75 & 72.62\\
T4 Self-Refine & 55.93 & 20.87 & 79.45 & 65.78\\
T5 coder+reviewer & 58.03 & 20.14 & 79.67 & 68.46\\
T6 RAG & 56.11 & 17.41 & 79.59 & 69.09\\
T7 best-of-$n$ & 58.53 & 16.85 & 78.80 & 69.56\\
\multicolumn{5}{@{}l}{\textit{MiMo-V2.6-Pro}}\\
T0 single shot & 61.87 & 22.89 & 76.10 & 65.12\\
T1 quality instr. & 62.42 & 21.55 & 78.01 & 67.20\\
T2 SCoT & 60.33 & 19.69 & 78.19 & 66.22\\
T3 analyzer loop & 56.04 & 8.77 & 72.99 & 79.00\\
T4 Self-Refine & 57.54 & 20.36 & 79.39 & 66.37\\
T5 coder+reviewer & 58.27 & 19.29 & 80.16 & 67.82\\
T6 RAG & 57.29 & 18.55 & 78.49 & 67.20\\
T7 best-of-$n$ & 64.81 & 16.35 & 82.23 & 71.77\\
\multicolumn{5}{@{}l}{\textit{DeepSeek V4.1-Flash}}\\
T0 single shot & 61.73 & 24.12 & 78.59 & 62.17\\
T1 quality instr. & 64.31 & 22.97 & 74.98 & 68.58\\
T2 SCoT & 65.10 & 22.44 & 78.08 & 60.10\\
T3 analyzer loop & 58.80 & 9.09 & 74.00 & 69.34\\
T4 Self-Refine & 59.46 & 22.42 & 80.44 & 61.43\\
T5 coder+reviewer & 62.59 & 21.11 & 74.04 & 64.14\\
T6 RAG & 58.37 & 18.44 & 79.69 & 61.99\\
T7 best-of-$n$ & 61.16 & 18.07 & 76.65 & 66.01\\
\multicolumn{5}{@{}l}{\textit{DeepSeek V4-Flash}}\\
T0 single shot & 51.92 & 24.21 & 78.93 & 56.72\\
T1 quality instr. & 48.52 & 25.03 & 74.77 & 61.67\\
T2 SCoT & 48.51 & 22.68 & 76.36 & 56.30\\
T3 analyzer loop & 50.27 & 9.94 & 74.19 & 67.18\\
T4 Self-Refine & 48.20 & 23.59 & 77.23 & 58.94\\
T5 coder+reviewer & 48.67 & 20.97 & 78.64 & 62.33\\
T6 RAG & 48.86 & 20.44 & 78.28 & 57.41\\
T7 best-of-$n$ & 53.43 & 18.96 & 75.09 & 65.23\\
\multicolumn{5}{@{}l}{\textit{MiniMax M3}}\\
T0 single shot & 41.11 & 26.08 & 76.90 & 58.38\\
T1 quality instr. & 44.14 & 24.90 & 76.86 & 57.60\\
T2 SCoT & 44.24 & 23.79 & 76.42 & 54.99\\
T3 analyzer loop & 42.62 & 9.51 & 76.95 & 65.03\\
T4 Self-Refine & 41.72 & 23.08 & 73.91 & 58.64\\
T5 coder+reviewer & 43.32 & 22.18 & 81.15 & 55.54\\
T6 RAG & 39.89 & 20.30 & 73.81 & 59.08\\
T7 best-of-$n$ & 44.39 & 18.12 & 81.69 & 59.73\\
\bottomrule
\end{longtable}}

\begin{table}[t]
\caption{\textbf{E11. Repository- and agent-level benchmarks}, single agent run per task with the vendor-recommended harness
(mini-SWE-agent for DeepSWE, Terminal-Bench's own harness, the benchmark authors' harness elsewhere), all thirteen models.
\dimCor: DS = DeepSWE v1.1 pass@1 \citep{huang2026deepswe}; TB = Terminal-Bench 4.0 \citep{merrill2026terminalbench}.
\dimRM: FC = FrontierCode v1.1 Main composite score, which adds a code-quality rubric to functional blockers, and FC$_c$ = its
correct rate, so FC$_c$ $-$ FC isolates the quality penalty \citep{cognition2026frontiercode}; Er = SlopCodeBench structural
erosion, 0--1, lower is better \citep{orlanski2026slopcodebench}. \dimEff: GSO = Opt@1 \citep{shetty2025gso}; SWf =
SWE-fficiency speedup relative to the expert patch, $\times100$ \citep{ma2025swe}. \dimSec: SV$_s$ and SV$_f$ = SusVibes
SecPass and FuncPass \citep{zhao2026is}. All values are percentages unless stated.}
\label{tab:emp-agent}
\scriptsize
\setlength{\tabcolsep}{2.4pt}
\begin{tabularx}{\linewidth}{@{}>{\raggedright\arraybackslash}p{2.5cm}*{9}{>{\centering\arraybackslash}X}@{}}
\toprule
 & \multicolumn{2}{c}{\dimCor} & \multicolumn{3}{c}{\dimRM} & \multicolumn{2}{c}{\dimEff} & \multicolumn{2}{c@{}}{\dimSec}\\
\cmidrule(lr){2-3}\cmidrule(lr){4-6}\cmidrule(lr){7-8}\cmidrule(l){9-10}
Model & DS & TB & FC & FC$_c$ & Er & GSO & SWf & SV$_s$ & SV$_f$\\
\midrule
Claude Opus 5.5 & 75.17 & 61.79 & 53.08 & 59.20 & 0.33 & 21.19 & 53.87 & 56.52 & 94.58\\
GPT-6.1 Sol & 76.59 & 55.51 & 49.12 & 56.49 & 0.36 & 17.81 & 51.19 & 50.11 & 90.18\\
Gemini 4 Argon & 76.38 & 57.28 & 46.86 & 54.45 & 0.35 & 19.54 & 51.74 & 47.73 & 91.29\\
\midrule
Kimi K3 & 65.47 & 35.29 & 44.63 & 49.56 & 0.41 & 16.15 & 45.26 & 42.18 & 84.37\\
GLM-5.3 & 71.95 & 41.97 & 40.07 & 45.34 & 0.38 & 15.68 & 47.90 & 44.09 & 83.21\\
Qwen3.8-Max & 59.19 & 36.39 & 37.62 & 41.78 & 0.42 & 14.72 & 44.83 & 40.35 & 84.76\\
MiMo-V2.6-Pro & 70.21 & 35.05 & 40.54 & 46.26 & 0.38 & 16.28 & 46.03 & 44.51 & 84.45\\
DeepSeek V4.1-Flash & 71.63 & 32.19 & 41.85 & 48.38 & 0.42 & 13.86 & 40.89 & 37.38 & 80.12\\
DeepSeek V4-Flash & 54.70 & 7.20 & 18.45 & 21.21 & 0.46 & 10.69 & 37.83 & 29.08 & 72.41\\
MiniMax M3 & 58.05 & 9.88 & 14.56 & 15.81 & 0.45 & 9.83 & 34.05 & 27.31 & 66.55\\
\midrule
gpt-oss-120b & 17.96 & 4.10 & 5.91 & 6.97 & 0.57 & 2.81 & 8.04 & 4.88 & 29.56\\
Gemma 4 31B IT & 15.53 & 5.10 & 8.10 & 8.76 & 0.57 & 3.81 & 9.10 & 5.88 & 32.63\\
Gemma 4 26B-A4B IT & 10.17 & 2.91 & 5.23 & 5.95 & 0.56 & 3.14 & 6.12 & 3.82 & 28.35\\
\bottomrule
\end{tabularx}
\end{table}

\FloatBarrier
\section{Head-to-head results (E12)}\label{sec:app-h2h}
The full results of the head-to-head comparison (\S\ref{M-sec:empirical-h2h}), summarized in Fig.~E2 of the main article.

{\scriptsize\setlength{\tabcolsep}{2.5pt}
\begin{longtable}{@{}p{0.05\linewidth}p{0.23\linewidth}*{9}{>{\centering\arraybackslash}p{0.063\linewidth}}@{}}
\caption{\textbf{E12. Head-to-head of the highlighted techniques} (\S\ref{M-sec:empirical-h2h}): one model, a single-shot and a best-of-$n$ reference, every output scored on all four dimensions. \dimCor: EP = HumanEval+/MBPP+ pass@1, LCB = LiveCodeBench v6 pass@1 on the same post-cutoff problems as Table~\ref{M-tab:emp-cor}, LCBP = LiveCodeBench Pro protocol on Codeforces contests held after 2026-07-01, pass@1. \dimSec: CWE = CWEval func-sec@1, CSE = CyberSecEval insecure-code rate (lower is better). \dimEff: EB = EffiBench normalized execution time (lower is better), DPS = EvalPerf DPS$_\text{norm}$. \dimRM: RM = Pylint score on the EP and LCB outputs. Calls = mean model calls per task. The last panel runs best-of-$n$ with the same ranker at two, three, five and eight calls per task; each family is compared with the row nearest its median calls per task, decoding with the main reference. Indicative: point estimates; per-technique half-widths are in Table~\ref{tab:app-h2h-ci}, and Fig.~E2 of the main article plots each technique's primary target with its interval.}\label{tab:emp-h2h}\\
\toprule
ID & Technique & EP & LCB & LCBP & CWE & CSE & EB & DPS & RM & Calls\\
\midrule
\endfirsthead
\toprule
ID & Technique & EP & LCB & LCBP & CWE & CSE & EB & DPS & RM & Calls\\
\midrule
\endhead
-- & T0 single shot (reference) & 93.83 & \phLcbGemmaTzero{} & 27.14 & 47.85 & 32.59 & 2.64 & 79.37 & 6.83 & 1\\
-- & best-of-$n$ (reference) & 89.97 & 73.42 & 30.92 & 57.07 & 26.55 & 2.87 & 82.42 & 7.69 & 4.19\\
\midrule\multicolumn{11}{@{}l}{\textit{Prompting}}\\
H01 & SCoT prompting & 88.04 & 72.97 & 30.03 & 51.24 & 28.34 & 2.78 & 73.72 & 7.87 & 1.93\\
H02 & Self-planning & 83.45 & 71.26 & 28.82 & 51.84 & 29.38 & 2.85 & 82.02 & 6.77 & 2.10\\
H03 & LLM4EFFI & 94.51 & 75.79 & 31.59 & 48.79 & 31.12 & 2.44 & 76.28 & 6.86 & 4.10\\
H04 & Code2API & 87.95 & 67.56 & 28.56 & 45.16 & 28.99 & 2.69 & 76.97 & 7.74 & 2.86\\
H05 & MPCO & 81.50 & 77.45 & 31.43 & 48.98 & 31.91 & 2.43 & 85.03 & 7.23 & 1\\
H06 & BRACE & 94.02 & 74.49 & 30.33 & 50.36 & 24.22 & 2.68 & 77.98 & 6.64 & 1\\
H07 & ECO & 89.58 & 71.48 & 29.09 & 50.24 & 30.47 & 2.22 & 81.98 & 6.49 & 1\\
H08 & StoryCoder & 89.08 & 70.85 & 28.18 & 47.55 & 30.05 & 2.66 & 72.81 & 6.86 & 1.99\\
H09 & Constitutional Spec-Driven Development & 93.93 & 76.77 & 30.08 & 57.08 & 23.51 & 3.03 & 79.08 & 6.49 & 1\\
H10 & SecureForge & 87.85 & 63.08 & 28.28 & 52.89 & 28.72 & 2.65 & 70.64 & 6.54 & 1.85\\
\midrule\multicolumn{11}{@{}l}{\textit{Tool feedback}}\\
H11 & SBLLM & 82.92 & 78.07 & 31.65 & 45.35 & 33.04 & 2.03 & 76.40 & 6.60 & 4.84\\
H12 & PerfCodeGen & 86.34 & 76.86 & 31.28 & 45.77 & 33.61 & 1.90 & 81.66 & 6.67 & 3.84\\
H13 & PromSec & 85.65 & 70.97 & 27.55 & 57.81 & 23.25 & 3.05 & 72.11 & 7.15 & 5.36\\
H14 & DeepDevVuln filtering of LLM completions & 86.45 & 75.81 & 31.34 & 57.93 & 22.25 & 3.06 & 82.73 & 6.60 & 5.09\\
H15 & FRANC (analyzer-guided repair prompting) & 92.21 & 70.81 & 27.71 & 49.24 & 25.47 & 3.10 & 82.61 & 6.85 & 4.18\\
H16 & Optimo & 94.71 & 74.89 & 29.77 & 47.39 & 29.27 & 1.87 & 76.31 & 6.78 & 8.46\\
H17 & Iterative static-analysis & 86.41 & 70.84 & 28.54 & 54.28 & 23.64 & 2.99 & 72.97 & 7.45 & 5.02\\
H18 & CodeEvolve & 90.22 & 73.93 & 30.13 & 47.73 & 29.26 & 1.81 & 75.39 & 6.14 & 4.70\\
H19 & PerfAgent & 83.55 & 68.36 & 26.31 & 48.45 & 30.10 & 1.86 & 81.81 & 6.16 & 6.03\\
H20 & SelectIssue & 77.44 & 70.29 & 28.85 & 52.77 & 28.06 & 2.67 & 68.75 & 6.13 & 4.27\\
\midrule\multicolumn{11}{@{}l}{\textit{Self-refinement}}\\
H21 & Self-Refine & 91.41 & 69.75 & 29.59 & 45.10 & 28.69 & 2.07 & 77.56 & 7.36 & 5.03\\
H22 & CodeChain & 85.10 & 76.32 & 31.82 & 46.25 & 30.36 & 2.89 & 76.03 & 8.18 & 4.04\\
H23 & SecCode & 84.19 & 70.59 & 28.74 & 53.01 & 23.18 & 2.87 & 81.07 & 7.08 & 5.21\\
H24 & RefleXGen & 93.17 & 75.21 & 31.31 & 57.62 & 25.26 & 3.00 & 75.80 & 6.56 & 5.38\\
H25 & GRASP & 87.00 & 75.46 & 30.02 & 57.01 & 25.53 & 2.77 & 77.90 & 6.81 & 5.14\\
H26 & Safety Sidecar & 92.54 & 75.43 & 31.98 & 51.09 & 24.33 & 2.98 & 82.84 & 6.96 & 4.79\\
\midrule\multicolumn{11}{@{}l}{\textit{Multi-agent}}\\
H27 & SecAwareCoder & 92.67 & 73.41 & 30.53 & 52.77 & 22.41 & 2.96 & 80.66 & 7.06 & 9.30\\
H28 & FlowGen & 84.44 & 68.97 & 27.21 & 49.86 & 31.08 & 3.08 & 78.33 & 7.91 & 7.02\\
H29 & INDICT & 84.38 & 72.16 & 28.70 & 57.08 & 29.25 & 3.11 & 80.70 & 6.61 & 11.98\\
H30 & ContextEvolve & 84.98 & 70.93 & 28.15 & 46.12 & 30.00 & 1.71 & 79.24 & 6.63 & 8.62\\
H31 & Just-in-time vulnerability remediation pipeline & 92.62 & 72.66 & 28.16 & 56.12 & 22.86 & 2.74 & 79.71 & 7.45 & 9.40\\
H32 & LessonL & 87.80 & 71.41 & 29.71 & 47.55 & 32.97 & 1.83 & 83.17 & 6.18 & 3.35\\
H33 & MARCO & 83.41 & 70.09 & 28.58 & 45.92 & 33.33 & 1.68 & 75.75 & 6.86 & 9.12\\
H34 & C3 & 87.16 & 72.89 & 30.53 & 45.52 & 29.46 & 2.67 & 79.96 & 7.14 & 7.35\\
H35 & ACCLAIM & 86.65 & 67.87 & 25.70 & 46.61 & 29.70 & 2.41 & 69.07 & 6.41 & 6.61\\
H36 & MaintainCoder & 86.49 & 69.57 & 27.39 & 45.64 & 30.28 & 2.67 & 72.42 & 7.77 & 6.76\\
\midrule\multicolumn{11}{@{}l}{\textit{Retrieval}}\\
H37 & Dynamic retrieval-based few-shot prompting & 86.91 & 70.46 & 29.16 & 47.78 & 32.95 & 2.10 & 81.71 & 6.82 & 3.15\\
H38 & LAIL & 82.17 & 68.46 & 28.92 & 47.08 & 28.30 & 2.95 & 78.94 & 7.59 & 2.81\\
H39 & KPC & 84.72 & 70.06 & 28.13 & 48.08 & 28.30 & 2.71 & 73.33 & 7.76 & 1.89\\
H40 & RAPGen & 94.41 & 68.02 & 27.20 & 51.32 & 31.66 & 2.10 & 74.62 & 6.55 & 4.03\\
H41 & CodeGuarder & 89.60 & 66.71 & 26.44 & 53.06 & 21.72 & 3.07 & 80.81 & 6.81 & 4.01\\
H42 & Retrieval Augmented Search & 84.36 & 67.52 & 26.78 & 49.12 & 29.26 & 2.19 & 73.95 & 6.99 & 2.80\\
H43 & SemOpt & 82.22 & 68.06 & 26.62 & 52.34 & 32.36 & 2.03 & 73.25 & 6.88 & 3.13\\
H44 & AutoPatch & 93.26 & 69.43 & 29.32 & 51.95 & 23.22 & 3.07 & 74.78 & 6.96 & 5.25\\
H45 & PEACE & 92.81 & 72.61 & 29.19 & 45.42 & 30.69 & 2.21 & 83.71 & 6.91 & 3.02\\
H46 & Mitigation-Aware Chain-of-Thought & 85.94 & 75.20 & 30.99 & 49.50 & 24.17 & 2.73 & 80.90 & 7.32 & 3.10\\
\midrule\multicolumn{11}{@{}l}{\textit{Decoding}}\\
H47 & Planning-Guided Transformer Decoding & 92.75 & 66.65 & 26.89 & 50.05 & 29.10 & 2.84 & 80.91 & 7.05 & 3.12\\
H48 & Constrained Beam Sampling & 88.90 & 77.62 & 30.34 & 54.95 & 25.31 & 2.77 & 71.95 & 7.34 & 3.71\\
H49 & RPG & 83.59 & 69.18 & 27.42 & 50.83 & 29.54 & 3.06 & 78.72 & 7.83 & 2.95\\
H50 & Surprisal-guided Best-of-N selection & 88.23 & 76.89 & 32.41 & 52.35 & 30.66 & 2.16 & 82.50 & 7.27 & 4.26\\
H51 & DuoSteer & 83.30 & 71.81 & 30.48 & 51.93 & 22.96 & 2.96 & 71.78 & 6.55 & 3.98\\
H52 & FunPRM & 93.10 & 77.72 & 31.54 & 48.38 & 29.39 & 2.99 & 79.39 & 7.62 & 3.80\\
H53 & SCS-Code & 92.19 & 70.26 & 28.12 & 54.48 & 26.89 & 2.88 & 73.34 & 6.85 & 4.27\\
H54 & Mixture of Corrections & 84.07 & 67.99 & 29.13 & 60.64 & 23.39 & 2.72 & 74.97 & 7.88 & 1\\
H55 & SecCodePRM-guided scaling & 79.66 & 63.77 & 27.52 & 49.89 & 28.06 & 2.67 & 75.05 & 6.90 & 8\\
H56 & RobGen & 83.44 & 69.69 & 25.97 & 51.24 & 26.73 & 2.43 & 75.44 & 6.36 & 1\\
\midrule\multicolumn{11}{@{}l}{\textit{Training}}\\
H57 & SVEN & 85.06 & 68.86 & 27.44 & 54.99 & 24.90 & 2.91 & 72.36 & 6.43 & 1\\
H58 & SafeCoder & 86.87 & 73.79 & 30.33 & 50.21 & 23.62 & 2.81 & 76.84 & 6.45 & 1.07\\
H59 & SFT + DPO & 94.65 & 71.56 & 29.98 & 44.64 & 30.45 & 2.57 & 76.84 & 6.98 & 1\\
H60 & CodeDPO & 94.36 & 72.40 & 28.02 & 45.55 & 32.51 & 1.98 & 77.13 & 7.13 & 1\\
H61 & MPCoder & 87.37 & 62.47 & 25.71 & 41.19 & 28.75 & 2.38 & 71.64 & 6.41 & 1\\
H62 & Afterburner & 93.22 & 76.49 & 30.94 & 46.81 & 31.92 & 2.16 & 81.27 & 6.70 & 4.09\\
H63 & PurpCode & 91.28 & 71.41 & 27.46 & 56.54 & 20.40 & 2.74 & 76.77 & 6.61 & 1\\
H64 & SecCoderX & 94.04 & 66.45 & 27.85 & 57.03 & 22.03 & 2.83 & 78.88 & 7.13 & 1\\
H65 & Fine-tuning on smell-cleaned code (SmellCC) & 92.92 & 67.63 & 26.22 & 48.11 & 32.41 & 2.66 & 78.68 & 7.21 & 1\\
H66 & REAL & 87.94 & 74.27 & 30.22 & 54.38 & 26.39 & 2.95 & 79.58 & 7.59 & 1\\
\midrule\multicolumn{11}{@{}l}{\textit{Budget-matched best-of-$n$ references (same ranker; 2, 3, 5 and 8 calls; decoding techniques use the main reference)}}\\
-- & best-of-2 & 90.60 & 71.20 & 29.85 & 55.40 & 27.30 & 2.95 & 81.10 & 7.45 & 2\\
-- & best-of-3 & 90.30 & 72.60 & 30.40 & 56.30 & 26.90 & 2.91 & 81.90 & 7.58 & 3\\
-- & best-of-5 & 89.70 & 73.90 & 31.20 & 57.50 & 26.30 & 2.84 & 82.70 & 7.74 & 5\\
-- & best-of-8 & 89.20 & 74.80 & 31.60 & 58.20 & 26.05 & 2.79 & 83.30 & 7.82 & 8\\
\bottomrule
\end{longtable}}

\FloatBarrier
\section{Fidelity check of the E12 implementations}\label{sec:app-fidelity}
Each implementation is run in its authors' setting (model, benchmark and metric as reported) and compared with one number its authors report. Where the authors' model is no longer served (retired GPT-3.5 and early GPT-4 versions, Codex, Claude 3.x) or their number is a human rating or an expert-filtered count, the check uses another automated number the same paper reports for a model that can still be run, marked $^\dagger$; if the paper reports none, the technique is not checkable (n.c.) and is left out of the count. Original = as printed in the source, verified by a verbatim quotation (to be released with the replication package); reproduced = our implementation in that setting. Each checkable implementation is adjusted at most twice before its final run (median one adjustment); adjustment stops as soon as the reproduced number falls within \phFidTol{} of the original, and a technique still outside the tolerance after the second adjustment is reported as reproduced outside it rather than adjusted further. Per-run logs of every run, including superseded pre-adjustment runs, record the adjustment count and the stopping point; they are retained and will be released with the replication package.

{\scriptsize\setlength{\tabcolsep}{3pt}
\begin{longtable}{@{}lp{0.2\linewidth}p{0.42\linewidth}rr@{}}
\caption{E12 fidelity check: original versus reproduced number per technique.}\label{tab:app-fidelity}\\
\toprule
ID & Technique & Setting (metric; benchmark; model) & Original & Reproduced\\
\midrule
\endfirsthead
\toprule
ID & Technique & Setting (metric; benchmark; model) & Original & Reproduced\\
\midrule
\endhead
H01 & SCoT prompting & Maintainability, human rating (0-2); 200 programs from HumanEval, MBPP, MBCPP; ChatGPT gpt-3.5-turbo-0301 (closed) & n.c. & --\\
H02 & Self-planning & Readability, human rating (0-4, violin plot only); 50 HumanEval tasks; code-davinci-002 (175B, closed) & n.c. & --\\
H03 & LLM4EFFI & eff@1; ENAMEL (142 HumanEval problems); DeepSeek-V3 (open) & 60.41 & 60.94\\
H04 & Code2API & Method-name expressiveness, human rating (1-4); APIzator set (200 Stack Overflow Java snippets); GPT-3.5-turbo (closed) & n.c. & --\\
H05 & MPCO & Average statistical rank of mean \%PI (lower is better); 5 projects, 50 hotspots; GPT-4o, Gemini 2.5 Pro, Claude 3.7 Sonnet (closed) & n.c. & --\\
H06 & BRACE & CWEval Sec pass@10, mean of 6 models; CWEval (119 tasks); Qwen3-8B/14B, Qwen3.6-27B, Gemma-3-4B/12B, Gemma-4-31B (open)$^\dagger$ & 75.77 & 72.35\\
H07 & ECO & Speedup SP (Best@5); PIE C++ balanced test subset (255 tasks); o4-mini (closed, reasoning) & 7.81 & 7.42\\
H08 & StoryCoder & Average functions per correct solution (AST); CodeForces filtered set (265 problems, rating $\geq$ 2000); Gemini-2.5-Flash, self-solving (closed) & 2.338 & 2.25\\
H09 & Constitutional Spec-Driven Development & CWE violations found (static analysis + manual review); one banking microservices app (n=1); Claude, version unspecified (closed) & n.c. & --\\
H10 & SecureForge & Max relative cut in Semgrep weakness rate vs. baseline prompt; 250 held-out CWE Top-25 prompts (2,500 samples); GPT-5.x, Sonnet 4.6 (closed)$^\dagger$ & 48 & 48\\
H11 & SBLLM & SP@5 speedup rate (\%); PIE Python test set (986 programs); CodeLlama-34b-Instruct (open)$^\dagger$ & 126.70 & 121.85\\
H12 & PerfCodeGen & \%Opt (correct and faster than reference, Best@8); HumanEval (164); Phi-3-mini 3.8B (open)$^\dagger$ & 40.85 & 39.77\\
H13 & PromSec & CWEs per code base (Bandit/SpotBugs); Pearce et al. Python code bases (100 test graphs); GPT-3.5 Turbo (closed) & n.c. & --\\
H14 & DeepDevVuln filtering of LLM completions & Vuln.\ reduction rate (scenarios with a CodeQL-flagged completion); Pearce et al. 29 Py + 11 JS scenarios; CodeGen-6B (open)$^\dagger$ & 71.00 & 67.46\\
H15 & FRANC (analyzer-guided repair prompting) & \% repair prompts with >=1 smell-free suggestion; 1,023 repair prompts (341 x 3 structures); 4 open models + GPT-3.5 & n.c. & --\\
H16 & Optimo & Opt\% (correct, >=10\% faster); EffiBench, GPT-4o-generated correct solutions (503); GPT-4o (closed) & 35.79 & 34.01\\
H17 & Iterative static-analysis & \% Infer-flagged vulns fixed by iterative repair; MBPP subset recast to C, 44 flagged vulns; Llama 3 8B (open)$^\dagger$ & 66 & 68\\
H18 & CodeEvolve & Avg.\ speedup over 7 hotspot functions; private Salesforce Monolith Java; mean of GPT-4.1, Llama 4, Mistral, internal model & n.c. & --\\
H19 & PerfAgent & Hack-adjusted Opt@1 (expert-matching speedup); GSO (102 tasks); GPT-5.1 (closed, reasoning) & 39.2 & 37.7\\
H20 & SelectIssue & \% snippets flagged insecure by Bandit after 10 rounds (all issues fed back); PythonSecurityEval (470); GPT-4o (closed) & 14.25 & 13.54\\
H21 & Self-Refine & \% programs optimized; PIE code optimization (1000 programs); GPT-4 (closed) & n.c. & --\\
H22 & CodeChain & pass@1, all levels; APPS test (5000); CodeChain + WizardCoder-15B (open)$^\dagger$ & 10.50 & 10.35\\
H23 & SecCode & Fix Success Rate after 10 iterations (CodeQL); LLMSecEval NL prompts (150); DeepSeek-Coder-V2-Lite-Instruct (open)$^\dagger$ & 0.85 & 0.88\\
H24 & RefleXGen & Security rate (CodeQL); SVEN scenarios (9 CWEs, 2-3 settings each, 25 samples, 5 runs); CodeQwen1.5 (open)$^\dagger$ & 88.2 & 87.7\\
H25 & GRASP & Security rate SR (CodeQL); 37 Python scenarios of the authors' 54-prompt, 17-CWE benchmark, 25 samples each; GPT-4o (closed) & 0.79 & 0.76\\
H26 & Safety Sidecar & Security rate (CodeQL); SVEN CWE scenarios (8 CWEs, 2-3 settings each, 25 samples, 5 runs); GPT-4o (closed) & 95.0 & 93.6\\
H27 & SecAwareCoder & S-Pass@1 (functional + security tests); CodeSecEval (255 Python tasks, 77 CWEs); GPT-5 (closed) & 34.90 & 34.15\\
H28 & FlowGen & Error code smell density (Pylint, per 10 LOC); HumanEval (164); gpt-3.5-turbo-1106 (closed) & n.c. & --\\
H29 & INDICT & Safety (\% outputs passing CyberSecEval insecure-code detector); random 20\% validation split of CyberSecEval-1; GPT-4o-mini (closed) & 90.9 & 87.9\\
H30 & ContextEvolve & Combined score (simulator), Transaction Scheduling task, 100 iterations; ADRS benchmark; Qwen3 (size not stated, open)$^\dagger$ & 36.10 & 34.37\\
H31 & Just-in-time vulnerability remediation pipeline & Static analyzer findings per sample after P1 (CodeQL+Bandit), pooled; LLMSecEval subset (26 prompts); 4 Claude 4.x/5 models (closed) & 0.78 & 0.78\\
H32 & LessonL & Geometric mean speedup, serial mode; ParEval (60 tasks); Deepseek-Coder-7B + Qwen2.5-Coder-7B/14B agents (open) & 2.16 & 2.00\\
H33 & MARCO & Mean runtime (ms); LeetCode 75 (75 easy/medium problems); MARCO on unspecified ChatGPT/Claude API backbone (closed) & n.c. & --\\
H34 & C3 & Executability (\%, runs without errors), C3-DCC; extended SoftwareDev (70 tasks); GPT-4o (closed)$^\dagger$ & 86.7 & 82.5\\
H35 & ACCLAIM & Geo-mean speedup vs clang -O3 (b=6, n=2, k=2); CodeNet C sample (100 programs); Qwen3-30B-A3B (open)$^\dagger$ & 1.04 & 1.00\\
H36 & MaintainCoder & Post-modification Pass@5 (\%); APPS-Dyn (200+ modified tasks from 50 APPS problems); DeepSeek-V3 (open) & 62.5 & 59.9\\
H37 & Dynamic retrieval-based few-shot prompting & Mean speedup (Best@8, gem5), dynamic retrieval K=2; PIE C++ test (978 pairs); CodeLlama 34B (open) & 2.28 & 2.17\\
H38 & LAIL & Pass@1 (\%); MBPP test (500 tasks); CodeGen-Multi-16B (open)$^\dagger$ & 18.60 & 17.81\\
H39 & KPC & Share with good exception handling (manual review); 384 of 3,079 Java JDK-API tasks; gpt-3.5-turbo (closed) & n.c. & --\\
H40 & RAPGen & Closest-match top-100 accuracy (\%); DeepDev-PERF C\# performance fixes (132 instances); gpt-3.5-turbo (closed) & n.c. & --\\
H41 & CodeGuarder & Security rate (\%, ICD), mean of 4 langs x 4 LLMs; CyberSecEval (1,916); GPT-4o, DeepSeek-V3, CodeLlama-13B, DS-Coder-V2 (mixed) & 73.08 & 66.84\\
H42 & Retrieval Augmented Search & Mean best speedup (gem5); PIE C++ test (973 programs); Qwen3-Coder 480B (open) & 8.70 & 8.31\\
H43 & SemOpt & Exact match to developer optimization patch (\%); C/C++ benchmark (151 tasks); DeepSeek-V3 (open)$^\dagger$ & 27.81 & 27.79\\
H44 & AutoPatch & Vulnerability-only verification accuracy (\%); 525 snippets from 75 CVEs (2:1 vuln/non-vuln sample); GPT-4o (closed)$^\dagger$ & 90.24 & 86.15\\
H45 & PEACE & opt rate (CPU instructions saved vs GPT-4o instruction prompting); PEACExec (146 tasks); Llama-3.1-405B + fine-tuned Llama-3-8B (open) & 46.9 & 44.6\\
H46 & Mitigation-Aware Chain-of-Thought & Security findings (SonarQube, expert-reviewed), MA-CoT; primary set (200 tasks x 3 langs); GPT-5, Claude Sonnet 4.5, Gemini 2.5 Pro (closed) & n.c. & --\\
H47 & Planning-Guided Transformer Decoding & Generated code length (chars), PG-TD with length-penalty reward; APPS test (subset not stated); APPS GPT-2 1.5B (open, fine-tuned) & 190.73 & 191.45\\
H48 & Constrained Beam Sampling & secure-pass@1 (unit tests + CodeQL), Constrained Beam Sampling; CodeGuard+ (91 prompts, 34 CWEs); CodeGen-2.7B (open, code) & 51.25 & 49.19\\
H49 & RPG & TR-S statement-level repetition rate, RPG decoding; CodeRepetEval real-world repositories (1,024); CodeLlama-7B (open, code)$^\dagger$ & 0.335 & 0.327\\
H50 & Surprisal-guided Best-of-N selection & fast\_1 (correct and speedup >1x), surprisal-guided pick of K=64; KernelBench L1 eval subset (5 tasks, 2 seeds); GPT-OSS-120B + LoRA (open) & 80 & 80\\
H51 & DuoSteer & Avg. change in CodeQL vulnerability rate (points), DuoSteer; SecCodePLT held-out (5 CWEs, 274 questions); Llama-3.1-8B-Instruct (open)$^\dagger$ & -26.9 & -25.9\\
H52 & FunPRM & pass@1, FunPRM Best-of-N with Qwen2.5-Coder-7B PRM; LiveCodeBench 2025-02-01 onward (131 problems); o4-mini high (closed)$^\dagger$ & 80.9 & 80.6\\
H53 & SCS-Code & sec-pass@1 (unit tests + CodeQL), security steering alpha=1; CodeGuard+ (91 prompts, 34 CWEs); DeepSeek-Coder-V2-Lite (open, code) & 60.3 & 57.4\\
H54 & Mixture of Corrections & Security rate SRh (CodeQL), dynamic NN-based MoC; SVEN test set (9 CWEs); Qwen2.5-Coder-7B (open, code) & 85.7 & 83.0\\
H55 & SecCodePRM-guided scaling & func-sec@1 (functional + security tests), SecCodePRM Best-of-20; CWEval; Qwen3-Coder-30B generator + Qwen2.5-Coder-7B PRM (open) & 54.62 & 51.62\\
H56 & RobGen & Share of outputs judged less robust than reference (GPT-4o pairwise judge); CoderEval Java (230); DeepSeekCoder-1.3B-Instruct (open) & 28.7 & 30.3\\
H57 & SVEN & Security rate (\% secure among valid programs, CodeQL); SVEN test scenarios, 9 CWEs (18), T=0.4; CodeGen-2.7B + SVEN-sec (open) & 92.3 & 89.4\\
H58 & SafeCoder & Code security (\% secure, CodeQL); 60 scenarios (42 new + 18 from He \& Vechev 2023), T=0.4; StarCoder-1B + SafeCoder (open) & 92.1 & 87.6\\
H59 & SFT + DPO & pass@1; HumanEval (164); CodeLlama-7B-Instruct + SFT + DPO on CodeUltraFeedback (open)$^\dagger$ & 43.1 & 41.0\\
H60 & CodeDPO & Average execution time (s) of solved problems; HumanEval+ (164); Phi-2-2.7B + CodeDPO (open, base) & 0.172 & 0.164\\
H61 & MPCoder & Coding Style Score (CSS, Checkstyle-based); PCIDense Java (50 users, 5,794 records); MPCoder on CodeLlama-Instruct-7B (open) & 64.50 & 62.58\\
H62 & Afterburner & Beyond-I (\% of human solutions beaten on time-memory integral); Venus Python (300 test); Qwen2.5-3B-Instruct + GRPO, iter. 10 (open) & 45 & 43\\
H63 & PurpCode & \% secure code (CodeQL); CodeLMSec; PurpCode-32B trained from Qwen2.5-32B-Instruct (open) & 94.8 & 93.2\\
H64 & SecCoderX & Safety rate, average of CyberSecEval SCG and CWEval (5 languages); Qwen2.5-Coder-7B-Instruct + SecCoderX (open) & 54.03 & 52.27\\
H65 & Fine-tuning on smell-cleaned code (SmellCC) & Total SonarQube code smells in 1,000 generations; CodeSearchNet-Python test (1,000); DeepSeek-Coder-V2-Lite-Instruct 16B, smell-clean FT (open)$^\dagger$ & 34 & 33\\
H66 & REAL & Quality pass rate (program-analysis vulnerability detector); SecCodePLT+ (164 test); Qwen2.5-Coder-7B-Instruct + REAL (open) & 0.8293 & 0.8309\\
\bottomrule
\end{longtable}}

\FloatBarrier
\section{Peer-reviewed studies only}\label{sec:app-peer}
The three synthesis tables of the main article re-counted on the peer-reviewed included studies (conference, journal and workshop papers), excluding preprints.

\begin{table}[h]
\centering
\footnotesize
\caption{Peer-reviewed studies only (sensitivity re-count of the main article's table of the same name). Direction of effect on correctness (rows) against direction of effect on the targeted quality dimension (columns), counted per technique row of \texttt{extraction.csv} (vote counting by direction, SWiM; no effect sizes are pooled). Counts from LLM-assisted extraction reviewed by the author.}\label{tab:effects-peer}
\begin{tabular}{lrrrrrr}
\toprule
\dimCor{} \textbackslash{} target & \up & \same & \down & mixed & \nr & total \\
\midrule
\up & 102 & 0 & 0 & 14 & 10 & 126 \\
\same & 18 & 1 & 0 & 1 & 0 & 20 \\
\down & 12 & 1 & 1 & 9 & 0 & 23 \\
mixed & 30 & 0 & 1 & 13 & 1 & 45 \\
\nr & 58 & 1 & 2 & 13 & 6 & 80 \\
\midrule
total & 220 & 3 & 4 & 50 & 17 & 294 \\
\bottomrule
\end{tabular}
\\[2pt]{\scriptsize \up{} improved, \same{} no clear change, \down{} worse, mixed = differs across settings, \nr{} = not reported.}
\end{table}

\begin{table}[h]
\centering
\footnotesize
\caption{Peer-reviewed studies only (sensitivity re-count of the main article's table of the same name). How often included studies measured anything beyond their targeted dimension: share of studies per family that measured at least one non-targeted quality dimension, and share that measured correctness. Counts from LLM-assisted extraction reviewed by the author.}\label{tab:nontargeted-peer}
\begin{tabular}{lrrrrr}
\toprule
Family & Studies & \multicolumn{2}{c}{Non-targeted dim.\ measured} & \multicolumn{2}{c}{\dimCor{} measured} \\
\cmidrule(lr){3-4}\cmidrule(lr){5-6}
 & & $n$ & \% & $n$ & \% \\
\midrule
Prompting & 58 & 9 & 16 & 40 & 69 \\
Tool feedback & 43 & 4 & 9 & 32 & 74 \\
Self-refinement & 6 & 1 & 17 & 5 & 83 \\
Multi-agent & 20 & 1 & 5 & 17 & 85 \\
Retrieval & 15 & 1 & 7 & 12 & 80 \\
Decoding & 6 & 0 & 0 & 6 & 100 \\
Training & 56 & 7 & 12 & 51 & 91 \\
Hybrid/other & 77 & 11 & 14 & 72 & 94 \\
\midrule
All studies & 256 & 31 & 12 & 215 & 84 \\
\bottomrule
\end{tabular}
\\[2pt]{\scriptsize A study can appear in more than one family when it evaluates several techniques; the last row counts each study once. Percentages are our computation over the extracted rows.}
\end{table}

\begin{table}[h]
\centering
\footnotesize
\caption{Peer-reviewed studies only (sensitivity re-count of the main article's table of the same name). Evidence quality by family: median QA score (0--10, items QA1--QA10 of Appendix~\ref{sec:app-studies}) and the share of studies that report compute or cost (QA7 $>0$), consider contamination (QA9 $>0$), use human evaluation, and release artifacts (QA8 $>0$). LLM-assisted extraction reviewed by the author.}\label{tab:qa-peer}
\begin{tabular}{lrrrrrr}
\toprule
Family & $n$ & Med.\ QA & Comp.\ \% & Cont.\ \% & Human \% & Artif.\ \% \\
\midrule
Prompting & 58 & 6.5 & 31 & 34 & 34 & 79 \\
Tool feedback & 43 & 6.5 & 49 & 33 & 23 & 74 \\
Self-refinement & 6 & 5.8 & 17 & 67 & 33 & 67 \\
Multi-agent & 20 & 5.5 & 45 & 15 & 10 & 60 \\
Retrieval & 15 & 6.0 & 53 & 53 & 40 & 60 \\
Decoding & 6 & 6.2 & 67 & 17 & 33 & 100 \\
Training & 56 & 6.5 & 46 & 62 & 30 & 73 \\
Hybrid/other & 77 & 5.5 & 68 & 27 & 23 & 65 \\
\midrule
All studies & 256 & 6.0 & 50 & 37 & 27 & 70 \\
\bottomrule
\end{tabular}
\end{table}

\FloatBarrier
\section{Per-family detail: studies, highlighted techniques, effects, compute fairness and limitations}\label{sec:app-family-detail}
For each technique family: the study-by-study grouped comparison and the caveats of each highlighted technique, which the family sections of the main article summarize, followed by effects on correctness and on non-targeted dimensions, compute fairness, and limitations (\S\ref{M-sec:discussion} of the main article compares the families).

\subsection{Prompting}\label{sec:fd-prompting}
\paragraph{Explicit non-functional requirements.}
Stated requirements help less than expected (Table~\ref{tab:prompting-results}). On RACE, five simultaneous requirements drive almost all models near zero on the joint correct-and-compliant metric, and models resist requests contradicting their stylistic priors \citep{zheng2024beyond}. ISO/IEC 25010-grounded requirements lowered unreadability density for all four requirements and reduced wording sensitivity \citep{pereira2026does}. Adding ``secure'' to a GPT-4 prompt cut weakness density by \rep{42.85\%} (not significant); the best zero-shot variant was CWE-specific for GPT-3 and generic ``comprehensive'' for GPT-3.5 and GPT-4 \citep{tony2025prompting}. Newer security prompts only shift the CWE mix \citep{kharma2026empirical} or trade high- for low-severity findings \citep{urmi2026prompt}. Energy prompts saved nothing consistently across six models and three hardware platforms \citep{apsan2026generating} and helped only some models and languages \citep{rekhi2026improving}; no single efficiency prompt transfers.

\paragraph{Personas, exemplars and style specifications.}
In CodePromptEval, a persona who ``follows best coding practices for maintainability'' produced fewer Pylint smells and lower complexity at slightly lower correctness, while few-shot and signature prompts raised correctness most but added smells, over a third more when combined with chain-of-thought and persona prompts \citep{khojah2025impact}. In ECCO, both effects of few-shot examples are model-dependent \citep{waghjale2024ecco}; retrieved exemplars are in \S\ref{M-sec:retrieval}. Personality-guided generation improves pass rates in \rep{23 out of 28} model--dataset combinations but measures no non-functional dimension \citep{guo2025personalityguided}. Instruction-plus-example system prompts controlled verbosity most stably over a two-turn revision; example-only prompts lost effect once code was extended \citep{bohr2025show}. Signatures, constraints and style descriptions are the prompt elements most tied to readability, though prompt design's overall role is bounded \citep{ye2026characterizing}.

\paragraph{Reasoning scaffolds.}
SCoT and self-planning (Table~\ref{tab:hl-prompting}) target correctness (relative pass@1 gains up to \rep{13.79\%} and \rep{25.4\%}) but also report human-judged quality gains \citep{li2025structured,jiang2024selfplanning}. LLM4EFFI carries plan-first to efficiency \citep{ye2025llm4effi}. ``Think step by step'' was dominated by simpler zero-shot security prompts \citep{tony2025prompting}, and WA-0CoT increased code size and maintainability warnings for C \citep{kharma2026empirical}. Step-by-step prompting could yield more efficient code on complex problems \citep{niu2024on}.

\paragraph{Reasoning effort and prompt sensitivity.}
Reasoning effort is the main knob frontier APIs expose, yet on A.S.E slow-thinking configurations tended to trail fast-thinking ones \citep{lian2026ase}, and no study we found varies effort within one model while measuring \dimRM{} or \dimEff{}. Sensitivity is better documented: one changed character can flip code from secure to vulnerable \citep{sternfeld2026minimal}; benign but less normative prompts make insecure code likelier, mitigated by chain-of-thought and self-correction prompts \citep{wang2026is}; two security prompts built on similar principles differed markedly \citep{bruni2025benchmarking}; and under-specification costs almost no correctness on the richer LiveCodeBench \citep{akli2026when}.

\paragraph{Highlighted techniques.}

In Table~\ref{tab:hl-prompting} (selection rule in \S\ref{M-sec:highlighted}), landmarks mostly structure reasoning before coding; recent leaders condition the prompt on a task or model diagnosis from another LLM or analyzer (ECO, BRACE, CSDD) optimize the prompt itself (MPCO, SecureForge), or restate the problem before coding (StoryCoder retells it as a narrative). SecureForge mines benign prompts whose code a static analyzer flags, broadens them by MCMC rephrasing, and evolves a system prompt against them with GEPA, reporting weakness rates up to \rep{48\%} lower on test-passing outputs with test pass rates essentially unchanged \citep{liu2026secureforge}.

\textbf{Caveats.} Landmarks' quality evidence is human-rated and partly unquantified; Code2API's ratings judge method names produced from human-written snippets. Several baselines are weak. ECO is compared with instruction-only prompting at best of five samples and lowers 3B-model accuracy from \rep{55.22} to \rep{37.80}. StoryCoder's quality evidence is structural: its correct CodeForces solutions contain more functions (\rep{1.820} to \rep{2.338} with Gemini-2.5-Flash), which its authors read as modularity without measuring maintainability. MPCO reports an aggregate rank, yet on one of its five projects few-shot prompting's mean improvement (\rep{19.29}) exceeds MPCO's (\rep{19.06}). BRACE's Safe Code Rate is a trained judge classifier's verdict; its average HumanEval pass@1 barely changes but drops on Gemma3-4B. CSDD rests on one application built twice by one developer with the same assistant. MPCO and CSDD omit correctness, and no row reports an effect on a non-targeted dimension.

\paragraph{Effects on correctness and on non-targeted dimensions.}

Where correctness is measured, signs conflict. RCI's security gain cost about ten pass@1 points (ten samples, one model) \citep{bruni2025benchmarking}; the PIE retrieval prompts show the same tension (\S\ref{M-sec:retrieval}); security-prompting studies more often report a drop or no change than a gain, and many skip correctness. Effects on non-targeted dimensions can be adverse (WA-0CoT's warnings); SCoT's human study rates correctness, smells and maintainability (0--2 scale, 200 programs per method) \citep{li2025structured}.

\paragraph{Compute fairness.}

Variants are rarely token-matched. RCI adds at least two calls per sample, yet \citet{tony2025prompting} and \citet{bruni2025benchmarking} compare it with single-call prompts rather than a resampled security prompt or an analyzer loop (\S\ref{M-sec:feedback}). No highlighted technique adding calls (meta-prompter, risk selector, primer synthesis, synthetic tests) has a budget-matched baseline either, and A.S.E compares across model variants.

\paragraph{Limitations.}

(i) \emph{Old or closed models}: the main security studies use GPT-3 to GPT-4o; GPT-3.5-turbo was hurt by prompts that helped GPT-4o \citep{bruni2025benchmarking}. (ii) \emph{Static analyzers as sole oracle}: these overestimate security by \rep{7 to 21 times}, with \rep{37 to 60\%} of ``secure'' outputs non-functional \citep{tessa2026how}. (iii) \emph{Small samples}: nine problems in \citet{apsan2026generating}, no inferential tests in \citet{bruni2025benchmarking}. (iv) \emph{Inconsistency}: SCoT gives both \rep{15.90\%} and \rep{16.17\%} as its maintainability gain \citep{li2025structured}. (v) \emph{Uncontrolled baselines}: the ISO study's baseline came from an earlier batch \citep{pereira2026does}. (vi) \emph{Prompt fragility}: one wording per condition, the norm, confounds technique and phrasing.

\subsection{Tool-feedback loops}\label{sec:fd-feedback}
\paragraph{Security-scanner loops.}
FDSP~\citep{alrashedy2023can} returns the Bandit report, requests several candidate fixes, and iterates each until Bandit is clean; LLM-verbalized reports add only \rep{1\% to 2\%}. With CodeQL, raw diagnostics help every model, while a stronger model's explanations help mainly strong instruction followers and not consistently more than raw feedback on real-world tasks~\citep{yan2026guiding}. Codexity~\citep{kim2024codexity} returns Infer and CppCheck findings for C; \citet{dolcetti2026helping} return one Infer finding at a time, including known false positives, which models also ``fixed''. SecureFixAgent~\citep{gajjar2025securefixagent} re-runs Bandit after each patch from small LoRA-tuned models; FRANC~\citep{siddiq2024franc} turns Bandit messages into repair prompts; PromSec~\citep{nazzal2024promsec} rewrites the prompt, not the code; \citet{sriram2026improving} append GCC, CodeQL and KLEE diagnostics plus retrieved repairs. Filtering returns nothing: a fine-tuned vulnerability detector discards flagged completions~\citep{chan2023transformerbased}. LLMSecGuard~\citep{kavian2024llm} proposes a Semgrep/Weggli loop without results. AutoSafeCoder's~\citep{nunez2024autosafecoder} ``Static Analyzer Agent'' is an LLM (Bandit only evaluates), so its signal source is other-LLM.

\paragraph{Maintainability and multi-dimension loops.}
\citet{blyth2025static} is the closest quality-loop template: Bandit and Pylint issues are tagged inline, security issues ordered by severity weights, and a candidate kept only if its fitness ($-\infty$ on any test failure) does not worsen. CORE~\citep{wadhwa2024core} filters: a proposer samples revisions from the CodeQL or SonarQube check description and vendor fix advice, the analyzer discards failures, and a ranker LLM scores the rest against a reviewer rubric. CodeCureAgent~\citep{joos2026codecureagent} repairs SonarQube warnings agentically; its change approver requires a build, the warning removed without new ones, and passing tests. \citet{abtahi2025augmenting} send SonarQube issue tables to GPT-3.5 Turbo, then leftovers to GPT-4o; \citet{islam2026from} feed \rep{14} metrics from Pylint, Radon, JSCPD, Bandit and others to several LLMs; LLMLOOP~\citep{ravi2025llmloop} chains compiler, test, PMD and mutation loops for Java but reports no PMD violation counts.

\paragraph{Compiler and linter guardrails.}
Compiler loops target \dimCor: ProCoder~\citep{bi2024iterative} turns project-context compiler errors into retrieval queries until the code compiles. SWE-agent~\citep{yang2024sweagent} reverts edits that a restricted \texttt{flake8} check flags for syntax or undefined names, raising resolved SWE-bench Lite issues from \rep{15.0} to \rep{18.0}.

\paragraph{Profiler and performance loops.}
EffiLearner~\citep{huang2024effilearner} returns per-line \texttt{line\_\allowbreak profiler} and \texttt{memory\_\allowbreak profiler} output; format matters, as aggregate time and memory alone \emph{increased} CodeLlama-70B's execution time by \rep{51.9\%} while the profile cut it by \rep{9.6\%}. SysLLMatic~\citep{peng2026sysllmatic} adds async-profiler or \texttt{perf} flame-graph hotspots and an optimization-pattern catalog; Optimo~\citep{peng2026multilevel} finds bottlenecks by differential profiling; PerfAgent~\citep{deng2026perfagent} gives a repository-level agent a \texttt{py-spy} hotspot summary; PerfCodeGen~\citep{peng2025perfcodegen} returns only the slowest unit test.

\paragraph{Evolutionary search.}
AlphaEvolve ``orchestrates an autonomous
pipeline of LLMs, whose task is to improve an algorithm by making direct changes to the code'' and keeps
the variants one or more evaluators score best; at Google it yielded a more efficient
data-centre scheduling heuristic and multiplied two $4\times4$ complex-valued matrices with
\rep{48} scalar multiplications \citep{novikov2025alphaevolve}. Open successors aim the search
at program efficiency: CodeEvolve selects targets from Java Flight Recorder profiles \citep{borra2026codeevolve} and
ContextEvolve compresses the search context into agents' summaries \citep{su2026contextevolve}; SBLLM
applies genetic-operator-style prompts to a population of optimized programs \citep{gao2025searchbased}. All inherit the evaluator's weaknesses: a timing evaluator rewards noise
(\S\ref{M-sec:meas-critiques}), and none of the studies we found scores evolved code for \dimRM{}.

\paragraph{Highlighted techniques.}
Table~\ref{tab:hl-feedback} lists the highlighted techniques (selection rule in \S\ref{M-sec:highlighted}). \citet{blyth2025static} tag a few Bandit and Pylint issues, chosen by severity weight, inline in GPT-4o's code each round and keep a revision only if its test-gated fitness does not fall; over ten rounds Bandit-flagged snippets fell from \rep{45.05\%} to \rep{13.41\%} while the pass rate rose from \rep{36.7\%} to \rep{46.59\%}.

\paragraph{Caveats.}
Several headline gains rest on weak comparisons. PerfCodeGen is compared only with the base model, not other refinement methods. CodeEvolve's baseline is a single LLM call evaluated with a different model (Claude 4 Sonnet), and one function at \rep{90.34$\times$} drives its average speedup. Optimo uses only GPT-4o and consumes more tokens and API calls than its baselines. Security results are tool-scored and small: Dolcetti et al.'s best model fixed \rep{39 out of 44} findings. FRANC reports no pre-repair value; PromSec shows CWE counts only in plots, and the unit-test re-evaluation of \citet{dai2026rethinking} contradicts its \dimCor{} gain. Filtering also shrinks output: for text-davinci-003, scenarios with a valid completion fell from \rep{27} to \rep{24}. Correctness is unreported for the filtering and FRANC results and unchecked after the Infer repairs.

\paragraph{Effects on correctness and non-targeted dimensions.}
Correctness held or rose mainly behind a test or fitness gate. Blyth et al.\ raise pass@1, LLMLOOP's PMD stage leaves it essentially unchanged (\rep{79.51\%} to \rep{79.57\%}), and the profiler loops keep only test-passing candidates, though EffiLearner, checking only open tests, still loses a little pass@1 for most models. Ungated, correctness is mostly unmeasured (FDSP, Codexity, SecureFixAgent, \citet{yan2026guiding}, \citet{islam2026from}) or falls: AutoSafeCoder's abstract claims ``no compromise in functionality'' while its table shows pass@1 dropping, and PromSec, validated mainly by code-graph similarity, lost Pass@1 under the unit tests of \citet{dai2026rethinking}. Exceptions run both ways: a gate is a safeguard, not a guarantee.

Goodhart's law also recurs. Reviewers rejected \rep{55.11\%} of CORE's CodeQL-passing revisions, and without its change approver CodeCureAgent would have admitted \rep{249} incorrect fixes. \citet{islam2026from} discarded later iterations in which local models removed the code, since empty output scored a perfect Maintainability Index and no code smells; \citet{dai2026rethinking} find secure-generation techniques often delete vulnerable lines or emit ``garbage code''; PerfAgent's authors saw agents cache results inside benchmark timing loops. Ruff appears only as an RL reward (\S\ref{M-sec:training}), where alone it pushed models toward shorter completions without improving correctness~\citep{skopin2026improving}. Scanner gates miss structural regressions: Semgrep gating raised latent security degradation from \rep{12.5\%} to \rep{20.8\%} because removed defensive logic triggers no rule~\citep{chen2026scaffoldcegis}, and runtime exploits were confirmed for \rep{14.53\%} of Bandit-Semgrep-clean samples~\citep{pourleyli2026beyond}. Yet most security loops are scored by their driving tool; when FDSP adds CodeQL, GPT-4's gain shrinks from \rep{32.7} points (Bandit) to \rep{10.2} (CodeQL).

Loops also introduce issues: \rep{5\%} of Blyth et al.'s iterations introduced new security vulnerabilities, convention issues most often; Checkov-driven Terraform repair regressed security in \rep{3.3\%} (strict) to \rep{13.8\%} (inclusive) of scenarios~\citep{agyekum2026does}. Among efficiency loops, only SysLLMatic reports a non-targeted dimension (cyclomatic complexity \rep{-0.95\%} to \rep{+21.05\%}). Few security-only loops report readability or efficiency, and no maintainability-only loop we found reports runtime.

\paragraph{Compute fairness.}
Every loop spends extra calls, but few face an equal-budget baseline. Blyth et al.\ allow up to \rep{10} repair calls plus one to infer function names, against one generation. FDSP's ablation credits most of its gain to sampling $J$ candidate fixes (\rep{10.0\%} vulnerable with a single solution, \rep{7.4\%} with full FDSP), confounding its comparison with single-response Bandit feedback. CORE draws \rep{10} proposals plus ranker calls; SecureFixAgent averages \rep{3} to \rep{4} iterations and AutoSafeCoder adds \rep{150} fuzzing iterations, both against a single pass. Codexity needs \rep{1.8} repair iterations and \rep{98.6 seconds} per generation with a found vulnerability; CodeCureAgent reports \rep{139K} tokens and about \rep{2.9 cents} per warning; PerfAgent, in a rare budget-aware comparison, beats an oracle \rep{best-of-five} baseline at lower cost. Counting repair cost, self-repair gains are often modest~\citep{olausson2024is}.

\paragraph{Limitations.}
Several loops use one model (Blyth et al., AutoSafeCoder, LLMLOOP, CodeCureAgent); the multi-model static-analysis loops (FDSP, \citet{yan2026guiding}, \citet{islam2026from}) lack a test gate. Benchmarks are small or narrow (SecurityEval has \rep{121} prompts; Abtahi and Azim study one project), and some preprints report carelessly: \citet{sriram2026improving} misattribute Self-Debug. CORE's human review used one label per revision.

\subsection{Self-refinement}\label{sec:fd-selfrefine}
\paragraph{Intrinsic self-feedback.}
Self-Refine alternates \textsc{feedback} and \textsc{refine} prompts on one model \citep{madaan2023selfrefine}; it is the only canonical method evaluated on readability, but that task rewrites existing CodeNet programs, scores the fraction of variables GPT-4 judges appropriately named, and checks no behavior. Outside code, LLMs struggle to correct their reasoning without external feedback, sometimes getting worse \citep{huang2024large}; a TACL survey finds self-correction works mainly with reliable external feedback and many positive reports rest on unfair evaluations \citep{kamoi2024when}. On EffiBench, an unsupervised Self-Refine prompt \emph{increased} CodeLlama-70B execution time \citep{huang2024effilearner}.

\paragraph{Execution-grounded self-repair.}
Self-Debug adds unit-test results and model-written code explanations \citep{chen2024teaching}; Self-Edit passes the example-test result as a comment to a fault-aware editor \citep{zhang2023selfedit}; Reflexion keeps verbal reflections in an episodic memory \citep{shinn2023reflexion}; LDB exposes intermediate variable values per basic block \citep{zhong2024debug}; CRITIC critiques with tools, but its program-synthesis tasks are mathematical \citep{gou2024critic}. All five are correctness-only, and the oracle is their weak point: Reflexion's self-generated MBPP Python tests have a \rep{16.3\%} false-positive rate \citep{shinn2023reflexion}, and self-generated tests bias self-debugging toward misleading feedback \citep{chen2025revisit}.

\paragraph{Self-refinement toward a quality target.}
Quality-targeted variants replace the test oracle with a non-functional signal (Table~\ref{tab:sr-results}). EffiLearner, on the border with \S\ref{M-sec:feedback}, feeds execution-time and memory profiles back to the model \citep{huang2024effilearner}. Controlled Self-Evolution adds diversified initial plans, feedback-guided mutation and a cross-task memory on EffiBench-X \citep{hu2026controlled}. A three-round Reflexion protocol returning the CWEs flagged by the Insecure Code Detector (ICD) raised the secure rate of five open code models \citep{datta2025secure}.

\paragraph{Open-ended ``improve this code'' loops.}
Untargeted iteration can degrade security. Over \rep{40} rounds of GPT-4o ``improvements'' to \rep{10} secure C and Java programs, critical vulnerabilities rose by \rep{37.6\%} after five iterations, and code complexity correlated with vulnerability counts \citep{shukla2025security}. In a 2026 study of three models, \rep{43.7 \%} of GPT-4o chains had more vulnerabilities than the baseline after ten rounds, and a static-analysis gate raised the latent degradation rate from \rep{12.5\%} to \rep{20.8 \%}, as rules miss removed defensive logic \citep{chen2026scaffoldcegis}.

\paragraph{Reasoning models as built-in refinement.}
Large reasoning models critique and revise inside a single call. Their traces contain flaw-detection and style-check actions, analyzed only against correctness \citep{halim2025study}. The cost is ``overthinking'', verbose and redundant reasoning \citep{sui2025stop}, and a 2026 audit finds more capable models generate more bloated and coupled code \citep{zhu2026aigenerated}. No study we found compares explicit refinement with internal revision at matched tokens on any quality dimension.

\paragraph{Highlighted techniques.}
Table~\ref{tab:hl-selfrefine} lists all \cnt{nHlSelSelfRefinement} eligible general-purpose techniques (selection rule in \S\ref{M-sec:highlighted}). Besides Self-Refine, they ground revision in structure (CodeChain, GRASP), in memory of past repairs (RefleXGen and Safety Sidecar, which share a first author) or in CodeQL findings (SecCode, partly a tool loop, \S\ref{M-sec:feedback}).

\textbf{Caveats.} Loop and instrument overlap: SecCode feeds CodeQL findings into the loop, and Safety Sidecar uses CodeQL in both pipeline and evaluation, computing rates only over compilable samples. RefleXGen and Safety Sidecar report a digit-for-digit identical base row under different model names (CodeQwen, qwen3-coder-plus). Correctness can suffer: RefleXGen lowers CodeQwen's pass rate from \rep{86.7} to \rep{69.8}, and GRASP reports an occasional trade-off at $k=1$ and compares with baselines only in Python. CodeChain's headline result is correctness (relative pass@1 gain of \rep{35\%} on APPS); its modularity gain is a GPT-4 Likert judgement shown only as a figure. No row reports an effect on a non-targeted dimension.

\paragraph{Effects on correctness and non-targeted dimensions.}
EffiLearner optimizes only code that already passes the public tests; its pass@1 on hidden tests fell by \rep{0\% to 0.5\%}. The security loops in Table~\ref{tab:sr-results} run a detector but no functional tests. Few studies measure a non-targeted dimension, one being the complexity analysis of \citet{shukla2025security}, so a loop may degrade one dimension while optimizing another, unnoticed.

\paragraph{Compute fairness.}
\citet{olausson2024is} set the standard: against i.i.d.\ sampling at an equal number of programs, self-repair gains on HumanEval and APPS are often modest and sometimes absent, and feedback quality is the bottleneck. Self-Debug reports matching baselines that draw over ten times as many candidates \citep{chen2024teaching}; a 2026 study finds self-repair more token-efficient than resampling, but repair decodes greedily and resampling at temperature \rep{0.8} \citep{arimbur2026how}. No quality-targeted study we found, the highlighted ones included, is budget-matched; they compare the final iteration with the initial output, and Self-Refine's no-refinement baseline makes fewer calls.

\paragraph{Limitations.}
Among canonical methods, the \dimRM{} evidence is one LLM-judged naming metric. In the Reflexion-plus-ICD study the detector that writes the feedback also scores the result, so part of the gain may reflect fitting its rules rather than more secure code. The degradation studies use few models and seed programs. Most correctness-only methods were evaluated on near-saturated HumanEval and MBPP.

\subsection{Multi-agent pipelines}\label{sec:fd-multiagent}
\paragraph{Process-mimicking role pipelines.}
ChatDev chains design, coding and testing through role-agent chats \citep{qian2024chatdev}, MetaGPT encodes standardized operating procedures in role prompts \citep{hong2024metagpt}, and Self-collaboration uses analyst, coder and tester roles \citep{dong2024selfcollaboration}. Quality is prompted, not evaluated: Self-collaboration's tester reports on ``functionality, readability, and maintainability'' but results are pass@1, and ChatDev's ``Quality'' multiplies completeness, executability and requirement consistency. Among highlighted role pipelines (Table~\ref{tab:hl-multiagent}), removing FlowGen's testing raised error-smell density by \rep{702.2\%} on average \citep{lin2025soen101}, and C3's role competition cut ``collaboration degeneration'' from \rep{32.4\%} to \rep{12.8\%} \citep{gu2026mitigating}.

\paragraph{Coder--tester and coder--reviewer loops.}
AgentCoder (programmer, test designer, test executor) \citep{huang2023agentcoder} and MapCoder (similar-problem retrieval, planning, coding, debugging on sample I/O) \citep{islam2024mapcoder} are correctness-only. The simplest quality-oriented design, coder plus reviewer, is re-evaluated in \S\ref{M-sec:empirical}. CudaForge, typical of the efficiency group, grounds its Judge reviewer in Nsight Compute hardware metrics; with OpenAI-o3 it reaches a \rep{1.68}$\times$ average speedup over PyTorch on KernelBench at \rep{97.6\%} correctness \citep{zhang2025cudaforge}.

\paragraph{Security-specialist agents.}
AutoSafeCoder adds a ``Static Analyzer Agent'' and a fuzzing agent to a coder \citep{nunez2024autosafecoder}, but the analyzer is an LLM prompted to find CWEs, with Bandit only scoring the result, so static-analysis feedback goes untested. MACGen passes structured artifacts, not a shared dialogue, between planner, security advisor, coder and reviewer \citep{yu2026macgen}. Multi-LLMSecCodeEval (preprint) compares ten single-model, ensemble, collaborative and hybrid pipelines \citep{sabir2026does}: on SecLLMEval, ensembles with static analysis beat a single LLM at secure generation by up to \rep{47.3\%}, LLM-only collaboration by \rep{8.9\% to 22.3\%}. Likewise, feeding CodeQL and Bandit findings to the generator in \citet{surikov2026securing} deepened the pooled drop in analyzer findings from \rep{42\%} to \rep{57\%}.

\paragraph{Repository-level agents.}
SWE-agent's agent-computer interface \citep{yang2024sweagent}, the OpenHands platform \citep{wang2025openhands} and Agentless's fixed localization, repair and validation phases (\rep{32.00\%} on SWE-bench Lite at \rep{\$0.70} per issue \citep{xia2024agentless}) are all scored by resolved issues, yet SWE-agent's case study notes a test-passing patch that ignores an existing utility function, ``which could lead to less maintainable or portable code''.

\paragraph{Quality of agent-produced code.}
Controlled studies point to complexity growth: six analyst--coder--tester--debugger configurations (GPT-4o family, HumanEval) form two complexity clusters \rep{50-130\%} apart, the heavier with no pass@1 advantage \citep{ashrafi2026how}. On SlopCodeBench, where \rep{15} agents repeatedly extend their own solutions, structural erosion rose in \rep{77\%} of trajectories and agent code was \rep{2.3x} more verbose than code in \rep{473} open-source Python repositories; quality guidance lowered initial verbosity and erosion, not the degradation rate \citep{orlanski2026slopcodebench}. A technical-debt audit finds code volume tracks structural degradation \citep{zhu2026aigenerated}. Field evidence comes mostly from AIDev's \rep{932,791} agent-authored pull requests (PRs) \citep{li2026aidev}. In another sample, \rep{83.8\%} of \rep{567} Claude Code PRs were merged \citep{watanabe2026on}; rejected agent PRs are larger and often fail CI \citep{ehsani2026where}; in merged bug-fix PRs SonarQube finds mostly code smells, with per-agent differences largely vanishing after churn normalization \citep{cynthia2026beyond}. On security, an LLM judge flags at least one security smell in \rep{38.9\%} of PRs from five agents \citep{sakib2026trust}, but a matched comparison finds agent code \emph{less} likely than human code to have one (odds ratio \rep{0.63}), with revert rates varying by vendor \citep{kraishan2026not}.

\paragraph{Highlighted techniques.}
Table~\ref{tab:hl-multiagent} lists the highlighted techniques (selection rule in \S\ref{M-sec:highlighted}). They differ less in agent count than in what agents exchange: development roles, security critiques or, for \dimEff, several agents' candidates and, recently, shared search memory such as LessonL's banked lessons. ACCLAIM lets a guiding agent interleave LLVM compiler passes with LLM rewriting agents at source, IR and assembly level under one call budget, reaching a \rep{1.25}$\times$ geometric-mean speedup over clang -O3 on C programs against \rep{1.18}$\times$ for a source-only agent at equal budget \citep{mikek2026agentic}. MaintainCoder has requirements-analysis, design-pattern and framework agents agree on an architectural blueprint before a coding agent and an execution-driven optimizer write the code, raising post-modification Pass@5 on APPS-Dyn, a test of how well code absorbs new requirements, from \rep{59.5} for plain DeepSeek-V3 to \rep{62.5} \citep{wang2025maintaincoder}.

\textbf{Caveats.} Headline numbers often overstate typical gains. ContextEvolve's \rep{33.3\%} comes from one task; its average margin over the top baseline is \rep{6.5\%}, using over three times OpenEvolve's API calls (but fewer tokens). On LessonL's benchmark, OpenAI o3 alone reaches \rep{2.21}, above LessonL, and incorrect or slower code counts as speedup 1. MARCO matched or beat its baseline on \rep{44} of \rep{75} problems. INDICT's strongest baseline, Reflexion+, reaches \rep{86.9}, and its safety score comes from a rule-based detector on a \rep{20\%} validation split. SecAwareCoder is evaluated on its own benchmark, CodeSecEval. \citet{surikov2026securing} use \rep{26} prompts, score their second configuration with the analyzers supplying its feedback, and their remediation introduced new vulnerabilities in \rep{15-22\%} of cases. C3's maintainability gain is a human rating on which it ties two other variants; four of the ten techniques report no correctness result.

\paragraph{Effects on correctness and non-targeted dimensions.}
AutoSafeCoder shows the trade-off this survey seeks: with GPT-4o it cut Bandit-flagged vulnerable SecurityEval samples from \rep{59/121 (49\%)} to \rep{44/121 (36\%)} but lowered HumanEval pass@1 from \rep{0.9085} to \rep{0.8780}, though its abstract reports ``no compromise in functionality'' \citep{nunez2024autosafecoder}. Multi-LLMSecCodeEval does not evaluate correctness \citep{sabir2026does}; MACGen uses the stronger joint functional-and-secure metric, raising BaxBench F\&S@1 from \rep{21.17\%} (direct prompting, GPT-4o) to \rep{31.12\%} \citep{yu2026macgen}. Few security or efficiency pipelines also report \dimRM, and in the field merge success does not reflect post-merge quality \citep{cynthia2026beyond}.

\paragraph{Compute fairness.}
Few pipelines are compared with a single agent at equal cost. MapCoder averages \rep{16.7} API calls and \rep{21.25} thousand tokens per problem against \rep{1} call and \rep{0.64} thousand for direct prompting, a limitation it acknowledges \citep{islam2024mapcoder}. AgentCoder's \rep{56.9K} HumanEval tokens (vs.\ \rep{138.2K} for MetaGPT) are compared only with multi-agent frameworks, and its abstract pairs MetaGPT's token count with another baseline's pass@1 \citep{huang2023agentcoder}. Neither MACGen's \rep{16\%} of INDICT's tokens \citep{yu2026macgen} nor CudaForge's roughly \rep{\$0.3} per kernel (compared with another agentic system) \citep{zhang2025cudaforge} is matched to a single agent. In ChatDev with a reasoning model, code review, where quality checks would sit, took \rep{59.4\%} of tokens \citep{salim2026tokenomics}. The multi-agent benefit shrinks as base models improve \citep{gao2025singleagent}, and a failure taxonomy of \rep{1600+} traces notes often minimal multi-agent gains on popular benchmarks \citep{cemri2025why}.

\paragraph{Limitations.}
The security pipelines use one or two benchmarks, often one model, and AutoSafeCoder is a preprint. Most field studies share one AIDev snapshot, use different instruments (SonarQube, LLM judges, rule-based smell detectors) and cannot fully separate agent output from human edits; one finds humans introduced most leaked secrets in agent-assisted PRs \citep{sakib2026trust}.

\subsection{Retrieval}\label{sec:fd-retrieval}
\paragraph{Security knowledge.}
Designs differ in the retrieved unit (Tables~\ref{tab:retrieval-results} and~\ref{tab:hl-retrieval}). RESCUE finds SecCoder's raw demonstrations \citep{zhang2024seccoder} do not significantly improve SecurePass@1, suggesting underused security knowledge, and instead hierarchically retrieves distilled guidelines and sliced examples \citep{shi2026rescue}. SEEK scopes CWE-graph retrieval with planner and selector agents \citep{gadey2026seek}. CodeGuarder (\rep{20.12\%} relative gain in security rate) and MA-CoT (security principles plus a language-specific checklist) retrieve before generation \citep{lin2025give,kharma2026enhancing}; AutoPatch retrieves after it \citep{seo2025autopatch}, SOSecure after a first draft, a post-generation loop \citep{mukherjee2025sosecure}.

\paragraph{Efficiency exemplars.}
Stand-alone \dimEff{} retrieval is rarer and mostly optimizes existing programs; hybrids with profiling or search loops are more common (\S\ref{M-sec:feedback}). PIE retrieves slow--fast pairs as few-shot examples (\rep{1.34$\times$} speedup at Best@1) \citep{shypula2024learning}; RAPGen retrieves an instruction mined from past fixes \citep{garg2025rapgen}. RAS builds on PIE; SemOpt and PEACE use RAPGen as a baseline \citep{anupam2026llm,zhao2026semopt,ren2025peace}. An unrelated AutoPatch adds control-flow-graph analysis to historical examples \citep{acharya2025optimizing}. A few retrieve similar solved problems for generation from a specification.

\paragraph{Reuse versus reimplementation.}
Without repository retrieval, GPT-3.5 in A3-CodGen reuses no local functions and writes redundant new ones \citep{liao2024a3codgen}. Conversely, in CodeRAG-Bench models copy context functions, sometimes overwriting the queried one, and over-complicate programs \citep{wang2025coderagbench}. Agent-authored pull requests miss reuse opportunities and are more redundant than human ones \citep{huang2026more}. Otherwise, \dimRM{} of RAG output is rarely measured (except by KPC and LAIL), and few methods retrieve style guides or project conventions.

\paragraph{Risks of retrieved content.}

One poisoned knowledge-base example can compromise up to \rep{48\%} of generated code \citep{lin2026exploring}, and a backdoored retriever makes GPT-4o generate vulnerable code in \rep{over 40\%} of targeted scenarios \citep{li2025exploring}. Staleness is the non-adversarial analogue: obsolete repository snapshots induced stale helper references in \rep{15/17} Qwen2.5-Coder-7B-Instruct samples \citep{weng2026when}. Few methods besides CodeGuarder are evaluated under poisoning \citep{lin2025give}.

\paragraph{Highlighted techniques.}

Table~\ref{tab:hl-retrieval} lists the highlighted techniques (selection rule in \S\ref{M-sec:highlighted}).

\textbf{Caveats.} RAPGen and SemOpt score agreement with developer fixes (expert and author judgment), not measured speed. PEACE's opt rate is relative to GPT-4o prompting, not the original code. RAS normalizes LLM calls across baselines. PIE's authors note that speedup ``may, however, come with some cost of correctness''. KPC's rate comes from the authors' manual review. LAIL is evaluated mainly by Pass@1, with quality a secondary human rating. AutoPatch uses its authors' dataset, and its abstract gives \rep{95.0} percent accuracy versus \rep{94.12\%} in its results table. MA-CoT only compiles or syntax-checks outputs, draws one sample per configuration on its own dataset and gains over a plain prompt; its authors report that generic CoT and zero-shot security prompts may themselves increase vulnerabilities.

\paragraph{Effects on correctness, compute fairness and limitations.}

Security RAG methods that measure correctness mostly report it preserved or improved, unlike security prompting (\S\ref{M-sec:prompting}), though a minority of extracted studies report a drop; efficiency retrieval, as in PIE, may cost correctness. Budgets are rarely matched: RESCUE's distillation and SEEK's agents face single-call baselines, and only RESCUE (input tokens) and SEEK (latency) report cost in Table~\ref{tab:retrieval-results}. Limitations: SecCoder uses models up to 6.7B parameters; SEEK is a recent preprint; security metrics rest on CodeQL and benchmark-specific checks, which can overestimate security (\S\ref{M-sec:prompting}); A3-CodGen counts calls to existing functions, not whether code is better.

\subsection{Decoding}\label{sec:fd-decoding}
\paragraph{Constrained decoding and steering.}
Synchromesh enforces syntax, scope and typing on partial programs in three domain-specific languages
\citep{poesia2022synchromesh}; type-constrained decoding extends this well-formedness target to
TypeScript, more than halving compilation errors \citep{mundler2025typeconstrained}. For \dimRM{},
PG-TD swaps its test reward for a length or comment reward \citep{zhang2023planning} and RPG penalizes
structural repetition \citep{dong2025rethinking}. For \dimSec{}, constrained beam sampling writes
constraints for each of the 91 CodeGuard+ prompts from the CWE it might trigger
\citep{fu2024constrained}; co-decoding replaces the rules with a small security model
\citep{hu2026cogate}; and 2026 steering methods (SCS-Code, DuoSteer)
use the target's own activations. MRepE, the only \dimRM{} steering method we found, combines
comment-density, naming and cyclomatic-complexity vectors while bounding correctness loss
\citep[preprint]{gao2026towards}. \citet{sandoval2026surgical} trace insecure generation to one late
layer; per-vulnerability steering vectors ``reduce insecure generation by up to \rep{74\%}'' across
five models.

\paragraph{Sample-and-select.}
MBR-exec keeps the candidate whose execution results agree most with the others
\citep{shi2022natural}; CodeT adds model-generated tests \citep{chen2023codet}; Coder-Reviewer rescores
by instruction likelihood because likelihood-only reranking is ``prone
to preferring degenerate solutions'' \citep{zhang2023coder}; LEVER learns a verifier over program
and execution result \citep{ni2023lever}. Without execution feedback, a code process reward model
is no better than a random reward in best-of-$N$ \citep{li2025codeprm}; FunPRM meta-learns
step-reward corrections from unit-test outcomes \citep{zhang2026funprm}. Quality-aware
selection needs a quality preference model: CodeFavor learns pairwise preferences from commits and
critiques, improving preference accuracy for correctness, efficiency
and security ``by up to \rep{28.8\%}'' \citep{liu2024learning}; Priority Sampling, a deterministic
confidence-ordered search over unique samples, raises a compiler model's code-size gain over
\texttt{-Oz} ``from \rep{2.87\%} to \rep{5\%}'' \citep{grubisic2024priority}. Test-time scaling adds
sequential revision to execution-grounded selection \citep{li2025s}; CodeMonkeys selects among
repository edits with model-generated tests \citep{ehrlich2025codemonkeys}. \emph{The
quality-verifier gap is now partly closed:} SecCodePRM (ICML 2026), a step-level security reward
model trained on analyzer labels, outranks probability ranking for best-of-$N$ on CWEval
\citep{yu2026seccodeprm}. For \dimEff{}, surprisal-guided selection is a preprint
\citep[preprint]{barnes2026surprisalguided} and PIE's Best@$k$ (fastest program passing all tests in
a simulator) an oracle bound \citep{shypula2024learning}. We found no peer-reviewed \dimRM{} or
\dimEff{} selector, and none reranking by static maintainability scores or profiled instruction counts.

\paragraph{Highlighted techniques.}

Table~\ref{tab:hl-decoding} lists the highlighted techniques (selection rule in
\S\ref{M-sec:highlighted}). MoC trains per-CWE linear probes on a code model's hidden states and, only when a probe flags a vulnerable state, adds a decaying mix of correction vectors, raising Qwen2.5-Coder-7B's CodeQL security ratio on the SVEN test set from \rep{76.8} to \rep{85.7} with HumanEval pass@1 \rep{79.9} $\to$ \rep{81.2} \citep{yu2025mixture}. RobGen uses a small classifier to predict whether the next line should open an \texttt{if} guard and, when \texttt{if} is among the three most likely tokens, raises its logit, cutting the share of DeepSeekCoder-1.3B's CoderEval Java outputs that GPT-4o judges less robust than the reference from \rep{42.2\%} to \rep{28.7\%} with Pass@1 slightly higher \citep{li2025preliminary}.

\paragraph{Caveats.}
Correctness effects are mixed. PG-TD's length penalty slightly lowers the public-test pass rate
(\rep{23.14} to \rep{22.82}); constrained beam sampling lowers Llama3-8B pass@1 from \rep{82.20} to
\rep{72.21}, and an independent study found, and repaired by script before testing, indentation issues
in most CodeGuard+ outputs for CodeLlama and Mistral \citep{dai2026rethinking}; SCS-Code lowers
CodeLlama pass@1 from \rep{68.1} to \rep{65.7}, whereas DuoSteer reports an average correctness change
of \rep{+7.5\%}. RPG improves on repetition penalty, which also cuts repetition but lowers HumanEval
pass@1 to \rep{0.092} (greedy: \rep{0.301}; RPG: \rep{0.325}). Evidence is thin for several recent
rows: surprisal selection rests on \rep{10} task--seed pairs
(sign test \rep{$p = 0.125$}); FunPRM targets correctness, and its readability gain comes from three
raters on \rep{100} pairs, read off a figure; DuoSteer reports both a test-swept and a held-out
configuration protocol.

\paragraph{Effects on correctness and non-targeted dimensions.}
Correctness selectors measure no other dimension. With the QC3-30B generator, SecCodePRM raises func@1 by \rep{11.09} points alongside
security \citep{yu2026seccodeprm}. The ICSE 2026 re-evaluation finds CodeGuard+'s security gains
limited or negative on Mistral and notes the original never measured the indentation defect above
\citep{dai2026rethinking}; its implementation also returns no code without its constraints
\citep{wendlinger2026securitybydesign}. No steering study we found measures \dimSec{} after \dimRM{}
steering or the reverse.

\paragraph{Compute fairness.}
Selectors compared at equal $N$ are budget-matched; comparisons with one greedy sample are not. CodeT
samples \rep{100} test generations per HumanEval problem \citep{chen2023codet}; CodeMonkeys spends
\rep{approximately 2300 USD} and resolves \rep{57.4\%} of issues from a pool whose oracle coverage is
\rep{69.8\%} \citep{ehrlich2025codemonkeys}; type constraints add \rep{39.1\%} average HumanEval
runtime \citep{mundler2025typeconstrained}. Surprisal selection is free only because every compared
strategy already tests all samples \citep{barnes2026surprisalguided}. Best-of-$n$ at equal verifier
and call budget is thus the baseline for iterative methods
(\S\ref{M-sec:feedback}--\S\ref{M-sec:multiagent}), as in \S\ref{M-sec:empirical}.

\paragraph{Limitations.}
Logit- and activation-level methods need white-box access (surprisal selection needs sequence
log-probabilities); of the highlighted techniques only FunPRM does not. The ICSE 2026 study could not
run three of four security techniques on commercial models \citep{dai2026rethinking}, and none applies
to the frontier APIs of \S\ref{M-sec:empirical}. Per-prompt CWE constraints do not transfer to unseen
tasks, and CodeQL labels miss vulnerabilities \citep{dai2026rethinking}.

\subsection{Training}\label{sec:fd-training}
\paragraph{Data curation and supervised fine-tuning.}

phi-1 filtered web code with a classifier trained on GPT-4 ``educational value'' ratings of about \rep{100k} files, added GPT-3.5-generated textbooks, and reached \rep{50.6\%} HumanEval pass@1 at 1.3B \citep{gunasekar2023textbooks}. Open pipelines followed: The Stack v2 removes low-quality files heuristically \citep{lozhkov2024starcoder}; OpenCoder applies over \rep{130} language-specific rules and reports that filtering by GitHub stars reduced diversity and worsened the model \citep{huang2025opencoder}; Seed-Coder distills an LLM oracle's readability, modularity, clarity and reusability scores into a 1.3B regression scorer \citep{seed2025seedcoder}; and SwallowCode combines pylint-based filtering with LLM rewriting that ``enforces style conformity'' \citep{fujii2026rewriting}. Some studies clean the code itself: \citet{jain2024llmassisted} rename variables and modularize solutions with an LLM, gaining up to \rep{30\%} correctness. Few curation studies measure \dimRM{} of generated code: beyond the SmellCC row of Table~\ref{tab:hl-training}, \citet{xue2026clean} report \rep{79.6\%} and \rep{83.1\%} fewer smells for DeepSeek-V2 and Qwen-Coder (their refactoring tool kept only \rep{91.3\%} correctness in its own test verification), and \citet{kuang2026on} find that data refactoring, selection and cleaning raise the Maintainability Index of generated code while synthesis and augmentation lower it. Large pretraining pipelines filter for readability-like properties but report only functional benchmarks.

\paragraph{Preference optimization with quality signals.}

Preference optimization learns from pairs of better and worse programs. CodeUltraFeedback draws responses from \rep{14} LLMs and brought CodeLlama-7B-Instruct to a CUF-Bench judge score of \rep{7.08} (GPT-3.5-Turbo: \rep{7.18}) and pass@1 from \rep{37.9} to \rep{43.1}, though DPO variants scored slightly below SFT alone \citep{weyssow2026codeultrafeedback}. CodeDPO reaches a \rep{1.45x} speedup on HumanEval+ with MagiCoder-S-DS-6.7B at \rep{83.5\%} correctness \citep{zhang2025codedpo}. PLUM \citep{zhang2024textbfplum} and AceCoder \citep{zeng2025acecoder} (unrelated to ACECode below) use synthesized tests for correctness only. For efficiency, Mercury found DPO stronger than SFT \citep{du2024mercury}, while Afterburner found both ``rapidly saturate'' \citep{du2025afterburner}.

Several security results share a mechanism: secure and insecure programs often differ in a few tokens (a sanitizer call, a parameterized query), which a sequence-level loss dilutes. Hence SVEN and SafeCoder mask their losses to the changed regions \citep{he2023large,he2024instruction}, Localized Preference Optimization (LPO) masks security-relevant tokens after finding SimPO ineffective \citep{hasan2025teaching}, and SRCode assigns token-level RL rewards \citep{quan2026learning}.

\paragraph{Security-targeted training.}

Security training has moved from curated fixes (SVEN, SafeCoder) to synthesized data and then RL. HexaCoder synthesizes vulnerable and repaired pairs with a CodeQL oracle \citep{hajipour2024hexacoder}; ProSec synthesizes vulnerability-inducing instructions from CWEs and trains on preference pairs \citep{xu2025prosec}; LPO refines its distilled pairs with CodeQL and Bandit feedback, reporting \rep{31\%} and \rep{64\%} fewer insecure files on Phi-2 and StarCoder2-7B \citep{hasan2025teaching}. RL methods include PurpCode (rule-referencing SFT plus multi-objective RL) \citep{liu2025purpcode}, SecCoderX (a learned vulnerability reward model in place of the analyzer) \citep{wu2026secure}, SRCode \citep{quan2026learning}, and pilots such as SecureCodeRL, whose evaluation had no Bandit findings at all \citep{sijwali2026securecoderl}. Training has also reached coding agents: SecureVibe applies SFT and execution-feedback RL to the agent's security planning and testing, raising security pass@1 by \rep{6.9} points on BaxBench and \rep{11.5} on SusVibes, and functionality pass@1 by \rep{4.1} on SWE-bench Verified \citep{wang2026securevibe}.

\paragraph{Efficiency-targeted training.}

Efficiency work uses profilers or simulators as the signal. PIE fine-tunes on \rep{77,967} human performance-improving C++ edit pairs timed in the gem5 simulator, adding performance-conditioned generation and self-play data \citep{shypula2024learning}; EffiCoder fine-tunes on the fastest correct candidate from several LLMs \citep{huang2025efficoder}. ACECode trains with proximal policy optimization (PPO) on a combined correctness-runtime reward \citep{yang2024acecode}; Afterburner runs Group Relative Policy Optimization (GRPO) in an iterative refinement loop \citep{du2025afterburner}; SuperCoder rewards assembly speedup over \texttt{gcc -O3} \citep{wei2026supercoder}; \citet{nichols2026performancealigned} align models to serial and OpenMP code performance; and RLPF ranks failed programs by execution progress and correct ones by improvement toward an expert reference \citep{jing2026rlpf}. Making execution time a usable reward is itself studied: \citet{chambon2026reinforcement} report that ``once timing drives the reward, small problems in measurement noise, reward sparsity, or GRPO instability overwhelm the signal and make RL fail''; with a calibrated sandbox and adapted GRPO they raise strict top-50\% pass@1 on their optimization tests from \rep{18.0\%} to \rep{31.3\%} on Qwen 2.5 7B and from \rep{30.7\%} to \rep{50.4\%} on CWM 32B ``while preserving pure-correctness scores''. The largest RL efficiency gains are on GPU kernels: Kevin's multi-turn RL raises its base model's kernel correctness from \rep{56\%} to \rep{82\%} and mean speedup over PyTorch Eager from \rep{0.53x} to \rep{1.10x}, finding serial refinement more useful than parallel sampling \citep{baronio2025kevin}; CUDA-L1 reports an average speedup of \rep{x3.12} (median \rep{x1.42}) over default KernelBench baselines \citep{li2025cuda}; and CUDA Agent reports \rep{100\%}, \rep{100\%} and \rep{92\%} faster rates over \texttt{torch.compile} on the three KernelBench levels \citep{dai2026cuda}. Kernels fall outside the general-purpose scope of the highlighted techniques (\S\ref{M-sec:highlighted}) but show what a precise, cheap-to-measure efficiency reward can do.

\paragraph{Reinforcement learning with program-analysis rewards.}

REAL is the clearest attempt to train for \dimSec{} and \dimRM{} jointly with \dimCor{} (its maintainability detector is MyPy type checking) \citep{yao2025training}. Its motivating negative result is that ``optimizing for code quality alone leads to reward hacking'', with the model emitting empty code or comments that maximize quality scores; its ablation confirms that a quality-only reward yields safer but less functional code. In 2026, \citet{jolfaei2026domainadaptable} map Ruff violations to the tokens causing them, and \citet{skopin2026improving} find that static-analysis penalties alone bias small models toward shorter completions. Themis trains code reward models with readability and maintainability criteria, citing ``the limited proficiency of current RMs beyond scoring for functional correctness'' \citep{paul2026themis}; we found no well-documented study using such a model as an RL reward and measuring \dimRM{} with independent instruments.

\paragraph{Side effects of reward optimization.}

The first side effect is \emph{length}. In RL from human feedback, ``a purely length-based reward reproduces most downstream RLHF improvements'' \citep{singhal2023long}; DPO also exploits verbosity \citep{park2024disentangling}; ODIN counters it by discarding a dedicated length head of the reward model \citep{chen2024odin}; and in RL with verifiable rewards (RLVR), GRPO has an optimization bias that ``artificially increases response length (especially for incorrect outputs)'' \citep{liu2025understanding}. Since length moves with lines of code, comment density and complexity metrics, judge-based \dimRM{} rewards face the same confound, which no code study we found controls for. Analyzer-only rewards pull the other way, toward short or empty code \citep{skopin2026improving,yao2025training}; PurpCode reports that with binary rewards ``models learn that generating no code is always `secure'\,'' \citep{liu2025purpcode}.

The second is \emph{exploitation of the test oracle} behind every correctness term. In a frontier RL run, an agent planned to ``circumvent the tests by making verify always return true'' \citep{baker2025monitoring}; on ImpossibleBench, whose tests contradict the task, GPT-5 cheats on \rep{76\%} of Oneoff-SWEbench tasks \citep{zhong2026impossiblebench}; \rep{28.5\%} of a SWE-bench Verified sample have tests accepting an incorrect patch \citep{rajan2026auditing}; in CATCH, a chain-of-thought monitor's protection erodes as the policy learns to mislead it with code comments \citep{wang2026catch}; and hacking learned in production coding environments generalizes to broader misalignment \citep{macdiarmid2025natural}. A correctness term thus does not prevent degenerate code: a special-cased solution is ``correct'' yet unmaintainable.

\paragraph{Highlighted techniques.}

Table~\ref{tab:hl-training} lists the highlighted techniques (selection rule in \S\ref{M-sec:highlighted}). Landmarks curate or mask \emph{data}; recent leaders shape the \emph{objective}, mostly through reinforcement-learning rewards. MPCoder learns Checkstyle style attributes, per-user style vectors and a shared, contrastively trained style adapter so that one CodeLlama-7B model writes Java in each user's own style, raising its Coding Style Score on PCISparse from \rep{46.29} to \rep{66.18}, while HumanEval-X Java correctness falls from \rep{30.49\%} to \rep{26.18\%} on that dataset \citep{dai2024mpcoder}.

\textbf{Caveats.} Several gains are measured with the training instrument: CodeUltraFeedback's GPT-3.5 judge also labelled the training data (with a GPT-4-Turbo judge the authors report ``more mitigated effects''), and REAL evaluates with its reward's program-analysis detectors. SafeCoder ``does not achieve strong generalization across different CWEs''. CodeDPO times only problems solved both before and after training. PurpCode's baseline is its own untrained base model, SecCoderX's the unaligned model, and Afterburner is evaluated on the same authors' Venus benchmark. Correctness costs are uneven (\S\ref{sec:training-effects}). No highlighted technique reports an effect on an untargeted dimension.

\paragraph{Effects on correctness and non-targeted dimensions.}\label{sec:training-effects}

\textbf{Correctness.} The security literature reports small correctness costs, which the evidence partly supports: SVEN ``closely matches the original LMs in functional correctness'' \citep{he2023large}, although its CodeGen-2.7B pass@1 falls from \rep{14.0} to \rep{11.7}, and ProSec and PurpCode report no or minimal utility loss \citep{xu2025prosec,liu2025purpcode}. SafeCoder leaves StarCoder-1B's correctness unchanged, but on Phi-2, against standard instruction tuning, raises security from \rep{69.9} to \rep{90.9} while lowering HumanEval pass@1 from \rep{48.3} to \rep{46.1} \citep{he2024instruction}. LPO lowers Mistral-7B's HumanEvalX pass@1 from \rep{37.1} (SFT) to \rep{28.7}, and without its regularizer gains security ``at a steep cost of decreased general coding ability'' \citep{hasan2025teaching}. On SecCodePLT+ the SFT baseline's function score (\rep{0.8659}) exceeds REAL's (\rep{0.7561}) \citep{yao2025training}. SecCoderX, whose metric weights safety rate by functional correctness, reports that prior methods ``often degrade ESR by \rep{14-54\%}'' \citep{wu2026secure}. Efficiency tuning shows the same tension: ACECode describes PIE as ``improving efficiency but compromising correctness'' \citep{yang2024acecode}, ECCO finds that ``no existing methods can improve time/space efficiency without sacrificing functional correctness'' \citep{waghjale2024ecco}, and a stricter test-coverage reward on hard problems ``reduces optimization failures but increases correctness failures'' \citep{zheng2026extrapolative}. Conversely, correctness-only RLVR neglects efficiency: \rep{36.6\%} of the RLVR baseline's correct programs were slow, versus \rep{15.0\%} for RLPF \citep{jing2026rlpf}. The correctness gains of EffiCoder, ACECode, Afterburner and SuperCoder cannot be separated from ordinary training on verified solutions.

\textbf{Non-targeted dimensions} are rarely measured (only \cnt{nNonTargTraining} of \cnt{nFamTraining} studies): no security study in Tables~\ref{tab:training-results} and~\ref{tab:hl-training} reports readability or runtime, nor any efficiency study security or readability. LPO's data pipeline hints at the risk: repeated security refinement produced ``overengineered and low quality code, training on which reduced models' code utility'' \citep{hasan2025teaching}. Terminology blurs the picture: LPO's ``code quality'' means pass@$k$, and REAL measures security and maintainability in separate scenarios, not on the same outputs.

\paragraph{Compute fairness.}

Training-time methods trade a one-off training cost for no extra inference calls, so per-query comparison with the loops of \S\ref{M-sec:feedback} is not meaningful; most studies report training cost in no form permitting amortization (phi-1's \rep{8} A100s for \rep{4} days excepted). Some headline numbers include inference-time search (PIE's \rep{6.86} speedup is the best of eight generations; Afterburner iterates with sandbox execution), and comparisons with inference-time baselines, such as REAL against constrained decoding and prompt optimization, are not budget-matched.

\paragraph{Limitations.}

\emph{Base-model dependence.} Most studies train models of at most 7B, REAL notes that ``reinforcement learning often requires a reasonably strong base model'' \citep{yao2025training}, and transfer to frontier models is untested. \emph{Oracle circularity.} Training signal and evaluation often share an analyzer family or judge (\S\ref{M-sec:hl-training}), and analyzers disagree: REAL found CodeQL ``unexpectedly ineffective'' on SecCodePLT+, and PurpCode found CodeGuru detects more vulnerabilities than CodeQL. \emph{Contamination.} Synthetic data from GPT-3.5, GPT-4 or frontier LLMs may overlap with public benchmarks; phi-1 studied contamination and PurpCode excluded benchmark sources, but many studies report neither. \emph{Reproducibility.} Data are often partly closed (phi-1 omits synthetic-data details ``for proprietary reasons''), and RL results depend on checkpoint selection, as RLPF notes.

\subsection{Hybrid and other techniques}\label{sec:fd-hybrid}

The \cnt{nFamOther} studies of the hybrid and other group (\S\ref{M-sec:hybrid}) are grouped here by the mechanisms their signal sources map to: retrieval (retrieved knowledge), self-refinement (the model's own critique), another LLM (a critic, reviewer or second agent), tool feedback (static analyzer, compiler, test execution or profiler), a learned verifier, training (training-time stage or a reward model) and decoding (in-decoding stage). Human input, used by 17 studies, is not counted as a mechanism. For each combination, the number of studies and up to four representative studies are listed; studies without a bibliography entry are named by their technique and record identifier (\S\ref{sec:app-studies}).

{\scriptsize\setlength{\tabcolsep}{3pt}
\begin{longtable}{@{}p{0.3\linewidth}rp{0.6\linewidth}@{}}
\caption{Hybrid and other studies by combination of mechanisms (derived from their signal sources and stage), with representative studies.}\label{tab:fd-hybrid}\\
\toprule
Mechanisms & Studies & Representative studies\\
\midrule
\endfirsthead
\toprule
Mechanisms & Studies & Representative studies\\
\midrule
\endhead
retrieval + self-refinement + tool feedback & 33 & Retrieval Augmented Search \citep{anupam2026llm}; SCGAgent \citep{saul2025scgagent}; ECO \citep{kim2025eco}; Safety Sidecar \citep{wang2026safety}\\
tool feedback & 27 & LLM4PLC (R00632); RustMap (R02061); LLMigrate (R02100); DeRep (R02183)\\
retrieval + tool feedback & 26 & SBLLM \citep{gao2025searchbased}; RAG + multi-tool \citep{sriram2026improving}; LLM-driven HLS repair framework (R01058); APILOT (R01272)\\
another LLM + tool feedback & 16 & Codexity Preshot Repair \citep{kim2024codexity}; Hybrid pipeline P10 \citep{sabir2026does}; CodeEvolve \citep{borra2026codeevolve}; SecureForge \citep{liu2026secureforge}\\
retrieval + another LLM + tool feedback & 15 & MARCO \citep{rahman2025marco}; SysLLMatic \citep{peng2026sysllmatic}; HLSPilot (R01152); VeriOpt (R03540)\\
self-refinement + tool feedback & 13 & CodeChain \citep{le2024codechain}; SecCode \citep{liu2024from}; LLM4EFFI \citep{ye2025llm4effi}; ArchCode (R00592)\\
retrieval + self-refinement + another LLM + tool feedback & 7 & LASSI-EE (R02311); FM Agent (S03419); Rustine (S03494); Bespoke OLAP synthesis pipeline (S03781)\\
tool feedback + training & 5 & Afterburner-GRPO within the Iterative~\ldots{} \citep{du2025afterburner}; TurduckenGen (R00265); opt-r1 / opt-r1-mini (R03479); AlgoSkill (R06051)\\
retrieval + self-refinement & 5 & KPC \citep{ren2023from}; RefleXGen \citep{wang2025reflexgenthe}; RAG with valid-package statements~\ldots{} (R00983); RAG-assisted segmented two-stage C-to-Rust~\ldots{} (R05241)\\
retrieval & 4 & AceCoder (R00277); Prototype2Code (R01185); Semantic patch learning from CVE patches (R04636); LLM-based AI agent with request classifier~\ldots{} (R05154)\\
learned verifier & 4 & SecCodePRM-guided inference-time scaling \citep{yu2026seccodeprm}; AutoParLLM (R00452); Fine-tuning on ACEOB efficient-inefficient~\ldots{} (R01180); PreMaQ (R06857)\\
self-refinement + another LLM + tool feedback & 4 & Adaptive self-improvement learning (S02612); Kernel-Smith (S03867); Feedback-grounded explicit planning in a~\ldots{} (S04124); SparseCraft (S04639)\\
learned verifier + training & 3 & FunPRM \citep{zhang2026funprm}; E-code (R01179); DeepGuard (R05109)\\
retrieval + self-refinement + another LLM & 3 & PEACE \citep{ren2025peace}; HCAG (R04905); RTL-Obliger (R06574)\\
retrieval + tool feedback + training & 3 & IDIOMS (S00303); TPU-Gen (S02710); daVinci-kernel (S04225)\\
retrieval + another LLM + tool feedback + training & 2 & HexaCoder \citep{hajipour2024hexacoder}; StitchCUDA (R04757)\\
tool feedback + learned verifier & 2 & SecureCoder (R01618); Surrogate-assisted kernel search (S04142)\\
none of these signals & 2 & iProg (R02038); CliffCompaction (S04576)\\
another LLM + tool feedback + learned verifier & 2 & T-LLM Compiler (R06491); Macro Thinking Micro Coding (S03493)\\
tool feedback + decoding & 1 & Vulnerability-tuning with~\ldots{} (R00457)\\
training & 1 & WaDec (R00998)\\
self-refinement + tool feedback + learned verifier & 1 & PromSec \citep{nazzal2024promsec}\\
retrieval + another LLM + tool feedback + learned verifier & 1 & AutoPatch \citep{seo2025autopatch}\\
another LLM + training & 1 & RefactorCoder-MoE (R03154)\\
retrieval + self-refinement + training & 1 & Stream of Revision (R04484)\\
retrieval + self-refinement + tool feedback + training & 1 & DepTrans (R05047)\\
self-refinement & 1 & Accumulated behavioral rules (R06233)\\
another LLM & 1 & Security-requirement appending /~\ldots{} (R06284)\\
retrieval + another LLM & 1 & Pipeline 1 \citep{surikov2026securing}\\
another LLM + learned verifier & 1 & ProbeDebias (R06886)\\
tool feedback + learned verifier + decoding & 1 & BetterV (S01747)\\
self-refinement + decoding & 1 & EfficientUICoder (S03256)\\
retrieval + another LLM + tool feedback + learned verifier + training & 1 & SwiftTrans (S04226)\\
self-refinement + tool feedback + decoding & 1 & Explicit agent-first (S04598)\\
\bottomrule
\end{longtable}}

\paragraph{Domains.} By keywords in their languages, benchmarks and titles, 111 of the studies work on specialized code: GPU or accelerator kernels (66), hardware design in Verilog, VHDL, Chisel or high-level synthesis (24), C-to-Rust migration (20), parallel code with OpenMP, MPI or OpenACC (9), smart contracts (3) and PLC programs (2); a study can fall under several. 90 evaluate on Python. The keyword match is approximate; the head-to-head excludes domain-specific techniques by the pass-2 rule of \S\ref{M-sec:highlighted}, not by these keywords.

\paragraph{Targets and effects.} By technique row, the target improves in 112 of 122 efficiency rows, 39 of 46 security rows and 33 of 35 readability/maintainability rows; correctness is unreported in 31, 9 and 3 of them. Efficiency rows thus carry most of the unreported correctness, as in the review overall.

\FloatBarrier
\section{Per-family result tables and highlighted techniques}\label{sec:app-family-tables}
The reported-effect and design tables of each technique family and its highlighted techniques, referenced from the family sections of the main article. Values are as reported by the original authors; settings differ across rows, so values are not directly comparable.

\subsection*{Prompting}
\begin{table}[H]
\caption{Reported effects of prompting strategies on quality and correctness. \up/\down/\same: improvement/degradation/no significant change in the stated metric relative to the study's baseline. Values as reported; settings differ, so rows are not directly comparable.}
\label{tab:prompting-results}
\footnotesize
\begin{tabularx}{\linewidth}{@{}l>{\raggedright\arraybackslash}X>{\raggedright\arraybackslash}X>{\raggedright\arraybackslash}X>{\raggedright\arraybackslash}X@{}}
\toprule
Study & Strategy & Models; benchmark & Target-dimension effect & \dimCor{} effect \\
\midrule
Tony~\citep{tony2025prompting} & RCI (one critique--improve round) & GPT-3/3.5/4; LLMSecEval, Bandit & \dimSec{} \up: weakness density \rep{77.55\%} lower (GPT-4) & \nr \\
Bruni~\citep{bruni2025benchmarking} & Security-aware persona prefix & GPT-4o; LLMSecEval + SecurityEval, Semgrep/CodeQL & \dimSec{} \up: \rep{56\%} fewer vulnerabilities & HumanEval pass@1 \rep{86.77\%}\,$\to$\,\rep{86.65\%} (GPT-4o-mini) \\
Bruni~\citep{bruni2025benchmarking} & RCI on baseline code & GPT-4o; as above & \dimSec{} \up: \rep{64.7\%} fewer vulnerabilities & \down{} pass@1 \rep{86.77\%}\,$\to$\,\rep{76.65\%} (GPT-4o-mini) \\
Kharma~\citep{kharma2026empirical} & WA-0CoT (CWE-aware CoT) & 5 LLMs; Java, C++, C, Python & \dimSec{} \same{} frequency and density; CWE mix shifts; \dimRM{} \down{} more warnings (C) & \nr \\
Urmi~\citep{urmi2026prompt} & Structured + security guidance & GPT-4o; 424 Python tasks, Bandit/CodeQL & \dimSec{} high-severity \rep{20.8\%}\,$\to$\,\rep{13.6\%}; low-severity \rep{32\%}\,$\to$\,\rep{43.5\%} & \nr{} (requested constructs sometimes removed) \\
Pereira~\citep{pereira2026does} & ISO/IEC 25010-grounded requirements & GPT-5.4; HumanEval/ET & \dimRM{} \up{} unreadability density \rep{0.88}\,$\to$\,\rep{0.69} (performance requirement) & not reliably improved; ET pass rate \down{} (error handling) \\
Khojah~\citep{khojah2025impact} & Few-shot vs.\ CoT+persona+packages & GPT-4o; CoderEval, Pylint & \dimRM{} missing docstring \rep{71\%} vs.\ \rep{22\%} & few-shot \up; persona/CoT \down \\
RACE~\citep{zheng2024beyond} & Explicit style requirement & 28 LLMs; RACE & \dimRM{} camel-case compliance \rep{below 30\%} (CodeLlama, WizardCoder) & joint metric only \\
Apsan~\citep{apsan2026generating} & Energy-oriented prompts & 6 LLMs; 9 EvoEval problems & \dimEff{} \same{} no consistent energy saving & \nr \\
Rekhi~\citep{rekhi2026improving} & 8 energy prompts & 10 LLMs; EffiBench & \dimEff{} \up{} up to \rep{25\%} (Python), \rep{17\%} (C++) & \nr{} (correct subset only) \\
ECCO~\citep{waghjale2024ecco} & Few-shot exemplars & open code LLMs, GPT-4o; ECCO & \dimEff{} mixed by model & mixed by model \\
\bottomrule
\end{tabularx}
\end{table}

\begin{table}[H]
\caption{Highlighted prompting techniques: landmarks (L, most cited) and recent leaders (R, 2025--2026, largest reported gain). Year = year of the cited version. Cit.\ = Semantic Scholar citations, October 2026. \dimCor{} = direction of the effect on correctness as coded in our extraction. Values as reported by the original authors; settings differ across rows, so values are not directly comparable.}
\label{tab:hl-prompting}
\scriptsize
\begin{tabularx}{\linewidth}{@{}>{\raggedright\arraybackslash}p{1.6cm}lrc>{\raggedright\arraybackslash}X>{\raggedright\arraybackslash}p{0.75cm}>{\raggedright\arraybackslash}X cc@{}}
\toprule
Technique & Year & Cit. & L/R & Core idea & Target & Reported result & \dimCor & Code \\
\midrule
SCoT \citep{li2025structured} & 2025 & 405 & L & Constrains the chain of thought to sequence, branch and loop structures, then implements code from it. & \dimRM & Maintainability, human rating 0--2, ChatGPT: CoT \rep{1.616} $\to$ \rep{1.873} & \up & -- \\
Self-planning \citep{jiang2024selfplanning} & 2024 & 306 & L & Few-shot prompts make the model write a concise step plan, then code guided by that plan. & \dimRM & Readability, human rating: gain reported without numbers (\nr) & \up & -- \\
LLM4EFFI \citep{ye2025llm4effi} & 2025 & 15 & L & Explores algorithms in a logic domain, optimizes the implementation in a code domain, then verifies with synthetic tests. & \dimEff & eff@1, ENAMEL, DeepSeek-V3: Effi-Learner \rep{52.22} $\to$ \rep{60.41} & \up & \ding{51} \\
Code2API \citep{mai2024are} & 2024 & 12 & L & Role, chain-of-thought and few-shot prompting turn Stack Overflow snippets into reusable, well-named API methods. & \dimRM & Method-name expressiveness, human rating 1--4, GPT-3.5: APIzator \rep{2.23} $\to$ \rep{3.65} & mixed & \ding{51} \\
MPCO \citep{gong2025tuning} & 2025 & 10 & L & A meta-prompter LLM writes project-, task- and model-aware code-optimization prompts. & \dimEff & Mean rank of \%PI over five projects (lower is better): contextual prompting \rep{2.00} $\to$ \rep{1.00} & \nr & -- \\
BRACE \citep{xu2026activating} & 2026 & 1 & R & A separate LLM predicts the task's likely CWEs; after catalog validation they are given to the target model as a cue. & \dimSec & Safe Code Rate (judge classifier), CyberNative, open-weight models (avg.): Secure Prompt \rep{33.9\%} $\to$ \rep{91.2\%} & mixed & -- \\
ECO \citep{kim2025eco} & 2025 & 2 & R & Prompt combines a rule-based static bottleneck diagnosis with retrieved optimization instructions distilled from slow--fast pairs. & \dimEff & Speedup, PIE, best of five, GPT-o4-mini: instruction-only \rep{1.99}$\times$ $\to$ \rep{7.81}$\times$ & mixed & -- \\
CSDD \citep{marri2026constitutional} & 2026 & 3 & R & A versioned constitution of CWE-mapped security principles constrains specifications and AI generation, traced to code. & \dimSec & CWE violations, one banking application, Claude: unconstrained \rep{11} $\to$ \rep{3} & \nr & \ding{51} \\
SecureForge \citep{liu2026secureforge} & 2026 & 3 & R & Mines benign prompts whose code Semgrep flags, broadens them by MCMC rephrasing, then evolves the system prompt with GEPA using the Semgrep findings as reflection feedback. & \dimSec & Semgrep weakness rate of test-passing outputs, ten LLMs, held-out CWE Top-25 scenarios: up to \rep{48\%} lower than the security-aware prompt & \same & \ding{51} \\
StoryCoder \citep{jang2026storycoder} & 2026 & 0 & R & Retells the problem as a narrative (task overview, constraints, example input/output) in a genre matched to the algorithm it picks, then generates code from the narrative. & \dimRM & Functions per correct solution, CodeForces, Gemini-2.5-Flash: repeated sampling \rep{1.820} $\to$ \rep{2.338} & \up & \ding{51} \\
\bottomrule
\end{tabularx}
\end{table}

\clearpage
\subsection*{Tool-feedback loops}
\begin{table}[H]
\caption{Design space of tool-feedback loops. Gate: acceptance check other than the driving tool. \dimCor: correctness measured, with direction. Author-reported values, not directly comparable across rows.}
\label{tab:fb-design}
\footnotesize
\begin{tabularx}{\linewidth}{@{}>{\raggedright\arraybackslash}p{1.95cm}>{\raggedright\arraybackslash}p{1.95cm}>{\raggedright\arraybackslash}X>{\raggedright\arraybackslash}p{1.3cm}>{\raggedright\arraybackslash}p{1.1cm}>{\raggedright\arraybackslash}p{0.85cm}>{\raggedright\arraybackslash}p{2.15cm}@{}}
\toprule
Study & Tool(s) & Feedback format & Max iter. & Gate & \dimCor & Lang.; models \\
\midrule
\citet{blyth2025static} & Bandit, Pylint & inline tags, severity order & \rep{10} & fitness & \up & Py; GPT-4o \\
\citet{wadhwa2024core} & CodeQL, Sonar & rule + fix advice; ranker & one pass, \rep{10} samples & ranker & review & Py, Java; GPT-3.5/4 \\
\citet{abtahi2025augmenting} & SonarQube & issue table + RAG & \rep{two} phases & manual & manual & one project; GPT-3.5/4o \\
\citet{joos2026codecureagent} & SonarQube & agent + rejection context & \rep{40} cycles & build, tests & manual & Java; GPT-4.1 mini \\
\citet{islam2026from} & Pylint, Radon, Bandit, \dots & metric diagnostics & \rep{five} & none & no & Py; several \\
\citet{ravi2025llmloop} & javac, PMD, tests & errors, violations & \rep{five}/loop & tests & \up & Java; GPT-4o-mini \\
\citet{alrashedy2023can} & Bandit & raw vs.\ verbalized; $J$ fixes & $K$ & none & no & Py; \rep{three} LLMs \\
\citet{nazzal2024promsec} & Bandit, SpotBugs & graph fix $\to$ new prompt & until clean & similarity & \down$^a$ & Py, Java; GPT-3.5 \\
\citet{kim2024codexity} & Infer, CppCheck & type + location & \rep{3} & none & no & C; \rep{five} LLMs \\
\citet{gajjar2025securefixagent} & Bandit & report + LLM check & to clean & none & no & Py; $<$\rep{8B} tuned \\
\citet{yan2026guiding} & CodeQL & raw vs.\ explained & \nr & none & no & Py; \rep{eight} LLMs \\
\citet{dolcetti2026helping} & Infer; tests & one finding per prompt & \rep{6} & separate & \up & C; \rep{4} open \\
\citet{sriram2026improving} & GCC, CodeQL, KLEE & diagnostics + RAG & \rep{three} & KLEE & \up & C/C++; \rep{two} small \\
\citet{nunez2024autosafecoder} & LLM ``analyzer'' & CWE + advice & \rep{four} & fuzzing & \down & Py; GPT-4o \\
\citet{huang2024effilearner} & line/mem.\ profiler & per-line profile & \rep{0 to 5} & open tests & small \down & Py; \rep{16}+\rep{6} LLMs \\
\citet{peng2025perfcodegen} & runtime & slowest test & \rep{one} & tests & \up & Py; open, GPT-3.5/4 \\
\citet{peng2026sysllmatic} & async-prof., perf & hotspots + catalog & \rep{T = 2} & tests & held & C++, Java; GPT-4o/4.1 \\
\citet{deng2026perfagent} & py-spy & hotspots, timing & \rep{5} & tests & \up & Py; GPT-5.1, Kimi-K2 \\
\bottomrule
\end{tabularx}
\par\smallskip{\scriptsize $^a$ functionality proxied mainly by code-graph similarity; Pass@1 fell when \citet{dai2026rethinking} ran unit tests.}
\end{table}

\begin{table}[H]
\caption{Reported effects on the target dimension and on correctness ($\to$: before to after). Author-reported values, not directly comparable across rows.}
\label{tab:fb-results}
\footnotesize
\begin{tabularx}{\linewidth}{@{}>{\raggedright\arraybackslash}p{2.05cm}>{\raggedright\arraybackslash}p{2.75cm}>{\raggedright\arraybackslash}X>{\raggedright\arraybackslash}p{2.6cm}@{}}
\toprule
Study & Setting & Target-dimension effect & \dimCor\ effect \\
\midrule
\citet{blyth2025static} & GPT-4o; PythonSecurityEval & insecure \rep{$>$40\%}$\to$\rep{13\%}; readability viol.\ \rep{$>$80\%}$\to$\rep{11\%} & pass@1 \rep{38.16\%}$\to$\rep{47.81\%} \\
\citet{wadhwa2024core} & GPT-3.5/4; CodeQL, Python & pass checks \rep{81.57\%}; reviewer rejection \rep{55.11\%} (\rep{47.55\%} ranked) & review only \\
\citet{joos2026codecureagent} & GPT-4.1 mini; SonarQube & plausible \rep{96.8\%}; correct \rep{86.3\%} & gated \\
\citet{abtahi2025augmenting} & GPT-3.5$\to$4o; one project & issues \rep{7,599}$\to$\rep{4,340}$\to$\rep{1,058} & manual \\
\citet{alrashedy2023can} & GPT-4; PythonSecurityEval & vulnerable: Bandit \rep{40.2\%}$\to$\rep{7.4\%}; CodeQL \rep{17.9\%}$\to$\rep{7.7\%} & \nr \\
PromSec in \citep{dai2026rethinking} & GPT-3.5; BigCodeBench & Secure-Pass@1 \rep{0.3219}$\to$\rep{0.2447} & Pass@1 \down \\
\citet{kim2024codexity} & gpt-3.5-turbo; C & vulnerable \rep{75.9\%}$\to$\rep{15.9\%} & \nr \\
\citet{gajjar2025securefixagent} & tuned $<$\rep{8B}; Bandit & fix accuracy \rep{87.83} vs.\ \rep{79.72} single-pass & \nr \\
\citet{yan2026guiding} & GPT-4o; CodeQL & target-vuln.\ rate $-$\rep{7.3\%}, $-$\rep{15.7\%}, $-$\rep{8.7\%} & \nr \\
\citet{dolcetti2026helping} & best models; Infer, C & \rep{89\%} of vulnerabilities fixed & \rep{62\%} of incorrect repaired \\
\citet{sriram2026improving} & CodeLlama-7B; C/C++ & CodeQL defects \rep{58.55\%}$\to$\rep{22.19\%} & KLEE \rep{19.98\%}$\to$\rep{12.18\%} \\
\citet{nunez2024autosafecoder} & GPT-4o; SecurityEval & vulnerable \rep{59/121 (49\%)}$\to$\rep{44/121 (36\%)} & pass@1 \rep{0.9085}$\to$\rep{0.8780} \\
\citet{ravi2025llmloop} & GPT-4o-mini; HumanEval-X & PMD \nr & pass@1 \rep{71.65\%}$\to$\rep{80.85\%} \\
\citet{huang2024effilearner} & StarCoder2-15B; EffiBench & time \rep{0.93 (s)}$\to$\rep{0.12 (s)} & e.g.\ \rep{5.8}$\to$\rep{5.4}$^b$ \\
\citet{peng2026sysllmatic} & GPT-4.1; DaCapo & latency \rep{1.54x} (compiler \rep{1.01x}) & gated \\
\citet{deng2026perfagent} & GPT-5.1; GSO / SWE-fficiency-Lite & expert-level \rep{19.6\%}$\to$\rep{39.2\%} / \rep{26\%}$\to$\rep{74\%} & gated \\
\bottomrule
\end{tabularx}
\par\smallskip{\scriptsize $^b$ pass@1 of OpenCodeInterpreter-DS-1.3B.}
\end{table}

\begin{table}[H]
\caption{Highlighted tool-feedback loop techniques: landmarks (L, most cited) and recent leaders (R, 2025--2026, largest reported gain). Year = year of the cited version. Cit.\ = Semantic Scholar citations, October 2026. \dimCor{} = direction of the effect on correctness as coded in our extraction. Values as reported by the original authors; settings differ across rows, so values are not directly comparable.}
\label{tab:hl-feedback}
\scriptsize
\begin{tabularx}{\linewidth}{@{}>{\raggedright\arraybackslash}p{1.75cm}>{\raggedright\arraybackslash}p{0.55cm}>{\raggedright\arraybackslash}p{0.4cm}>{\raggedright\arraybackslash}p{0.3cm}>{\raggedright\arraybackslash}X>{\raggedright\arraybackslash}p{0.75cm}>{\raggedright\arraybackslash}X>{\raggedright\arraybackslash}p{0.55cm}>{\raggedright\arraybackslash}p{0.4cm}@{}}
\toprule
Technique & Year & Cit. & L/R & Core idea & Target & Reported result & \dimCor & Code \\
\midrule
SBLLM \citep{gao2025searchbased} & 2025 & 76 & L & Evolves optimized programs: selects representative executed samples, retrieves optimization patterns, combines them with genetic-operator-style prompts. & \dimEff & SP@5 (\%), GPT-4, PIE Python: ICL \rep{181.53} $\to$ \rep{209.59} & \up & \ding{51} \\
PerfCodeGen \citep{peng2025perfcodegen} & 2025 & 58 & L & After a correctness phase, returns the most time-consuming unit test and asks for a faster program. & \dimEff & \%Opt Best@8, GPT-4, MBPP: base \rep{43.48} $\to$ \rep{56.52} & \up & -- \\
PromSec \citep{nazzal2024promsec} & 2024 & 55 & L & A contrastively trained graph GAN repairs the vulnerable code graph; the repaired code is turned into an improved prompt, iteratively. & \dimSec & \nr{} (CWE counts shown only in plots) & \up & \ding{51} \\
DeepDevVuln filter \citep{chan2023transformerbased} & 2023 & 41 & L & A fine-tuned transformer vulnerability detector flags incomplete code; completions it flags are discarded. & \dimSec & vulnerable scenarios, text-davinci-003: unfiltered \rep{78.00\%} $\to$ filtered \rep{8.0\%} & \nr & -- \\
FRANC repair \citep{siddiq2024franc} & 2024 & 32 & L & Builds a repair prompt from Bandit or SpotBugs messages (generated-knowledge prompting, three prompt structures) and re-queries the model. & \dimSec, \dimRM & prompts with a good suggestion after repair: up to \rep{80\%}; no pre-repair value & \nr & \ding{51} \\
Optimo \citep{peng2026multilevel} & 2026 & 0 & R & Differential profiling finds bottlenecks, routed to multi-level, pattern-specific optimization prompts with shared and special strategies and correctness reflection. & \dimEff & Opt\%, GPT-4o code, EffiBench: SBLLM \rep{3.58\%} $\to$ \rep{35.79\%} & \up & \ding{51} \\
Infer repair loop \citep{dolcetti2026helping} & 2026 & 27 & R & Feeds one Infer warning at a time, with fix instructions, back to the model for iterative repair. & \dimSec & Infer findings fixed, C, best of four open models: \rep{39 out of 44} (\rep{89\%}) & \nr & -- \\
CodeEvolve \citep{borra2026codeevolve} & 2026 & 1 & R & Selects targets from runtime profiles; evolves edits with MCTS-augmented search, a refinement agent and cascaded validation. & \dimEff & avg.\ speedup, 7 Java hotspots: single-call RPBD \rep{7.12$\times$} $\to$ \rep{15.22$\times$} & \up & -- \\
PerfAgent \citep{deng2026perfagent} & 2026 & 3 & R & Curated profiler hotspot summary, an objective-driven loop that keeps optimizing, selective test validation, and best-patch selection. & \dimEff & hack-adjusted Opt@1, GPT-5.1, GSO: OpenHands \rep{19.6\%} $\to$ \rep{39.2\%} & \up & -- \\
SelectIssue loop \citep{blyth2025static} & 2025 & 17 & R & Each round tags a few Bandit and Pylint issues, picked by severity weight, inline at their code lines; a revision is kept only if its test-gated fitness does not fall. & \dimSec, \dimRM & Bandit-flagged insecure snippets, GPT-4o, PythonSecurityEval, two issues per round, ten rounds: \rep{45.05\%} $\to$ \rep{13.41\%} & \up & -- \\
\bottomrule
\end{tabularx}
\end{table}

\clearpage
\subsection*{Self-refinement}
\begin{table}[H]
\caption{Reported effects of self-refinement on non-functional quality (Self-Refine on PIE: Table~\ref{tab:hl-selfrefine}). Values as reported by the original authors; settings differ across rows, so values are not directly comparable.}
\label{tab:sr-results}
\footnotesize
\begin{tabularx}{\linewidth}{@{}>{\raggedright\arraybackslash}p{3.2cm} >{\raggedright\arraybackslash}X l l@{}}
\toprule
Study (model) & Setting and metric & Before $\rightarrow$ after & \dimCor \\
\midrule
Self-Refine (GPT-4) & CodeNet rewrite; GPT-4-judged meaningful names (\dimRM) & \rep{27.4} $\rightarrow$ \rep{56.2} & \nr \\
EffiLearner (StarCoder2-15B) & EffiBench; execution time, s (\dimEff) & \rep{0.93} $\rightarrow$ \rep{0.12} & \down{} \rep{0\% to 0.5\%} \\
EffiLearner (StarCoder2-15B) & EffiBench; total memory usage, Mb$\cdot$s (\dimEff) & \rep{22.02} $\rightarrow$ \rep{2.03} & as above \\
Self-Refine in EffiLearner (CodeLlama-70B) & EffiBench; execution time change (\dimEff) & $+$\rep{51.9\%} & \nr \\
Reflexion + ICD (five open code LLMs) & Instruct Prime; ICD secure rate (\dimSec) & \rep{70.74\%} $\rightarrow$ \rep{79.43\%} & \nr \\
\bottomrule
\end{tabularx}
\end{table}

\begin{table}[H]
\caption{Highlighted self-refinement techniques: landmarks (L, most cited) and recent leaders (R, 2025--2026, largest reported gain). Year = year of the cited version. Cit.\ = Semantic Scholar citations, October 2026. \dimCor{} = direction of the effect on correctness as coded in our extraction. Values as reported by the original authors; settings differ across rows, so values are not directly comparable.}
\label{tab:hl-selfrefine}
\scriptsize
\begin{tabularx}{\linewidth}{@{}>{\raggedright\arraybackslash}p{1.6cm}lrc>{\raggedright\arraybackslash}X>{\raggedright\arraybackslash}p{0.75cm}>{\raggedright\arraybackslash}X cc@{}}
\toprule
Technique & Year & Cit. & L/R & Core idea & Target & Reported result & \dimCor & Code \\
\midrule
Self-Refine \citep{madaan2023selfrefine} & 2023 & 5050 & L & One LLM alternates between feedback on its own output and refinement using that feedback, for several iterations. & \dimEff & \% programs optimized, PIE, GPT-4: no refinement \rep{27.3} $\to$ \rep{36.0} & \nr & \ding{51} \\
CodeChain \citep{le2024codechain} & 2024 & 91 & L & Modular chain of thought, then self-revisions conditioned on clustered representative sub-modules from earlier samples. & \dimRM & Modularity and reusability, GPT-4 Likert rating, \rep{20} APPS tasks: shown only as a figure (\nr) & \up & \ding{51} \\
SecCode \citep{liu2024from} & 2024 & 8 & L & Score-based ``encouragement'' prompts drive repeated vulnerability detection and fixing, cross-checked with CodeQL. & \dimSec & Fix success rate after ten iterations, LEval, GPT-3.5: CoT \rep{0.64} $\to$ \rep{0.96} & \up & -- \\
RefleXGen \citep{wang2025reflexgenthe} & 2025 & 7 & L & Self-reflection on code security; reflections and secure snippets accumulate in a knowledge base retrieved for later tasks. & \dimSec & CodeQL security rate, SVEN scenarios, GPT-3.5 Turbo: base \rep{75.5} $\to$ \rep{89.1} & mixed & -- \\
GRASP \citep{patir2025fortifying} & 2025 & 4 & L & Revises a seed solution by traversing a dependency graph (DAG) of the relevant secure coding practices. & \dimSec & CodeQL security rate, \rep{37} Python scenarios, GPT-4o: PromSec \rep{0.52} $\to$ \rep{0.79} & mixed & -- \\
Safety Sidecar \citep{wang2026safety} & 2026 & 1 & R & A lightweight self-check routes risky code to a reflective prompt fed by a memory of past secure repairs. & \dimSec & CodeQL security rate, SVEN scenarios, GPT-4o: base agent \rep{85.7} $\to$ \rep{95.0} & mixed & -- \\
\bottomrule
\end{tabularx}
\end{table}

\clearpage
\subsection*{Multi-agent pipelines}
\begin{table}[H]
\caption{Highlighted multi-agent techniques: landmarks (L, most cited) and recent leaders (R, 2025--2026, largest reported gain). Year = year of the cited version. Cit.\ = Semantic Scholar citations, October 2026. \dimCor{} = direction of the effect on correctness as coded in our extraction. Values as reported by the original authors; settings differ across rows, so values are not directly comparable.}
\label{tab:hl-multiagent}
\scriptsize
\begin{tabularx}{\linewidth}{@{}>{\raggedright\arraybackslash}p{1.6cm}lrc>{\raggedright\arraybackslash}X>{\raggedright\arraybackslash}p{0.75cm}>{\raggedright\arraybackslash}X cc@{}}
\toprule
Technique & Year & Cit. & L/R & Core idea & Target & Reported result & \dimCor & Code \\
\midrule
SecAwareCoder \citep{wang2026toward} & 2026 & 54 & L & Task-adaptive threat modeling yields vulnerability hypotheses that guide constraint-aware generation and security-test synthesis; execution feedback drives repair. & \dimSec & S-Pass@1, CodeSecEval, GPT-5: CoT \rep{27.45} $\to$ \rep{34.90} & \up & \ding{51} \\
FlowGen \citep{lin2025soen101} & 2025 & 51 & L & Agents in requirement-engineer, architect, developer, tester and scrum-master roles emulate Waterfall, TDD or Scrum, with self-refinement. & \dimRM & Pylint error-smell density, GPT-3.5: \rep{81.8\%} to \rep{96.0\%} lower than RawGPT & \up & \ding{51} \\
INDICT \citep{le2024indict} & 2024 & 4 & L & A safety critic and a helpfulness critic, grounded in web search and code execution, give preemptive and post-hoc feedback to the coder. & \dimSec & Safety, CyberSecEval-1 split, GPT-4o-mini: direct \rep{78.2} $\to$ \rep{90.9} & \nr & \ding{51} \\
ContextEvolve \citep{su2026contextevolve} & 2026 & 2 & R & Summarizer, Navigator and Sampler agents compress evolutionary-search context into semantic state, optimization direction and retrieved exemplars. & \dimEff & ADRS combined score vs.\ top baseline: \rep{6.5\%} higher on average & \same & \ding{51} \\
Just-in-time remediation (P1) \citep{surikov2026securing} & 2026 & 0 & R & A validator LLM reviews generated code; its findings, enriched with ATT\&CK techniques, CWE examples and best practices, drive remediation. & \dimSec & CodeQL+Bandit findings per sample, four Claude models pooled: initial \rep{1.34} $\to$ \rep{0.78} & \nr & \ding{51} \\
LessonL \citep{liu2025lessons} & 2025 & 0 & R & Agents solicit lessons from each other's successes and failures, bank them, and select relevant lessons for later rounds. & \dimEff & Geometric-mean speedup, ParEval serial: Reflexion (Qwen14B) \rep{1.88} $\to$ \rep{2.16} & \up & \ding{51} \\
MARCO \citep{rahman2025marco} & 2025 & 3 & R & An optimizer agent and a performance-evaluator agent iterate in a feedback loop; web search supplies recent optimization techniques. & \dimEff & Mean runtime (ms), LeetCode 75: Claude 3.5 Sonnet \rep{57.5} $\to$ \rep{49.1} & \nr & -- \\
C3 \citep{gu2026mitigating} & 2026 & 0 & R & Adds auction-based role allocation (CAB) and opponent-aware decentralized competition (DCC) to a collaborative development pipeline. & \dimRM & Human maintainability rating, extended SoftwareDev, GPT-4o: MetaGPT \rep{3.6} $\to$ \rep{3.8} & \up & -- \\
MaintainCoder \citep{wang2025maintaincoder} & 2025 & 18 & L & Requirements-analysis, design-pattern-selection, framework-design and framework-evaluation agents produce an architectural blueprint; a coding agent and an execution-driven optimizer then write and refine the code. & \dimRM & Post-modification Pass@5 (code changed for new requirements), APPS-Dyn, DeepSeek-V3: plain \rep{59.5} $\to$ \rep{62.5} & \up & \ding{51} \\
ACCLAIM \citep{mikek2026agentic} & 2026 & 3 & L & A guiding agent interleaves LLVM front-, middle- and back-end passes with LLM rewriting agents at source, IR and assembly level under one call budget; a test-generating agent checks every candidate. & \dimEff & Geometric-mean speedup over clang -O3, Claude 3.7 Sonnet, 100 CodeNet C programs, equal budget: source-only agent \rep{1.18}$\times$ $\to$ \rep{1.25}$\times$ & \nr & \ding{51} \\
\bottomrule
\end{tabularx}
\end{table}

\clearpage
\subsection*{Retrieval}
\begin{table}[H]
\caption{Reported effects of further retrieval-augmented methods on quality and correctness, as reported by the original authors; settings differ, so values are not directly comparable.}
\label{tab:retrieval-results}
\footnotesize
\begin{tabularx}{\linewidth}{@{}l>{\raggedright\arraybackslash}X>{\raggedright\arraybackslash}X>{\raggedright\arraybackslash}X>{\raggedright\arraybackslash}X@{}}
\toprule
Method & Retrieved unit & Models; benchmark & Target-dimension effect & \dimCor{} effect \\
\midrule
SecCoder~\citep{zhang2024seccoder} & Secure demonstrations & CodeGen, SantaCoder, InCoder; CodeQL & \dimSec{} \up{} \rep{7.20\%} avg.\ on unseen cases & \same{} ``almost the same'' \\
RESCUE~\citep{shi2026rescue} & Distilled guidelines, sliced examples & 6 LLMs; 4 benchmarks & \dimSec{} \up{} SecurePass@1 \rep{4.8} points & \same{} comparable \\
SEEK~\citep{gadey2026seek} & CWE-graph fix instances & 5 open LLMs; CodeGuard+ & \dimSec{} SecurePass@1 \rep{77.45\%} & in SecurePass@1 \\
SOSecure~\citep{mukherjee2025sosecure} & Stack Overflow critiques (post-draft) & GPT-4; LLMSecEval & \dimSec{} fix rate \rep{91.3\%} vs.\ \rep{56.5\%} & \nr \\
A3-CodGen~\citep{liao2024a3codgen} & Local functions & GPT-3.5; 13,784 PyPI functions & \dimRM{} reuse F1 \rep{0.589} (vanilla 0) & \nr \\
\bottomrule
\end{tabularx}
\end{table}

\begin{table}[H]
\caption{Highlighted retrieval techniques: landmarks (L, most cited) and recent leaders (R, 2025--2026, largest reported gain). Year = year of the cited version. Cit.\ = Semantic Scholar citations, October 2026. \dimCor{} = direction of the effect on correctness as coded in our extraction. Values as reported by the original authors; settings differ across rows, so values are not directly comparable.}
\label{tab:hl-retrieval}
\scriptsize
\begin{tabularx}{\linewidth}{@{}>{\raggedright\arraybackslash}p{1.6cm}lrc>{\raggedright\arraybackslash}X>{\raggedright\arraybackslash}p{0.75cm}>{\raggedright\arraybackslash}X cc@{}}
\toprule
Technique & Year & Cit. & L/R & Core idea & Target & Reported result & \dimCor & Code \\
\midrule
PIE dynamic retrieval \citep{shypula2024learning} & 2024 & 171 & L & Retrieves the slow--fast program pairs most similar to the input and uses them as few-shot examples for optimization. & \dimEff & Speedup (Best@8), PIE, CodeLlama 34B: CoT \rep{1.30$\times$} $\to$ \rep{2.28$\times$} & mixed & \ding{51} \\
LAIL \citep{li2025large} & 2025 & 92 & L & Labels candidate examples by how much they raise the LLM's likelihood of the reference program; trains a contrastive retriever on these labels. & \dimRM & Human code-quality score: best example selector \rep{1.738} $\to$ \rep{1.839} & \up & -- \\
KPC \citep{ren2023from} & 2023 & 50 & L & Builds an exception-handling knowledge base from Java API documentation; chained prompts check generated code for unhandled exceptions and rewrite it. & \dimRM & Good exception-handling practice (manual review), GPT-3.5: direct \rep{14.58\%} $\to$ \rep{98.86\%} & \up & \ding{51} \\
RAPGen \citep{garg2025rapgen} & 2025 & 36 & L & Retrieves an instruction from a knowledge base of past performance-bug fixes and builds a zero-shot fix prompt from it. & \dimEff & Expert-judged closest-match top-100 accuracy, C\# fixes, GPT-3.5: reasoning prompt \rep{43.2} $\to$ \rep{60.6} & \nr & -- \\
CodeGuarder \citep{lin2025give} & 2025 & 30 & L & Decomposes the query into sub-tasks, retrieves root causes and secure samples, re-ranks by model susceptibility, injects them into the prompt. & \dimSec & Security rate, CyberSecEval, average over LLMs: RACG without security knowledge \rep{60.84} $\to$ \rep{73.08} & \up & -- \\
RAS \citep{anupam2026llm} & 2026 & 6 & R & Beam search over candidate optimizations; slow--fast exemplars are retrieved by an LLM-written natural-language description of the program. & \dimEff & Average speedup, PIE, Qwen3-Coder: dynamic retrieval \rep{4.23$\times$} $\to$ \rep{8.70$\times$} & \up & \ding{51} \\
SemOpt \citep{zhao2026semopt} & 2026 & 6 & R & Builds a strategy library from optimization commits; LLM-written Semgrep rules locate applicable code; the LLM applies the matched strategy. & \dimEff & Equivalence to developer patch (manual), C/C++, DeepSeek-V3: BM25 RAG \rep{23.80\%} $\to$ \rep{42.38\%} & \nr & \ding{51} \\
AutoPatch \citep{seo2025autopatch} & 2025 & 7 & R & Matches generated code to post-cutoff CVEs in a retrieval database, verifies the vulnerability, and patches with the retrieved CVE context. & \dimSec & Patch accuracy, authors' CVE dataset: VSP \rep{55.56\%} $\to$ \rep{94.12\%} (GPT-4o) & \nr & \ding{51} \\
PEACE \citep{ren2025peace} & 2025 & 11 & R & Optimizes functions in dependency order, guided by filtered historical edits; a fine-tuned optimizer also uses similar internal and external fast implementations. & \dimEff & Opt rate (\%, CPU instructions vs.\ GPT-4o prompting), PeacExec: SBLLM \rep{+28.5} $\to$ \rep{+46.9} & \up & \ding{51} \\
MA-CoT \citep{kharma2026enhancing} & 2026 & 2 & R & Classifies the task into security domains with an LLM, retrieves matching CWE mitigation strategies from a knowledge base, and builds them into the prompt. & \dimSec & Total security findings (SonarQube, expert-validated), authors' primary dataset, three LLMs: Vanilla \rep{92} $\to$ \rep{39} & \nr & -- \\
\bottomrule
\end{tabularx}
\end{table}

\clearpage
\subsection*{Decoding}
\begin{table}[H]
\centering
\footnotesize
\caption{Further decoding-time results (highlighted techniques are in Table~\ref{tab:hl-decoding}).
Values as reported by the authors; settings differ, so values are not directly comparable.}
\label{tab:dec-results}
\begin{tabularx}{\linewidth}{@{}l l X l@{}}
\toprule
Method & Dim. & Setting (model; benchmark; metric) & Reported \\
\midrule
Type-constrained \citep{mundler2025typeconstrained} & \dimCor{} & 2B--34B; HumanEval (TypeScript); compiler-error reduction & \rep{74.8\%} \\
MRepE \citep{gao2026towards} & \dimRM{} & DeepSeek-R1 14B; MBPP; correctness loss & \rep{0.79\%} \\
CodeT \citep{chen2023codet} & \dimCor{} & code-davinci-002; HumanEval; pass@1 & \rep{65.8\%} \\
S* \citep{li2025s} & \dimCor{} & R1-Distill-Qwen-32B; LiveCodeBench; accuracy & \rep{85.7\%} \\
\bottomrule
\end{tabularx}
\end{table}

\begin{table}[H]
\caption{Highlighted decoding and selection techniques: landmarks (L, most cited) and recent leaders (R, 2025--2026, largest reported gain). Year = year of the cited version. Cit.\ = Semantic Scholar citations, October 2026. \dimCor{} = direction of the effect on correctness as coded in our extraction. Values as reported by the original authors; settings differ across rows, so values are not directly comparable.}
\label{tab:hl-decoding}
\scriptsize
\begin{tabularx}{\linewidth}{@{}>{\raggedright\arraybackslash}p{1.75cm}>{\raggedright\arraybackslash}p{0.55cm}>{\raggedright\arraybackslash}p{0.4cm}>{\raggedright\arraybackslash}p{0.3cm}>{\raggedright\arraybackslash}X>{\raggedright\arraybackslash}p{0.75cm}>{\raggedright\arraybackslash}X>{\raggedright\arraybackslash}p{0.55cm}>{\raggedright\arraybackslash}p{0.4cm}@{}}
\toprule
Technique & Year & Cit. & L/R & Core idea & Target & Reported result & \dimCor & Code \\
\midrule
PG-TD, length reward \citep{zhang2023planning} & 2023 & 272 & L & Tree search during decoding; replacing the test reward with a length penalty or comment reward controls code style. & \dimRM & code length, APPS: default reward \rep{248.42} $\to$ length penalty \rep{190.73} & \same & \ding{51} \\
Constrained beam sampling \citep{fu2024constrained} & 2024 & 61 & L & Beam sampling with hand-written positive and negative key-phrase constraints per prompt, enforcing secure API patterns. & \dimSec & secure-pass@1, CodeGen-2.7B, CodeGuard+: beam \rep{33.77} $\to$ \rep{51.25} & mixed & \ding{51} \\
RPG \citep{dong2025rethinking} & 2025 & 15 & L & A grammar-based pushdown automaton detects structural repetition during decoding and decays the likelihood of tokens causing it. & \dimRM & TR-S (lower is better), real-world repos: greedy \rep{0.631}, repetition penalty \rep{0.358}, RPG \rep{0.335} & \up & \ding{51} \\
Surprisal-guided selection \citep{barnes2026surprisalguided} & 2026 & 1 & R & Among correct samples, keeps the one with the highest surprisal (lowest model confidence) instead of the most confident. & \dimEff & fast and correct, GPT-OSS-120B, KernelBench L1 subset: most confident \rep{50\%} $\to$ \rep{80\%} & \nr & -- \\
DuoSteer \citep{yan2026interpreting} & 2026 & 0 & R & Adds safety and correctness steering vectors at causally identified attention heads during inference. & \dimSec & vulnerability rate, Llama-3.1-8B-Instruct, five CWEs: average change \rep{$-$26.9\%} vs.\ unsteered & \up & \ding{51} \\
FunPRM \citep{zhang2026funprm} & 2026 & 4 & R & Prompts for code written function by function, treats each function as a step, and selects best-of-$N$ with a meta-corrected process reward model. & \dimRM & readability preference, Qwen2.5-7B-Coder, correct pairs: base \rep{45.8} vs.\ FunPRM \rep{54.2} & \up & -- \\
SCS-Code \citep{wendlinger2026securitybydesign} & 2026 & 1 & R & Adds a difference-in-means secure-code concept vector to the residual stream at inference. & \dimSec & sec-pass@1, DeepSeek-Coder-V2-Lite, CodeGuard+: unsteered \rep{52.8} $\to$ \rep{60.3} & mixed & \ding{51} \\
MoC \citep{yu2025mixture} & 2025 & 6 & L & Per-CWE linear probes on hidden states flag a vulnerable state; only then a decaying mix of per-CWE correction vectors is added to the hidden state. & \dimSec & Security ratio (CodeQL), SVEN test set, Qwen2.5-Coder-7B: base \rep{76.8} $\to$ \rep{85.7} (learned dynamic corrections) & \up & \ding{51} \\
SecCodePRM \citep{yu2026seccodeprm} & 2026 & 1 & R & A process reward model, trained on step-level labels derived from vulnerability-fix diffs, scores each code step; best-of-$N$ keeps the candidate with the highest risk-weighted security reward. & \dimSec & func-sec@1, CWEval, QC3-30B generator, best of 20: probability ranking \rep{41.64} $\to$ \rep{54.62} (QwenPRM \rep{47.90}) & \up & \ding{51} \\
RobGen \citep{li2025preliminary} & 2025 & 8 & L & A small classifier predicts whether the next line should open an \texttt{if} guard; when \texttt{if} is among the three most likely tokens, a logits processor raises its logit. & \dimSec & Outputs judged less robust than the reference (GPT-4o), CoderEval Java, DeepSeekCoder-1.3B: greedy \rep{42.2\%} $\to$ \rep{28.7\%} (lower is better) & \up & \ding{51} \\
\bottomrule
\end{tabularx}
\end{table}

\clearpage
\subsection*{Training}
\begin{table}[H]
\caption{Reported effects of further training-time techniques; the highlighted ones are in Table~\ref{tab:hl-training} (signal source in parentheses; PO = preference optimization; $^{a}$with two-step generation; $^{b}$as stated by the authors). Values as reported by the original authors; settings differ across rows, so values are not directly comparable.}
\label{tab:training-results}
\footnotesize
\begin{tabularx}{\linewidth}{@{}l>{\raggedright\arraybackslash}p{1.75cm}>{\raggedright\arraybackslash}p{2.55cm}>{\raggedright\arraybackslash}X>{\raggedright\arraybackslash}p{1.75cm}@{}}
\toprule
Method & Signal & Base model & Target effect & \dimCor{} effect \\
\midrule
HexaCoder & SFT (CodeQL) & four LLMs & \dimSec{} \up{} up to \rep{85\%} fewer vulnerable$^{a}$ (3 benchmarks) & \same{}$^{b}$ \\
ProSec & PO (CWE tasks) & several code LLMs & \dimSec{} \up{} \rep{25.2\%}--\rep{35.4\%} vs.\ SafeCoder data (PurpleLlama) & \same{}$^{b}$ \\
PurpCode & SFT+RL & Qwen2.5-32B & \dimSec{} \up{} \rep{12\%} over next best (CyberSecEval) & \same{}$^{b}$ \\
SecCoderX & RL (learned RM) & CodeLlama-7B, Qwen2.5-Coder & \dimSec{}$\times$\dimCor{} \up{} ESR about \rep{10\%} & in ESR \\
PIE & SFT (gem5) & GPT-3.5 & \dimEff{} \up{} \rep{6.86} speedup, best of 8 (PIE) & filtered \\
EffiCoder & SFT (profiler) & Qwen2.5-Coder-7B-Instr. & \dimEff{} \up{} time \rep{48.4\%} lower (EffiBench) & \rep{44.8\%}$\to$\rep{57.7\%} \\
ACECode & RL (execution) & four code LLMs & \dimEff{} \up{} faster in \rep{65\%}--\rep{72\%} & \rep{1.84\%}--\rep{14.51\%} \up \\
Afterburner & RL (sandbox) & Qwen2.5-3B & \dimEff{} \up{} beats humans \rep{31\%}$\to$\rep{45\%} (Venus) & \rep{47\%}$\to$\rep{62\%} \\
SuperCoder & RL (speedup) & Qwen2.5-Coder-7B-Instr. & \dimEff{} \up{} \rep{1.10x}$\to$\rep{1.46x} vs.\ \texttt{-O3} & \rep{61.4\%}$\to$\rep{95.0\%} \\
RLPF & RL (runtime) & Qwen3-32B & \dimEff{} \up{} \rep{8.1\%}$\to$\rep{38.6\%} rel.\ efficiency (PerfCodeBench) & \rep{11.1\%}$\to$\rep{54.6\%} \\
\bottomrule
\end{tabularx}
\end{table}

\begin{table}[H]
\caption{Highlighted training techniques: landmarks (L, most cited) and recent leaders (R, 2025--2026, largest reported gain). Year = year of the cited version. Cit.\ = Semantic Scholar citations, October 2026. \dimCor{} = direction of the effect on correctness as coded in our extraction. Values as reported by the original authors; settings differ across rows, so values are not directly comparable.}
\label{tab:hl-training}
\scriptsize
\begin{tabularx}{\linewidth}{@{}>{\raggedright\arraybackslash}p{1.6cm}lrc>{\raggedright\arraybackslash}X>{\raggedright\arraybackslash}p{0.75cm}>{\raggedright\arraybackslash}X cc@{}}
\toprule
Technique & Year & Cit. & L/R & Core idea & Target & Reported result & \dimCor & Code \\
\midrule
SVEN \citep{he2023large} & 2023 & 337 & L & Learns property-specific prefixes with region-specific losses on vulnerability-fix diffs; LM weights unchanged. & \dimSec & Secure rate, CodeGen-2.7B: original LM \rep{59.1\%} $\to$ \rep{92.3\%} & mixed & \ding{51} \\
SafeCoder \citep{he2024instruction} & 2024 & 110 & L & Joint standard and security instruction tuning; masked likelihood on secure code, masked unlikelihood on vulnerable code. & \dimSec & Code security, StarCoder-1B: standard instr.\ tuning \rep{62.9} $\to$ \rep{92.1} & \same & \ding{51} \\
CodeUltra\-Feedback \citep{weyssow2026codeultrafeedback} & 2026 & 81 & L & GPT-3.5 judge rates responses of many LLMs on five coding preferences; SFT then DPO on the ratings. & \dimRM, \dimEff{} (judged) & Judge score, complexity/efficiency, CodeLlama-7B-Instr.: base \rep{6.16} $\to$ \rep{7.89} & \up & \ding{51} \\
CodeDPO \citep{zhang2025codedpo} & 2025 & 63 & L & Ranks self-generated code and tests by PageRank-style mutual validation, adds fastest-versus-slowest pairs, trains with DPO. & \dimEff & Mean time (s), HumanEval+, commonly solved: before \rep{0.250} $\to$ \rep{0.172} & \up & \ding{51} \\
Afterburner \citep{du2025afterburner} & 2025 & 20 & R & GRPO with execution-time rewards inside an iterative optimization loop that feeds back measured performance. & \dimEff & Beyond-I, Venus, Qwen2.5-3B: \rep{31\%} $\to$ \rep{45\%} & \up & \ding{51} \\
PurpCode \citep{liu2025purpcode} & 2025 & 23 & R & Rule-learning SFT on safety rules, then GRPO with multi-objective rewards from code analyzers, tests and a learned verifier. & \dimSec & Secure code, CodeLMSec, Qwen2.5-32B: base \rep{29.2} $\to$ \rep{94.8} & \same & \ding{51} \\
SmellCC data \citep{xue2026clean} & 2026 & 9 & R & Removes code smells from the fine-tuning corpus by LLM-based refactoring before training. & \dimRM & SonarQube smells, DeepSeek-V1: \rep{61.0\%} fewer than fine-tuning on the original data & \up & \ding{51} \\
SecCoderX \citep{wu2026secure} & 2026 & 5 & R & Online GRPO against a CWE-conditioned reasoning vulnerability reward model, with rewards that preserve functionality. & \dimSec & Safety rate vs.\ the unaligned model: \rep{11\%}--\rep{16\%} gain & mixed & \ding{51} \\
REAL \citep{yao2025training} & 2025 & 12 & R & RL whose reward combines program-analysis detection of security or maintainability issues with unit-test results. & \dimSec, \dimRM & Quality, SecCodePLT+, Qwen2.5-Coder-7B: SFT \rep{0.5122} $\to$ \rep{0.8293} & \up & \ding{51} \\
MPCoder \citep{dai2024mpcoder} & 2024 & 33 & L & Learns Checkstyle style attributes through residual prompts, per-user style vectors, and a contrastively trained multi-user style adapter gated into the output distribution. & \dimRM & Coding Style Score (similarity to the user's own Java style), CodeLlama-7B-Instr., PCISparse: base \rep{46.29} $\to$ \rep{66.18} (single-user L-LDB \rep{61.09}) & mixed & \ding{51} \\
\bottomrule
\end{tabularx}
\end{table}

\clearpage

\FloatBarrier
\section{The model landscape: sampling, reasoning, cutoffs and vendor reports}\label{sec:app-models-detail}
\subsection{Overview of the 2025--2026 models}\label{sec:models}

Most techniques in \S\S\ref{M-sec:prompting}--\ref{M-sec:training} of the main article were evaluated on now-superseded models. Table~\ref{tab:models-landscape} (\S\ref{sec:app-landscape}) lists frontier closed and major open-weight and code-specialized models released January 2025 to October 2026, under vendor names, with vendor-reported coding scores, architecture and training cutoffs; the Supplementary Material (\S\ref{sec:app-models}) lists every release with its month and source. The experiments (\S\ref{M-sec:empirical}) use a fixed subset.

\subsubsection{Frontier and open-weight models}\label{sec:models-families}

\paragraph{Frontier closed models.} Frontier vendors now ship tiered flagships every few months. Anthropic's fifth generation adds a top tier, Claude Fable, alongside Claude Mythos variants whose most sensitive capabilities go to vetted users via a trusted access program \citep{anthropic2026system}; current models are Claude Fable 5.1, Opus 5.5, Sonnet 5.5 and Haiku 5.5 \citep{anthropic2026models}. OpenAI moved to three tiers per generation: GPT-5.6 Sol, Terra and Luna \citep{openai2026gpt56}, then GPT-6 Astra, Sol and Luna \citep{openai2026gpt6}. Google's Gemini 2.5 and Gemini 4 Argon are sparse mixture-of-experts (MoE) models \citep{comanici2025gemini,googledeepmind2026geminid}, and xAI's recent cards name the vendor SpaceXAI, ``a doing-business-as (DBA) name of XAI LLC'' \citep{spacexai2026modelb}.

\paragraph{Open-weight and code models.} The largest open-weight models are MoEs from DeepSeek, Qwen, Z.ai, Moonshot, MiniMax and Xiaomi, up to \rep{2.4T} total and \rep{95B} active parameters for Qwen3.8 \citep{qwenteam2026qwenqwen3824ta95b}. Western releases are mostly smaller (gpt-oss, Gemma, Llama 4, Olmo, Muse Glimmer), except NVIDIA's Nemotron 3 Ultra and Mistral Large 4 (weights announced, not yet released). Technical reports describe DeepSeek-R1 \citep{deepseekai2025deepseekr1}, Kimi k1.5, K2 and K2.5 \citep{team2025kimib,team2025kimi,team2026kimi}, MiniMax-M1 \citep{minimax2025minimaxm1}, Magistral \citep{mistralai2025magistral}, OLMo 2 \citep{olmo20242}, Llama-Nemotron \citep{bercovich2025llamanemotron}, and Gemma 3 and Gemma 4 \citep{team2025gemma,team2026gemma}, the latter pairing dense and MoE variants; Ai2's Olmo 3 also releases its training data \citep{olmo2025olmo}. Code-specialized releases include Meta's CWM \citep{team2025cwm}, Qwen3-Coder and Qwen3-Coder-Next \citep{qwenteam2025qwen3coder,cao2026qwen3codernext}, Mistral's Devstral \citep{mistralai2025introducing} and Ai2's SERA coding agents \citep{shen2026sera}.

\paragraph{Benchmarks.} Vendor benchmarks changed as fast as the models. By September 2026 the top SWE-bench Verified entry on the vals.ai leaderboard (grey literature) reached \rep{97.00\%} \citep{valsai2026swebench}, and OpenAI ``stopped reporting SWE-bench Verified scores'' in February 2026 \citep{openai2026why}. Recent cards report SWE-bench Pro (a public set plus held-out and proprietary commercial repository sets) \citep{deng2026swebench}, Terminal-Bench 2.0 to 4.0 \citep{merrill2026terminalbench} and DeepSWE. On Terminal-Bench 2.1, open models report \rep{90.6\%} (DeepSeek-V4.1-Flash, above Claude Opus 5 at \rep{89.1\%}) \citep{deepseekai2026deepseekv41flash}, \rep{88.3} (Kimi K3) \citep{moonshotai2026moonshotaikimik3} and \rep{88.2} (GLM-5.3) \citep{zai2026zaiorgglm53}, against \rep{88.8\%} for GPT-5.6 Sol \citep{openai2026gpt56}. Harnesses differ, so the numbers are not comparable, but open weights, required for decoding- and training-time techniques (\S\ref{M-sec:decoding}, \S\ref{M-sec:training}), no longer trail far on correctness.

How vendors moved away from user-controlled sampling, what reasoning models change for code quality, the training cutoffs that define our contamination windows, and what vendors themselves report about code quality are detailed in the following subsections.

\subsection{The Retreat From Sampling Control}\label{sec:models-sampling}

From Claude Opus 4.7 onward, non-default \texttt{temperature}, \texttt{top\_p} or \texttt{top\_k} values on Anthropic's models ``return a 400 error on every request, regardless of whether thinking is used''; older models restrict them only with thinking on \citep{anthropic2026thinking}. OpenAI's GPT-6 guide says to drop \texttt{temperature} and \texttt{top\_p} unless reasoning effort is \texttt{none} \citep{openai2026using}. Google recommended Gemini 3's default temperature of 1.0 \citep{google2026gemini}, and for Gemini 4 Argon sampling parameters ``are ignored'' \citep{google2026whats}. xAI's reasoning models cannot turn reasoning off and reject penalty and stop parameters \citep{spacexai2026reasoning}. Only an effort level remains. Greedy decoding and temperature sweeps, the usual pass@$k$ setting in the surveyed papers, are not reproducible on these APIs; variance must come from repeated samples at a recorded effort, and logit-level techniques (\S\ref{M-sec:decoding}) need open weights.

\subsection{Reasoning Models and Code Quality}\label{sec:models-reasoning}

Most flagships since mid-2025 are reasoning models with adjustable effort. Vendors report effort's effect on correctness and cost, not quality; independent evidence covers only security. On CWEval, o4-mini produced secure code more often than the non-reasoning Claude 3.7 Sonnet, yet a non-reasoning model in an agentic workflow matched it \citep{saul2025scgagent}. SecPI reports that ``even state-of-the-art RLMs frequently introduce critical security vulnerabilities'' and that generic security reminders ``consistently degrade functional correctness'' \citep{wang2026secpi}. We found no study of reasoning effort's effect on readability, maintainability or efficiency. Hidden reasoning also complicates compute fairness: an added critique step competes with the model's own deliberation (\S\ref{M-sec:selfrefine}).

\subsection{Training Cutoffs and Contamination Windows}\label{sec:models-contamination}

Stated cutoffs range from \rep{June 2024} for gpt-oss \citep{openai2025gptoss120b} and GPT-4.1 \citep{openai2025introducing}, and \rep{August 2024} for Llama 4 and Gemma 3 \citep{meta2025metallamallama4maverick17b128einstruct,googledeepmind2025gemma}, to \rep{Jun 2026} for Anthropic's current models \citep{anthropic2026models} and \rep{June 2026} pretraining for Grok 4.7 \citep{spacexai2026model}. Two details widen these windows. Some vendors date data separately: Nemotron 3 Ultra's pre-training data runs to \rep{September 2025} and its post-training data to \rep{May 2026} \citep{nvidia2026nvidianvidianemotron3ultra550ba55bbf16}, and Grok 4.6 used ``data generated as late as June 2026 in its supplemental training'' \citep{spacexai2026modelb}. And DeepSeek, Qwen, Z.ai and Moonshot state no cutoff in the sources we examined, so date splits for them must first probe the cutoff; open training data, as in Olmo 3 \citep{olmo2025olmo}, permits a direct audit. Memorization of older benchmarks is documented: from issue text alone, models find the buggy file of a SWE-bench Verified issue with ``up to 76\% accuracy'', against ``up to 53\%'' on other repositories \citep{liang2026swebench}, and agents degrade without familiar repository cues \citep{chen2026schrodingers}. Date-tagged benchmarks allow evaluation ``only on time-windows after the model's cutoff date'' \citep{jain2025livecodebench}, but with frontier cutoffs in mid-2026 such windows hold only months of tasks. See \S\ref{M-sec:measurement} for benchmark choice under this constraint; \S\ref{M-sec:empirical} reports contaminated and post-cutoff splits separately.

\subsection{What Vendors Report About Code Quality}\label{sec:models-nfq}

The NF column of Table~\ref{tab:models-landscape} is mostly \nr. On security, Qwen reports SecCodeBench and CWEval \citep{cao2026qwen3codernext,qwenteam2026qwenqwen35397ba17b}, Meta CyberSecEval insecure-code tests \citep{meta2026muse}, and xAI an internal eval, SecureCodeReview \citep{spacexai2026modelb}. On efficiency, MiniMax reports SWE-Perf, SWE-fficiency and KernelBench Hard \citep{minimax2025minimaxaiminimaxm21,minimax2026minimax}. Closest to maintainability is FrontierCode, which ``measures whether AI-generated pull requests would be mergeable by open-source maintainers, measuring behavioral correctness, regression safety, scope discipline, style, and adherence to codebase conventions'' \citep{spacexai2026modelb}; Google, OpenAI and Anthropic also report it \citep{googledeepmind2026geminib,openai2026gpt6,anthropic2026introducing}, but since it includes correctness it does not isolate \dimRM. Mistral adds a blind human rating of ``coding quality'' \citep{mistralai2026introducing}, and OpenAI an extraneous-edit rate \citep{openai2025introducing}. No vendor reports complexity, code smells or lint violations, and system-card cyber-security sections measure offensive capability, not ordinary generated code's security. Independently, a SonarQube analysis of \rep{4,442} Java assignments found ``no direct correlation'' between pass@1 and ``the overall quality and security'' of the generated code \citep{sabra2025assessing}.

\clearpage
\section{Related surveys and publication trends}\label{sec:app-related}

\begin{table}[H]
\caption{Existing secondary studies compared with this review. Dims: quality dimensions treated as
a focus (\dimRM, \dimEff, \dimSec); ``--'' means none. Exp.: whether the study runs its own
experiments. Years and study counts as stated by each survey (\nr: not stated). Values as
reported by the original authors; settings differ across rows, so values are not directly
comparable.}
\label{tab:rel-surveys}
\footnotesize
\begin{tabularx}{\linewidth}{@{}>{\raggedright\arraybackslash}p{1.9cm}>{\raggedright\arraybackslash}X>{\raggedright\arraybackslash}p{1.35cm}>{\raggedright\arraybackslash}X>{\raggedright\arraybackslash}p{0.6cm}>{\raggedright\arraybackslash}p{1.35cm}>{\raggedright\arraybackslash}p{1.05cm}@{}}
\toprule
Study & Scope & Dims & Technique coverage & Exp. & Years & Studies \\
\midrule
\citet{hou2023large} & LLMs for software engineering & -- & Model use, data, tuning, evaluation by task & no & \rep{2017}--\rep{2024} & \rep{395} \\
\citet{jiang2024survey} & LLM code generation & -- & Data, training, prompting, retrieval, agents & no & \nr & \rep{235} \\
\citet{dong2025survey} & LLM-based code generation agents & -- & Single- and multi-agent architectures & no & \rep{2022}--\rep{2025} & \nr \\
\citet{yang2024robustness} & Non-functional properties of code models & \dimSec{}; efficiency of models, not code & Evaluation and enhancement methods per property & no & \nr & \rep{146} \\
\citet{kamoi2024when} & Self-correction of LLMs (all tasks) & -- & Intrinsic and external feedback, fine-tuning & no & \nr & \nr \\
\citet{chochlov2026tertiary} & Secondary studies of LLM code generation (tertiary) & -- & Effects, scenarios, integration challenges & no & \rep{2017}--\rep{2025} & \rep{30} \\
\midrule
\citet{tony2025prompting} & Prompting for secure code & \dimSec & Prompting only & yes & from \rep{2018} & \rep{13} \\
\citet{basic2024from} & LLMs in code security & \dimSec & Prompting; vulnerability detection and repair & no & to early \rep{2026} & \rep{102} \\
\citet{patir2026sok} & Secure code generation (SoK) & \dimSec & Prompting, fine-tuning, RL, agentic workflows & yes & \rep{2021}--\rep{2026} & \nr \\
\citet{matias2026survey} & Maintainability and evolvability & \dimRM & Mitigation strategies & no & \rep{2020}--\rep{2024} & \rep{87} \\
\citet{ehsani2026faster} & Technical debt (multivocal) & \dimRM & Mitigation strategies, metrics, tools & no & to \rep{2026} & \rep{104} \\
\citet{alomari2026using} & LLMs to enhance code quality & \dimRM & Refactoring, smell detection; prompting, fine-tuning & no & to \rep{2024} & \rep{49} \\
\citet{gong2025language} & LM-based code optimization & \dimEff & Code optimization techniques & no & \nr & \rep{53} \\
\citet{ali2026energyefficient} & Energy efficiency of LLM-generated code & \dimEff & Prompting and fine-tuning strategies & no & \nr & \rep{18} \\
\midrule
\citet{he2026bridging} & Training-data causes of code quality issues & \dimRM, \dimEff, \dimSec & Detection and mitigation across data, model and generation & no & \rep{2021}--\rep{2026} & \rep{114} \\
\citet{sun2026quality} & Non-functional quality of generated code (ISO/IEC~25010) & \dimRM, \dimEff, \dimSec & Prompting and feedback strategies & small & \rep{2022}--\rep{2025} & \rep{109} \\
This review & Techniques that improve LLM code quality & \dimRM, \dimEff, \dimSec{} (\dimCor{} as control) & Seven families, faceted by stage and signal source & yes & 2021--2026 & \cnt{nIncluded} \\
\bottomrule
\end{tabularx}
\end{table}

\begin{figure}[H]
\centering
\includegraphics[width=0.9\linewidth]{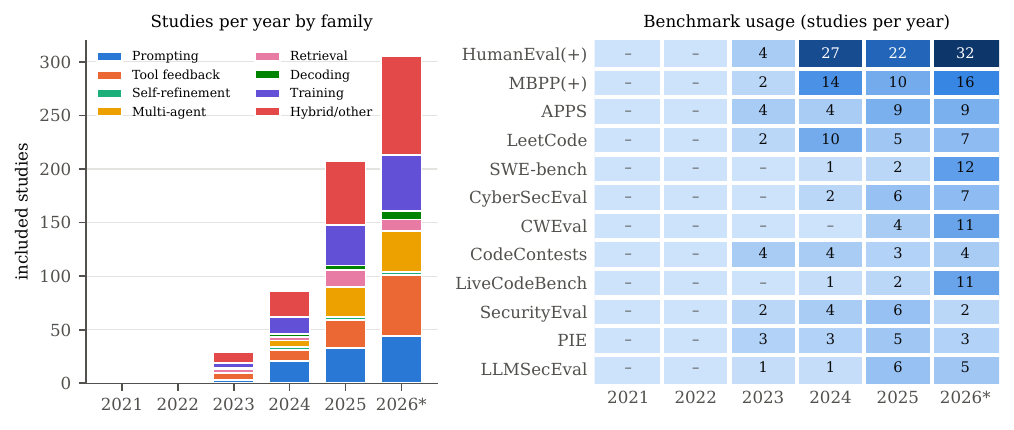}
\Description{Two stacked bar charts by year: included studies per technique family, and the twelve most used benchmarks.}
\caption{Left: included studies per year by technique family (2026 covers January to mid-October). Right: benchmarks used by included studies per year (top 12). Counts from LLM-assisted extraction reviewed by the author.}\label{fig:trends}
\end{figure}

\IfFileExists{ACM-Reference-Format.bst}{\bibliographystyle{ACM-Reference-Format}}{\bibliographystyle{plainnat}}
\bibliography{bib/core,bib/02_background,bib/05_measurement,bib/07_prompting,bib/08_feedback_loops,bib/09_self_refinement,bib/10_multi_agent,bib/11_retrieval,bib/12_decoding,bib/13_training,bib/14_models}